\documentclass[11pt]{article}
\usepackage{epsfig}
\usepackage{a4}
\usepackage{amssymb}
\usepackage{amsfonts}

\def\to{\rightarrow}

\newcommand\nutau{$\nu_\tau$}
\newcommand\anutau{$\bar{\nu}_\tau$}
\newcommand\numu{$\nu_\mu$}
\newcommand\anumu{$\bar{\nu}_\mu$}
\newcommand\nue{$\nu_e$}
\newcommand\anue{$\bar{\nu}_e$}
\newcommand\nux{$\nu_x$}
\def\sq2{sin^2(2\Theta)}

\newcommand{\hepph}[1]{\tt hep-ph/#1}
\newcommand{\hepex}[1]{\tt hep-ex/#1}
\newcommand{\astroph}[1]{\tt astro-ph/#1}

\def\dms2{\Delta m^2}
\def\D32{\Delta m^2_{32}}
\def\dm31{\Delta m^2_{31}}
\def\Dm32{$\Delta m^2_{32}$}
\def\t12{\theta_{12}}
\def\th13{$\theta_{13}$}
\def\t23{\theta_{23}}
\def\s2t13{$\sin^2\theta_{13}$}
\def\nuall{\nu^{^{\hbox{\hspace*{-3mm}{\tiny (---)}}}}}
\def\EB{$E_B$}
\def\avenue{$\langle E_{\nu_e} \rangle$}
\def\aveanue{$\langle E_{\bar{\nu}_e} \rangle$}
\def\avenux{$\langle E_{\nu_x} \rangle$}
\def\lelx{$L_e/L_x$}
\def\fEB{E_B}
\def\favenue{\langle E_{\nu_e} \rangle}
\def\faveanue{\langle E_{\bar{\nu}_e} \rangle}
\def\favenux{\langle E_{\nu_x} \rangle}
\def\flelx{L_e/L_x}

\def\Journal#1#2#3#4{{#1} {\bf #2} (#3) #4.}
\def\etal{{\it et\ al.}}

\def\NIMA{{\em Nucl. Instrum. Methods} A}

\def\NPB{{\em Nucl. Phys.} B}
\def\PLB{{\em Phys. Lett.}  B}
\def\PRL{{\em Phys. Rev. Lett.}}

\def\PRD{{\em Phys. Rev.} D}

\def\APJ{{\em Astrop. Phys. J.}}

\begin{document}
\thispagestyle{empty}

\begin{flushright}
{\tt ICARUS-TM/04-04}\\ 
\today
\end{flushright}
\vspace*{1cm}
\begin{center}
{\Large{\bf Decoupling supernova and neutrino oscillation physics \\ with
LAr TPC detectors}}\\ 

\vspace{1cm}
{\large I. Gil-Botella}\footnote{Ines.Gil.Botella@cern.ch} and
{\large A. Rubbia}\footnote{Andre.Rubbia@cern.ch}

Institut f\"{u}r Teilchenphysik, ETHZ, CH-8093 Z\"{u}rich,
Switzerland
\end{center}
\vspace{1.cm}
\begin{abstract}
\noindent
Core collapse supernovae are a huge source of all flavor
neutrinos. 
The flavor composition, energy spectrum and time structure of
the neutrino burst  from a galactic supernova can provide information
about the explosion mechanism
and the mechanisms of proto neutron star cooling. 
Such data can also give information
about the intrinsic properties of the neutrino such as flavor
oscillations. 
One important question is to understand to which extend can the
supernova and the neutrino physics be decoupled in the observation
of a single supernova. On one hand, the understanding of the supernova
explosion mechanism is still plagued by uncertainties which have an impact
on the precision with which one can predict time, energy and flavor-dependent
neutrino fluxes. On the other hand, the neutrino mixing properties are
not fully known, since the type of mass hierarchy and the value of
the  \th13 angle are unknown, and in fact large uncertainty still exists on
the prediction of the actual effect of neutrino oscillations in the event
of a supernova explosion.
In this paper we discuss the possibility to probe the neutrino mixing angle
\th13 and the type of mass hierarchy from the detection of supernova neutrinos
with a liquid argon TPC detector. 
Moreover, describing the supernova neutrino emission by a set of five parameters
(average energy of the different neutrino flavors, their relative
luminosity and the total supernova binding energy),
we quantitatively study how it is possible to constrain these parameters.
A characteristic feature of the liquid argon TPC is the accessibility to
four independent detection channels ( (1) elastic scattering off electrons, (2) charged neutrino
and (3) antineutrino and (4) neutral currents on argon nuclei) which have different sensitivities
to electron-neutrino, anti-electron-neutrino and other neutrino flavors 
(muon and tau (anti)neutrinos). This allows to over-constrain 
the five supernova and the flavor mixing parameters and to some 
extent disentangle neutrino from supernova
physics.
Numerically, we find that a very massive liquid argon detector (O(100~kton))
is needed to perform accurate measurements of these parameters, specially in supernova scenarios
where the average energies of electron and non-electron neutrinos are
similar (almost degenerate neutrinos) or if no information about the \th13 mixing angle and type of
mass hierarchy is available. 

\vspace{1cm}
\end{abstract}

\newpage
\pagestyle{plain} 
\setcounter{page}{1}
\setcounter{footnote}{0}

%%%%%%%%%%%%%%%%%%%%%%
\section{Introduction}
%%%%%%%%%%%%%%%%%%%%%%

Core collapse supernovae are a huge source of all flavor
neutrinos. Neutrino astrophysics entered a new phase with the
detection of neutrinos from the supernova SN1987A in the Large
Magellanic Cloud by the Kamiokande and IMB detectors \cite{SN1987}. In
spite of these fundamental neutrino observations, the 19 events
observed are not statistically significant enough to obtain
precise quantitative information on the neutrino spectrum. 
Currently running neutrino detectors like Superkamiokande \cite{Fukuda:2000fq}
or SNO \cite{Tanner:zr} have the capabilities to provide high
statistics information about supernova and neutrino properties if a
supernova collapse were to take place in the near future.

The flavor composition, energy spectrum and time structure of
the neutrino burst  from a galactic supernova can give information
about the explosion mechanism
and the mechanisms of proto neutron star cooling.

The neutrino signal from a galactic supernova can also give information
about the flavor
oscillation of neutrinos. Although new data from solar, atmospheric, reactor and
accelerator neutrinos \cite{nuobserv} have contributed to the
understanding of the neutrino properties, still the neutrino mixing
angle \th13 and the nature of the mass hierarchy, i.e., normal or
inverted, remain unknown. Only an upper limit on the mixing angle is available dominated
by negative measurements from reactor experiments \cite{reactors}.
These parameters can be probed by the
observation of supernovae neutrino bursts, since neutrinos will travel
long distances before reaching the Earth and will, as they travel through
the exploding star, in addition traverse
regions of different matter densities where matter enhanced oscillations \cite{msw}
will take place. In the case of the  \th13 angle, matter enhancement has 
the striking feature that  very small mixing angles, beyond any
value detectable by next generation accelerators, could in fact
alter significantly the neutrino spectrum. Hence, supernova neutrinos
provide indeed a complementary tool to study  the  \th13 angle.

The main question of course is to understand to which extend can the
supernova and the neutrino physics be decoupled in the observation
of a single supernova. On one hand, the understanding of the supernova
explosion mechanism is still plagued by uncertainties which have an impact
on the precision with which one can predict time, energy and flavor-dependent
neutrino fluxes. On the other hand, the intrinsic neutrino properties are
not fully known, since the type of mass hierarchy and the value of
the  \th13 angle are unknown.

The total event rates and
the energy spectra of the detected supernova neutrinos depend on the neutrino flavor
and on the \th13 mixing angle and the type of mass hierarchy. However,
it is not a priori easy to separate neutrino and supernova physics.
Different solutions to this problem have been proposed.
Global fits of data (\cite{Barger,Valle}) and analyses considering
qualitative features of the supernova neutrino fluxes
(\cite{Lunardini}) have been performed for the case of \v{C}erenkov
detectors. The potential of liquid scintillator detectors has also be
investigated (\cite{Choubey}). 

We have already qualitatively illustrated in our paper
\cite{IApaper} that four independent channels are a priori accessible
with a liquid argon TPC: elastic scattering on electrons, \nue ~charged
current absorption on argon, \anue~charged current absorption on argon
and neutral current interactions on argon. 
They have different sensitivities
to electron-neutrino, anti-electron-neutrino and other neutrino flavors 
(muon and tau (anti)neutrinos).
These events being sensitive to different combinations of neutrino and antineutrino
flavors can be combined to over-constrain the features of the supernova fluxes that
are at the moment not well constrained by theory and the mixing parameters and
allow neutrino and supernova physics effects
to be disentangled.

In the present paper we study
quantitatively the capability of LAr TPC detectors to simultaneously measure
the supernova and oscillation parameters, performing a global fit to
the event energy distribution for elastic and charged current
events and to the rate for the neutral current events.

This study is motivated by the 3 kton LAr ICARUS experiment
\cite{icarus3000} that was proposed at the Gran Sasso laboratory,
and by ideas for a very large liquid argon TPC with
a mass in the range of 100 ktons 
(see \cite{Cline:2001pt} and independently \cite{Rubbia:2004tz}).  

\section{Theoretical framework}

We briefly summarize in this section the neutrino features relevant
for this analysis. We essentially follow the general framework already
described in our previous paper \cite{IApaper}.

\subsection{Supernova physics}
\label{sec:supphys}
At the end of the life of a massive star with a core mass greater than
the Chandrasekhar mass, the energy generation in the core stops, and
the star undergoes a fast gravitational collapse. Within a duration of tens
of milliseconds, this collapse compresses the inner core beyond nuclear
matter densities. During this stage electron neutrinos are produced via
neutronisation processes. While the electron neutrinos produced could
escape the star during the first period (producing the ``neutrino burst''),
the very dense core becomes opaque to neutrinos which remain captured
within the supernova core producing the so-called inner core rebound.
This generates a shock wave which propagates radially outward through
the star remnants, supplying the ``explosion'' mechanism. During the shock
phase, thermal neutrinos and antineutrinos of all flavors are produced
due to high temperatures. It is believed that
the 99\% of the total binding energy of the
star, \EB ~$\approx$ 3 $\times$ 10$^{53}$ ergs, is emitted in the form
of neutrinos. 
These neutrinos are in equilibrium with their surrounding matter density
and their energy spectra can be described by a function close to a 
Fermi-Dirac distribution. The flux of a neutrino $\nu_{\alpha}$
emitted can then be written as \cite{Lunardini}:

\begin{equation}
\phi_{\alpha}(E_{\alpha},  L_{\alpha}, D, T_{\alpha}, \eta_{\alpha}) =
\frac{L_{\alpha}}{4\pi D^2 F_3(\eta_\alpha) T^4_{\alpha}}
\frac{E_{\alpha}^2}{e^{E_{\alpha}/T_{\alpha}-\eta_{\alpha}}+1} 
\end{equation}

\noindent
where $L_{\alpha}$ is the luminosity of the flavor $\nu_{\alpha}$
(\EB = $\sum L_{\alpha}$), $D$ is the distance to the supernova,
$E_{\alpha}$ is the energy of the $\nu_{\alpha}$ neutrino,
$T_{\alpha}$ is the neutrino temperature inside the neutrinosphere and
$\eta_{\alpha}$ is the ``pinching'' factor. We have taken
$\eta_{\alpha}$ = 0 and hence no deviations in the high energy tail
are considered. The neutrino average energy $\langle E_{\nu_\alpha}
\rangle$ depends on both $T_{\alpha}$ and $\eta_{\alpha}$. For
$\eta_{\alpha}$ = 0, $\langle E_{\nu_\alpha}\rangle$ $\approx$
3.15 $T_{\alpha}$. The normalization factor in absence of pinching is
equal to $F_3(0) \approx$ 5.68. 

The original \numu, \nutau, \anumu~and \anutau ~fluxes are
approximately equal and therefore we treat them as \nux. 
Out of all neutrinos, muon and tau neutrinos and their antiparticles
interact with matter through neutral currents only, whereas electron
neutrinos and their antiparticles feel both charged and neutral currents.
Therefore, muon and tau flavors should decouple first and hence
have the largest temperatures.
This energy hierarchy between the different neutrino flavors is
generally believed to hold and imply $\langle E_{\nu_e} \rangle < \langle
E_{\bar{\nu}_e} \rangle < \langle E_{\nu_x} \rangle$, due to the
different strength of neutrino interactions with the surrounding matter.  
However, the specific neutrino spectra remain a matter of detailed
calculations. In particular, the simulations of Raffelt et al.~\cite{Raffelt}
seem to indicate that the energy differences between flavors could be very 
small.
We use a range of typical values of the neutrino average energies and relative
luminosities provided by several simulations in the following
intervals \cite{Lunardini, Raffelt}:
\begin{eqnarray}
\begin{tabular}{ccc}
\avenue = (7--18) MeV; & \aveanue = (14--22) MeV; & \avenux = (15--35)
MeV \\
\end{tabular}
\end{eqnarray} 
\vspace{-1cm}
\begin{eqnarray}
\begin{tabular}{cc}
\lelx = (0.5--2); & $L_e$ = $L_{\bar{e}}$ \\
\end{tabular}
\end{eqnarray}
In practice, we will consider two specific scenarios (I \& II) in order
to understand the effects of a hierarchical versus non-hierarchical 
distribution of energies on our results, see below, and take luminosity
equipartition. Some detailed simulations may also give rise to cases
where the electron luminosities of the neutrinos vary between (0.5--2) times
the muon and tau luminosities. We did not explicitly consider this case, even
though our calculations could be extended to such a case.
We will be later concerned on how well one can determine the energies and
the relative luminosities of the neutrinos through the detection
of a single supernova collapse.

We consider core collapse supernovae at a distance $D$ of 10 kpc
(galactic supernovae).
We assume that in the actual event of a supernova, 
this distance will be precisely estimated by independent astronomical observations
and we hence take this value as precisely known in our discussion\footnote{We note
that the error on the distance will essentially affect the overall normalization of
the fluxes, parameterized by the binding energy in our discussion. This error will not
fundamentally affect our conclusions. On the other hand, the fact that a supernova
could actually occur at a closer or larger distance does affect the statistics of collected
event. We take 10~kpc as a reference value for easy comparison with other authors.}. 

In order to parameterize the supernova physics, we employ {\it five}
supernova parameters:
\begin{enumerate}
\item the total binding energy \EB $=\sum L_{\alpha}$ , 
\item the average
energies of the neutrinos emitted from the supernova \avenue,
\aveanue, \avenux ~and
\item the relative luminosities of the electron and
non-electron neutrinos \lelx.
\end{enumerate}

We always assume $L_e$ = $L_{\bar{e}}$ and $L_x = L_{\bar x}$, hence we obtain:
\begin{eqnarray}
E_B & = & L_e + L_{\bar e} + 2 L_x + 2 L_{\bar x} = 2 L_e + 4 L_x \nonumber \\
&  = &2 L_x(L_e/L_x + 2) 
\end{eqnarray}
from which $L_x =  L_{\bar x} = E_B/(2(L_e/L_x+2))$.

In order to take into account the uncertainties on the knowledge of
the supernova physics, we consider two different sets of reference
values for the supernova parameters from Ref. \cite{Langanke} and  \cite{Raffelt}, 
corresponding to one hierarchical energy scenario  (I) and a non-hierarchical energy scenario (II). They
differ in the assumptions on the average energies of the
neutrino flavors (Table \ref{tab:sncoolscenario}).  

\begin{table}[htbp]
\centering
\begin{tabular}{|c|ccccc|c|} \hline
SN & \EB & $\langle E_{\nu_e} \rangle$ & $\langle E_{\bar\nu_e} \rangle$  & $\langle
E_{\nu_{\mu,\tau}} \rangle$ = $\langle E_{\bar\nu_{\mu,\tau}} \rangle$ & Luminosity & Ref. \\
scenario & ($\times$ 10$^{53}$ erg) & (MeV) & (MeV) & (MeV) &  & \\
\hline
%\\
{\bf I} & 3 & 11 & 16 & 25 &  L$_{\nu_e}$ = L$_{\bar\nu_e}$ = 
L$_{\nu_x}$ & \cite{Langanke} \\
%\\
{\bf II} & 3 & 13 & 16 & 17.6 &  L$_{\nu_e}$ = L$_{\bar\nu_e}$ = 
L$_{\nu_x}$ & \cite{Raffelt}\\
%\\
\hline
\end{tabular}
\caption{Assumed supernova parameters: scenarios I and II.}   
\label{tab:sncoolscenario}
\end{table}

\subsection{Neutrino flavor oscillation physics}
\label{sec:oscphys}
We summarize here the supernova flux changes introduced by flavor oscillations
(see e.g. Ref. \cite{IApaper} for more details).
Before arriving at Earth the neutrino fluxes travel through the
supernova mantle undergoing two MSW \cite{msw} resonances: one at high density
$\rho\approx $  10$^3$--10$^4$ g cm$^{-3}$ (H--resonance) which is 
governed by the atmospheric parameters ($\Delta m^2_{31}$ and
$\theta_{13}$) and the other one at low density $\rho\approx$ 
10--30 g cm$^{-3}$ (L--resonance), characterized by the solar
parameters ($\Delta m^2_{21}$ and $\theta_{12}$). The exact propagation
of the neutrinos and their oscillation from one flavor to another depend
on the matter density profile. We refer the reader to Ref. \cite{IApaper} for more details
on our implementation of these effects.

The transitions in the two resonance layers can be considered
independently and each transition is reduced to a two neutrino oscillation
problem. 
The H--resonance lies in the neutrino channel for normal mass
hierarchy and in the antineutrino channel for the inverted
hierarchy. The L--resonance lies in the neutrino channel for both the
hierarchies \cite{Dighe}. 

The propagation through the resonance L is always adiabatic for the
LMA solar parameters. In the case of the H resonance, the ``jump''
probability depends on the \th13 angle as \cite{Dighe}:
\begin{equation}
P_H \propto \exp\left[-const ~ \sin^2\theta_{13} \left(\frac{\Delta
m^2_{31}}{E}\right)^{2/3}\right]
\label{eq:ph}
\end{equation} 

Considering P$_{ee}$ = P($\nu_e$ $\to$ $\nu_e$)  and
$\overline{P}_{ee}$ = P($\bar\nu_e$ $\to$ $\bar\nu_e$) the survival
probabilities, the neutrino fluxes arriving at Earth ($\phi_{\nu}$) can
be written in terms of the fluxes in absence of oscillations
($\phi^o_{\nu}$) as: 
\begin{eqnarray}
\begin{tabular}{l}
$\phi_{\nu_e} = \phi^o_{\nu_e} P_{ee} + \phi^o_{\nu_x} (1-P_{ee})$ \\
$\phi_{\bar\nu_e} = \phi^o_{\bar\nu_e} \overline{P}_{ee} + \phi^o_{\bar\nu_x}
(1-\overline{P}_{ee})$ \\
$\phi_{\nu_\mu} + \phi_{\nu_\tau} = \phi^o_{\nu_e} (1-P_{ee}) +
\phi^o_{\nu_x} (1+P_{ee})$ \\
$\phi_{\bar\nu_\mu} + \phi_{\bar\nu_\tau} = \phi^o_{\bar\nu_e} (1-\overline{P}_{ee}) +
\phi^o_{\bar\nu_x} (1+\overline{P}_{ee})$
\end{tabular}
\label{eq:flux}
\end{eqnarray}

\noindent with $\phi^o_{\nu_x}$ = $\phi^o_{\nu_\mu}$ =
$\phi^o_{\nu_\tau}$ = $\phi^o_{\bar\nu_x}$ = $\phi^o_{\bar\nu_\mu}$ =
$\phi^o_{\bar\nu_\tau}$.

Using a standard parametrization of the mixing matrix and considering
that the neutrino bursts do not cross the Earth, the
survival probabilities can be expressed in terms of the ``jump''
probability in the H resonance $P_H$:
\begin{eqnarray}
\begin{tabular}{l}
$P_{ee} = P_H \sin^2\theta_{12} \cos^2\theta_{13} + (1-P_H) \sin^2\theta_{13}$ \\
$\overline{P}_{ee} = \cos^2\theta_{12} \cos^2\theta_{13}$
\end{tabular}
\label{eq:peenh}
\end{eqnarray}
\noindent for normal hierarchy (\Dm32 $>$ 0) and
\begin{eqnarray}
\begin{tabular}{l}
$P_{ee} = \sin^2\theta_{12} \cos^2\theta_{13}$ \\
$\overline{P}_{ee} = P_H \cos^2\theta_{12} \cos^2\theta_{13} + (1-P_H)
\sin^2\theta_{13}$
\end{tabular}
\label{eq:peeih}
\end{eqnarray}
\noindent for inverted hierarchy (\Dm32 $<$ 0). 

The following neutrino oscillations parameters have been used,
obtained from other neutrino observations \cite{nuobserv,reactors}:
\begin{eqnarray}
\begin{tabular}{ll}
$\sin^22\theta_{23} = 1.$ & $|\Delta m^2_{32}| \approx |\Delta m^2_{31}| = 3 \times 10^{-3} eV^2$\\
$\sin^2 \theta_{12} = 0.3$ & $\Delta m^2_{21} = 7 \times 10^{-5} eV^2$\\
$\sin^2 \theta_{13} < 0.02$ (at 90\% C.L.)& $\Delta m^2_{31}>0$ or $\Delta m^2_{31}<0$\\
\end{tabular}
\end{eqnarray}

According to the adiabaticity conditions on the H resonance and the
value of the \th13 angle, we can consider two limits: the {\it small mixing
angle (S)} (\s2t13 $<$ 2 $\times$ 10$^{-6}$) and the {\it large mixing
angle (L)} case (\s2t13 $>$ 3 $\times$ 10$^{-4}$). The value of the
survival probabilities and the neutrino fluxes at Earth in these
regions are summarized in Table~\ref{tab:pee} for both mass
hierarchies.
For intermediate mixing angle 
values (2 $\times$ 10$^{-6}$ $<$ \s2t13 $<$ 3 $\times$ 10$^{-4}$), the
``jump'' probability $P_H$ depends on \th13 and on the neutrino energy 
(see e.g. Ref. \cite{IApaper}). 

We can distinguish four extreme cases
depending only on the type of mass
hierarchy and the value of the $\theta_{13}$ mixing angle:
\begin{enumerate}
\item {\bf n.h.-L} for normal mass hierarchy and large \th13,
\item {\bf n.h.-S} for normal hierarchy and small \th13,
\item {\bf i.h.-L}  for inverted mass hierarchy and large \th13, 
\item {\bf i.h.-S} for inverted mass hierarchy and small \th13.
\end{enumerate}
%and where ``small'' (resp. ``large'') mean \s2t13 $<$ $2\times 10^{-6}$ 
%(resp. \s2t13 $>$ $3\times 10^{-4}$). 
For instance, we select the mixing angle \s2t13 = 10$^{-3}$ for
the large \th13 case and \s2t13 = 10$^{-7}$ for the small \th13 case.

These results have been discussed in Ref.~\cite{IApaper}. We recall here
the salient feature that for normal hierarchy and large mixing angle,
the $\nu_e$ flux on Earth is fully dominated by $\nu_x$ neutrinos in the core of the
supernova due to the phenomenon of adiabatic conversion. Similarly,
for antineutrinos for inverted hierarchy and large mixing angle.

\begin{table}[htbp]
\centering
\begin{tabular}{|c|c|l|} 
\hline 
Limit oscillation  & Survival probabilities & \multicolumn{1}{|c|}{Neutrino fluxes at Earth}\\
scenarios &  P$_{ee}$ = P($\nu_e$ $\to$ $\nu_e$)  &
\multicolumn{1}{|c|}{$\phi^o_{\alpha}$=w/o osc, $\phi_{\alpha}$=with
osc} \\ \hline \hline
n.h.-L & $P_{ee}$ = 0. & $\phi_{\nu_e} = \phi^o_{\nu_x}$ \\
\s2t13 $>$ 3 $\times$ 10$^{-4}$         & $\overline{P}_{ee}$ = 0.7 & $\phi_{\bar\nu_e}$ = 0.7
$\phi^o_{\bar\nu_e}$ + 0.3 $\phi^o_{\bar\nu_x}$ \\ 
       &  & $\phi_{\nu_\mu + \nu_\tau + \bar\nu_\mu + \bar\nu_\tau} =
\phi^o_{\nu_e}$ + 0.3 $\phi^o_{\bar\nu_e}$ + 2.7 $\phi^o_{\nu_x}$  \\
\hline

n.h.-S or i.h.-S& $P_{ee}$ = 0.3 & $\phi_{\nu_e}$ = 0.3 $\phi^o_{\nu_e}$ + 0.7 $\phi^o_{\nu_x}$\\
\s2t13 $<$ 2 $\times$ 10$^{-6}$          & $\overline{P}_{ee}$ = 0.7 & $\phi_{\bar\nu_e}$ = 0.7
$\phi^o_{\bar\nu_e}$ + 0.3 $\phi^o_{\bar\nu_x}$\\
       &  & $\phi_{\nu_\mu + \nu_\tau + \bar\nu_\mu + \bar\nu_\tau}$ =
0.7 $\phi^o_{\nu_e}$ + 0.3 $\phi^o_{\bar\nu_e}$ + 3 $\phi^o_{\nu_x}$ \\
\hline

i.h.-L & $P_{ee}$ = 0.3 & $\phi_{\nu_e}$ =  0.3 $\phi^o_{\nu_e}$ + 0.7 $\phi^o_{\nu_x}$\\
\s2t13 $>$ 3 $\times$ 10$^{-4}$          & $\overline{P}_{ee}$ = 0. & $\phi_{\bar\nu_e}$ =
$\phi^o_{\nu_x}$\\ 
       &  & $\phi_{\nu_\mu + \nu_\tau + \bar\nu_\mu + \bar\nu_\tau}$ =
0.7 $\phi^o_{\nu_e}$ + $\phi^o_{\bar\nu_e}$ + 2.3 $\phi^o_{\nu_x}$ \\
\hline
\end{tabular}
\caption{Survival probabilities for \nue ~and \anue ~and neutrino
fluxes at Earth ($\phi_{\nu}$)  for normal and inverted mass hierarchies
and large (L) or small (S) mixing angle \th13 in terms of the fluxes 
($\phi^o_{\nu}$) in absence of oscillations.} 
\label{tab:pee}
\end{table}

%%%%%%%%%%%%%%%%%%%%%%%%%%%%%%%%%%%%%%%%%%
\section{Supernova neutrino signal in LAr}
%%%%%%%%%%%%%%%%%%%%%%%%%%%%%%%%%%%%%%%%%%

In a liquid argon TPC supernova neutrinos ($E_\nu <$ 100 MeV) can be detected through four
different channels (\cite{IApaper}):

\begin{enumerate}

\item {\bf charged-current (CC) interactions on argon}
\begin{equation}
\nu_e ~^{40}Ar \to e^- + A' + nN
\label{eq:ccnue}
\end{equation}
\begin{equation}
\bar{\nu}_e ~^{40}Ar \to e^+ + A' + nN
\label{eq:ccanue}
\end{equation}
where $A'$ is the remnant nucleus and $nN$ represent emitted nucleons or other debris ($\alpha$'s, ...)
(if any).
At low energy ($<$ 30 MeV) or at low $Q^2$, the dominant reactions are
$\nu_e ~^{40}Ar \to e^- ~^{40}K^*$ and $\bar{\nu}_e ~^{40}Ar \to e^+ ~^{40}Cl^*$.
The neutrino energy thresholds of these reactions are 1.5 MeV and 7.48
MeV, respectively. The associated photons emitted in the de-excitation of the
final state nucleus can be used to separate the two reactions.

\item {\bf neutral-current (NC) interactions on argon}
\begin{equation}
\nuall ~^{40}Ar \to \nuall + A' + nN
\label{eq:ncnu}
\end{equation}

where $A'$ is the remnant nucleus and $nN$ represent emitted nucleons or other debris ($\alpha$'s, ...) (if any).
At low energy ($<$ 30 MeV) or at low $Q^2$, the dominant reaction is
$\nuall ~^{40}Ar \to \nuall ~^{40}Ar^*$.
The energy threshold of this reaction is 1.46 MeV.
The associated photons emitted in the de-excitation of the
final state nucleus can be used to identify the reaction.

\item  {\bf elastic scattering (ELAS)} on electrons
\begin{equation}
\nuall ~e^- \to \nuall ~e^-
\label{eq:elas}
\end{equation}
\end{enumerate}

All these channels have been described in more details in reference
\cite{IApaper}. 
We use the nuclear cross sections of the CC and NC processes on argon that
have been calculated by Random Phase Approximation (RPA) for neutrino
energies up to 100 MeV including all the possible multipoles
(\cite{martinez}). These processes are very important on argon because
their high cross sections. As already mentioned, it is possible to separate the different
channels by measuring the associated photons from the $K$, $Cl$ or
$Ar$ de-excitation, or by the absence of them in the case of elastic
scattering. We recall here that the energy dependence of the cross-section
is $\propto E^2$ for the nuclear processes and $\propto E$ for the scattering
off electrons.

Table \ref{tab:rates100kton} shows the expected number of neutrino
events from a supernova at 10 kpc in a 100 kton detector. The four
detection channels are considered independently and oscillation and
non-oscillation cases are computed for normal and inverted
hierarchies. The supernova parameters correspond to scenario I.
No threshold on the electron
energy has been applied since the burst nature of the supernova signal
allows to set trigger and detection thresholds at the level of the MeV.

\begin{table}[htbp]
\centering
\begin{tabular}{|c||c|cc|cc|} \hline
 \multicolumn{6}{|c|}{\bf Scenario I: expected events in 100 kton detector} \\
 \multicolumn{6}{|c|}{$\langle E_{\nu_e}
\rangle$ = 11 MeV, $\langle E_{\bar\nu_e} \rangle$ = 16 MeV, $\langle
E_{\nu_{x}} \rangle$ = $\langle E_{\bar\nu_{x}} \rangle$
= 25 MeV}\\
 \multicolumn{6}{|c|}{ and luminosity equipartition}\\
\hline
Reaction & Without & \multicolumn{2}{c|}{Oscillation
(n.h.)} & \multicolumn{2}{c|}{Oscillation (i.h.)} \\
 & oscillation & Large $\theta_{13}$ & Small $\theta_{13}$ & Large
$\theta_{13}$ & Small $\theta_{13}$\\ \hline \hline
{\bf ELAS} & 1330 & 1330 & 1330 & 1330 & 1330 \\
\hline
{\bf \nue CC} & 6240 & 31320 & 23820 & 23820 & 23820 \\ 
{\bf \anue CC} & 540 &  1110 &  1110 &  2420 & 1110 \\ 
\hline
{\bf NC} & 30440 & 30440 & 30440 & 30440 & 30440 \\
\hline \hline
{\bf TOTAL} & 38550 & 64200 & 56700 & 58010 & 56700 \\\hline
\end{tabular}
\caption{Expected neutrino events in a 100 kton detector from a
supernova at a distance of 10 kpc.}   
\label{tab:rates100kton}
\end{table}

\subsection{Charged current events}
In order to understand the dependencies of the expected number of
events with the supernova parameters and the \th13 mixing angle,
we perform a scan of these parameters for the \nue CC, and \anue CC 
processes on argon. While we scan two parameters, the rest are fixed
according to the values of scenario I. 

\begin{figure}[htbp]
\epsfig{file=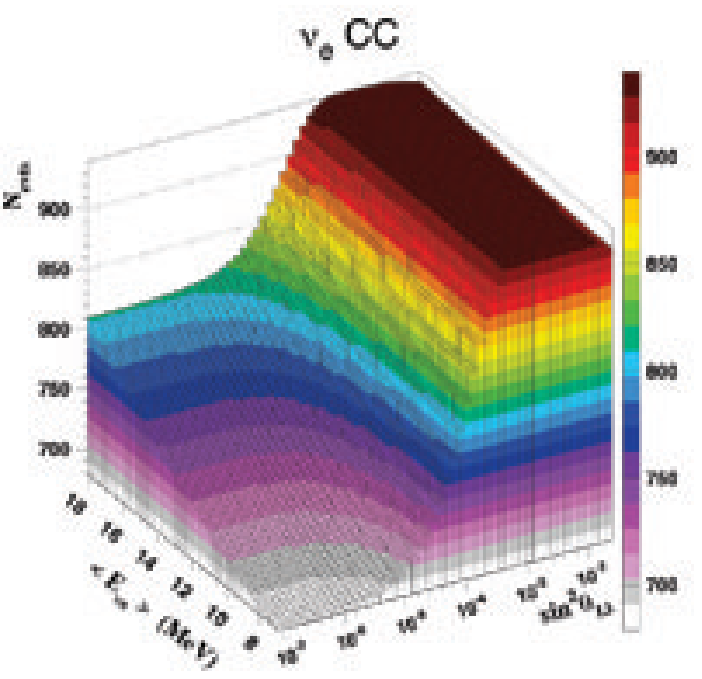,width=0.5\linewidth}
\epsfig{file=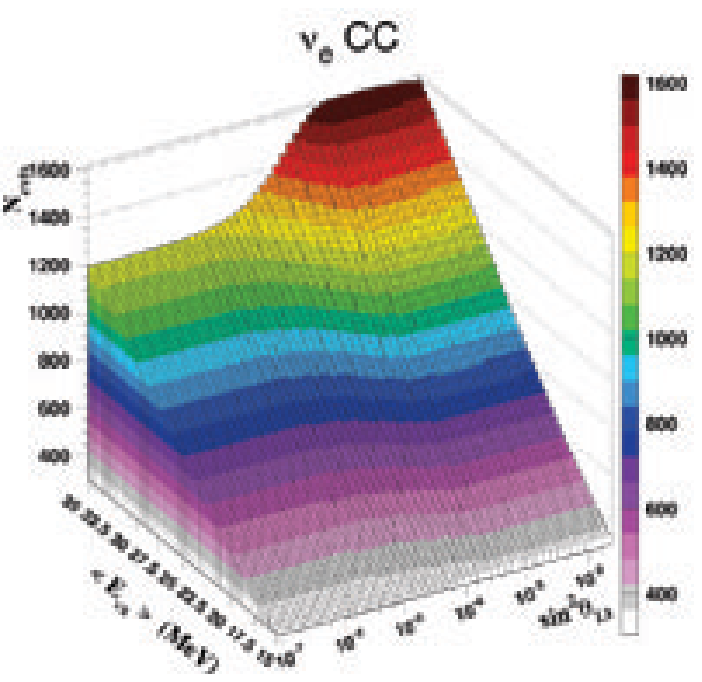,width=0.5\linewidth}
%\begin{center}
%\epsfig{file=EPS/3dpar_nh_eanue.eps,width=0.55\linewidth}
%\end{center}
\caption{{\bf Normal mass hierarchy}: Variation of the expected number of \nue CC events in a 3
kton detector as a function of \s2t13 and the original
supernova neutrino average energies \avenue ~and \avenux.}
\label{fig:3dparnuecc}
\end{figure}

Figure \ref{fig:3dparnuecc} shows the expected number of
\nue CC events in a 3 kton detector as a function of \s2t13 and the
\avenue ~and \avenux ~average energies
for a normal mass hierarchy. The range of the parameters is determined
by possible physical regions discussed in Section~\ref{sec:supphys}, namely
\avenue = (7--18) MeV and \avenux = (15--35)
MeV.
We can see from these figures that large and small mixing angle cases
are not symmetric and the results depend on the value of the neutrino
average energy. \nue CC events are very sensitive to the \avenue ~and
\avenux ~energies while they are of course independent on the \aveanue ~value.

For large \th13 mixing angle, maximal flavor conversion occurs and all
$\nu_{\mu,\tau}$ neutrinos convert into \nue's (see Table~\ref{tab:pee}). Therefore, the
number of expected events will only depend on the original average
energy of \nux ~neutrinos and not on \avenue, as can be seen in the
figures. For small \th13 mixing angle the expected flux of \nue's at
the detector is $\approx 30\%$ of original \nue ~and $\approx 70\%$ of \nux
~neutrinos produced in the supernova. In this case, the number of \nue
CC increases linearly with \avenue ~and \avenux.

\begin{figure}[htbp]
\epsfig{file=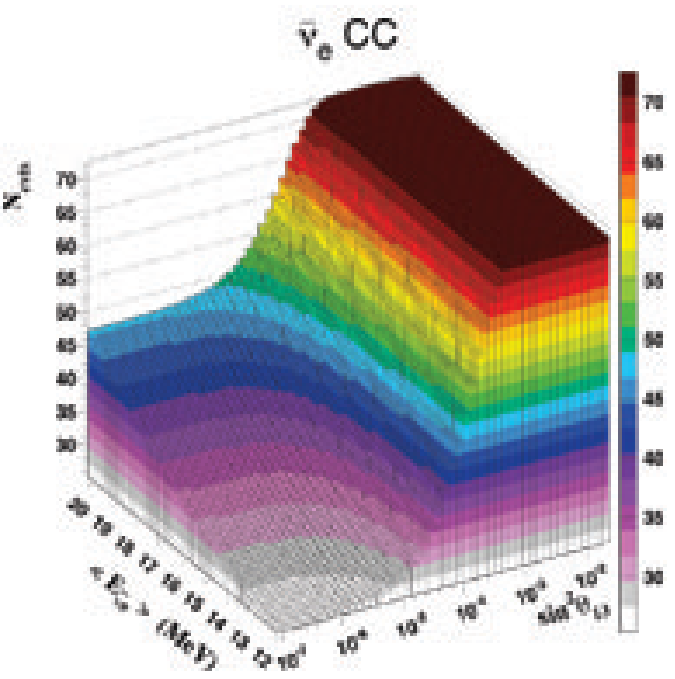,width=0.5\linewidth}
\epsfig{file=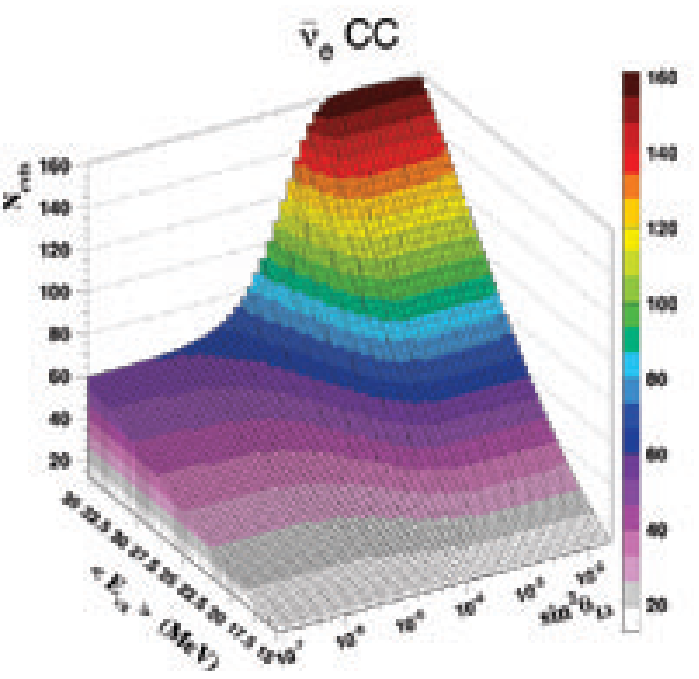,width=0.5\linewidth}
\caption{{\bf Inverted mass hierarchy}: Variation of the expected number of \anue CC events in a 3
kton detector as a function of \s2t13 and the original
supernova neutrino average energies \aveanue ~and \avenux.}
\label{fig:3dparanuecc}
\end{figure}

Figure \ref{fig:3dparanuecc} shows the number of \anue CC events
expected in a 3 kton detector as a function of \s2t13 and the
\aveanue ~and \avenux ~energies for the case of inverted hierarchy.
The H resonance lies in the antineutrino channel for the inverted
hierarchy and therefore the behavior of the \anue CC rates for i.h.~is
quite similar to the \nue CC for n.h. In the first case the variations
between small and large mixing angle are more pronounced than for \nue
CC but the event rates statistics are smaller. 

The two channels \nue CC and \anue CC are sensitive to the \th13
mixing angle and to the mass hierarchy but in different ways. This
will help to distinguish the oscillation scenarios.

\subsection{Neutral current events}

Similarly,  we perform a scan of the parameters for the NC
processes on argon. While we scan two parameters, the rest are fixed
according to the values of scenario I. 

\begin{figure}[htbp]
\begin{center}
\epsfig{file=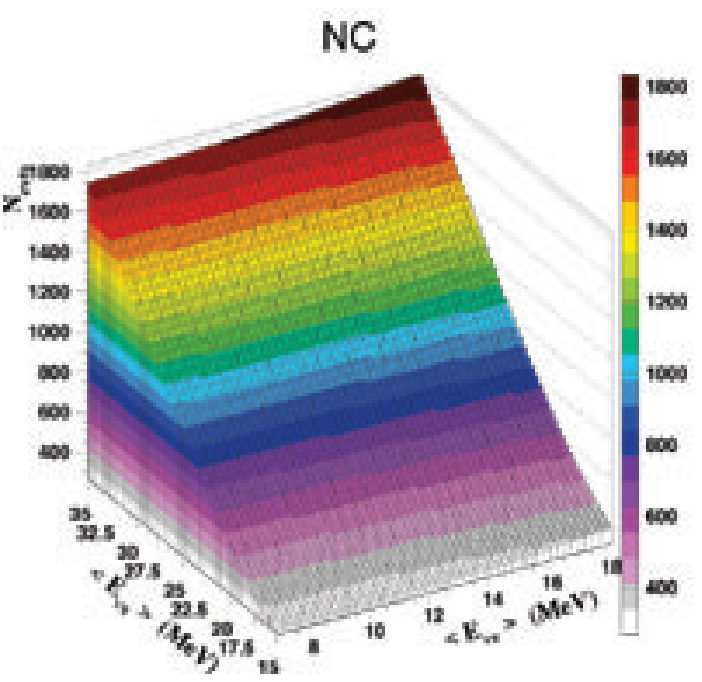,width=0.47\linewidth}
\epsfig{file=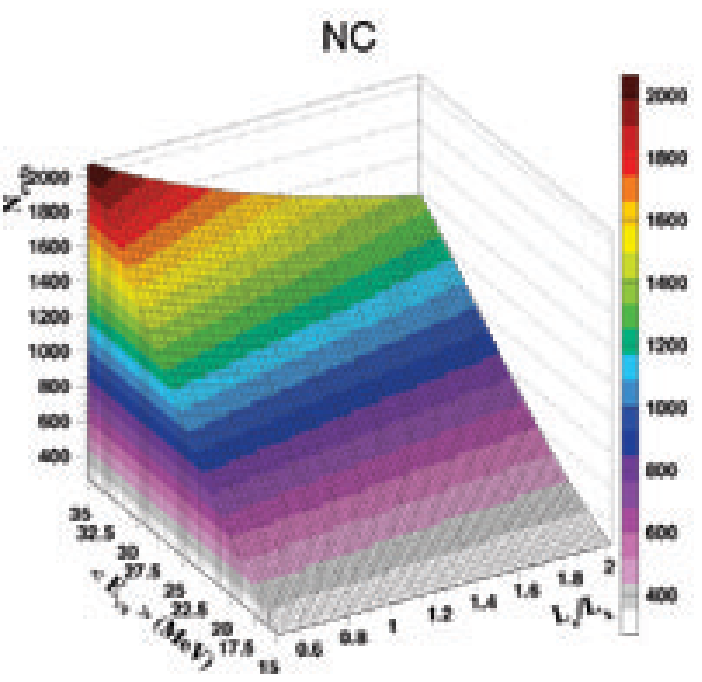,width=0.47\linewidth}
\epsfig{file=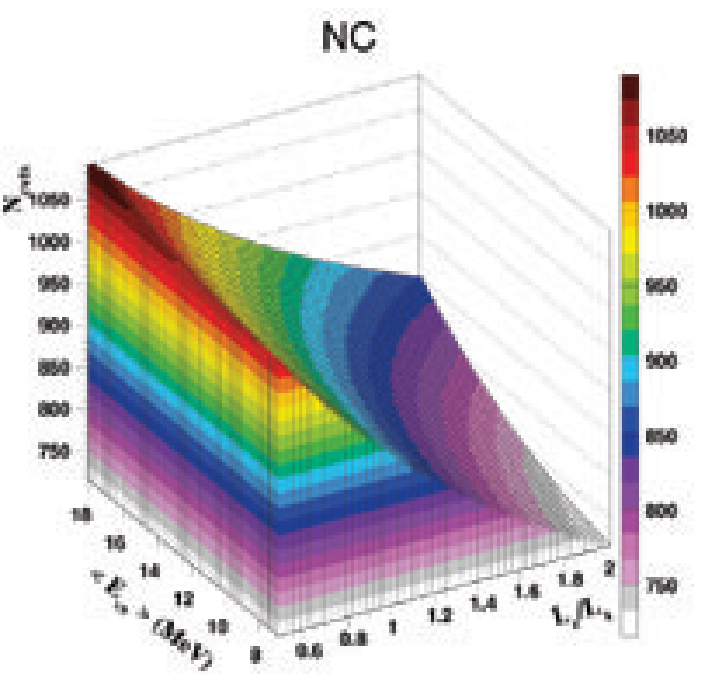,width=0.47\linewidth}
\epsfig{file=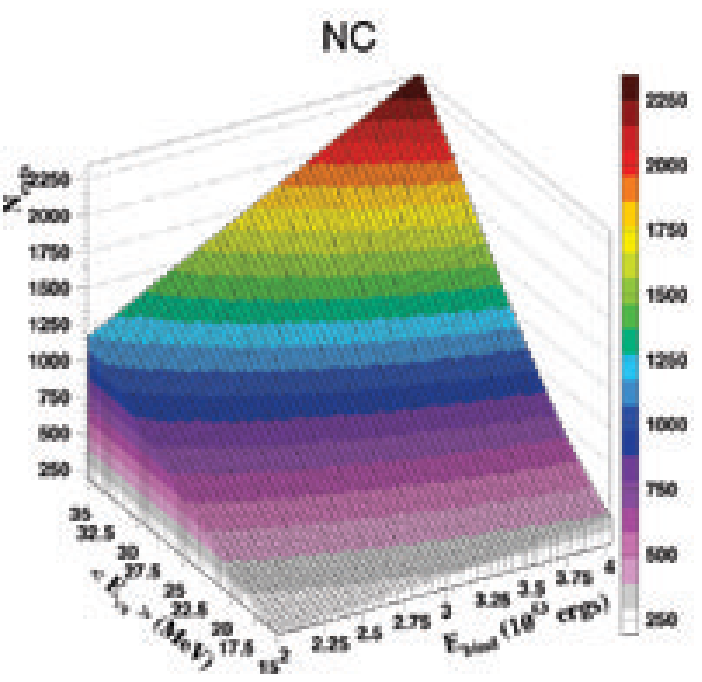,width=0.47\linewidth}
\end{center}
\caption{Variation of the expected number of NC events in a 3 kton
detector as a function of the original supernova \nue ~and \nux 
~average energies, the ratio between the electron and non-electron
neutrino luminosities (\lelx) and the total binding energy (\EB).} 
\label{fig:3dparnc}
\end{figure}

Figure~\ref{fig:3dparnc} shows the variation of the expected events
from the NC process with several supernova parameters normalized to
a 3 kton detector.

In the first plot we see the dependence of the NC events with the
original average energies \avenue ~and \avenux ~within the ranges
defined in Section~\ref{sec:supphys}, namely
\avenue = (7--18) MeV and \avenux = (15--35) MeV.
Since neutral currents are by definition flavor independent, they are
not sensitive to the \th13 mixing angle. The number of events
increases quickly with average energies because of the quadratic
dependence of the NC cross section on argon. The variations within
the range of considered energies of \nux ~are large while
within the range of the \avenue, the variation is much smaller.

In the second and third plots we scan the average energies as a
function of the ratio between the electron and non-electron neutrino
luminosities (\lelx). As \lelx ~increases, the
flux of \nue's is increasing with respect
to the \nux's. Accordingly, the number of NC events decreases because \nue's
are assumed to be less energetic than \nux's. We can see that the number of events
increases with \avenux ~and \avenue. 

The last plot shows the variations with \avenux ~and the total binding
energy \EB. For low \avenux ~energy, the number of events is almost
constant. But as soon as the \nux ~energy increases, the changes with
\EB ~are much bigger, given the quadratic dependence of the cross-section
on energy.

All these figures show that the NC events are very sensitive to the
different supernova parameters independently of the neutrino
oscillations. This process hence constitutes an
excellent probe for the supernova properties independent of the
neutrino oscillation physics.
The combination of the information extracted from other
interactions on argon with NC will allow to decouple supernova
from oscillation physics.

Elastic events off electrons are also in part sensitive to all neutrino and
antineutrino flavors.
We note however that the rate of NC events is much larger than that of
elastic events (see Table~\ref{tab:rates100kton}), hence, the statistical
power of the nuclear NC events is much larger than that of elastic events
on electrons.

\subsection{Energy distribution of the charged current and elastic events}
We have presented how the expected rates of the different processes
change depending on \th13, the mass hierarchy and the astrophysical
parameters. But also the energy spectra of the neutrinos is affected
by the value of these parameters. See our reference \cite{IApaper} for examples
of the variations of the spectra for different values of the \th13
parameter and mass hierarchies. 

\begin{figure}[htbp]
\begin{center}
\epsfig{file=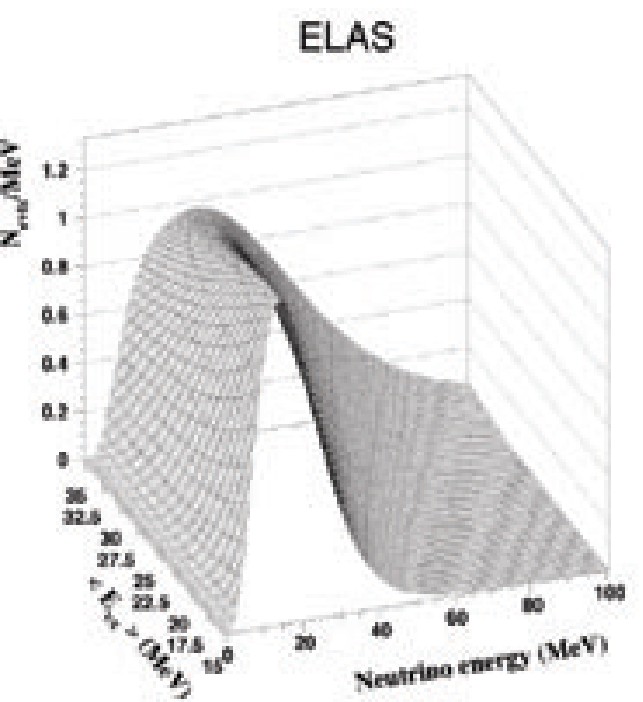,width=0.47\linewidth}
\epsfig{file=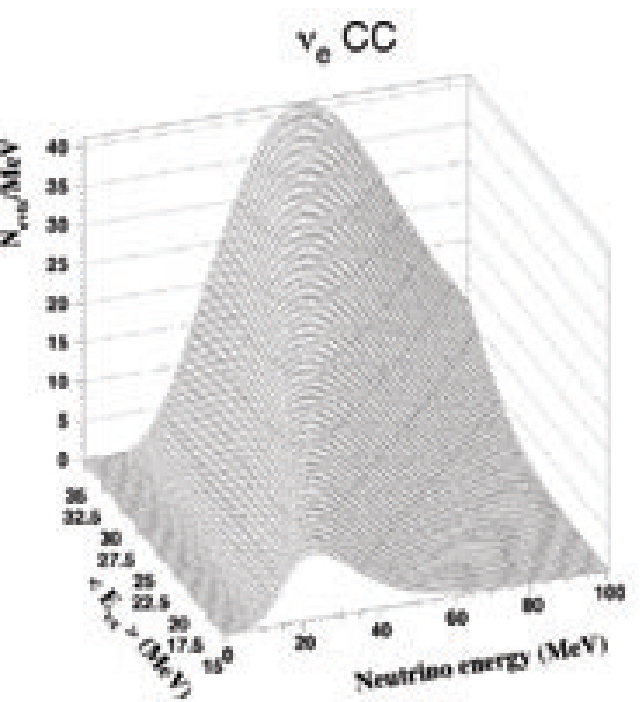,width=0.47\linewidth}
\epsfig{file=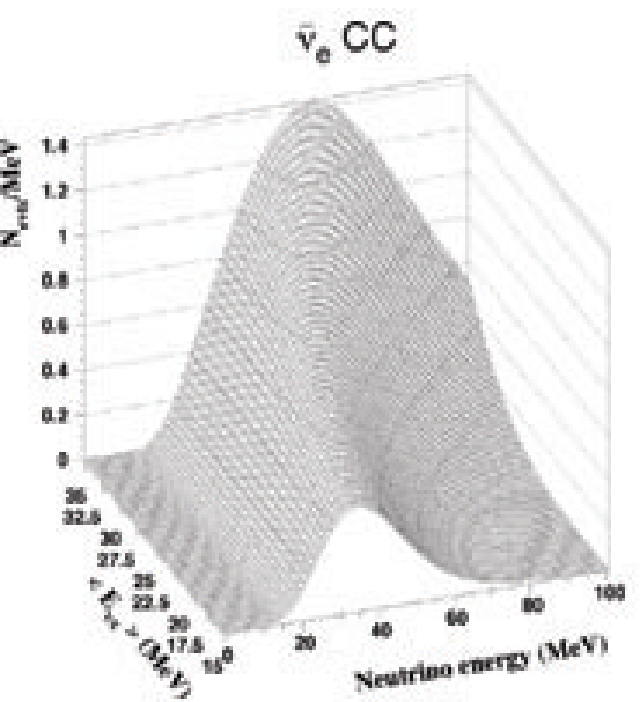,width=0.47\linewidth}
\end{center}
\caption{{\bf Large mixing angle (\s2t13 = 10$^{-3}$) and normal mass hierarchy:} Neutrino energy 
spectra for the elastic scattering, \nue CC
and \anue CC interaction processes on argon as a function of the
\avenux ~average energy. The
distributions are normalized to a 3 kton detector.} 
\label{fig:3dfunc}
\end{figure}

\begin{figure}[htbp]
\begin{center}
\epsfig{file=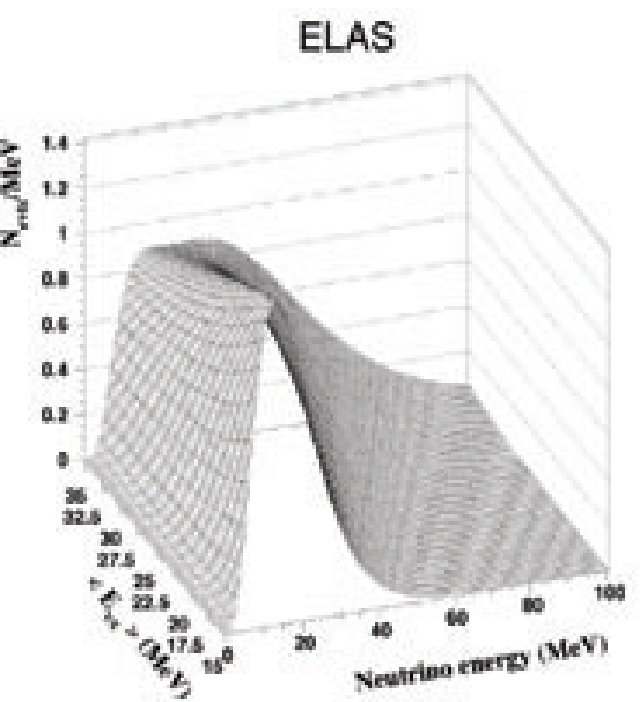,width=0.47\linewidth}
\epsfig{file=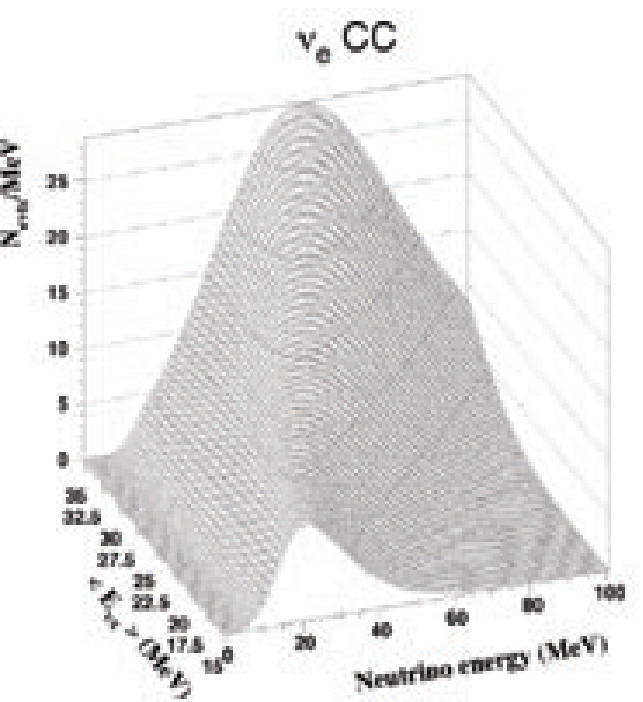,width=0.47\linewidth}
\epsfig{file=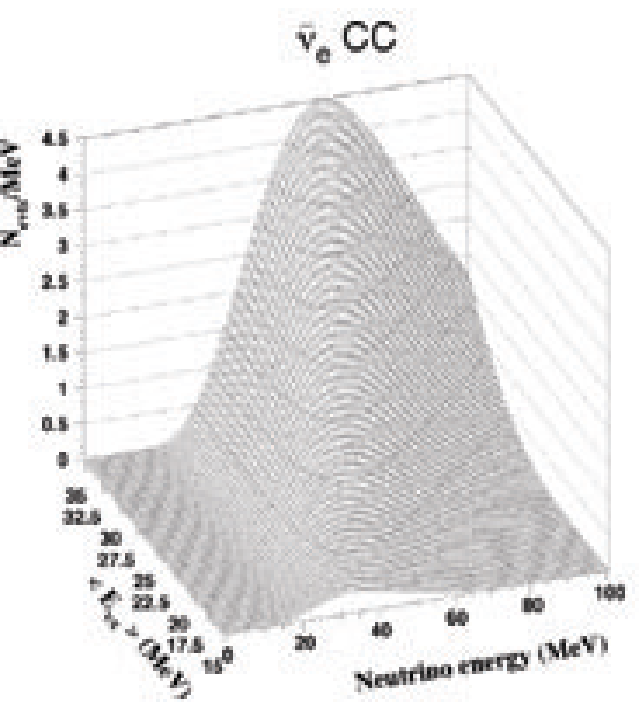,width=0.47\linewidth}
\end{center}
\caption{{\bf Large mixing angle (\s2t13 = 10$^{-3}$) and inverted mass hierarchy:} Neutrino energy 
spectra for the elastic scattering, \nue CC
and \anue CC interaction processes on argon as a function of the
\avenux ~average energy. The
distributions are normalized to a 3 kton detector.}  
\label{fig:3dfuncih}
\end{figure}

Figures \ref{fig:3dfunc} and \ref{fig:3dfuncih} show
 the variation of the neutrino energy spectra for the elastic
scattering, \nue CC and \anue CC processes as a function of the
\avenux ~average energy in the case of large mixing angle (\s2t13 =
10$^{-3}$) and normal or inverted mass hierarchies, respectively. The
total number of events corresponds to the expectations for a 3 kton
detector. We have taken energy bins of 1~MeV.

As we consider a large \th13 mixing angle, the \nue ~flux arriving at
Earth is $\phi_{\nu_e} \simeq \phi^o_{\nu_x}$(\avenux). The energy distribution of the \nue CC events depends
strongly on the \avenux ~parameter. For the inverted mass hierarchy,
the main variations as a function of \avenux ~are for the \anue CC
events.

%%%%%%%%%%%%%%%%%%%%%%%%%
\section{Analysis procedure}
%%%%%%%%%%%%%%%%%%%%%%%%%

We will consider three different situations. In all cases,
we exploit the information from the four
neutrino detection channels (ELAS, \nue CC, \anue CC and NC).
When relevant, we compare the results
with the case of not having access to NC events, which
is experimentally the most challenging signature.

\begin{enumerate}
\item
{\bf Study of the different oscillation scenarios without
knowledge of the supernova parameters}: We investigate the sensitivity to the \th13 and mass hierarchy
parameters without any assumption on
the supernova parameters. Quantitative results on the determination
of the \s2t13 mixing angle are
obtained. This is described in Section~\ref{sec:first}.
\item
{\bf Study of the supernova parameters assuming that the
oscillation parameters are known}: Assuming that we know the value of the oscillation parameters from
terrestrial experiments, we study the possible determination of the {\it five}
supernova parameters \EB,  \avenue, \aveanue, \avenux ~and \lelx.
Two cases are assumed: (a) the \th13 is large (\s2t13 $=$
10$^{-3}$) and has been measured with a 10\% precision from
long-baseline experiments; (b) the \th13 is small (\s2t13 $<$
10$^{-4}$) and an upper limit on its value has been determined. 
This is described in Section~\ref{sec:second}.
\item {\bf Study of the supernova parameters without any knowledge on the
neutrino oscillation parameters}: 
We investigate the information that can be extracted from supernova
neutrinos if a supernova explosion occurred nowadays when no 
information about the oscillation parameters \th13 and mass hierarchy
(sign[$\D32$]) is available. This is described in Section~\ref{sec:third}.
\end{enumerate}

As was shown in the previous section, the simultaneous observation of
ELAS, \nue CC, \anue CC and NC events on argon enables us to
extract information about the oscillation parameters \th13
and the type of mass hierarchy (sign[$\D32$]) and the five 
supernova parameters.

In order to determine all these parameters quantitatively we use a $\chi^2$ method.
This analysis procedure includes all the uncertainties and parameter
correlations.
We define a set of reference values for the relevant supernova and
oscillation parameters that we consider as ``true''. Different values of the \th13 angle are chosen in the
analysis for normal and inverted mass hierarchies.  
Two sets of reference values of the astrophysical parameters are
studied according to scenarios I and II described in Table
\ref{tab:sncoolscenario}. 

%The reference values of the oscillation parameters
%correspond to the four different oscillation scenarios that we
%consider as ``true'' and we want to test.

To see how well we can discriminate every ``true'' scenario,
we minimize the $\chi^2$ in the space of the five astrophysical
parameters and the two oscillation parameters. The $\chi^2$ function
is defined as: 
\begin{equation}
\chi^2(x) = \sum_{i=1}^{Nbins}
\frac{\left(N_i(x) - N_i(x^0) \right)^2}{\sigma_i^2(x)}
\end{equation} 

\noindent where $N_{bins}$ is the number of bins considered in every
histogram; $N_i$ is the number of events per bin $i$; $\sigma$ is
the statistical error associated to the bin $i$; $x$ corresponds
to the set of parameters to be determined $x$ $\equiv$ \{\EB, \avenue,
\aveanue, \avenux, \lelx, \s2t13, sign[$\D32$]\}, which are freely
varying and $x^0$ are the ``true'' values of these parameters. When
computing the $\chi^2$ for one parameter, the others are left free. In this
way, we can naturally take into account the uncertainty on the given 
parameter introduced by the lack of knowledge on the others.
This is particularly relevant in the discussion of the decoupling
of supernova and oscillation parameters.

The total $\chi^2$ is computed as the sum of the minimized $\chi^2$
corresponding to every detection channel:
\begin{equation}
\chi^2_{total} = \chi^2_{ELAS} + \chi^2_{\nu_e CC} + \chi^2_{\bar{\nu}_e CC}
+ \chi^2_{NC}
\end{equation}

The fit is performed with the MINUIT \cite{minuit} package, and is 
expected to get back the same values of the parameters, starting from the reference
distributions.  At each iteration, a different 
set of parameters is probed, and with the same procedure used to get 
the reference histograms.
The condition of the hierarchy between
the average energies, confirmed by various supernova
simulations, is enforced during the minimization:
\begin{equation}
\langle E_{\nu_e} \rangle \leq \langle E_{\bar{\nu}_e} \rangle \leq \langle E_{\nu_x} \rangle
\end{equation}
Otherwise, the ranges of the parameters are left free.

We compare the energy distributions of the elastic scattering, \nue CC
and \anue CC neutrino events from the ``true'' scenarios with those obtained
with all possible values of the parameters, taking into account the
oscillations inside the SN matter. A bin width of 1 MeV is considered
in the spectra. 
For the NC channel, the energy distribution cannot be
determined due to the neutrino presence in the final state. We only
consider the number of NC events in the fit.

In order to compute the precision of the determination of the parameters,
we consider two methods: (1) a one-dimensional ``scan'' of a given
parameter; the other variables are left free and minimized at each step;
the resp.~1, 2, 3 sigmas are given by resp. $\chi^2_{min}+1$, $+4$ and $+9$.
(2) a two-dimensional ``scan'' of a two-parameter plane; the other
variables are left free and minimized at each point in the plane;
the resp. 68\%, 90\%, 99\% C.L. are given by resp.
$\chi^2_{min}+2.4$, $+4.6$ and $+9.2$.

%%%%%%%%%%%%%%%%%%%%%%%%%%%%%%%%%%%%%%%%%%%%%%%%%%%%%%%%%%%%%
\section{Study of the different oscillation scenarios without
knowledge of the supernova parameters}
\label{sec:first}
%%%%%%%%%%%%%%%%%%%%%%%%%%%%%%%%%%%%%%

In this first section we investigate the possible
oscillation scenarios without any assumption on the supernova
properties. 

As was shown in table \ref{tab:pee}, the value of the survival
probabilities P$_{ee}$ and $\overline{P}_{ee}$ is constant in two
extreme regions of the \th13 parameter: the large mixing angle case
(\s2t13 $>$ $3\times 10^{-4}$) and the small mixing angle case
(\s2t13 $<$ $2\times 10^{-6}$). It means that the neutrino fluxes do
not change with \th13 in these regions.
On the other hand, if the angle is in an intermediate range 2 $\times$
10$^{-6}$ $<$ \s2t13 $<$ 3 $\times$ 10$^{-4}$, the fluxes depend on
the \th13 value and on the neutrino energy. Hence, different energy
spectra and event rates are expected for every \th13 value.

For this reason we first analyze the true small and large mixing angle
scenarios and study the possibility
to distinguish the mass hierarchy and put a bound on the \th13
value. Then, we investigate different values of the angle in the
intermediate region and we determine the statistical accuracy
with which it can be measured.

\subsection{Small and large mixing angle extreme cases}
%%%%%%%%%%%%%%%%%%%%%%%%%%%%%%%%%%%%%%%%%%%%%%%

We consider the four extreme oscillation cases (see Section~\ref{sec:oscphys}) as possible ``true'' values
and we use the information of the neutrino detection channels 
to discriminate between them. Because we assume no knowledge on the 
original supernova neutrino parameters, we let them free while performing the
$\chi^2$ minimization. For example, for every value of the \th13 and a
fixed hierarchy, we vary the supernova parameters in order to obtain
the set of values that minimize the $\chi^2$ function for this given value
of the mixing angle and mass hierarchy. The $\chi^2$ value
result of this fit can then be studied as a function of \th13 and the mass
hierarchy. 

\begin{figure}[htbp]
\epsfig{file=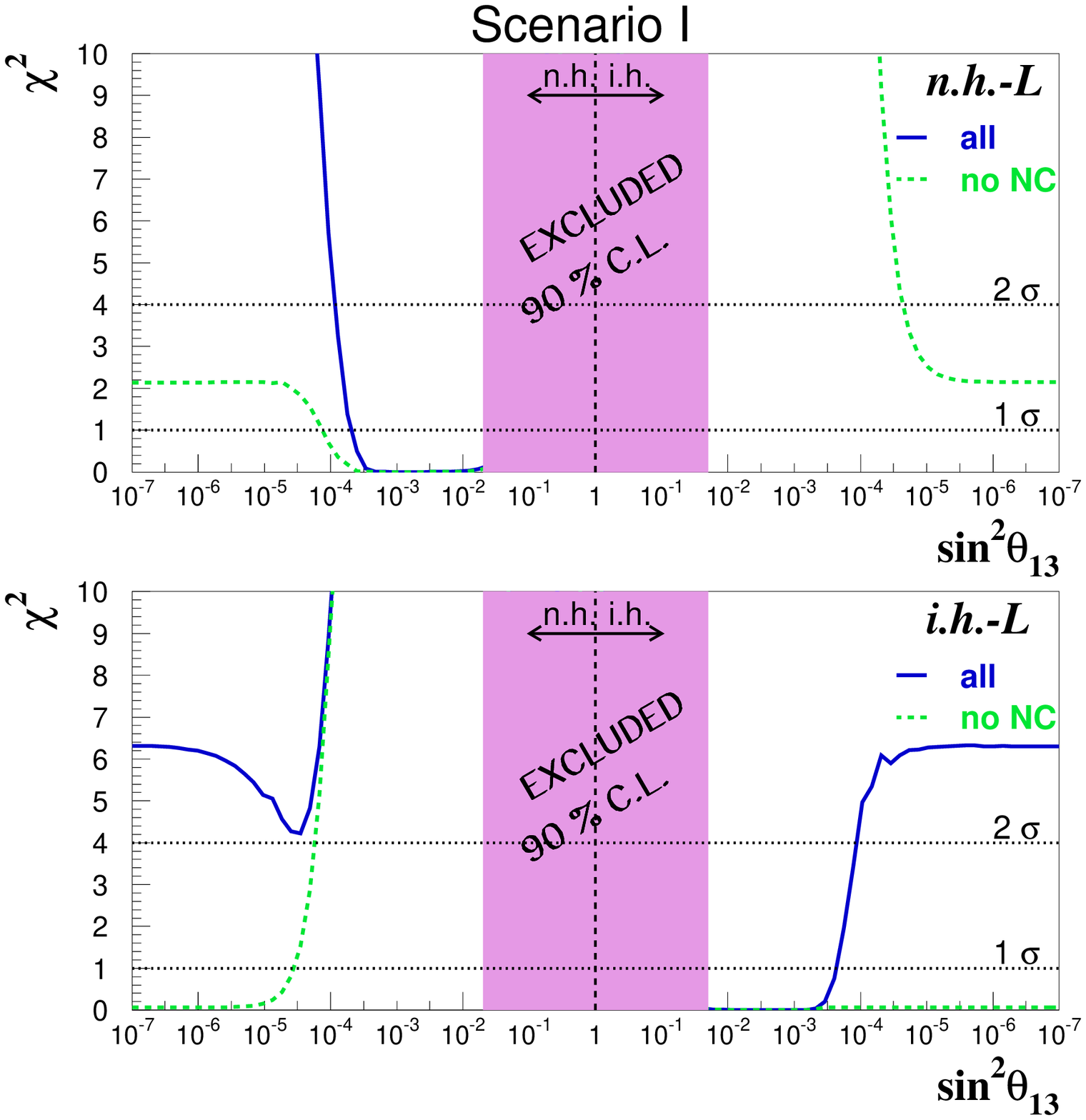,width=0.55\linewidth}
\hspace{-1cm}
\epsfig{file=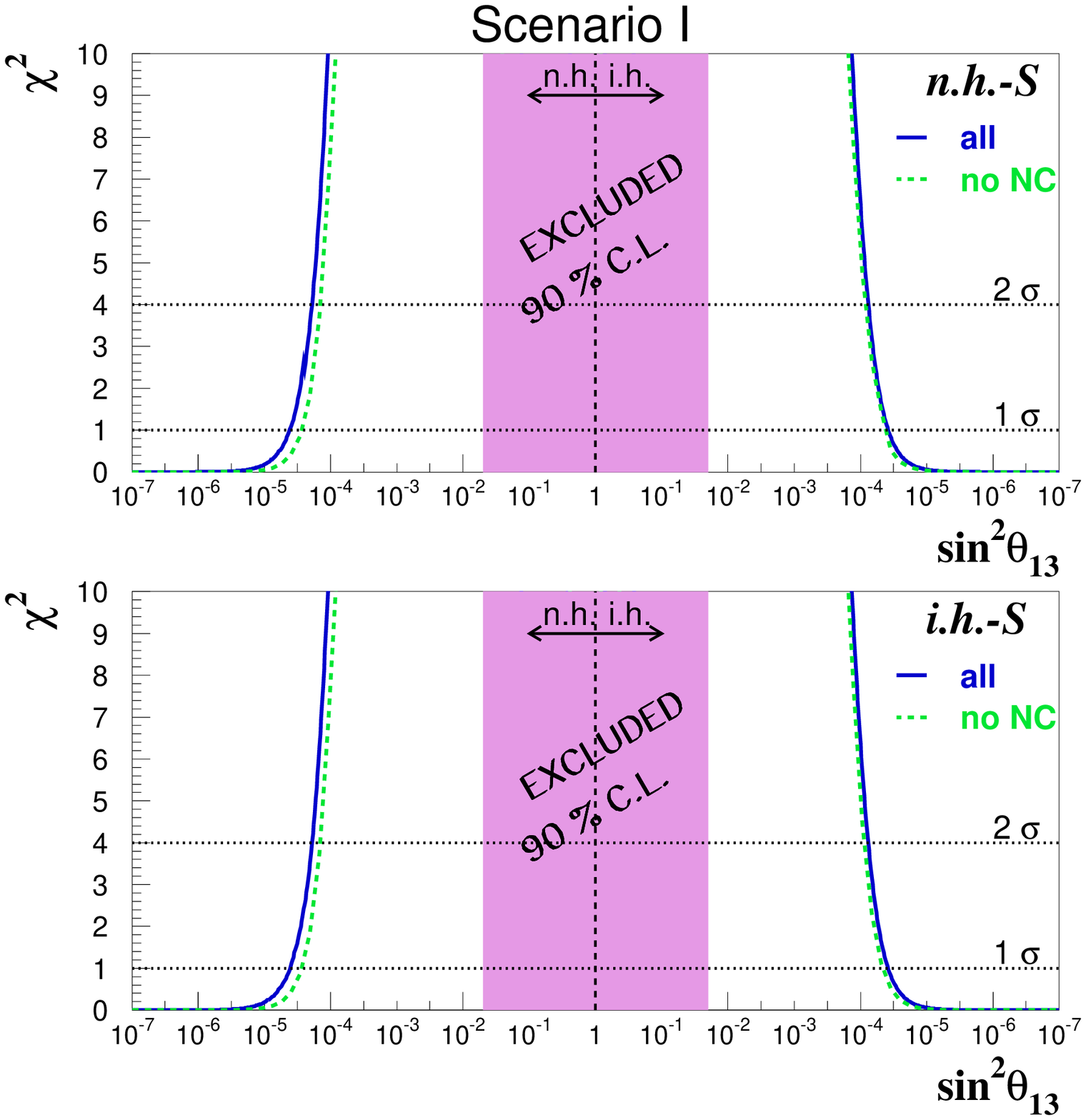,width=0.55\linewidth}
\epsfig{file=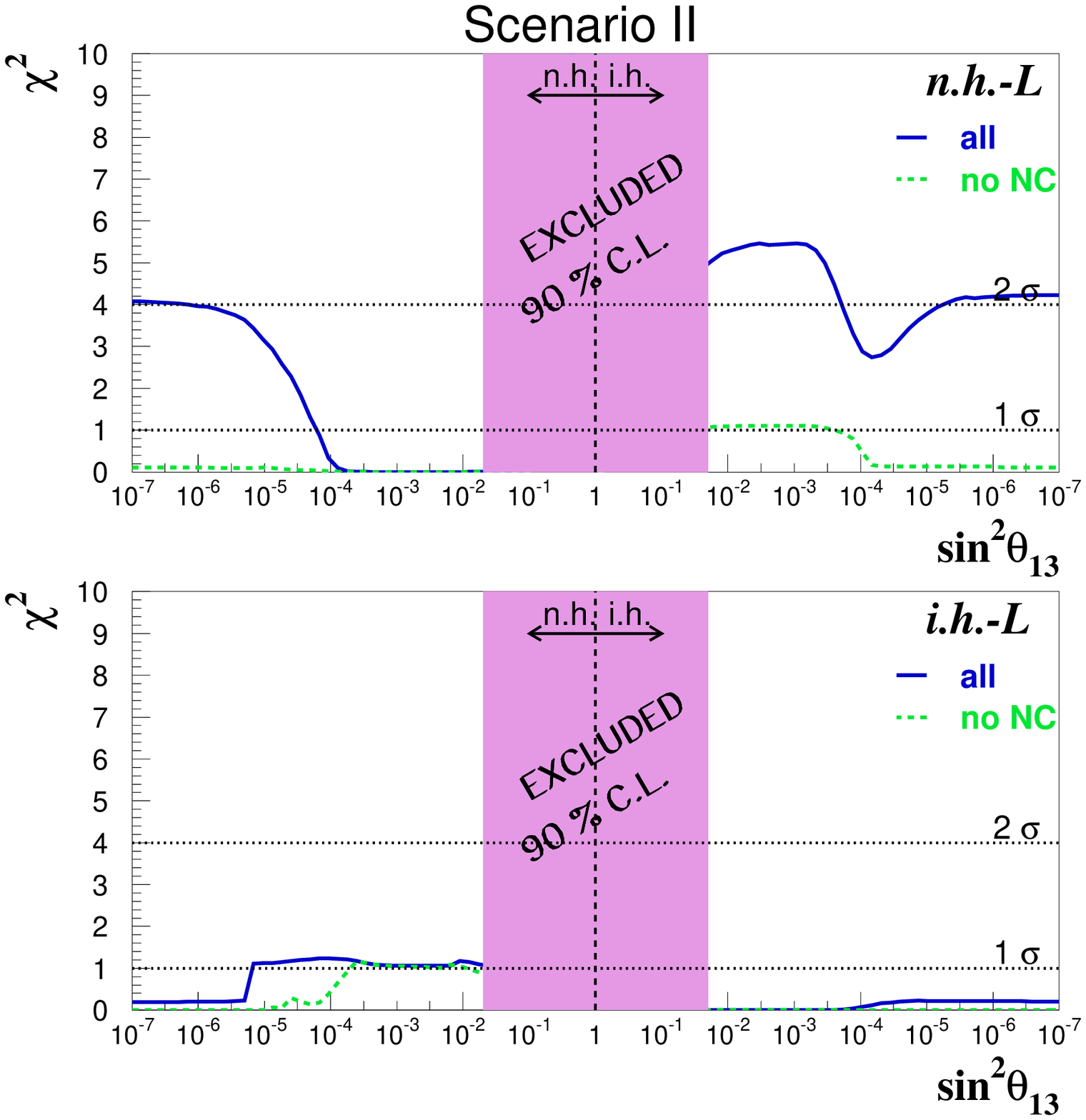,width=0.55\linewidth}
\hspace{-1cm}
\epsfig{file=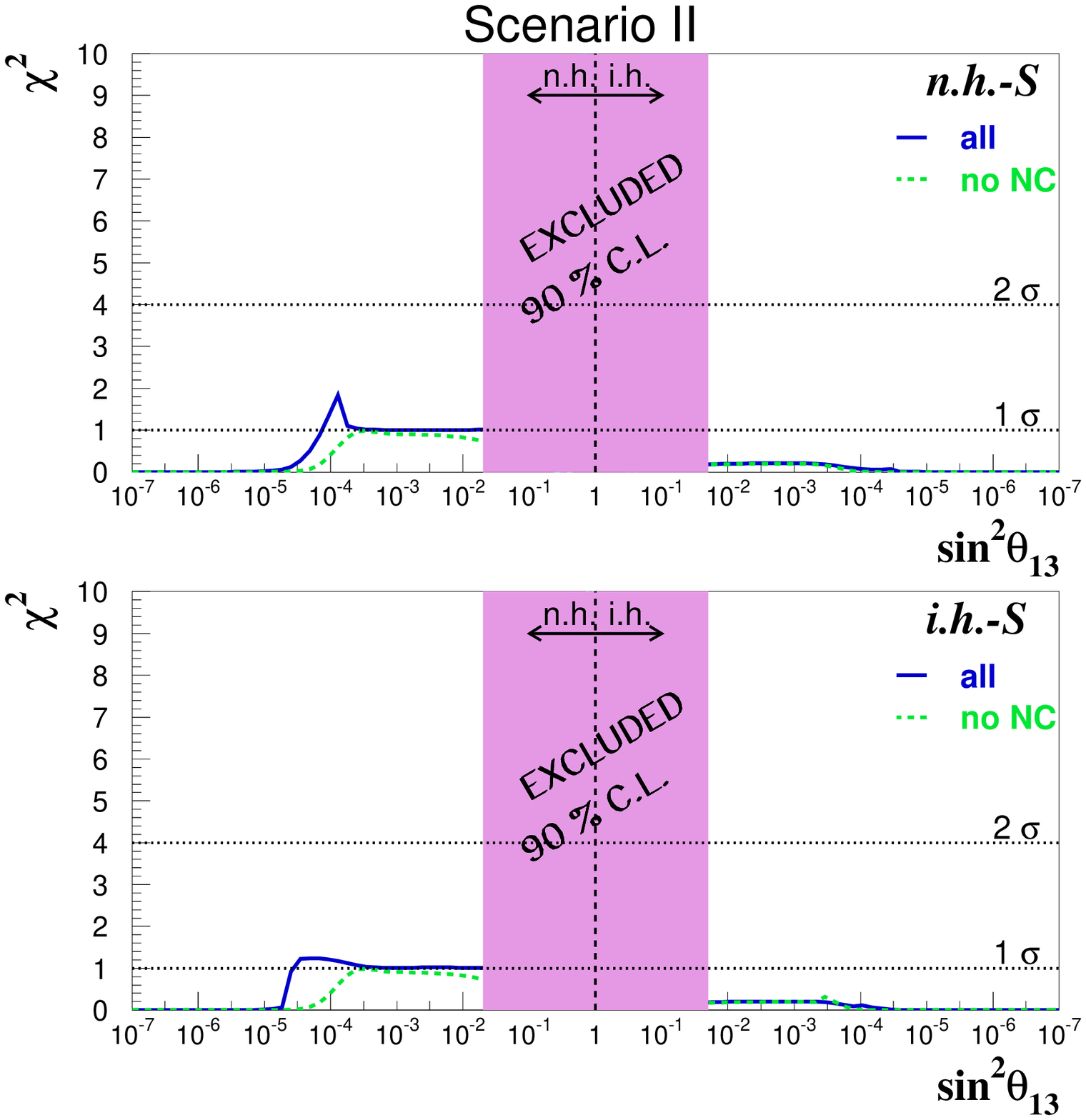,width=0.55\linewidth}
\caption{{\bf Extreme mixing angle cases and 3~kton detector}: 
$\chi^2$ value of the fit as a function of \s2t13 for a 3~kton detector. 
Curves on the left (right) part of every plot
correspond to the results of the fit under the assumption of normal (inverted)
hierarchy. Four
limiting oscillation scenarios ({n.h.-L, n.h.-S, i.h.-L, i.h.-S})
are considered as ``true'' and the reference values of the supernova
parameters correspond to scenario I (top) and II (bottom). Solid
curves are computed using the information of the four supernova
neutrino detection processes and the dashed line does not
take into account NC events.} 
\label{fig:3kton}
\end{figure}

A summary of the results on the determination on the \th13 mixing angle for the
different oscillation cases can be found in
Table~\ref{tab:th13limits}. We include the results for a 3 kton and a
100 kton detector and supernova scenarios I and II.

\begin{table}[htbp]
\small{
\centering
\begin{tabular}{|c|c|c|c|c|c|} 
\hline
\multicolumn{6}{|c|}{Determination of $\theta_{13}$ and mass hierarchy (SN parameters free)}\\
\hline
{\bf ``True''} & \multicolumn{2}{|c|}{3 kton LAr} &  \multicolumn{2}{|c|}{100 kton LAr} & \\
{\bf osc.} & \multicolumn{2}{c|}{\bf \s2t13 and assumed hierarchy}  &
\multicolumn{2}{c|}{\bf \s2t13 and assumed hierarchy}  & Events \\  
{\bf case} & \multicolumn{1}{|c}{SN scen I} & \multicolumn{1}{c|}{SN scen II} &
\multicolumn{1}{|c}{SN scen I} & \multicolumn{1}{c|}{SN scen II} & {type} \\ 
% & \multicolumn{3}{|c|}{}\\ 
\hline \hline
                 & $>$ 2.1$\times$10$^{-4}$ n.h. & $>$ 6.2$\times$10$^{-5}$ n.h. & $>$ 4.5$\times$10$^{-4}$ n.h. & $>$ 1.8$\times$10$^{-4}$ n.h. & all \\ 
{\sf \bf n.h.-L} & excluded i.h.                 & excluded i.h.                 & excluded i.h.                 & excluded i.h. & all \\ \cline{2-6}
                 & $>$ 7.5$\times$10$^{-5}$ n.h. & -- & $>$ 2.5$\times$10$^{-4}$ n.h. & $>$ 9.4$\times$10$^{-5}$ n.h. & no NC \\ 
	 	 & excluded i.h.                 & -- & excluded i.h.            & excluded i.h. & no NC \\ 
\hline \hline
		 & $<$ 2.4$\times$10$^{-5}$ n.h. & -- & $<$ 4.3$\times$10$^{-6}$ n.h. & $<$ 1.2$\times$10$^{-5}$ n.h. &all \\ 
{\sf \bf n.h.-S} & $<$ 3.8$\times$10$^{-5}$ i.h. & -- & $<$ 5.9$\times$10$^{-6}$ i.h. & $<$ 4.3$\times$10$^{-6}$ i.h. &all \\ \cline{2-6}
                 & $<$ 3.7$\times$10$^{-5}$ n.h. & -- & $<$ 7.3$\times$10$^{-6}$ n.h. & $<$ 3.2$\times$10$^{-5}$ n.h. &no NC \\ 
		 & $<$ 4.1$\times$10$^{-5}$ i.h. & -- & $<$ 7.3$\times$10$^{-6}$ i.h. & $<$ 1.1$\times$10$^{-4}$ i.h. &no NC \\ 
\hline \hline
		 & excluded n.h.                 & -- & excluded n.h.                 & excluded n.h.                 & all\\
{\sf \bf i.h.-L} & $>$ 2.3$\times$10$^{-4}$ i.h. & -- & $>$ 5.0$\times$10$^{-4}$ i.h. & $>$ 2.3$\times$10$^{-4}$ i.h. & all \\ \cline{2-6}
		 & $<$ 2.8$\times$10$^{-5}$ {\it n.h.} & -- & excluded n.h.                 & $<$ 2.0$\times$10$^{-5}$ {\it n.h.} & no NC \\ 
		 & --                                  & -- & $>$ 4.9$\times$10$^{-4}$ i.h. & --                                  & no NC \\
\hline \hline
		 & $<$ 2.4$\times$10$^{-5}$ n.h. & -- & $<$ 5.9$\times$10$^{-6}$ n.h. & $<$ 1.2$\times$10$^{-5}$ n.h. &all \\
{\sf \bf i.h.-S} & $<$ 3.8$\times$10$^{-5}$ i.h. & -- & $<$ 4.0$\times$10$^{-6}$ i.h. & $<$ 4.1$\times$10$^{-5}$ i.h. &all \\ \cline{2-6}
		 & $<$ 3.7$\times$10$^{-5}$ n.h. & -- & $<$ 7.2$\times$10$^{-6}$ n.h. & $<$ 3.2$\times$10$^{-5}$ n.h  &no NC \\
		 & $<$ 4.6$\times$10$^{-5}$ i.h. & -- & $<$ 7.4$\times$10$^{-6}$ i.h. & $<$ 1.1$\times$10$^{-4}$ i.h. &no NC \\
\hline
\end{tabular}
\caption{Estimated limit on the \th13 mixing angle at 1$\sigma$ for
different ``true'' oscillation cases with a 3 and a 100 kton detectors. 
We compare the results considering as reference values
for the supernova neutrino parameters scenarios I (hierarchical) and
II (non-hierarchical).} 
\label{tab:th13limits}
}
\end{table}

\subsubsection{Supernova scenario I and 3 kton detector}
We first discuss results for a 3~kton detector.
Plots in figure \ref{fig:3kton} show the variation of the minimized
$\chi^2$ value with the \s2t13 parameter and the mass hierarchy. In
the top right corner of the figure is indicated in bold the ``true''
scenario considered (the $x^0$ parameters). The left part of a given plot
corresponds to the results of the fit {\it under the assumption of normal hierarchy and the
right part for an inverted hierarchy}. 

Solid lines are computed considering the four supernova neutrino detection
channels. Dashed lines do not take into
account the NC processes, only ELAS and CC events. The one- and
two- sigma levels are shown by horizontal lines. The four plots on resp. the
top (bottom) are obtained using as true supernova
parameters those from resp. scenario I (II). 
The shaded region illustrates the
excluded region by the results of reactor experiments \cite{reactors}.

We see that normal and inverted hierarchies are indistinguishable for
small ``true'' \th13 mixing angle. The results of the fit are
similar and only an upper bound on \s2t13 can be set. This limit is
\s2t13 $<$ 2.4 $\times$ 10$^{-5}$ for assumed n.h. and \s2t13 $<$ 3.8 $\times$
10$^{-5}$ for assumed i.h.

If the ``true'' mixing angle is large, we are able to distinguish among
hierarchies. Assuming the ``true'' hierarchy is normal and the
value of \th13 is large (n.h.-L), we could put a lower limit on the
\th13 angle (\s2t13 $>$ 2.1 $\times$ 10$^{-4}$) for an assumed n.h. 
The assumed i.h. is excluded. The same is possible
for the case of inverted hierarchy being the limit \s2t13 $>$
2.3 $\times$ 10$^{-4}$.  The assumed n.h. is excluded.

We illustrate the importance of the neutral current events.
If we do not include the information given by NC events, we loose
sensitivity in the large mixing angle cases. The limit in the case of
n.h.-L can be set only at the level of 1$\sigma$ and no limit can be
determined for i.h.-L. Moreover, without NC events, i.h.-L could be
misidentified as n.h.-S. 

%For the small mixing cases, the NC channel does not
%help in adjusting the energy distributions obtained with large mixing
%angle values to the ones corresponding to the true n.h.-S. 
%For large mixing angle the neutrino flavor conversion is
%total and the \nue ~flux is only composed by oscillated \nux
%~neutrinos. A change on the energy or relative luminosity of \nux's is
%not sufficient to reproduce the low energy tail of the \nue CC energy
%distribution of the n.h.-S case.  
%On the other hand, the NC events are very important to exclude the
%small mixing angle region in the n.h.-L case. For small mixing the
%\nue ~flux arriving at Earth is a mixture of the original \nue ~and
%\nux's from the supernova. Decreasing the relative 
%contribution of \nue's (decreasing \lelx) and adjusting energies, the
%\nue CC energy distribution of the {\bf n.h.-L} case can be
%reproduced. However, the number of NC events will increase and they
%will give us the possibility to exclude the region. If we do not
%consider them, we loose this sensitivity. 

\subsubsection{Supernova scenario II and 100 kton detector}
If we consider the supernova scenario
II, we see that the results of the fit obtained with a 3 kton detector
are quite marginal. In this
case the average energies of \nux ~are very close to those \nue ~and
\anue. It is impossible to distinguish among hierarchies
and no bounds on the mixing angle can be set. 
Only a lower limit is obtained for the case of n.h.-L
(\s2t13 $>$ 6.2 $\times$ 10$^{-5}$).

\begin{figure}[htbp]
\epsfig{file=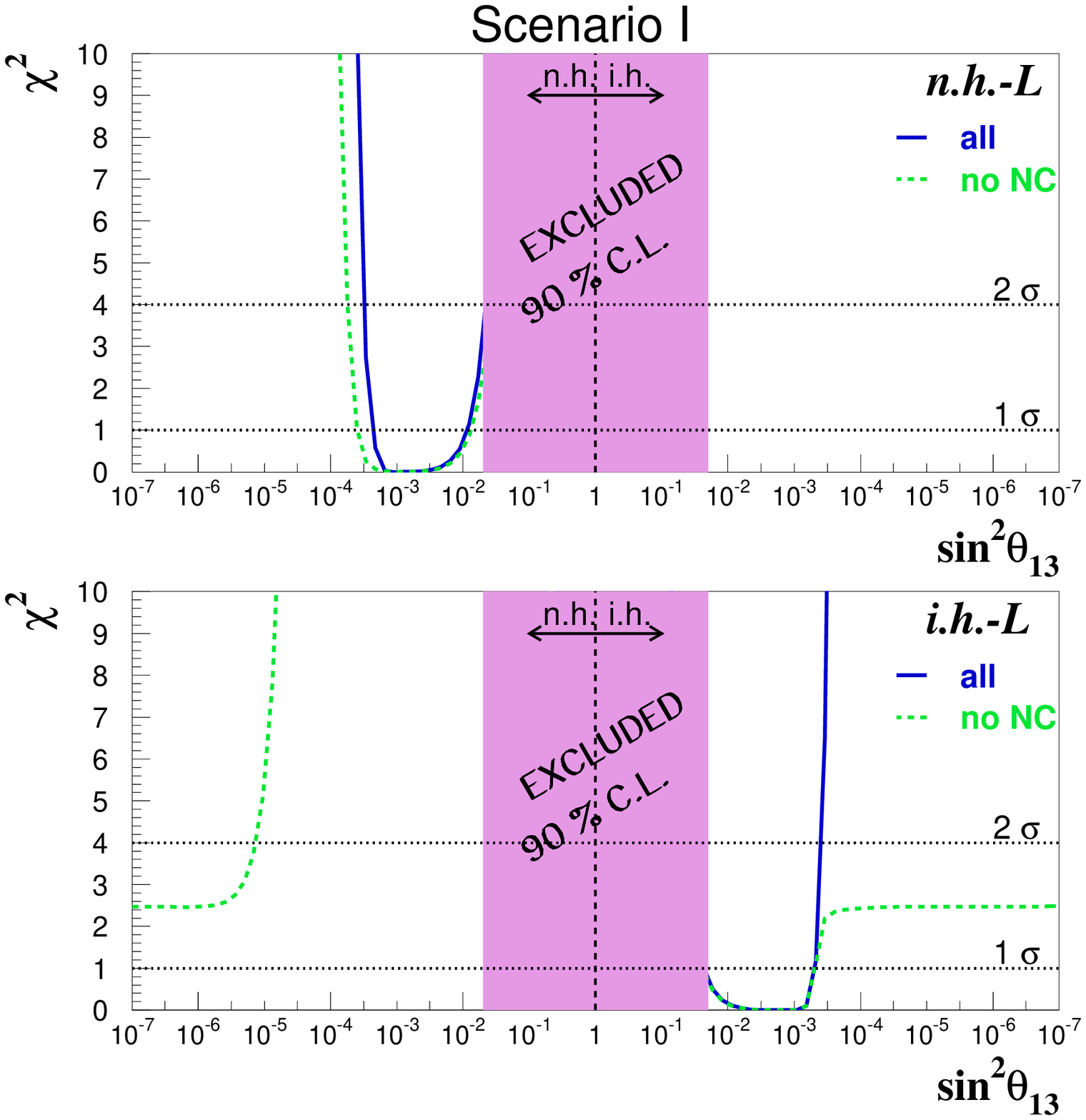,width=0.55\linewidth}
\hspace{-1cm}
\epsfig{file=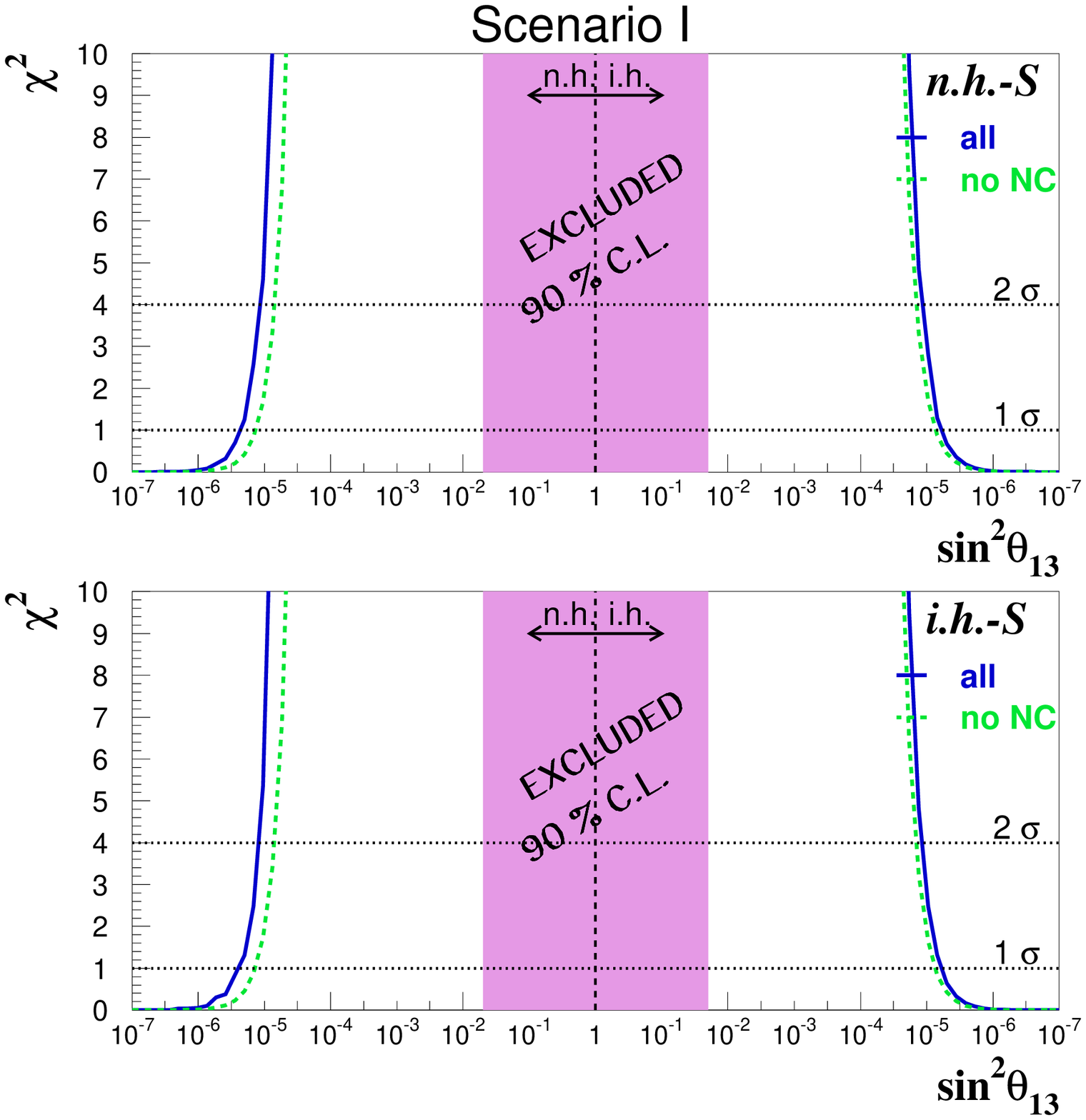,width=0.55\linewidth}
\epsfig{file=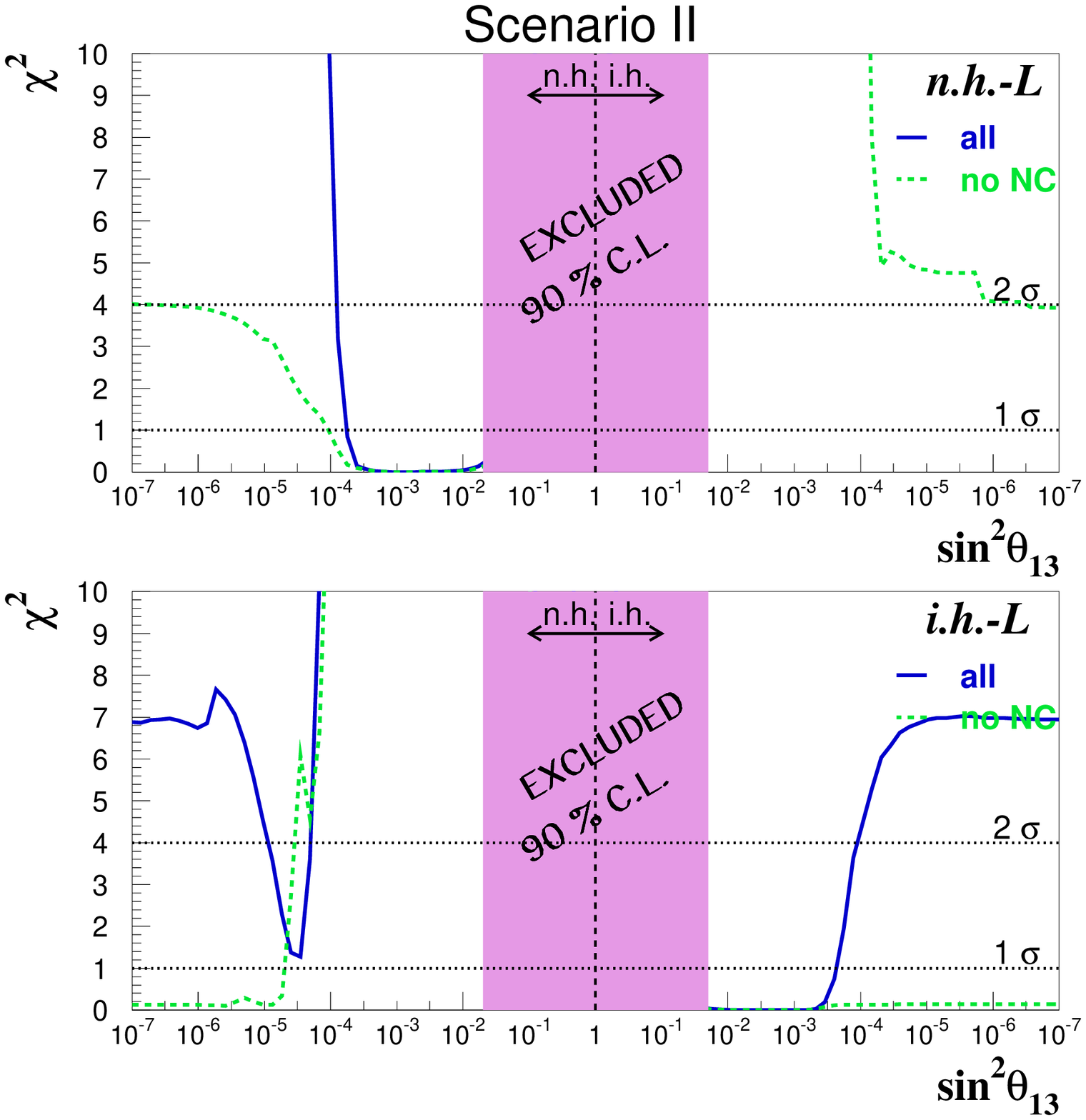,width=0.55\linewidth}
\hspace{-1cm}
\epsfig{file=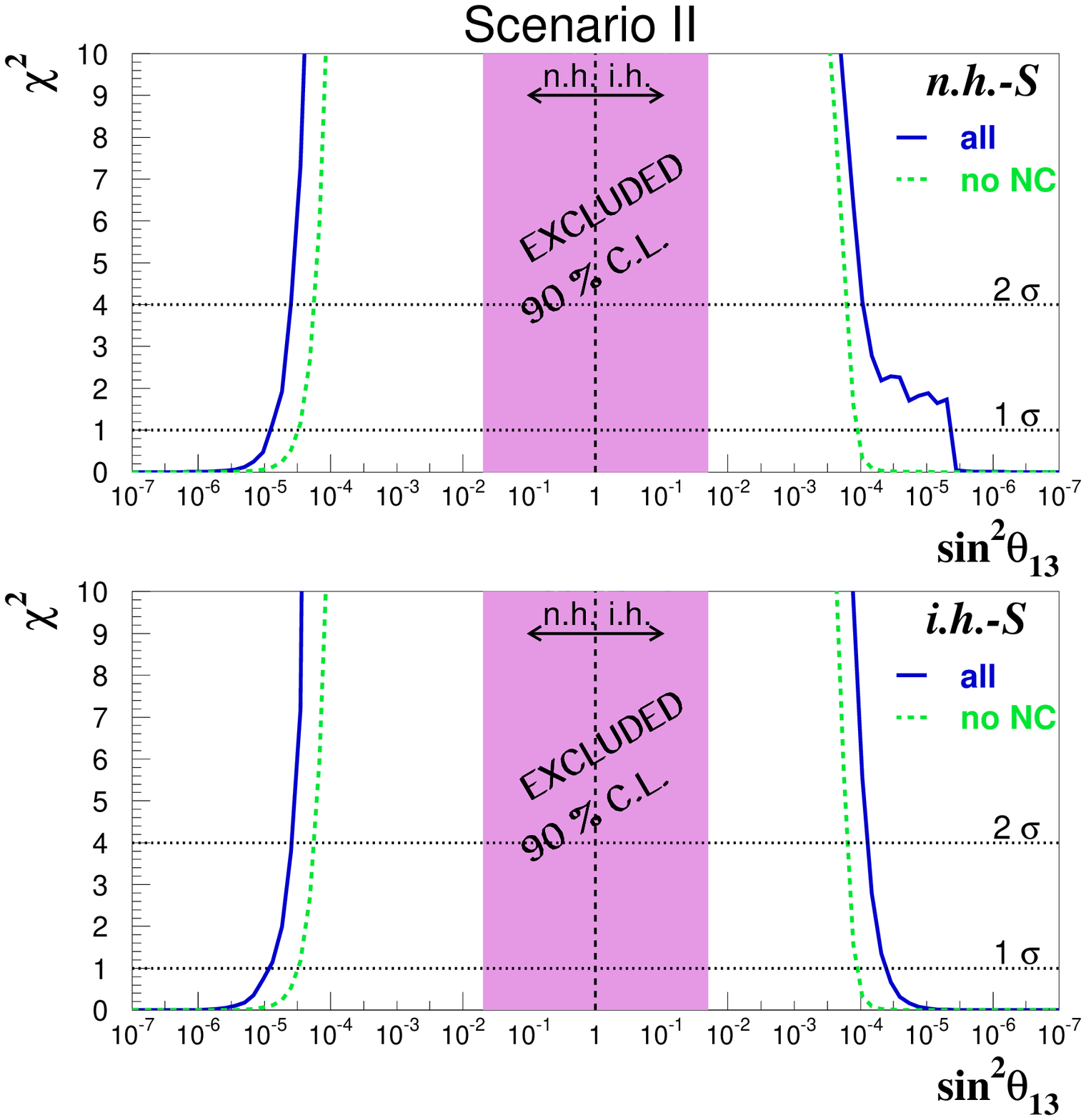,width=0.55\linewidth}
\caption{{\bf Extreme mixing angle cases and 100~kton detector}:
$\chi^2$ value of the fit as a function of \s2t13 for a 100~kton 
detector. Curves on the left (right) part of every plot
correspond to the results of the fit under the assumption of a normal (inverted)
hierarchy. Four limiting oscillation scenarios ({n.h.-L, n.h.-S, i.h.-L, i.h.-S})
are considered as ``true'' and the reference values of the supernova
parameters correspond to scenario I (top) and II (bottom). Solid
curves are computed using the information of the four supernova neutrino
detection processes and the dashed line does not take into
account NC events.} 
\label{fig:70kton}
\end{figure}

In order to be sensitive to the oscillation parameters even in the
scenario II we consider a 100~kton detector (see Figure~\ref{fig:70kton}).
In this case the sensitivity to the parameters is
almost the same for scenarios I and II.
We can appreciate that the neutral current events are crucial even
with such a big detector. They are specially important to discriminate
the large mixing angle regions if the average energies of the
neutrinos are close, as we already discussed. 

\subsection{Intermediate value of the mixing angle}
%%%%%%%%%%%%%%%%%%%%%%%%%%%%%%%%%%%%%%%%%%%%%%%%%%%

If the \th13 mixing angle is in the intermediate range (2 $\times$
10$^{-6}$ $<$ \s2t13 $<$ 3 $\times$ 10$^{-4}$), maximal sensitivity to
the angle is achieved and measurements of the value are possible in
this region, since one expects an energy dependent effect in the relevant
energy region of the supernova neutrinos (see e.g. Ref. \cite{IApaper}).

\begin{figure}[htbp]
\epsfig{file=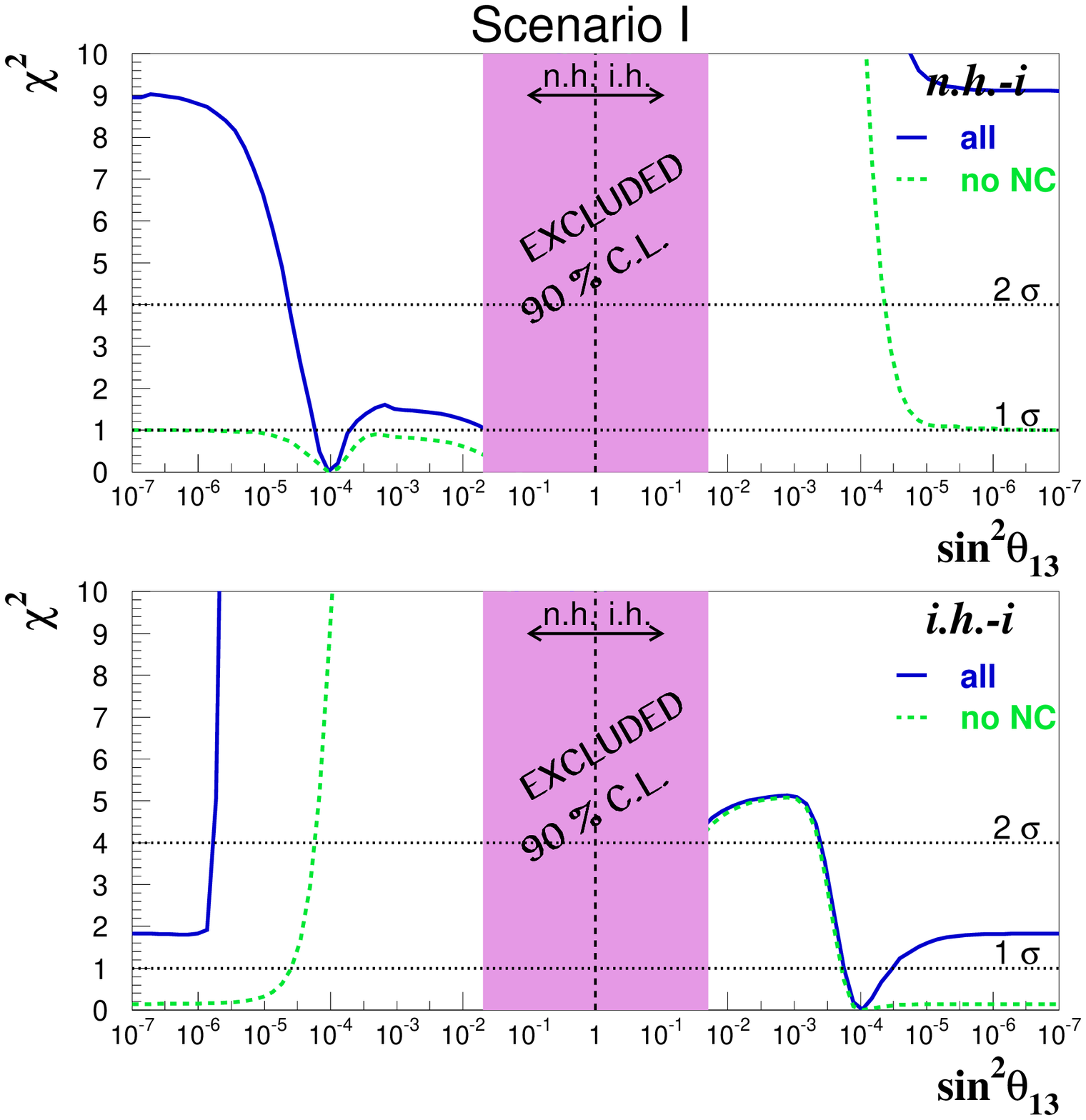,width=0.55\linewidth}
\hspace{-1cm}
\epsfig{file=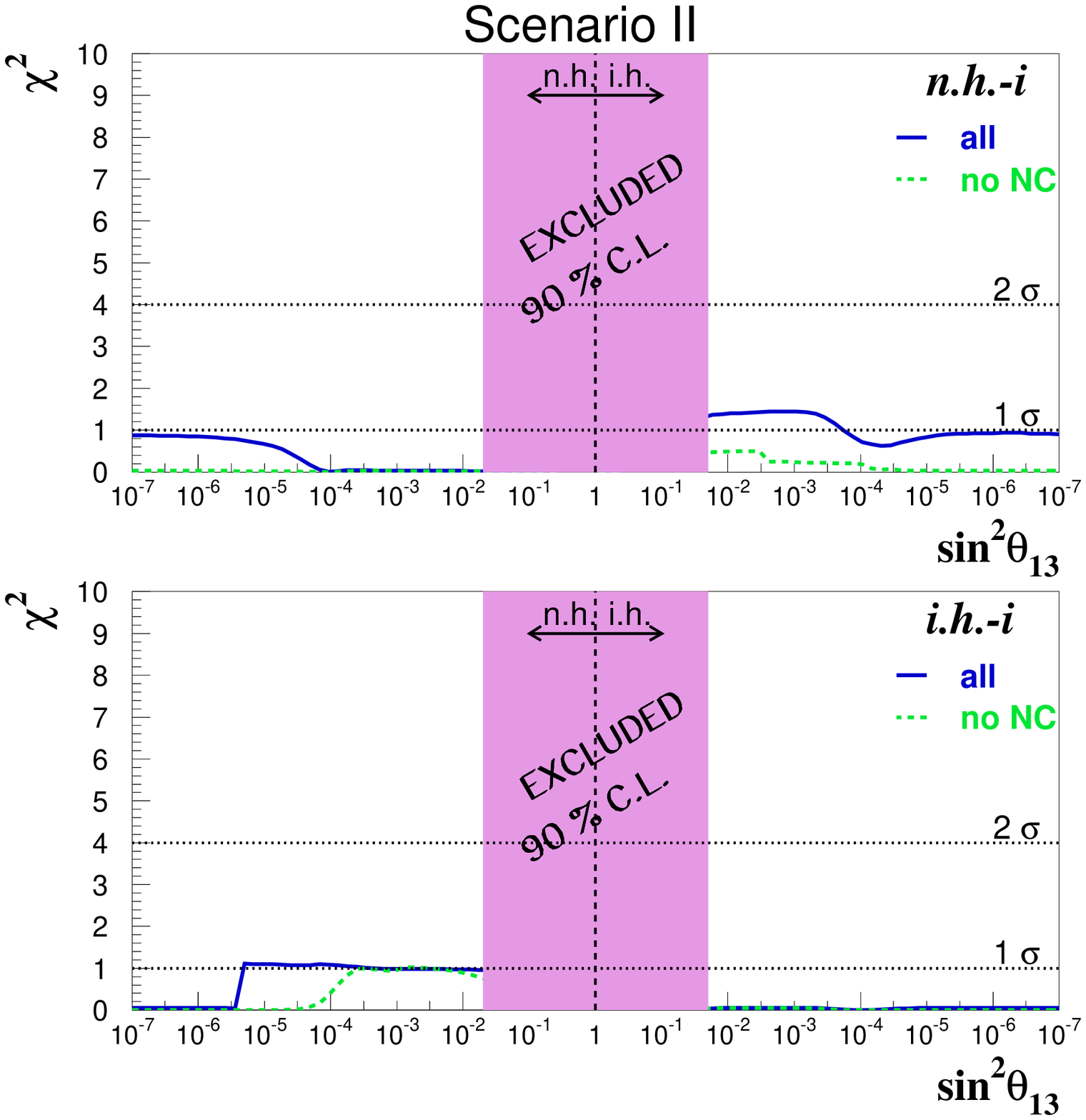,width=0.55\linewidth}
\caption{{\bf Intermediate mixing angle region (\s2t13 = 10$^{-4}$)}: 
$\chi^2$ value of the fit as a function of \s2t13 for a 3
kton detector. Curves on the left (right) part of every plot
correspond to results of the fit for assumed normal (inverted) hierarchy. 
Two true oscillation scenarios are 
considered (n.h.~and i.h.) and the true values of the supernova parameters
correspond to scenario I (left) and II (right). Solid curves are
computed using the information of the four supernova neutrino
detection processes and the dashed line does not take into
account NC events.}
\label{fig:exts2t13_inter}
\end{figure}

If we consider a mixing angle \s2t13 =
10$^{-4}$, the $\chi^2$ fit gives the results shown in
figure \ref{fig:exts2t13_inter}.
True normal (n.h.-i) and inverted (i.h.-i) hierarchies for true \s2t13 =
10$^{-4}$ and the two supernova scenarios are studied. Solid lines
correspond to the contribution of the four neutrino detection channels
while dashed lines only consider elastic and CC events.
The one- and two- sigma levels are shown by horizontal lines.

A determination of the angle is possible at 1$\sigma$ level for
scenario I with a 3 kton detector considering all neutrino
processes. We obtain the following
1$\sigma$ intervals and mass hierarchies for a 3~kton detector: 
\begin{equation}
5.8 \times 10^{-5} < \sin^2\theta_{13} < 1.9 \times 10^{-4}~~~(assumed~n.h.)
\end{equation}
\begin{equation}
3.3 \times 10^{-5} < \sin^2\theta_{13} < 1.8 \times 10^{-4}~~~(assumed~i.h.)
\end{equation}
\noindent The wrong hierarchies are excluded in both cases. 

\begin{figure}[htbp]
\epsfig{file=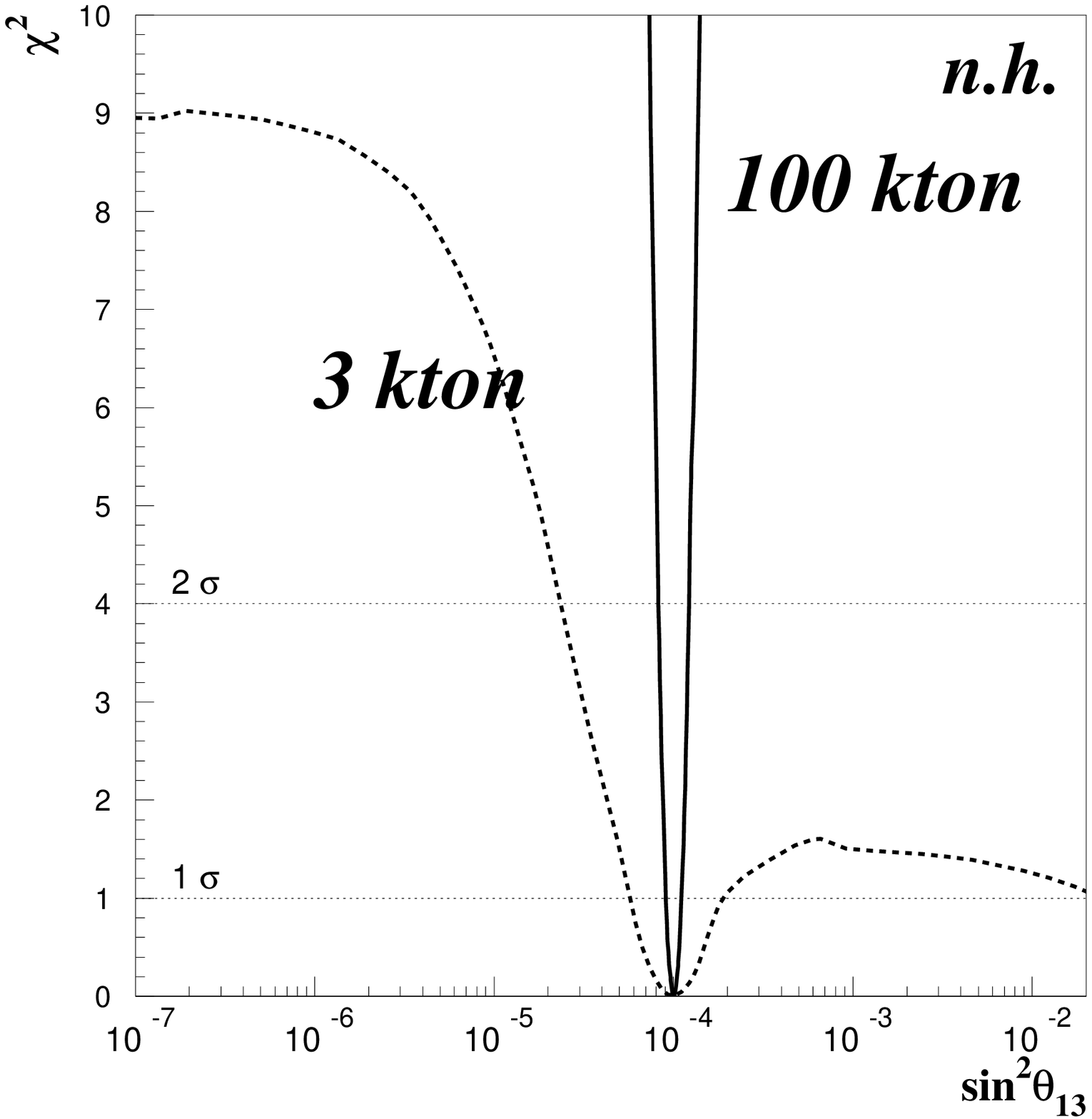,width=0.55\linewidth}
\hspace{-1cm}
\epsfig{file=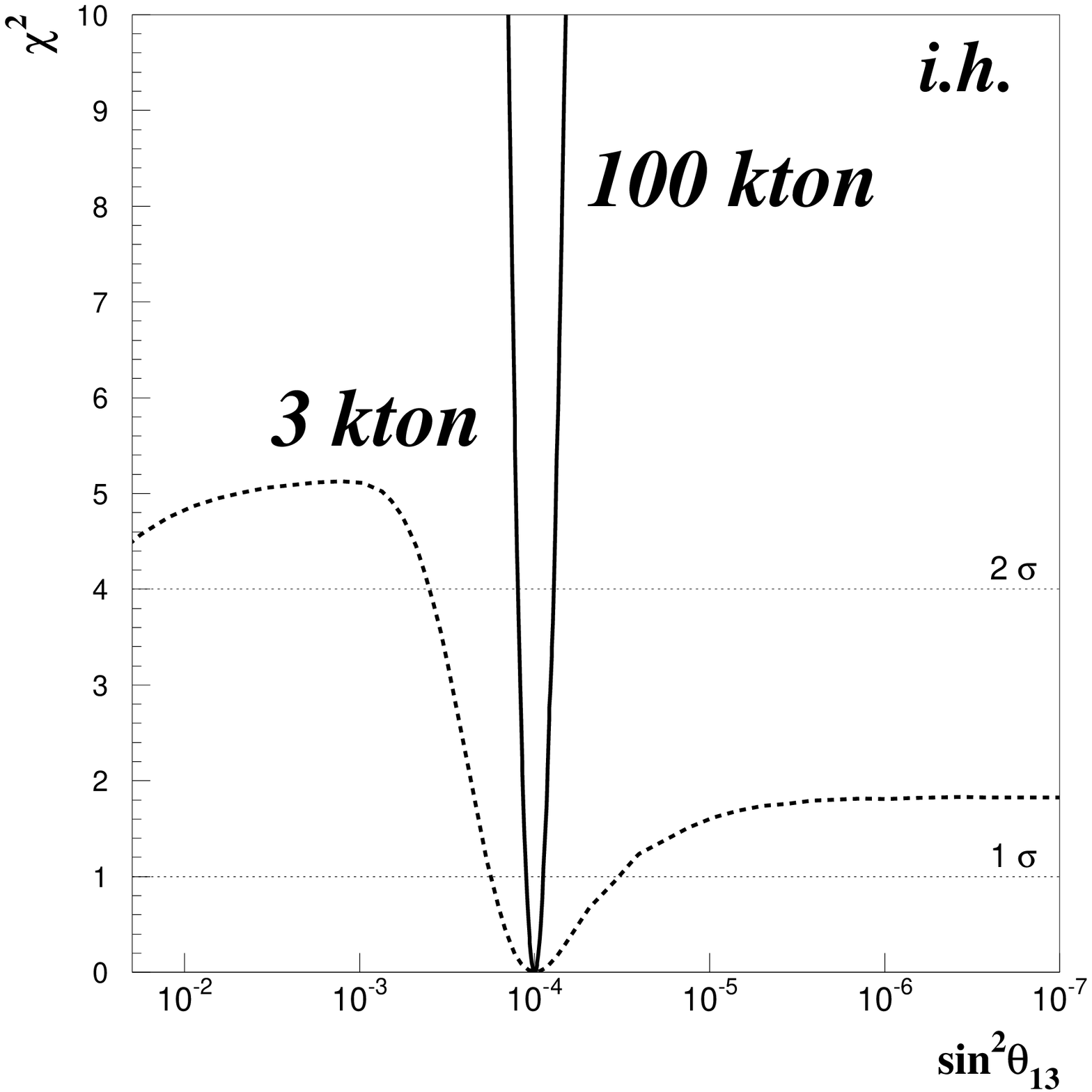,width=0.55\linewidth}
\caption{Determination of the \s2t13 parameter in the intermediate
region. We have considered as ``true'' values \s2t13 = 10$^{-4}$,
normal (left) and inverted (right) hierarchies and the supernova
parameters of scenario I. The solid curve corresponds to the results
of the $\chi^2$ fit obtained with a 100 kton detector and dashed line
shows the contribution of a 3 kton. The one- and two- sigma
levels are shown by horizontal lines.}  
\label{fig:compnh_inter}
\end{figure}

However, the statistics achieved with
such a mass are not enough to put any constraint on the \th13 angle
if the supernova parameters correspond to scenario II.
Figure~\ref{fig:compnh_inter} compares the determination of the \s2t13
parameter using a 3 and a 100 kton detector. The $\chi^2$
of the fit is plotted as a function of the \s2t13 value. 

If we consider a 100 kton detector, much more statistically accurate measurements can
be performed at 1$\sigma$ and 2$\sigma$ levels:
\begin{equation}
9.1 (8.1) \times 10^{-5} < \sin^2\theta_{13} < 1.1 (1.3) \times 10^{-4}
\hspace{0.5cm} 1\sigma (2\sigma)~~~(assumed~n.h.)
\end{equation}
\begin{equation}
8.9 (7.4) \times 10^{-5} < \sin^2\theta_{13} < 1.1 (1.3) \times 10^{-4}
\hspace{0.5cm} 1\sigma (2\sigma)~~~(assumed~i.h.)
\end{equation}
\noindent The wrong hierarchies are excluded in both cases. 

\begin{figure}[htbp]
\epsfig{file=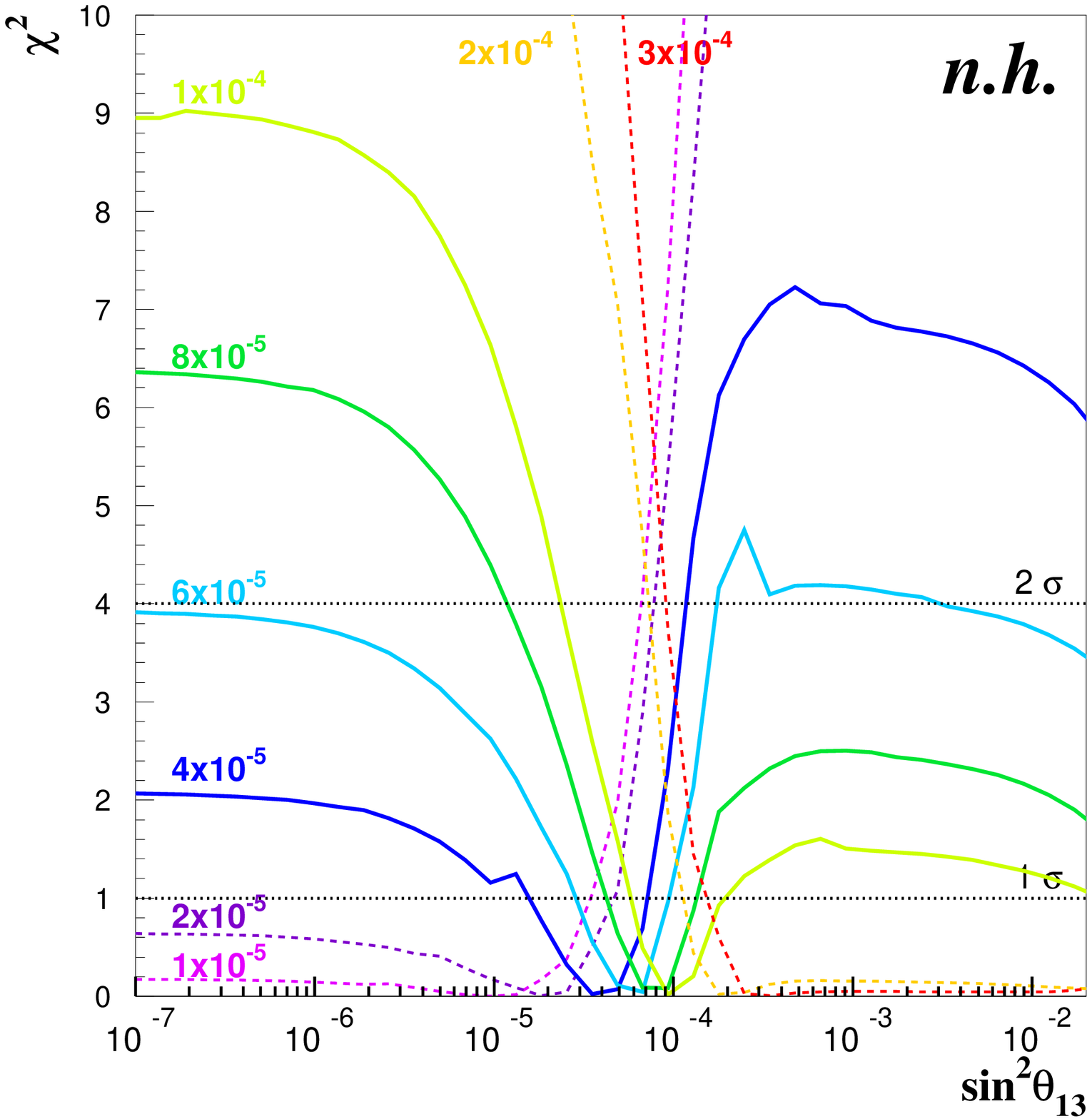,width=0.55\linewidth}
\hspace{-1cm}
\epsfig{file=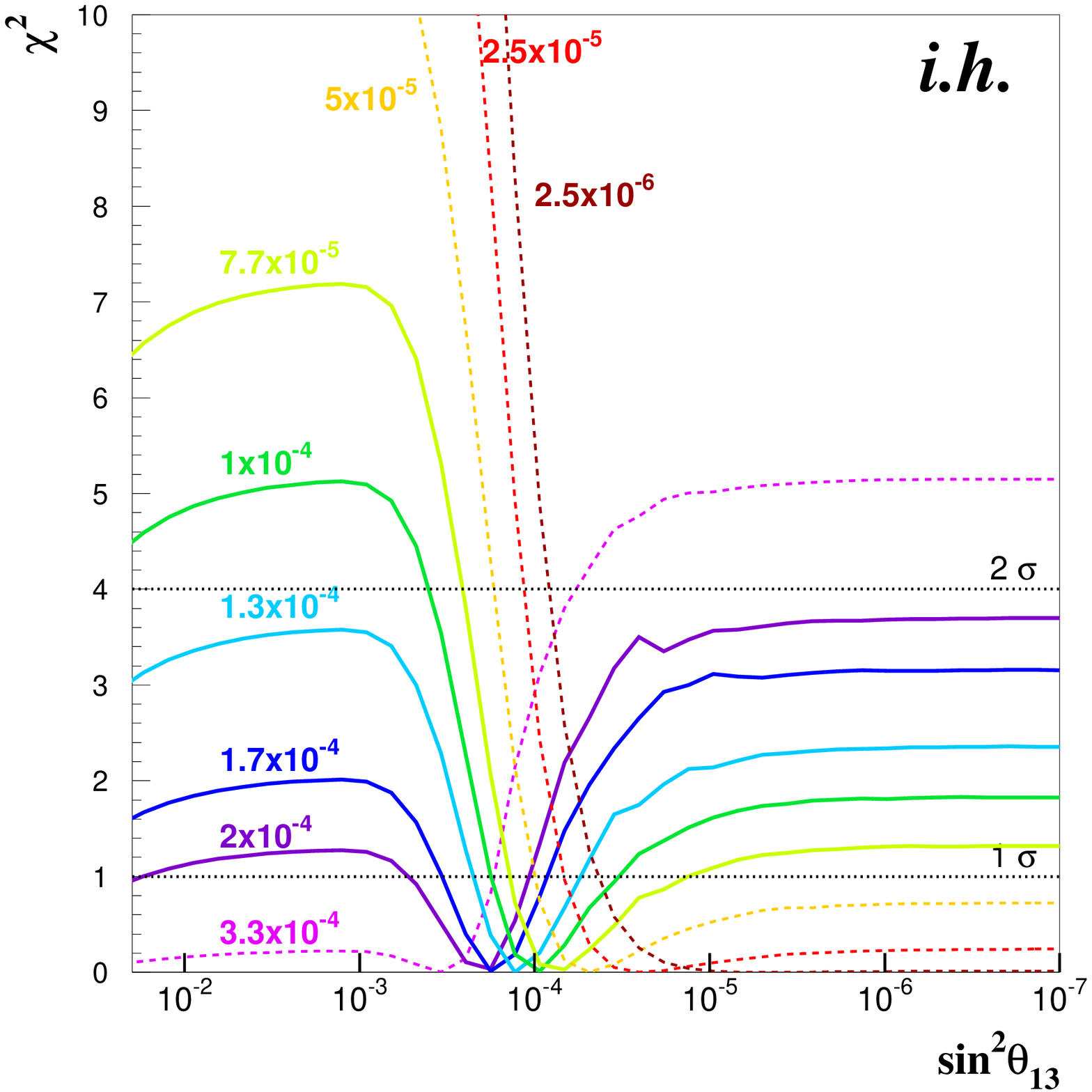,width=0.55\linewidth}
\caption{Determination of the \s2t13 parameter in the intermediate
region with a 3 kton detector. We have considered different
``true'' values for \s2t13 and normal and inverted hierarchies. The
reference supernova parameters correspond to scenario I. The
one- and two- sigma levels are shown by horizontal lines.} 
\label{fig:s2t13_inter_all}
\end{figure}

The precision on the measurement of the mixing angle will depend on
the exact value of \th13. Considering different reference values in
the intermediate region, 
Figure \ref{fig:s2t13_inter_all} shows the superposition of different determinations of the
\s2t13 parameter, for the cases of true normal and inverted
hierarchies normalized to a 3~kton detector. The angle can be measured at 1$\sigma$ in both cases.

The estimated limits and ranges of \s2t13 are summarized in table
\ref{tab:interthlimit} for true normal and inverted hierarchies. If
the \s2t13 angle is between $\approx$
(4$\times$10$^{-5}$ -- 1$\times$10$^{-4}$) (n.h.) or $\approx$
(7$\times$10$^{-5}$ -- 3$\times$10$^{-4}$) (i.h.), it could be
constrained in a certain range. Otherwise, upper or lower limits on
its value can be set.

\begin{table}[htbp]
\small{
\centering
\begin{tabular}{|c|c|}
\hline
\multicolumn{2}{|c|}{3 kton detector}\\
\hline
{\bf ``True'' value} & {\bf Limit or range of \s2t13} \\
{\bf \s2t13}     & {\bf at 1$\sigma$ level} \\ \hline \hline
{\bf True n.h.}       &  n.h. assumed \\ \hline
1. $\times$ 10$^{-5}$ & \s2t13 $<$ 3.4 $\times$ 10$^{-5}$ \\ \hline
2. $\times$ 10$^{-5}$ & \s2t13 $<$ 4.6 $\times$ 10$^{-5}$ \\ \hline
4. $\times$ 10$^{-5}$ & 1.6 $\times$ 10$^{-5}$ $<$ \s2t13 $<$ 7.2
$\times$ 10$^{-5}$ \\ \hline
6. $\times$ 10$^{-5}$ & 2.8 $\times$ 10$^{-5}$ $<$ \s2t13 $<$ 9.4
$\times$ 10$^{-5}$ \\ \hline 
8. $\times$ 10$^{-5}$ & 4.2 $\times$ 10$^{-5}$ $<$ \s2t13 $<$ 1.3
$\times$ 10$^{-4}$ \\ \hline
1. $\times$ 10$^{-4}$ & 5.8 $\times$ 10$^{-5}$ $<$ \s2t13 $<$ 1.9
$\times$ 10$^{-4}$ \\ \hline
2. $\times$ 10$^{-4}$ & \s2t13 $>$ 1.1 $\times$ 10$^{-4}$ \\ \hline
3. $\times$ 10$^{-4}$ & \s2t13 $>$ 1.5 $\times$ 10$^{-4}$ \\ \hline \hline
{\bf True i.h.} & i.h. assumed  \\ \hline
2.5 $\times$ 10$^{-6}$ & \s2t13 $<$ 4.0 $\times$ 10$^{-5}$ \\ \hline
2.5 $\times$ 10$^{-5}$ & \s2t13 $<$ 6.8 $\times$ 10$^{-5}$ \\ \hline
5.0 $\times$ 10$^{-5}$ & \s2t13 $<$ 9.9 $\times$ 10$^{-5}$ \\ \hline
7.7 $\times$ 10$^{-5}$ & 1.3 $\times$ 10$^{-5}$ $<$ \s2t13 $<$ 1.4
$\times$ 10$^{-4}$ \\ \hline
1.0 $\times$ 10$^{-4}$ & 3.3 $\times$ 10$^{-5}$ $<$ \s2t13 $<$ 1.8
$\times$ 10$^{-4}$ \\ \hline
1.3 $\times$ 10$^{-4}$ & 5.4 $\times$ 10$^{-5}$ $<$ \s2t13 $<$ 2.2
$\times$ 10$^{-4}$ \\ \hline
1.7 $\times$ 10$^{-4}$ & 8.5 $\times$ 10$^{-5}$ $<$ \s2t13 $<$ 3.4
$\times$ 10$^{-4}$ \\ \hline
2.0 $\times$ 10$^{-4}$ & 1.1 $\times$ 10$^{-4}$ $<$ \s2t13 $<$ 5.3
$\times$ 10$^{-4}$ \\ \hline
3.3 $\times$ 10$^{-4}$ & \s2t13 $>$ 1.7 $\times$ 10$^{-4}$ \\ \hline
\end{tabular}
\caption{Estimated limit on the \th13 mixing angle at 1$\sigma$ level
for different ``true'' values of the angle and mass hierarchies in the
intermediate range.}
\label{tab:interthlimit}
}
\end{table}

\newpage
%%%%%%%%%%%%%%%%%%%%%%%%%%%%%%%%%%%%%%%%%%%%%%%%%%%%%%%%%%%%%%%
\section{Study of the supernova parameters assuming that the
oscillation parameters are known}  
\label{sec:second}
%%%%%%%%%%%%%%%%%%%%%%%%%%%%%%%%%%%%%%%%%%%%%%%%%%%%%%%%%%%%%%%

In this section we study the information about the astrophysical
parameters that can be extracted from the detection of supernova
neutrinos assuming that the oscillation
parameters \s2t13 and the mass hierarchy have been determined from
terrestrial experiments, essentially from long-baseline experiments.

Considering the range of sensitivity of \th13 of future accelerator experiments
(see e.g.~\cite{Apollonio:2002en}), we can study two cases: 

\begin{enumerate}
\item {\bf If \th13 angle is measured by long-baseline experiments}:
If the true value of the \th13 angle is large (\s2t13 $\gtrsim$ 10$^{-3}$)
then future superbeams should be able to measure it (see e.g. \cite{Apollonio:2002en}). We assume that
the angle will be determined with a precision of 10\%. Therefore, we
study how well we can constrain the supernova parameters under the external constraint \s2t13 =
10$^{-3}$ $\pm$ 10$^{-4}$, considering that the mass hierarchy is also
known.

\item {\bf If an upper bound on \th13 angle is set by long-baseline experiments}:
If the angle is very small, then future long-baseline
experiments might not have the possibility to measure this angle and will only place an
upper limit on the value. As a second example, we consider
that the limit \s2t13 $<$ 10$^{-4}$ has been set and we
investigate the consequences of this limit into the determination of
the supernova parameters. 
\end{enumerate}

\subsection{If \th13 angle is measured by long-baseline experiments} 
%%%%%%%%%%%%%%%%%%%%%%%%%%%%%%%%%%%%%%%%%%%%%%%%%%%%%%%%%%%%%%%%

We assume that the value of the true \th13 mixing angle has been determined
by long-baseline experiments to be \s2t13 =
10$^{-3}$ $\pm$ 10$^{-4}$ and we study the cases of true normal and
inverted hierarchies. We will compare the precisions in the determinations
of the supernova parameters with a 3 and 100~kton detectors.
In the fits, we let free the supernova parameters (\EB, \avenue, \aveanue, \avenux,
\lelx) and we perform a $\chi^2$ minimization in order to obtain the allowed
regions in different 1D parameter regions and 2D parameter planes.
The expected accuracies
at 90\% C.L. to be achieved with a 3 and 100 kton detectors 
are summarized in Table~\ref{tab:accurasnpar}.

\begin{table}[htbp]
\centering
\begin{tabular}{|c|c|c||c|c|c|c|c|} 
\hline
\multicolumn{8}{|c|}{With constraint  \s2t13 = 10$^{-3}$ $\pm$
10$^{-4}$ from terrestrial experiment} \\
\hline
Detector mass & True hierarchy & SN scen. & $\frac{\Delta\favenue}{\favenue}$ &
$\frac{\Delta\faveanue}{\faveanue}$ &
$\frac{\Delta\favenux}{\favenux}$ & $\frac{\Delta\fEB}{\fEB}$ &
$\frac{\Delta(\flelx)}{(\flelx)}$ \\ \hline \hline 
3 kton & n.h. & I & -- & -- & $\sim$ 3\% & $\sim 23$\% & -- \\
3 kton & i.h. & I & $\sim$ 46\% & -- & $\sim$ 7\% & $\sim$ 17\% & $\sim 60$\% \\
\hline
3 kton & n.h. & II & -- & $\sim 26$\% & $\sim$ 4\% & $\sim 28$\% & -- \\
3 kton & i.h. & II & -- & $\sim 36$\% & $\sim$ 19\% & $\sim$ 20\% &
-- \\ \hline\hline
100 kton & n.h. & I & $\sim$ 14\% & $\sim$ 4\% & $<$ 1\% & $\sim$
2\% & $\sim$ 11\% \\
100 kton & i.h. & I & $\sim$ 5\% & $\sim$ 9\% & $\sim$ 1\% & $\sim$ 2\% &
$\sim$ 9\% \\ \hline 
100 kton & n.h. & II & $\sim$ 9\% & $\sim$ 3\% & $<$ 1\% & $\sim$
4\% & $\sim$ 12\% \\
100 kton & i.h. & II & $\sim$ 6\% & $\sim$ 6\% & $\sim$ 1\% & $\sim$ 2\% &
$\sim$ 32\% \\
\hline
\end{tabular}
\caption{Expected accuracies at 90\% C.L. in the determination of the
supernova parameters using the neutrinos measured with a 3 kton and a
100 kton detector. We have assumed that the mass hierarchy is
known and the \th13 mixing angle has been measured by long-baseline
experiments, being equal to \s2t13 = 10$^{-3}$ $\pm$
10$^{-4}$. Supernova scenarios I and II are tested.}    
\label{tab:accurasnpar}
\end{table}

\begin{figure}[htbp]
\begin{center}
\fbox{\LARGE \sf \s2t13 = 10$^{-3}$ $\pm$ 10$^{-4}$ and n.h.}
\end{center}
\begin{center}
{\bf \Large 3 kton LAr}
\end{center}
\vspace{-0.7cm}
%\epsfig{file=EPS/2dfitcase1p_2bis.eps,width=0.55\linewidth}
%\hspace{-1cm}
%\epsfig{file=EPS/2dfitcase1p_3.eps,width=0.55\linewidth}
%\vspace{-1cm}
\begin{center}
\epsfig{file=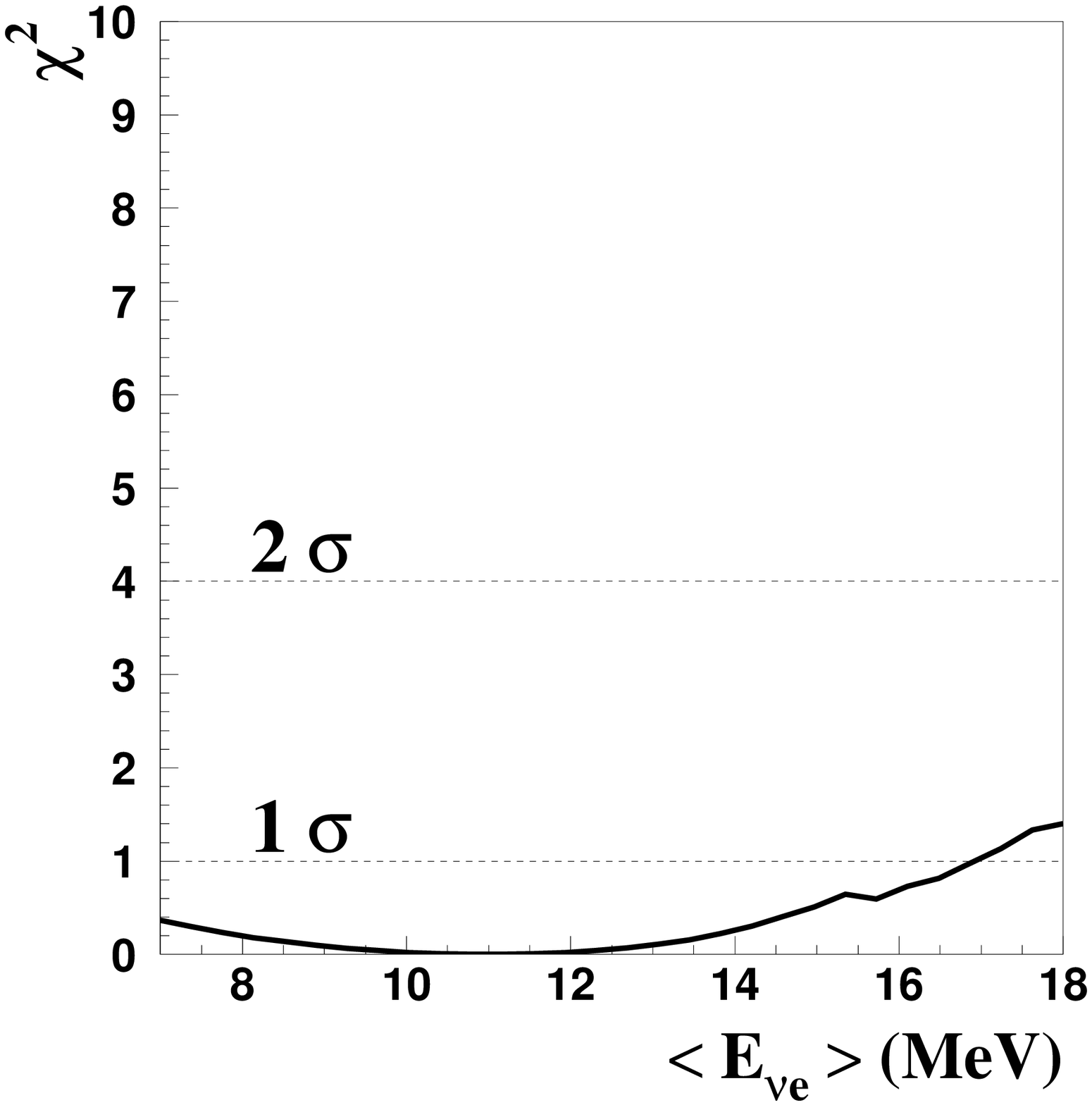,width=0.3\linewidth} \hspace{-0.5cm}
\epsfig{file=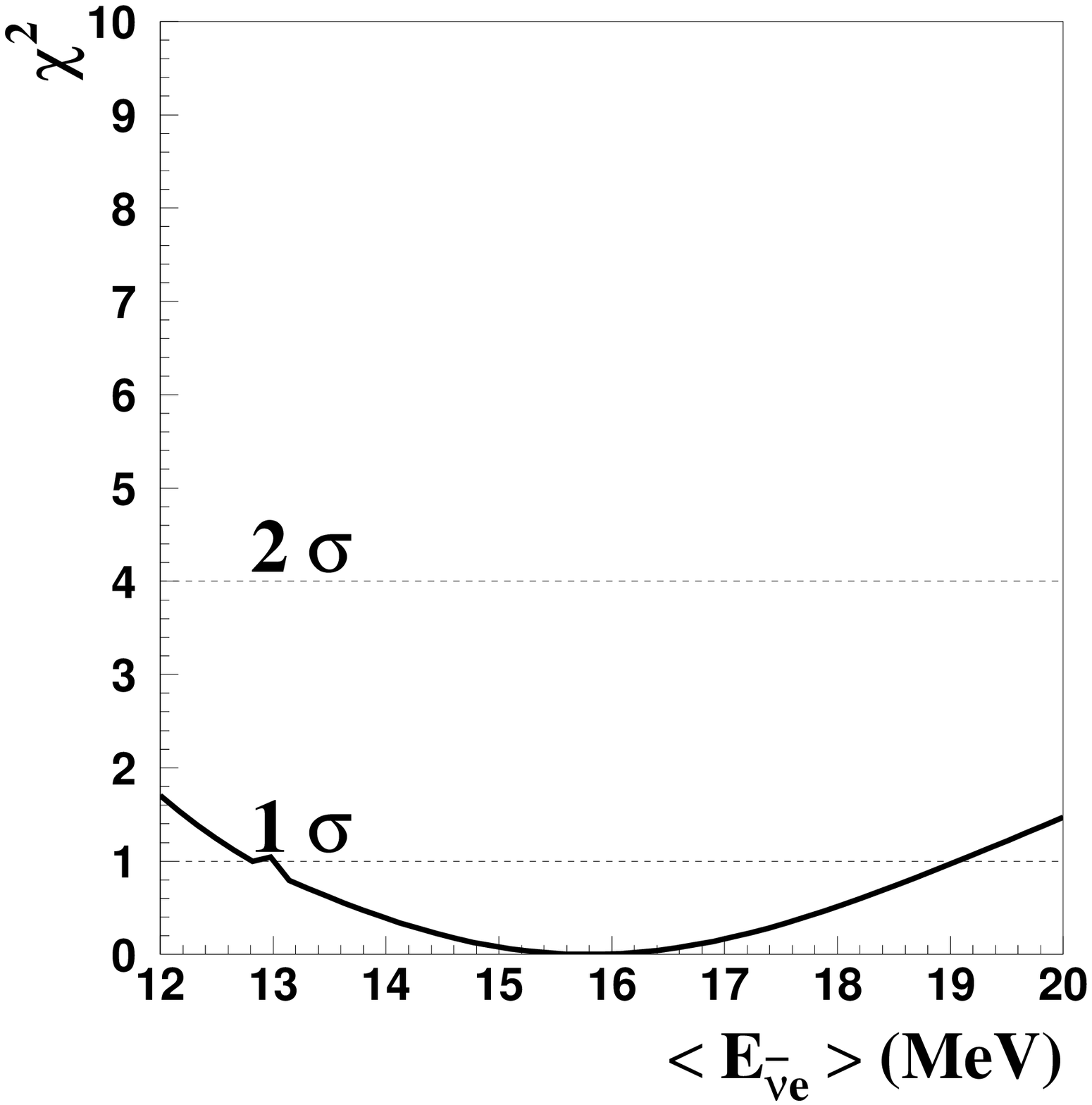,width=0.3\linewidth} \hspace{-0.5cm}
\epsfig{file=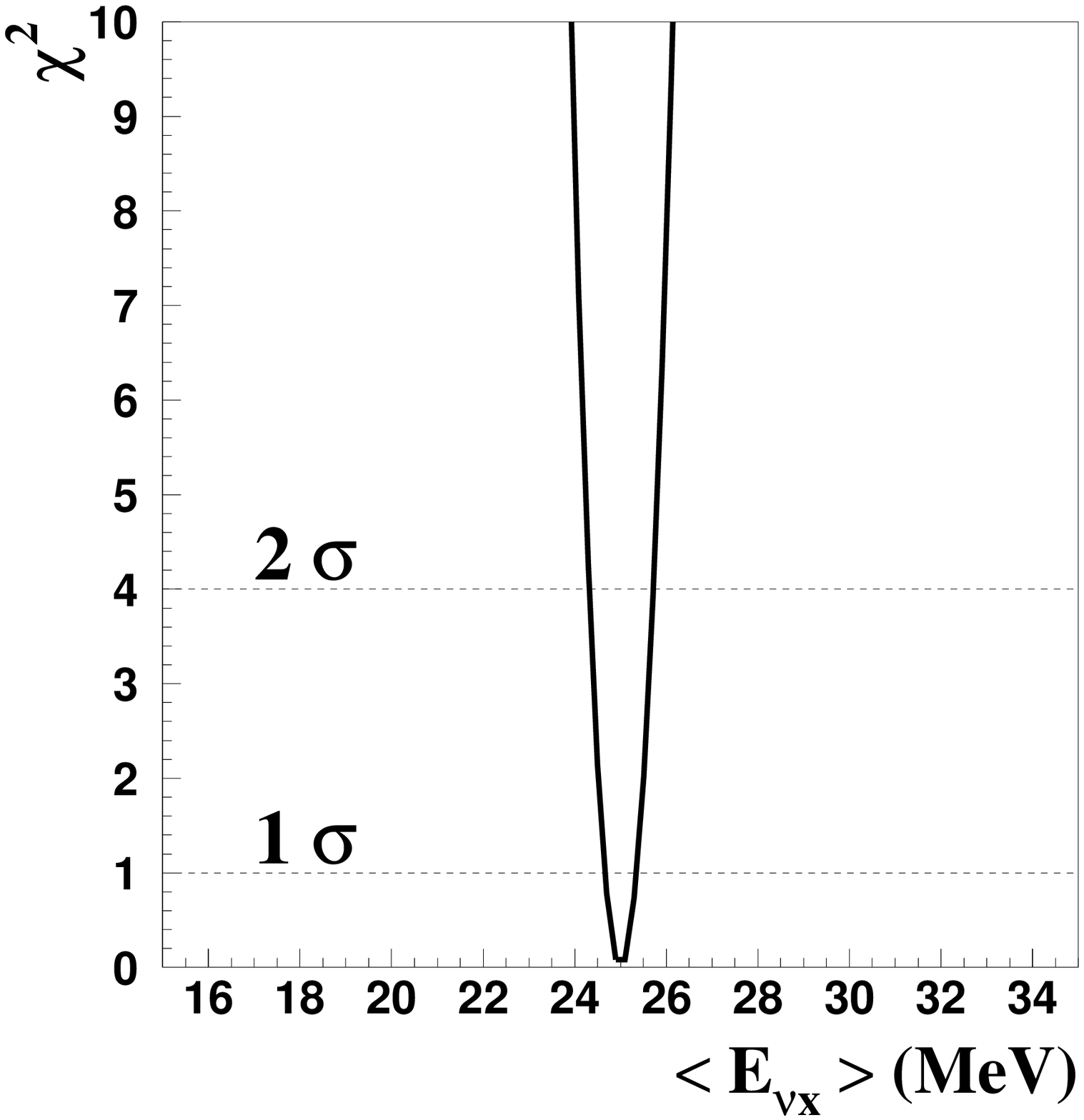,width=0.3\linewidth}
\end{center}
\vspace{-1cm}
\begin{center}
\epsfig{file=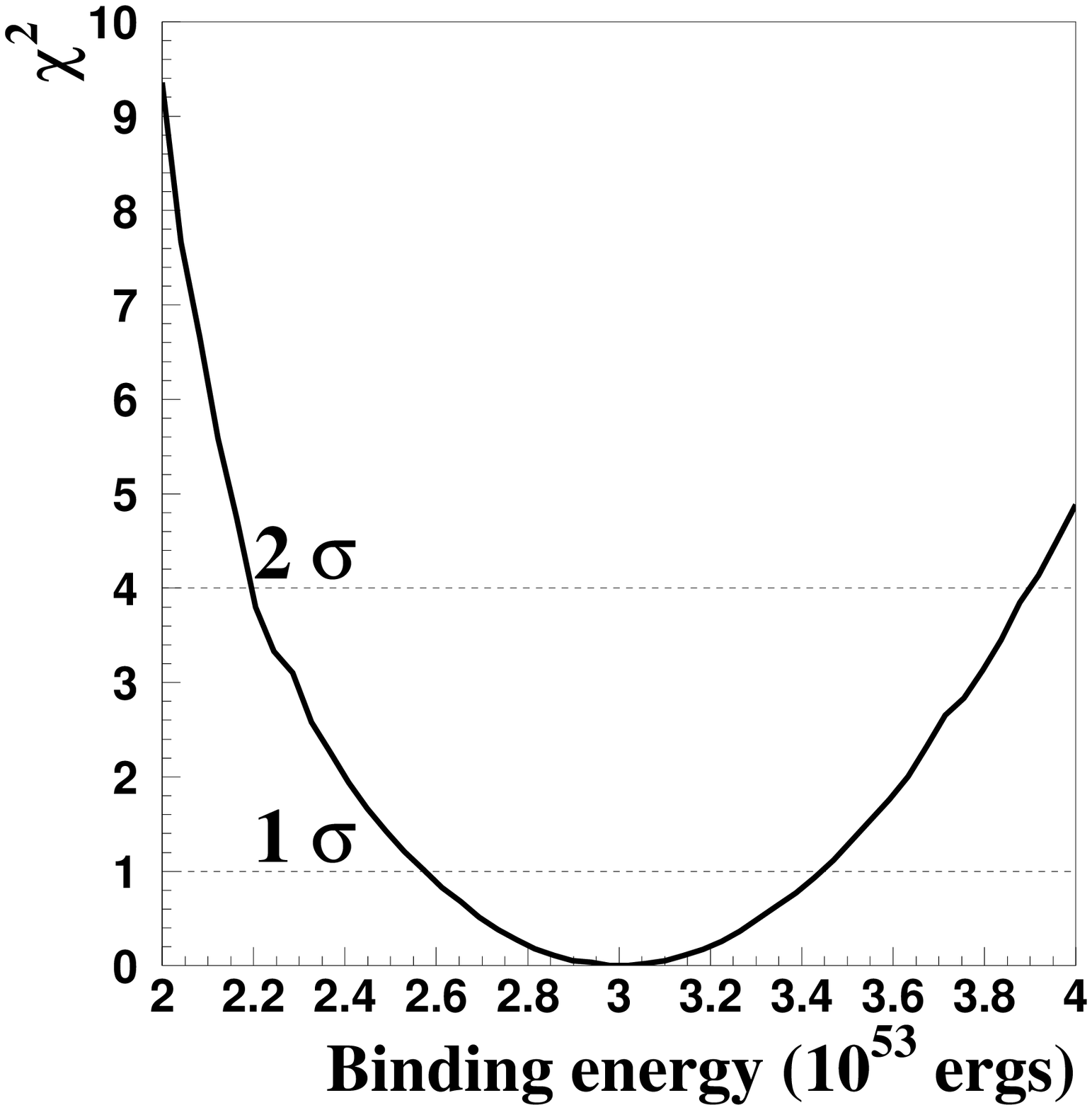,width=0.3\linewidth} \hspace{-0.5cm}
\epsfig{file=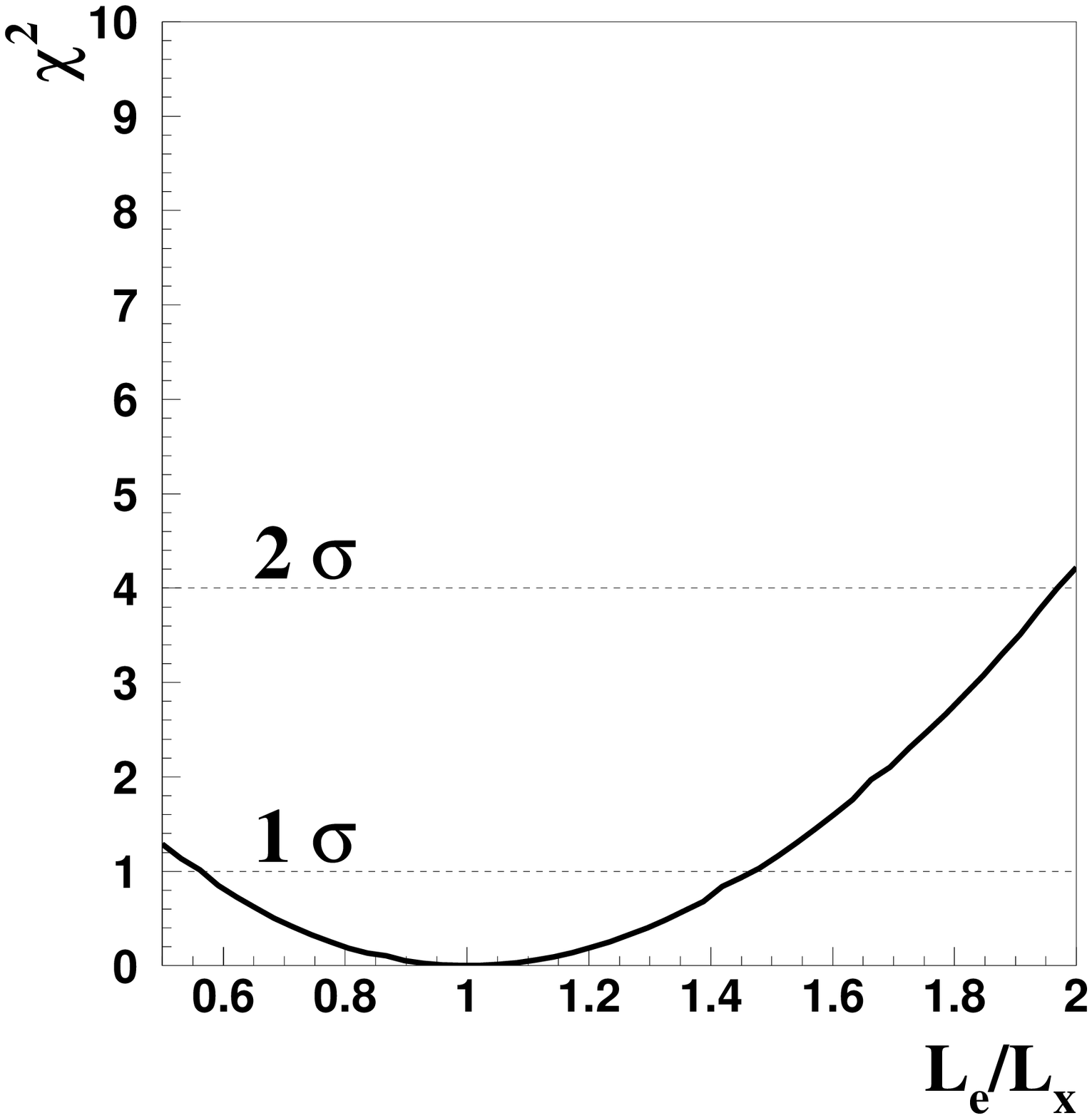,width=0.3\linewidth}
\end{center}

\begin{center}
\Large \bf 100 kton LAr
\end{center}
\vspace{-0.7cm}
\epsfig{file=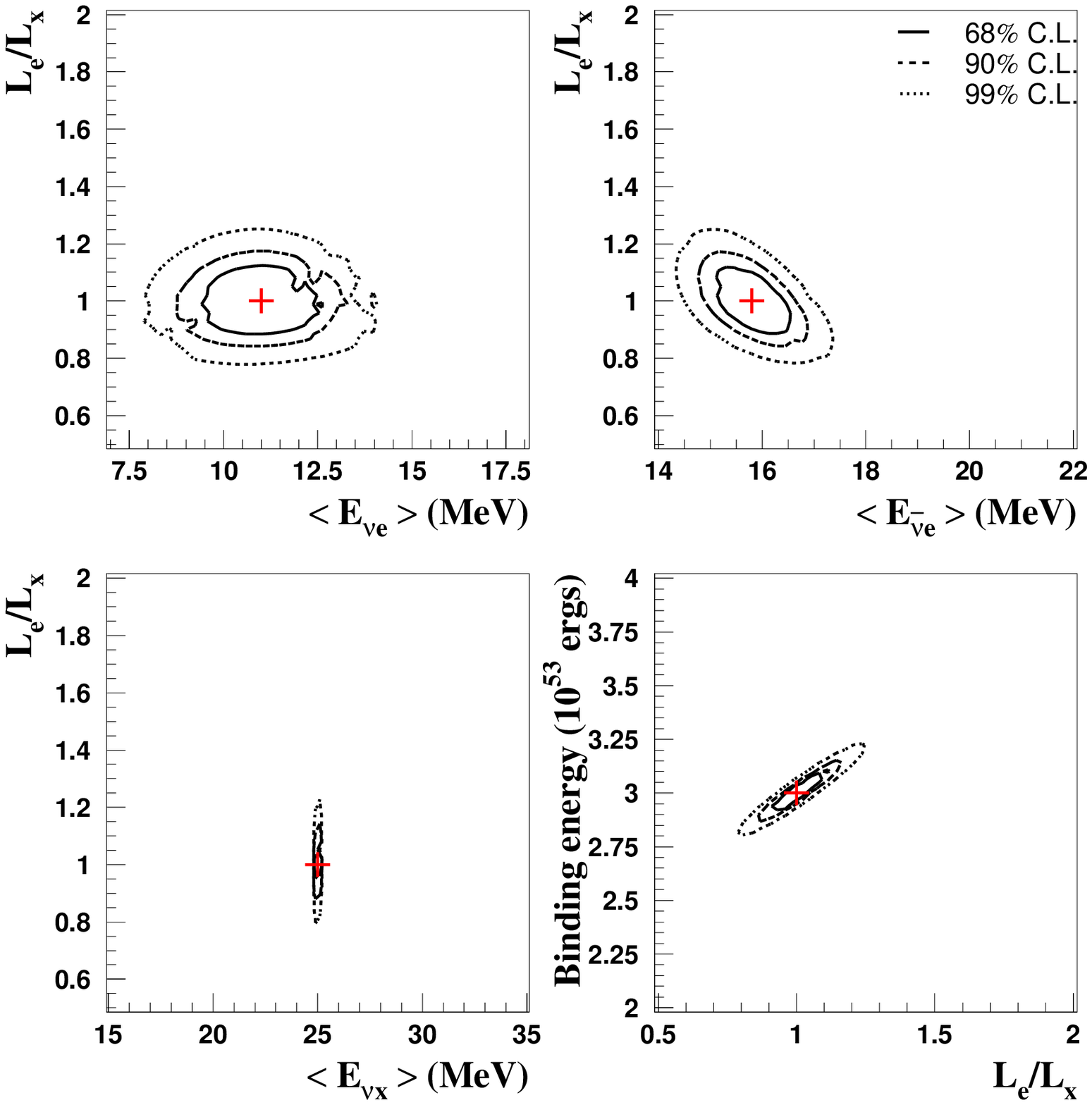,width=0.55\linewidth}
\hspace{-1cm}
\epsfig{file=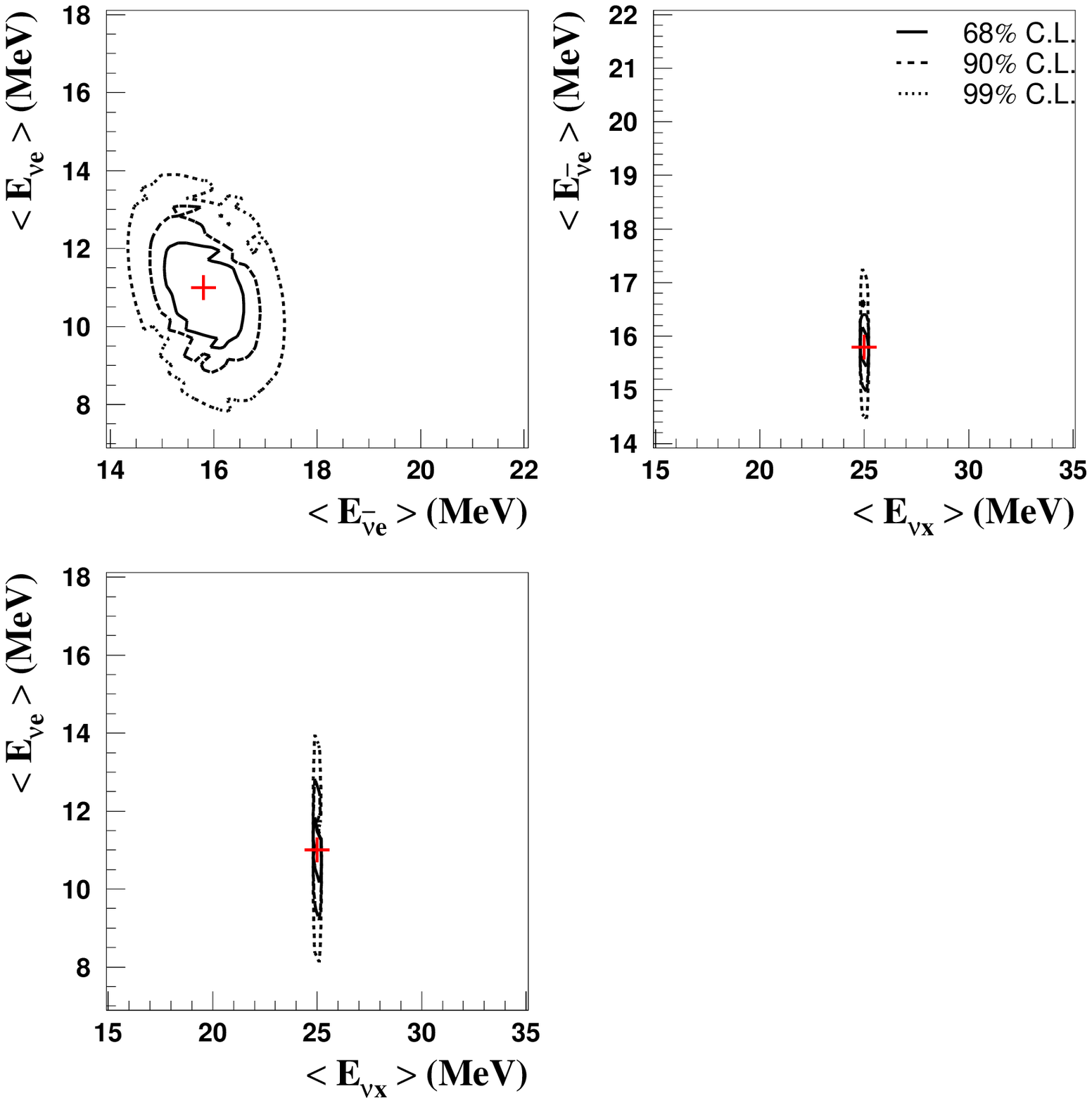,width=0.55\linewidth}
\caption{(Top) $\chi^2$ value of the fit as a function of the
supernova parameters for a 3 kton detector, assuming that the \th13
mixing angle has been measured with a precision of 10\% (\s2t13 =
10$^{-3}$ $\pm$ 10$^{-4}$) and the mass hierarchy is normal ($\dm31$
$>$ 0). (Bottom) 68\%, 90\% and 99\% C.L. allowed regions for the
supernova parameters with a 100 kton detector. Crosses indicate the
value of the parameters for the best fits.}  
\label{fig:2dfitcase1}
\end{figure}

\begin{figure}[htbp]
\begin{center}
\fbox{\LARGE \sf \s2t13 = 10$^{-3}$ $\pm$ 10$^{-4}$ and i.h.}
\end{center}
\begin{center}
\Large \bf 3 kton LAr
\end{center}
\vspace{-0.7cm}
%\epsfig{file=EPS/2dfitcase1ihp_2bis.eps,width=0.55\linewidth}
%\hspace{-1cm}
%\epsfig{file=EPS/2dfitcase1ihp_3.eps,width=0.55\linewidth}
%\vspace{-1cm}
\begin{center}
\epsfig{file=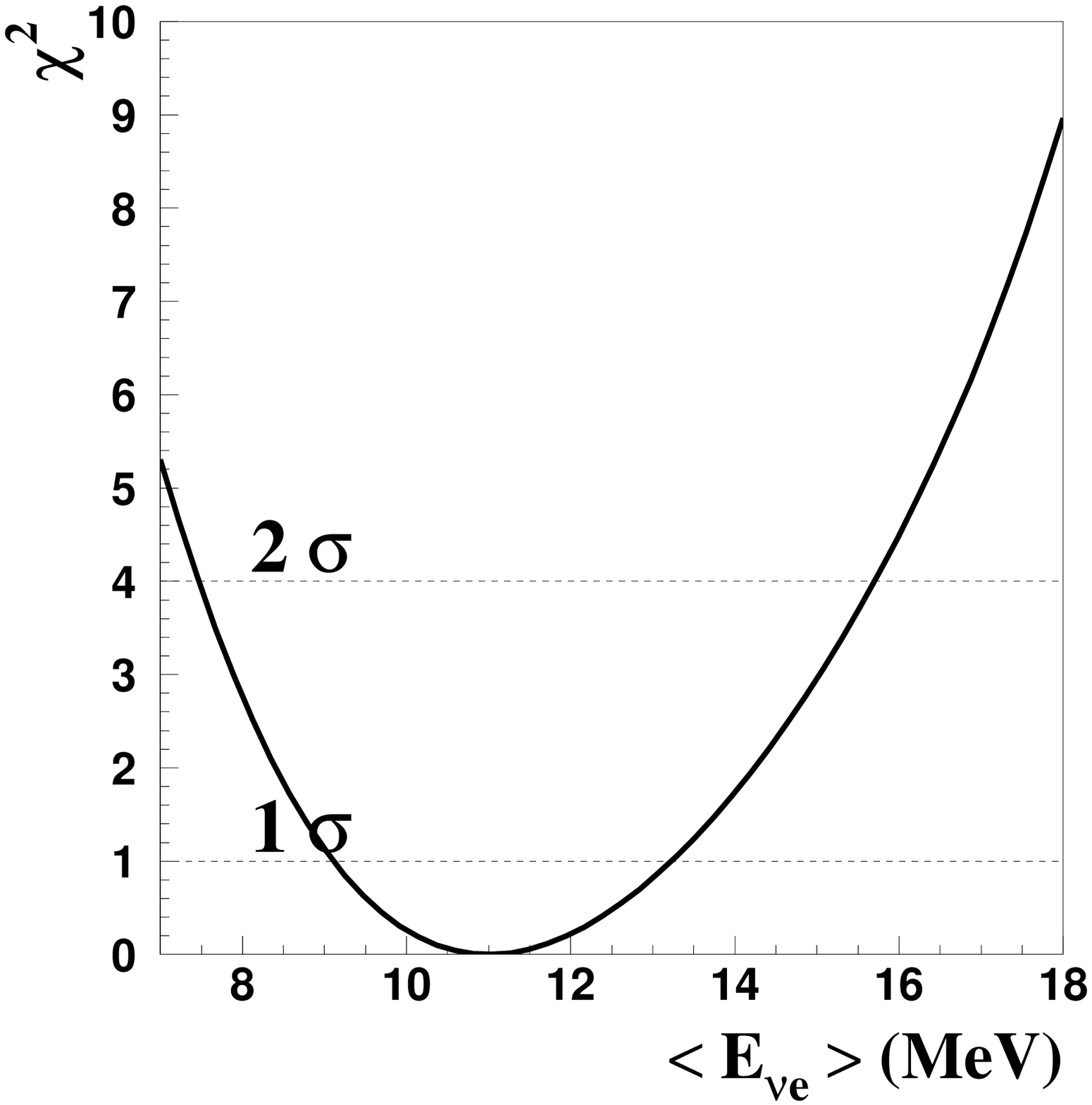,width=0.3\linewidth} \hspace{-0.5cm}
\epsfig{file=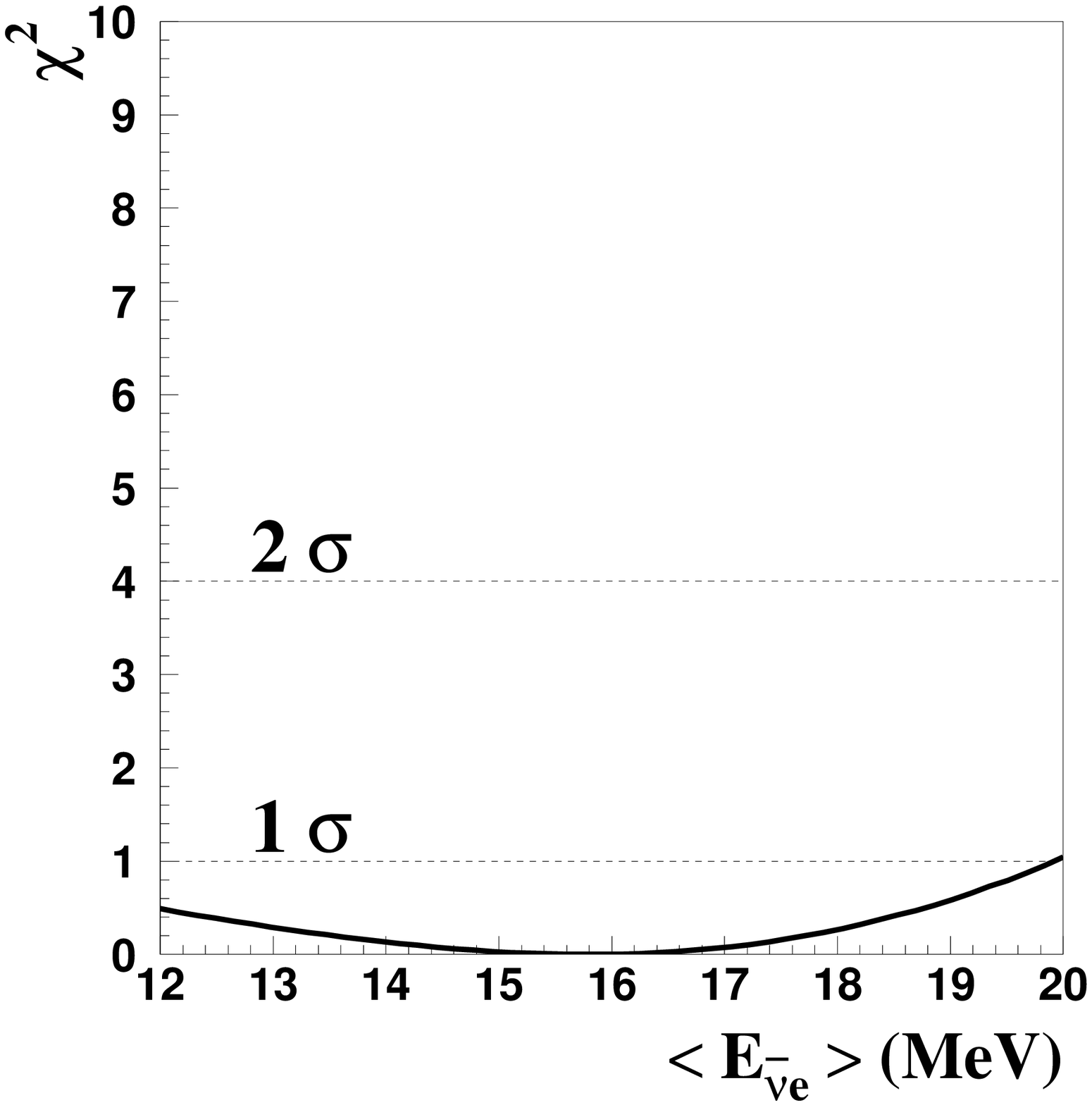,width=0.3\linewidth} \hspace{-0.5cm}
\epsfig{file=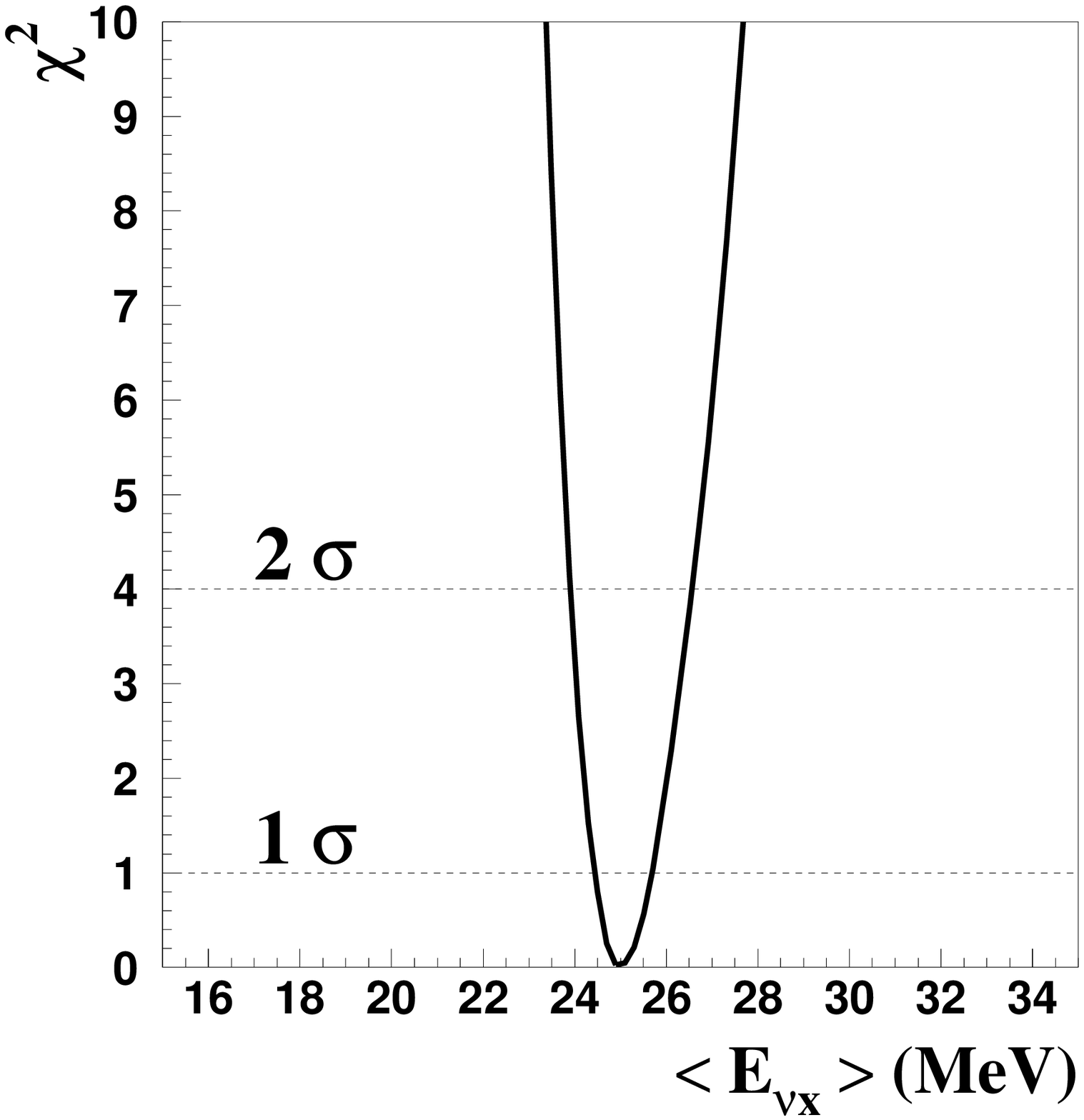,width=0.3\linewidth}
\end{center}
\vspace{-1cm}
\begin{center}
\epsfig{file=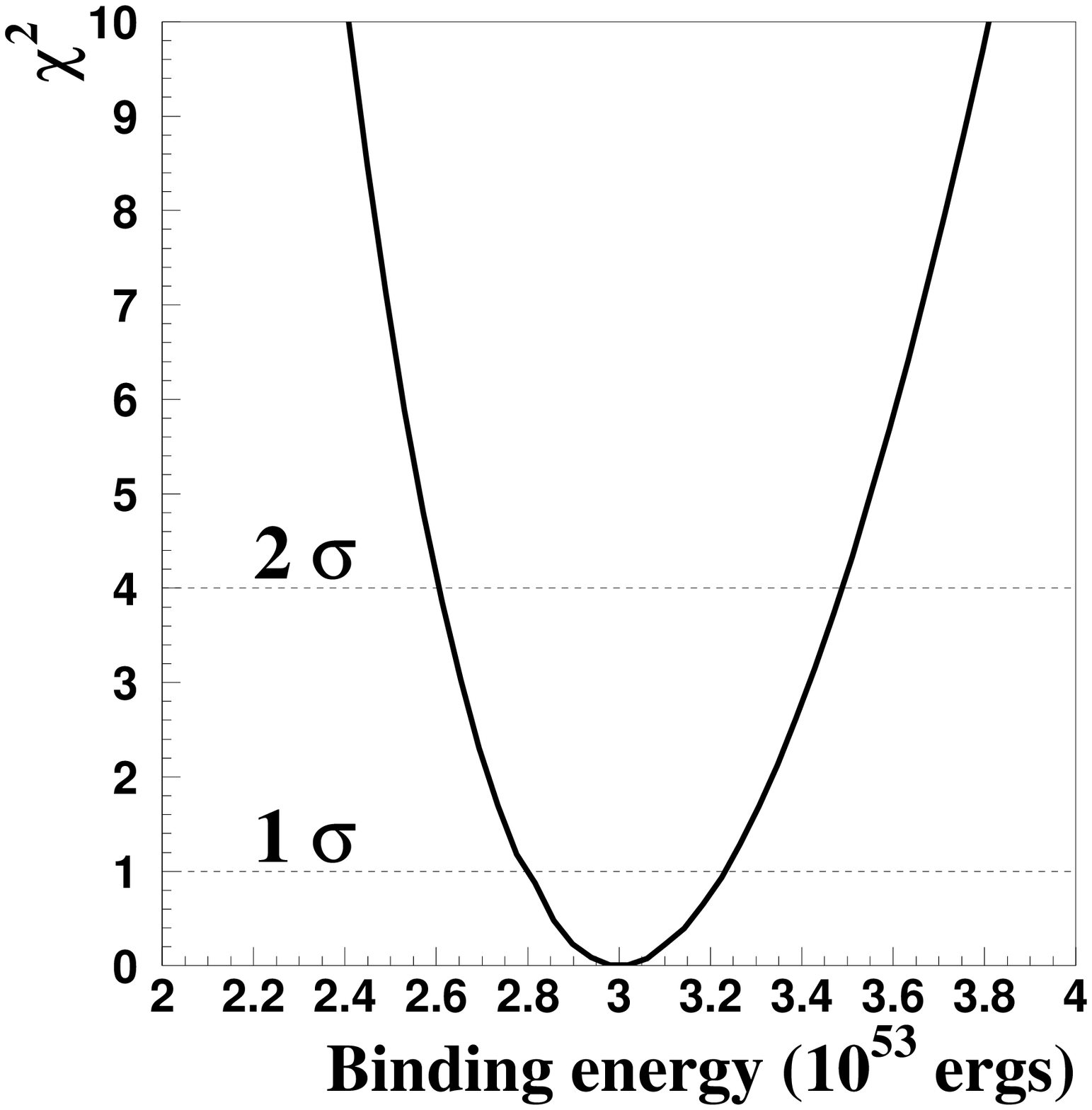,width=0.3\linewidth} \hspace{-0.5cm}
\epsfig{file=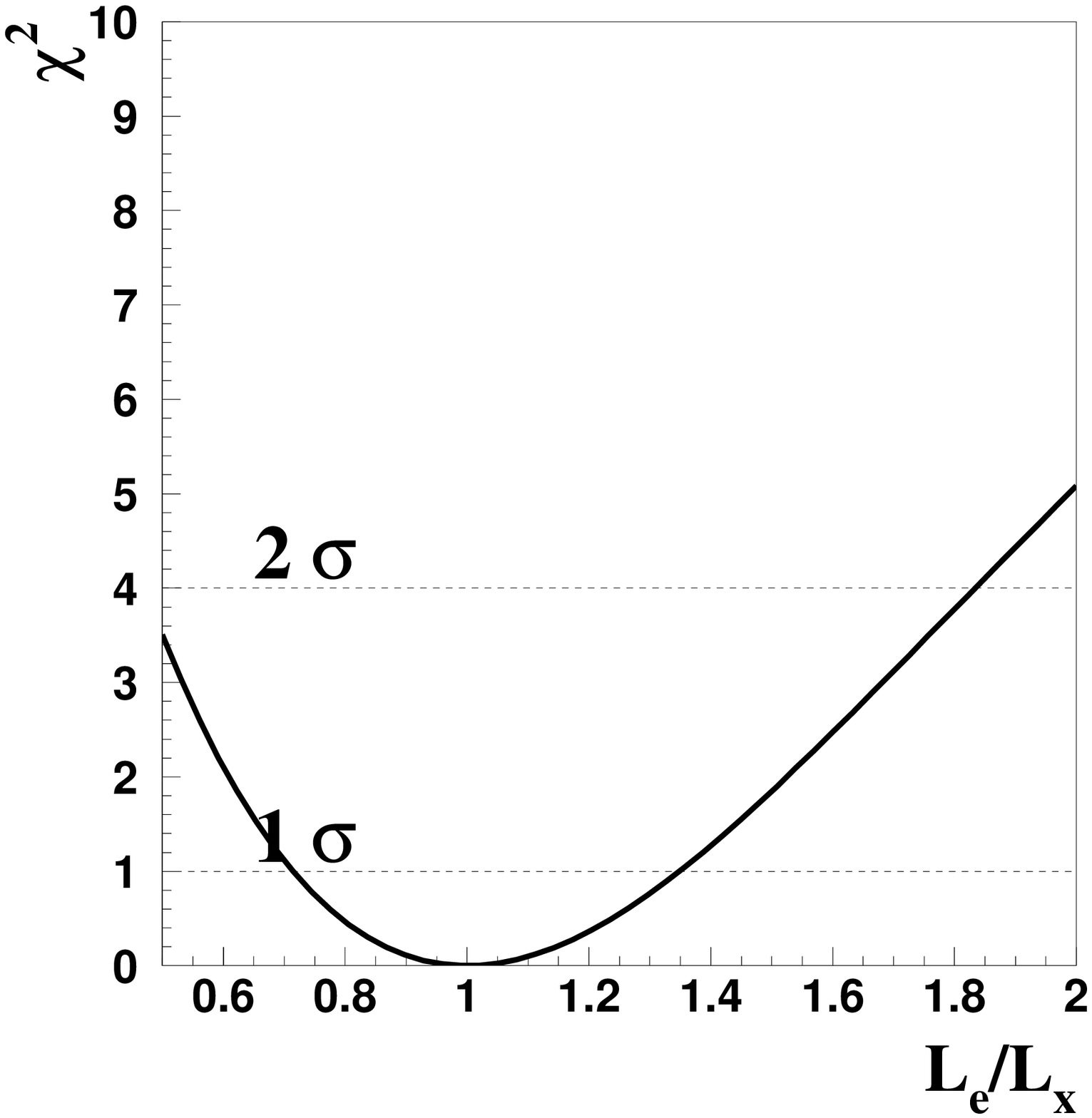,width=0.3\linewidth}
\end{center}

\begin{center}
\Large \bf 100 kton LAr
\end{center}
\vspace{-0.7cm}
\epsfig{file=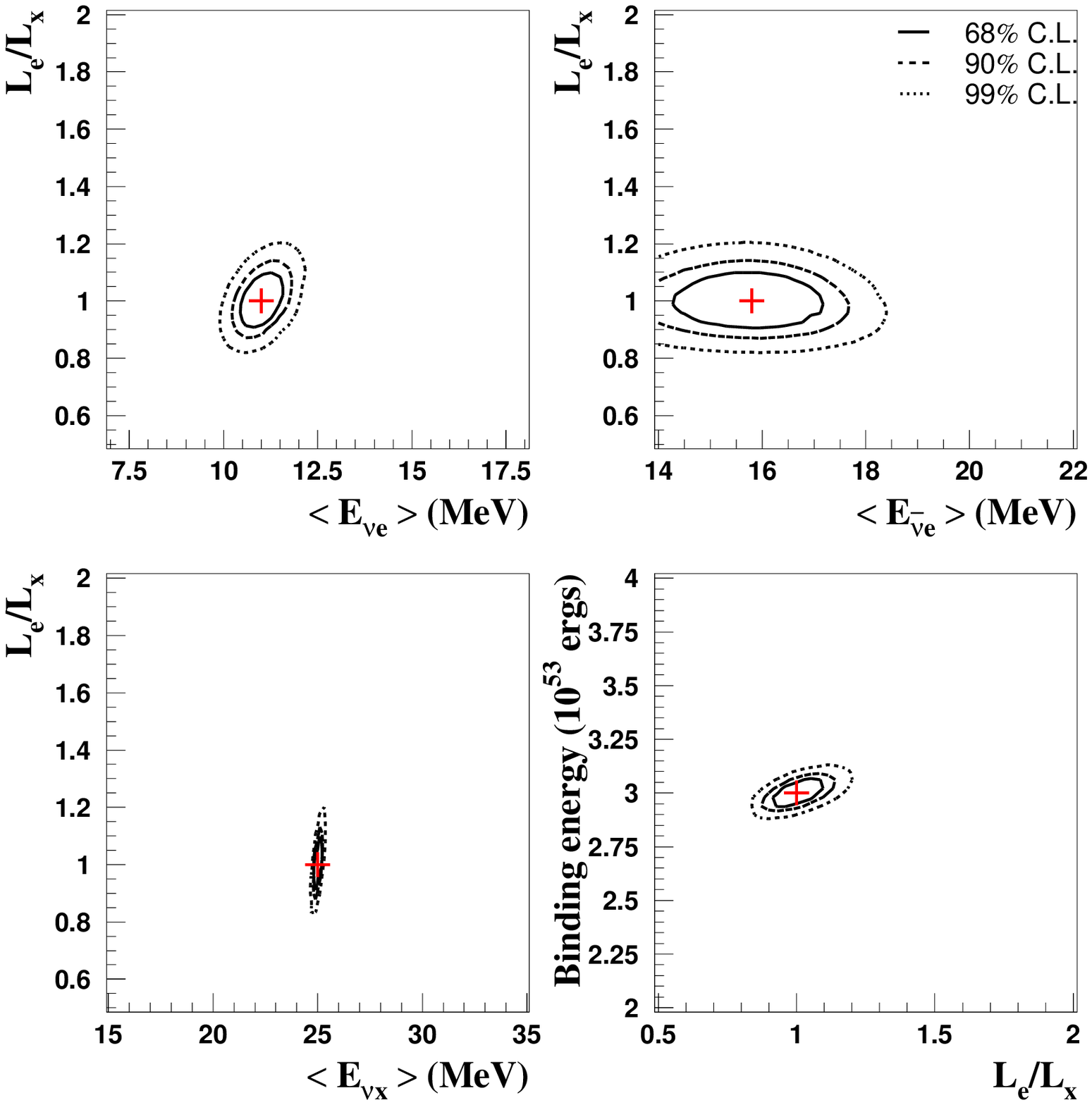,width=0.55\linewidth}
\hspace{-1cm}
\epsfig{file=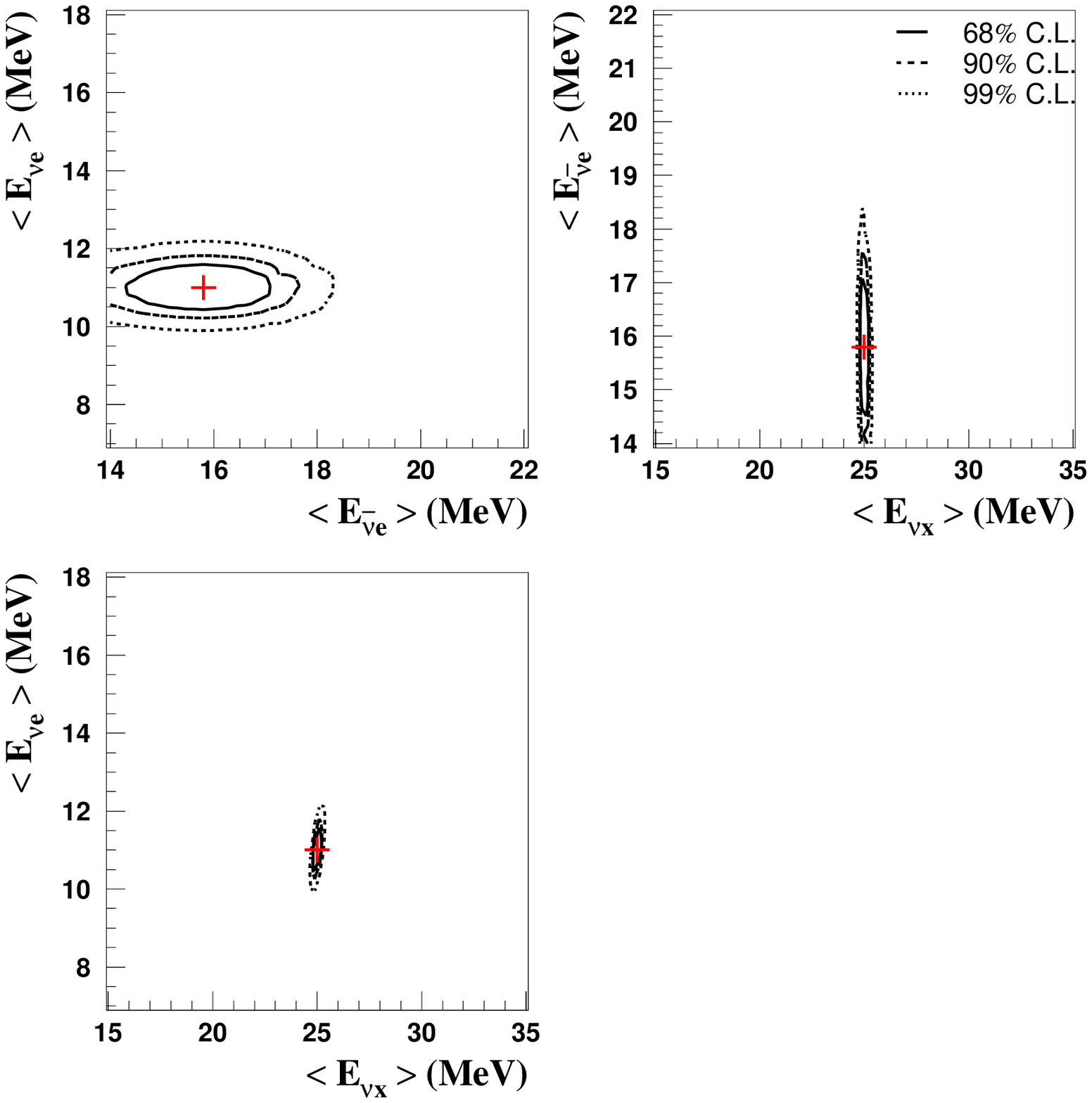,width=0.55\linewidth}
\caption{(Top) $\chi^2$ value of the fit as a function of the
supernova parameters for a 3 kton detector, assuming that the \th13
mixing angle has been measured with a precision of 10\% (\s2t13 =
10$^{-3}$ $\pm$ 10$^{-4}$) and the mass hierarchy is inverted ($\dm31$
$<$ 0). (Bottom) 68\%, 90\% and 99\% C.L. allowed regions for the
supernova parameters with a 100 kton detector. Crosses indicate the
value of the parameters for the best fits.}
\label{fig:2dfitcase1ih}
\end{figure}

Figure \ref{fig:2dfitcase1} shows on the top the $\chi^2$ value of the
fit as a function of the supernova parameters for a 3 kton detector
for true normal hierarchy. Figures \ref{fig:2dfitcase1ih} correspond
to the inverted hierarchy case. The bottom plots are the 68\%, 90\%
and 99\% C.L. allowed regions for the astrophysical parameters with a
100 kton detector. The crosses on the plots indicate the values of the
parameters taken as reference for the fit, which correspond to
scenario I.  

%the 68\%, 90\% and 99\%
%C.L. allowed regions for the astrophysical parameters assuming normal
%hierarchy. Figures \ref{fig:2dfitcase1ih} correspond to the inverted
%hierarchy case. We compare the results obtained with a 3 kton and 100 kton detector. 
%The crosses on the plots
%indicate the values of the parameters taken as reference for
%the fit, which correspond to scenario I.   

The plots \lelx ~vs $\langle E_{\nu} \rangle$ show the luminosity of
every flavor as a function of its average energy. We see that with the
statistics given by the 3 kton detector is not possible put any
constraint on the \avenue~and \aveanue~energies or on \lelx. However,
the \avenux~energy can be determined very precisely. The accuracy that
we can achieve at 90\% C.L. is $\Delta$\avenux /\avenux $\sim$ 3\%. 

The plot \EB ~vs \lelx ~shows the strong correlation between these two
variables. 
Using a 100 kton detector, all the variables can be determined in a
certain range. The best results are obtained for \avenux ~because we
are considering the large \th13 mixing angle.
In this case the \nue's have oscillated into \numu's and \nutau's (so \nux's) 
and vice versa. Therefore, the most important reaction on argon (\nue CC) will
be mainly sensitive to the original \avenux. 

In the inverted hierarchy case, a fraction of the \nue ~flux arrives at the detector
hence in this case we are sensitive to the \avenue ~energy and a bit less to
\avenux, but both variables can be estimated with accuracies of
$\sim$ 46\% and $\sim$ 7\%, respectively, for a 3 kton detector.  

%%%  SCENARIO II  %%%%

\begin{figure}[htbp]
\begin{center}
\fbox{\LARGE \sf \s2t13 = 10$^{-3}$ $\pm$ 10$^{-4}$ and n.h. (scen II)}
\end{center}
\begin{center}
\Large \bf 3 kton LAr
\end{center}
\vspace{-0.7cm}
%\epsfig{file=EPS/2dfitcase1scen2p_2bis.eps,width=0.55\linewidth}
%\hspace{-1cm}
%\epsfig{file=EPS/2dfitcase1scen2p_3.eps,width=0.55\linewidth}
%\vspace{-1cm}
\begin{center}
\epsfig{file=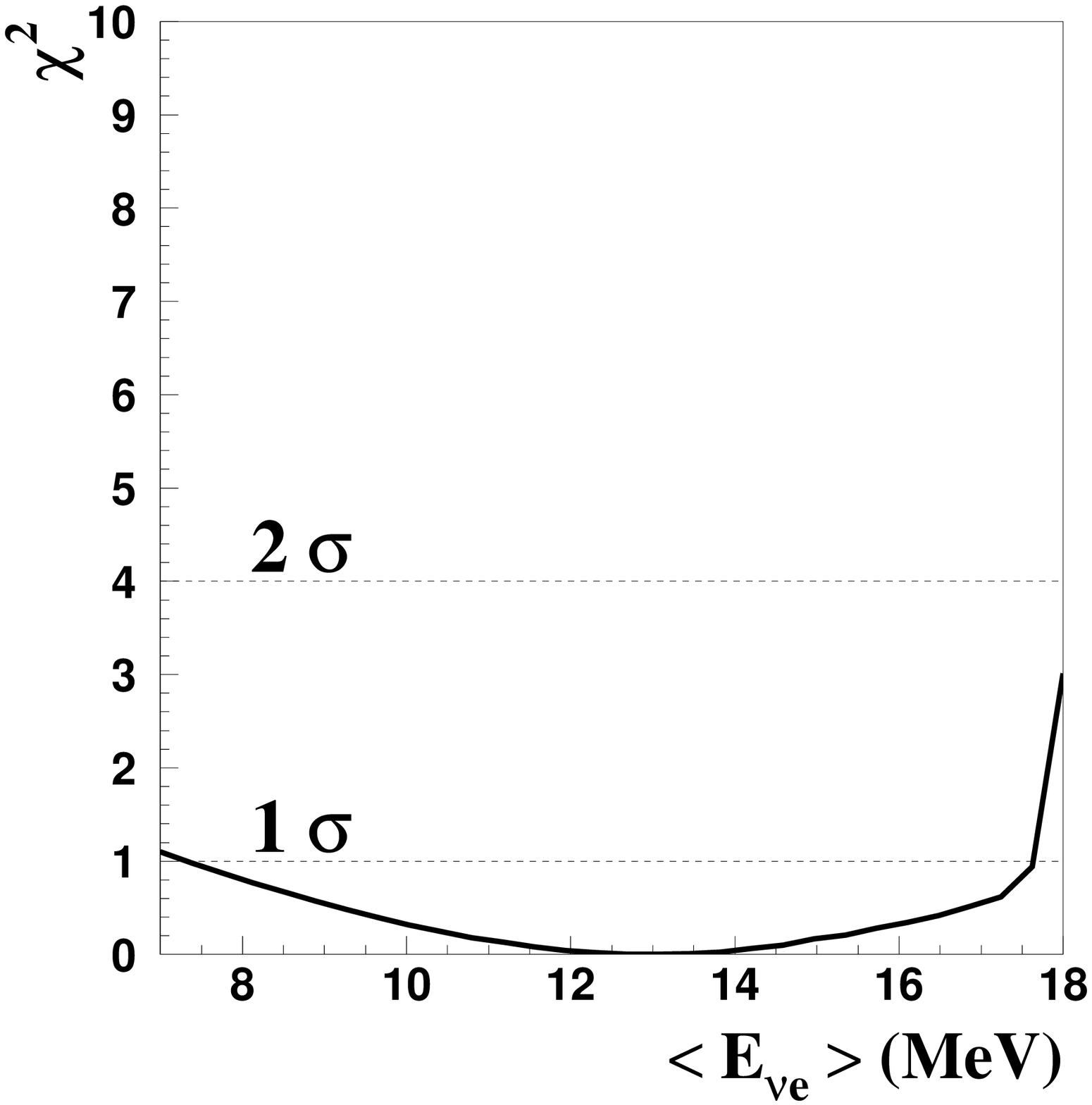,width=0.3\linewidth} \hspace{-0.5cm}
\epsfig{file=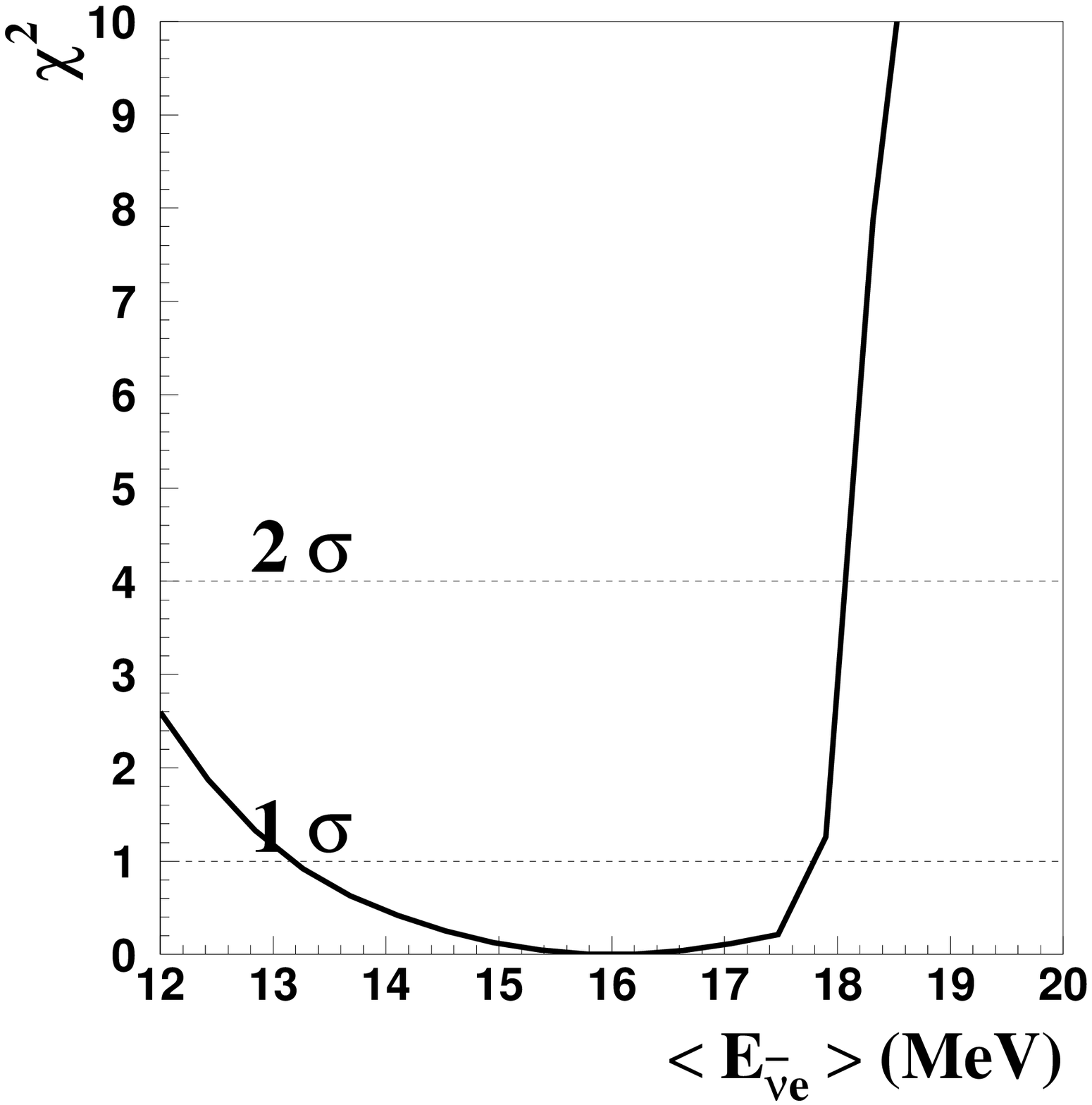,width=0.3\linewidth} \hspace{-0.5cm}
\epsfig{file=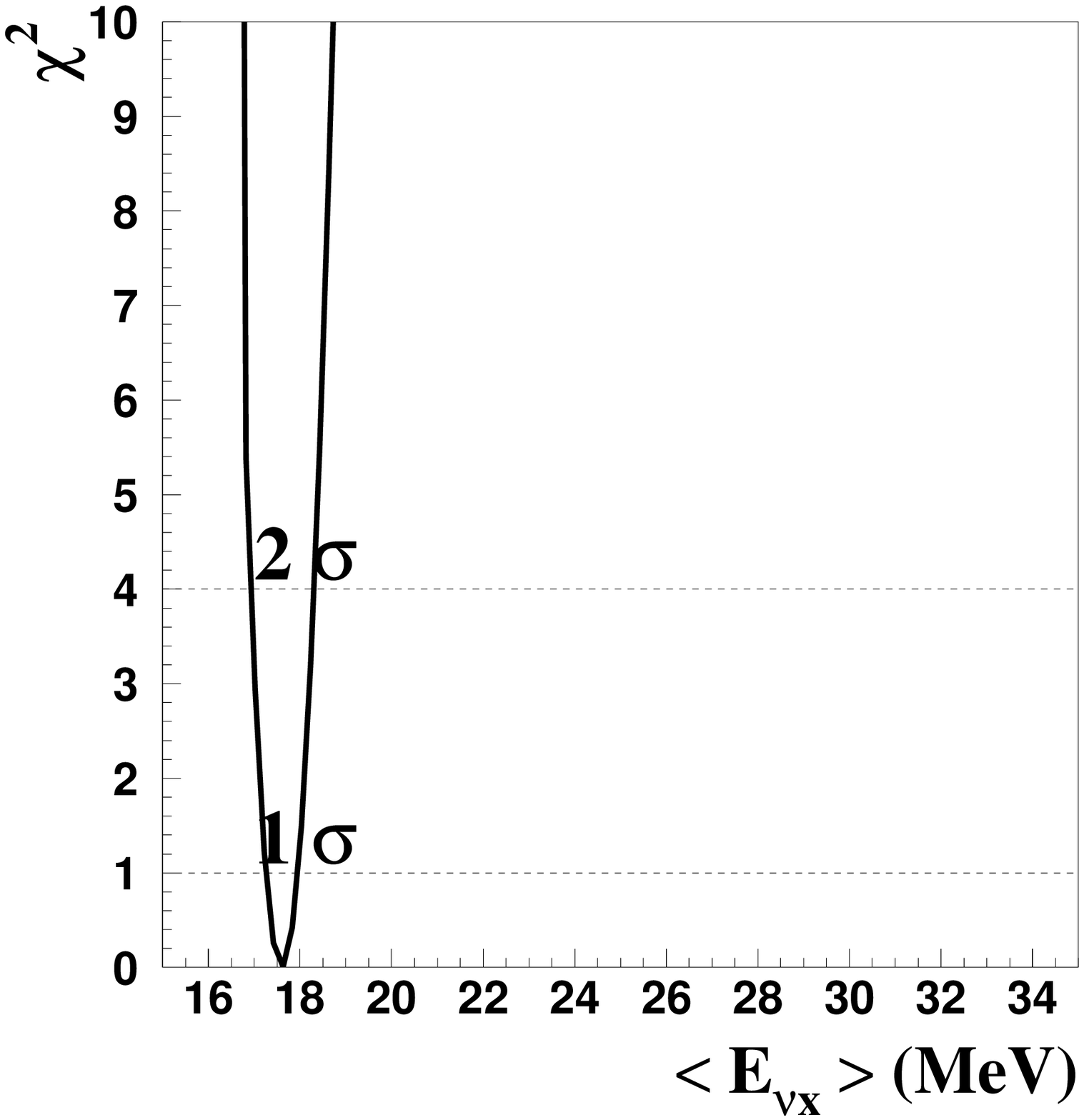,width=0.3\linewidth}
\end{center}
\vspace{-1cm}
\begin{center}
\epsfig{file=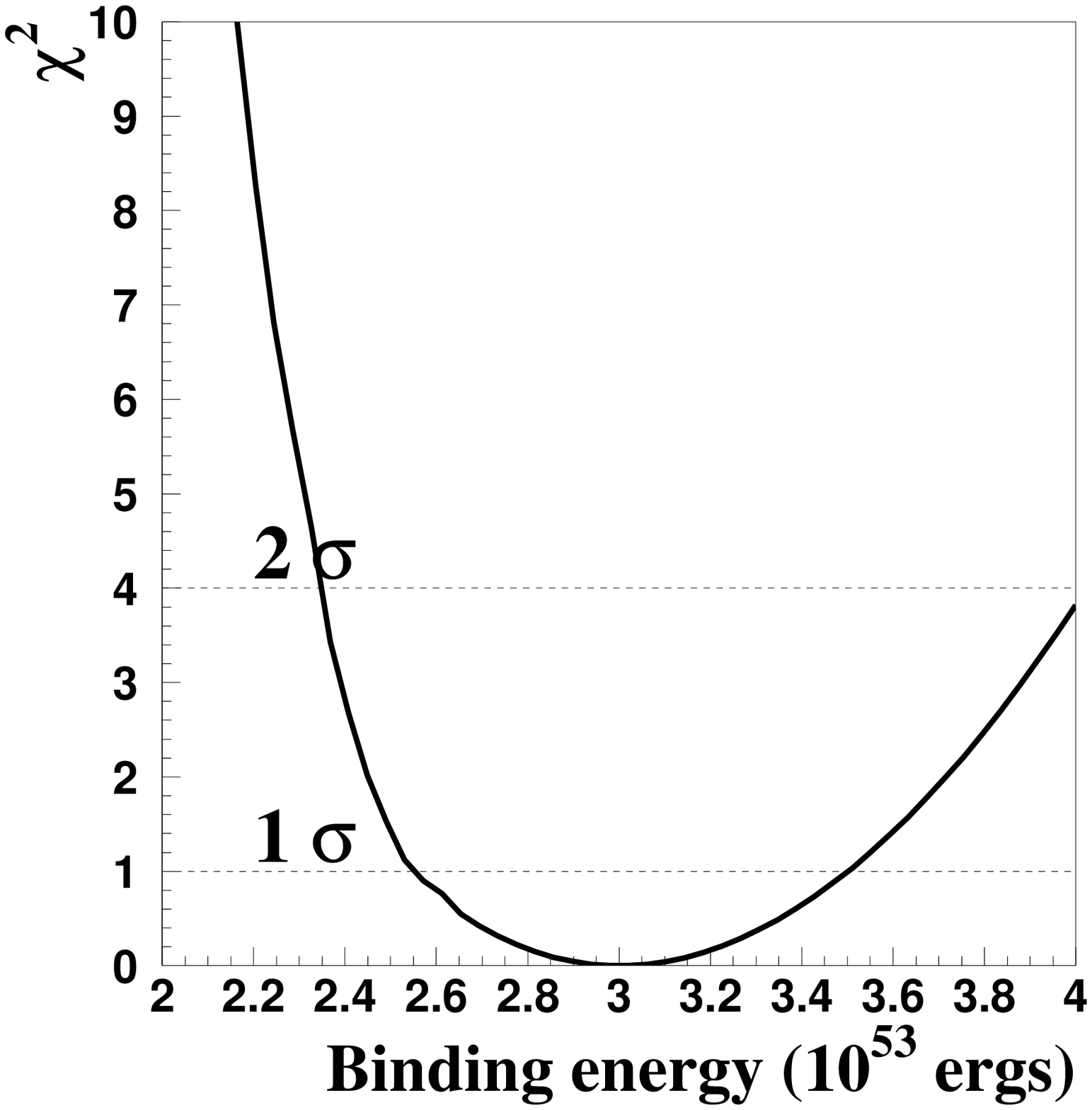,width=0.3\linewidth} \hspace{-0.5cm}
\epsfig{file=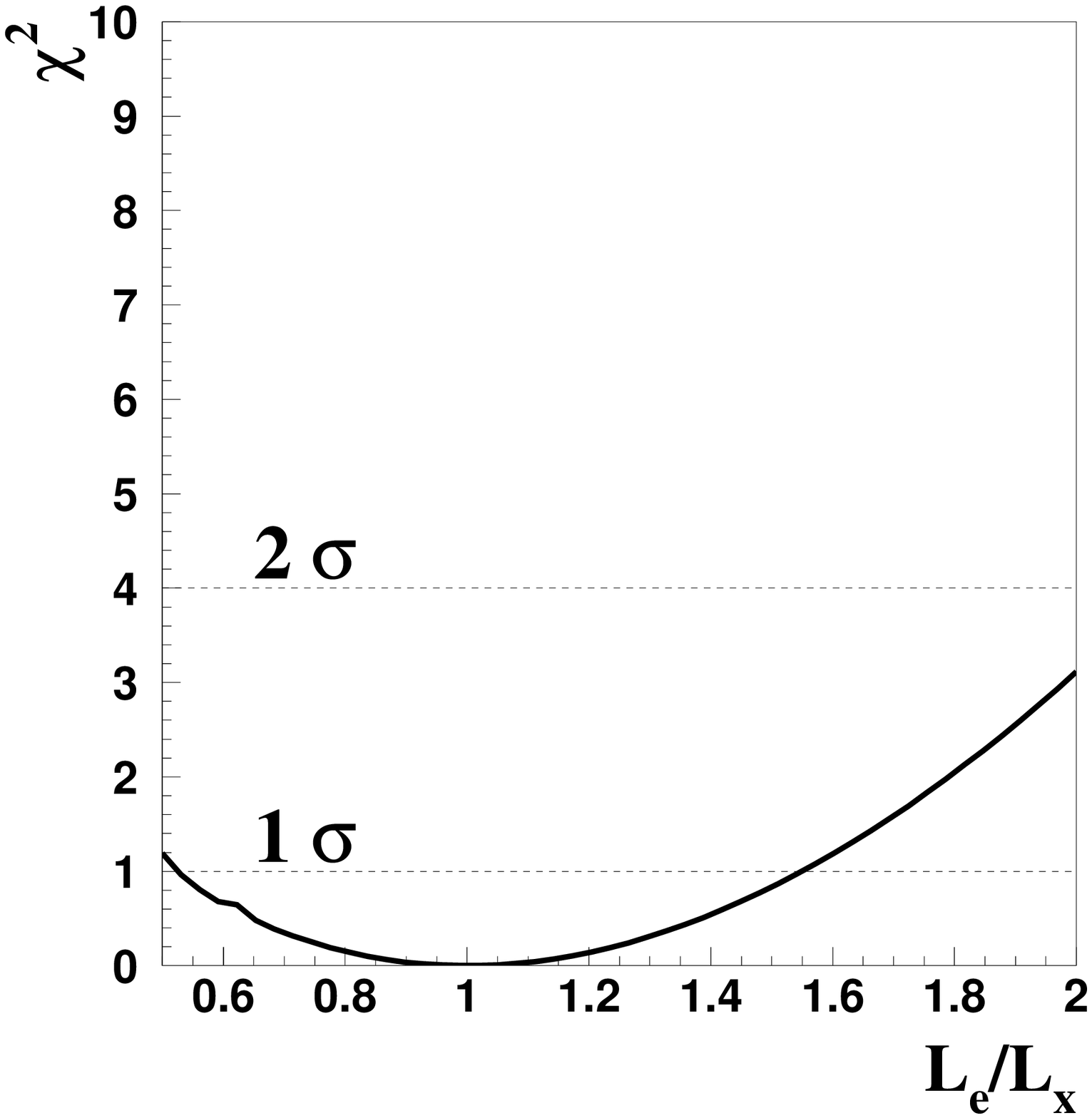,width=0.3\linewidth}
\end{center}

\begin{center}
\Large \bf 100 kton LAr
\end{center}
\vspace{-0.7cm}
\epsfig{file=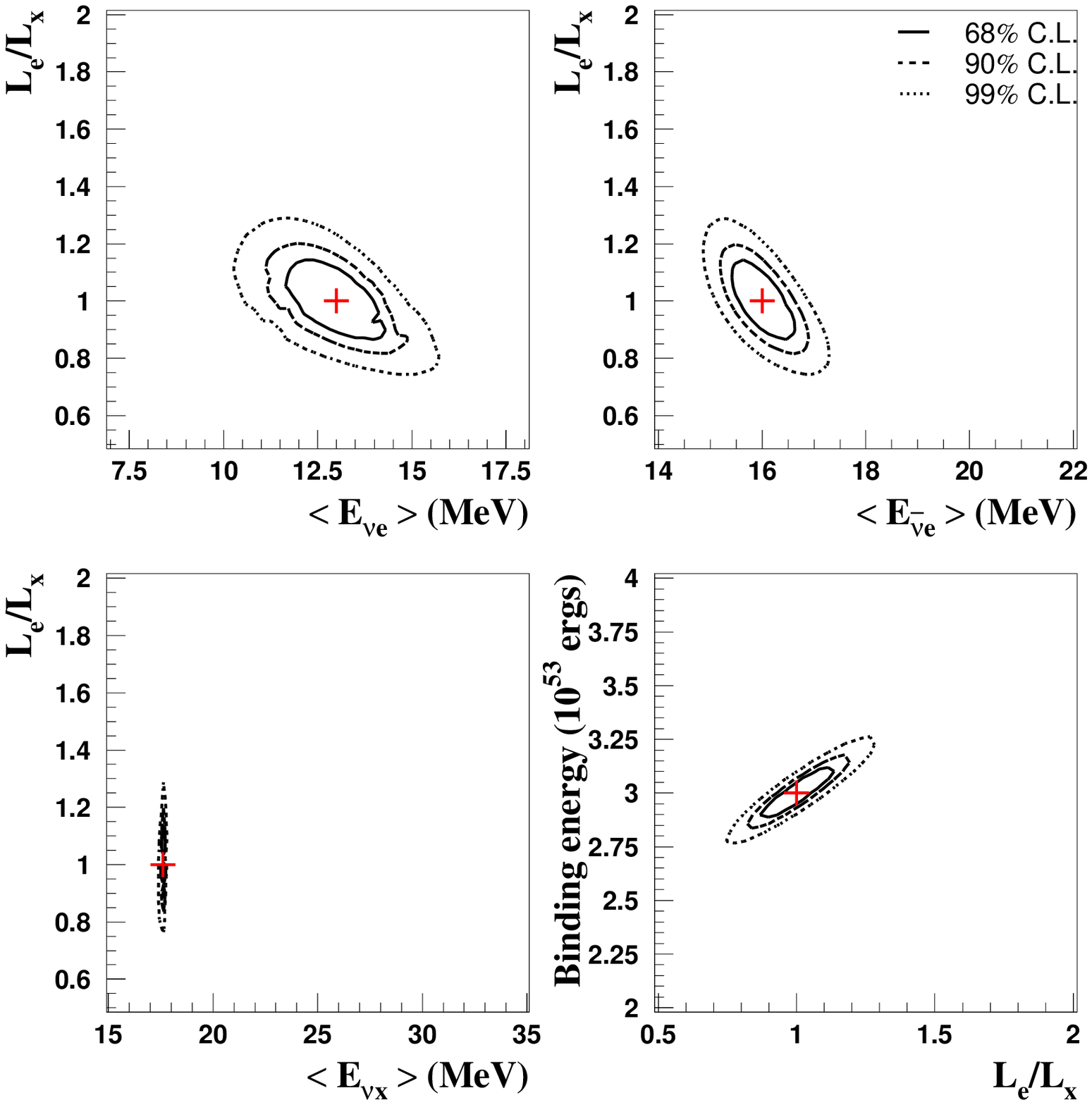,width=0.55\linewidth}
\hspace{-1cm}
\epsfig{file=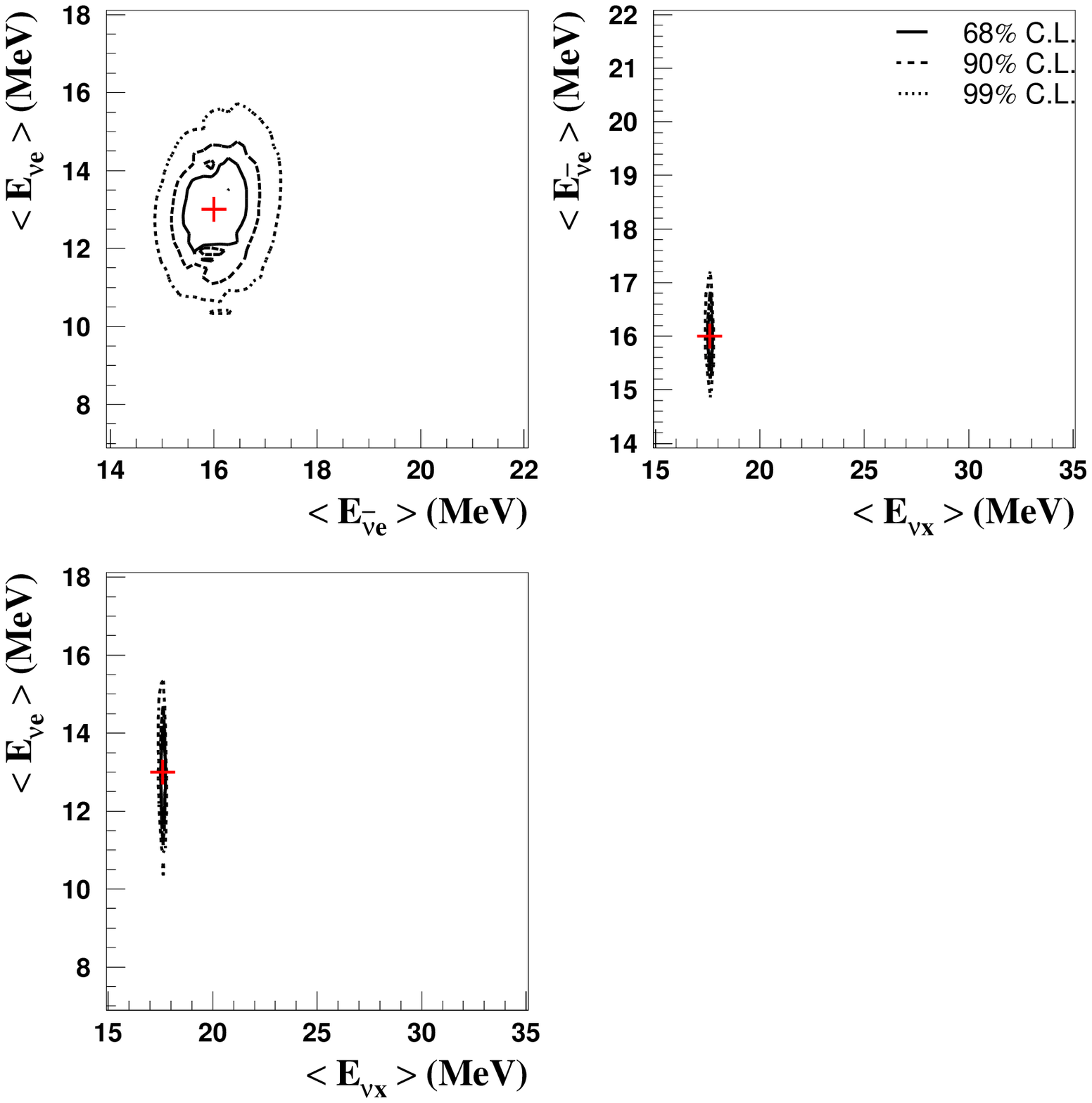,width=0.55\linewidth}
\caption{(Top) $\chi^2$ value of the fit as a function of the
supernova parameters for a 3 kton detector, assuming that the \th13
mixing angle has been measured with a precision of 10\% (\s2t13 =
10$^{-3}$ $\pm$ 10$^{-4}$) and the mass hierarchy is normal ($\dm31$
$>$ 0). The reference values taken for the neutrino average energies
are the ones of scenario II (see table
\ref{tab:sncoolscenario}). (Bottom) 68\%, 90\% and 99\% C.L. allowed
regions for the supernova parameters with a 100 kton detector. Crosses
indicate the value of the parameters for the best fits.}
\label{fig:2dfitcase1scen2}
\end{figure}

\begin{figure}[htbp]
\begin{center}
\fbox{\LARGE \sf \s2t13 = 10$^{-3}$ $\pm$ 10$^{-4}$ and i.h. (scen II)}
\end{center}
\begin{center}
\Large \bf 3 kton LAr
\end{center}
\vspace{-0.7cm}
%\epsfig{file=EPS/2dfitcase1ihscen2p_2bis.eps,width=0.55\linewidth}
%\hspace{-1cm}
%\epsfig{file=EPS/2dfitcase1ihscen2p_3.eps,width=0.55\linewidth}
%\vspace{-1cm}
\begin{center}
\epsfig{file=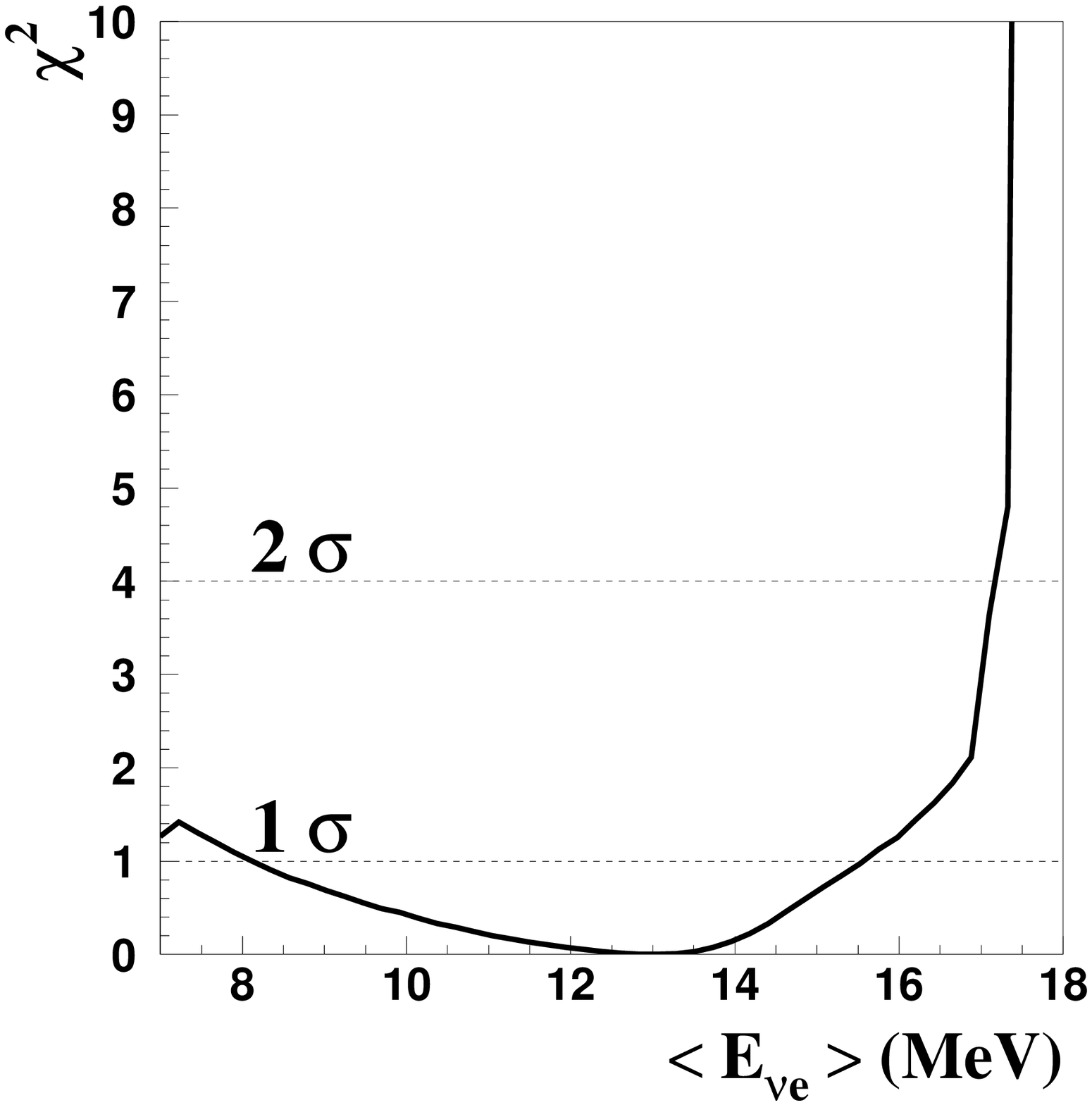,width=0.3\linewidth} \hspace{-0.5cm}
\epsfig{file=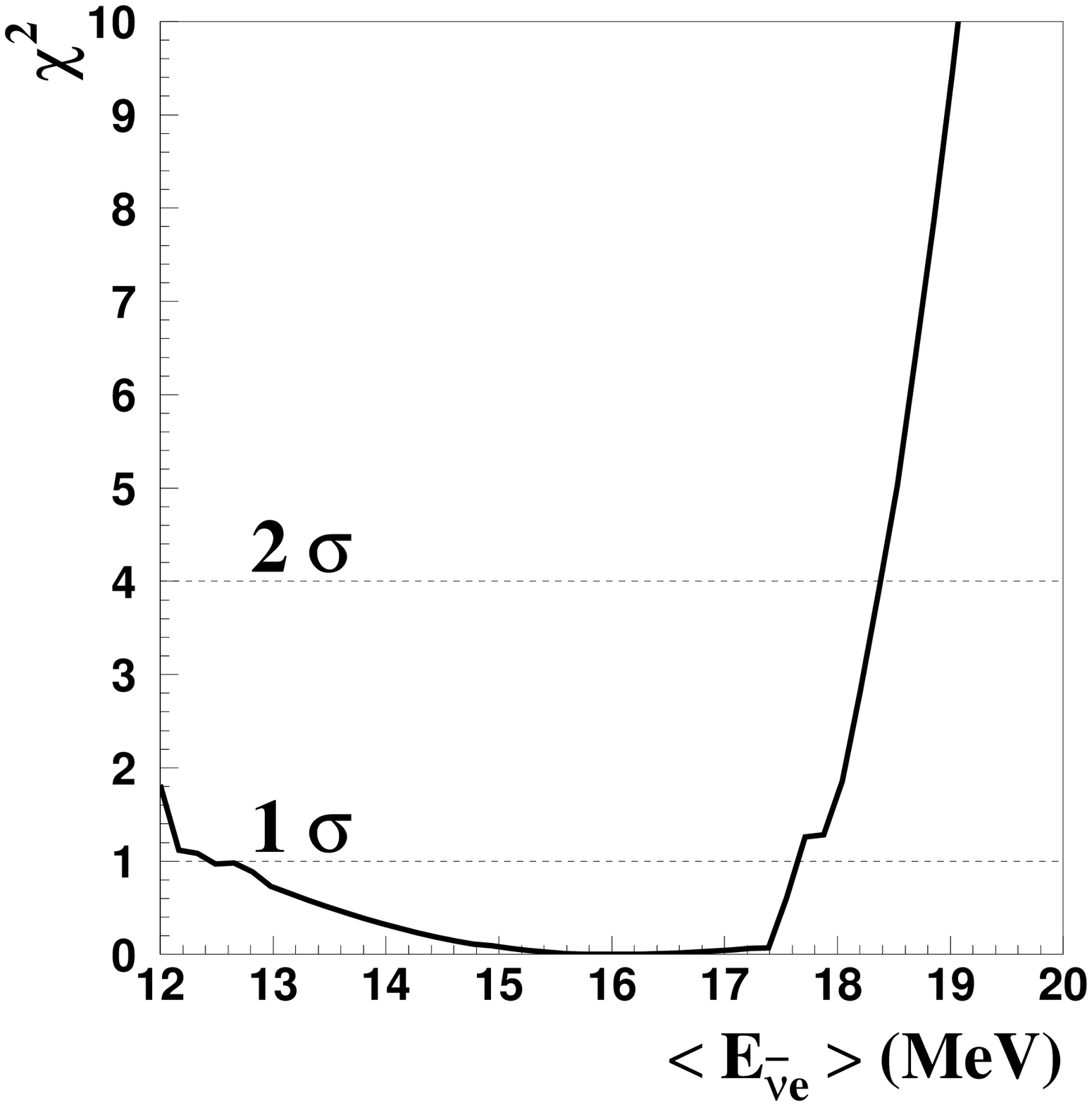,width=0.3\linewidth} \hspace{-0.5cm}
\epsfig{file=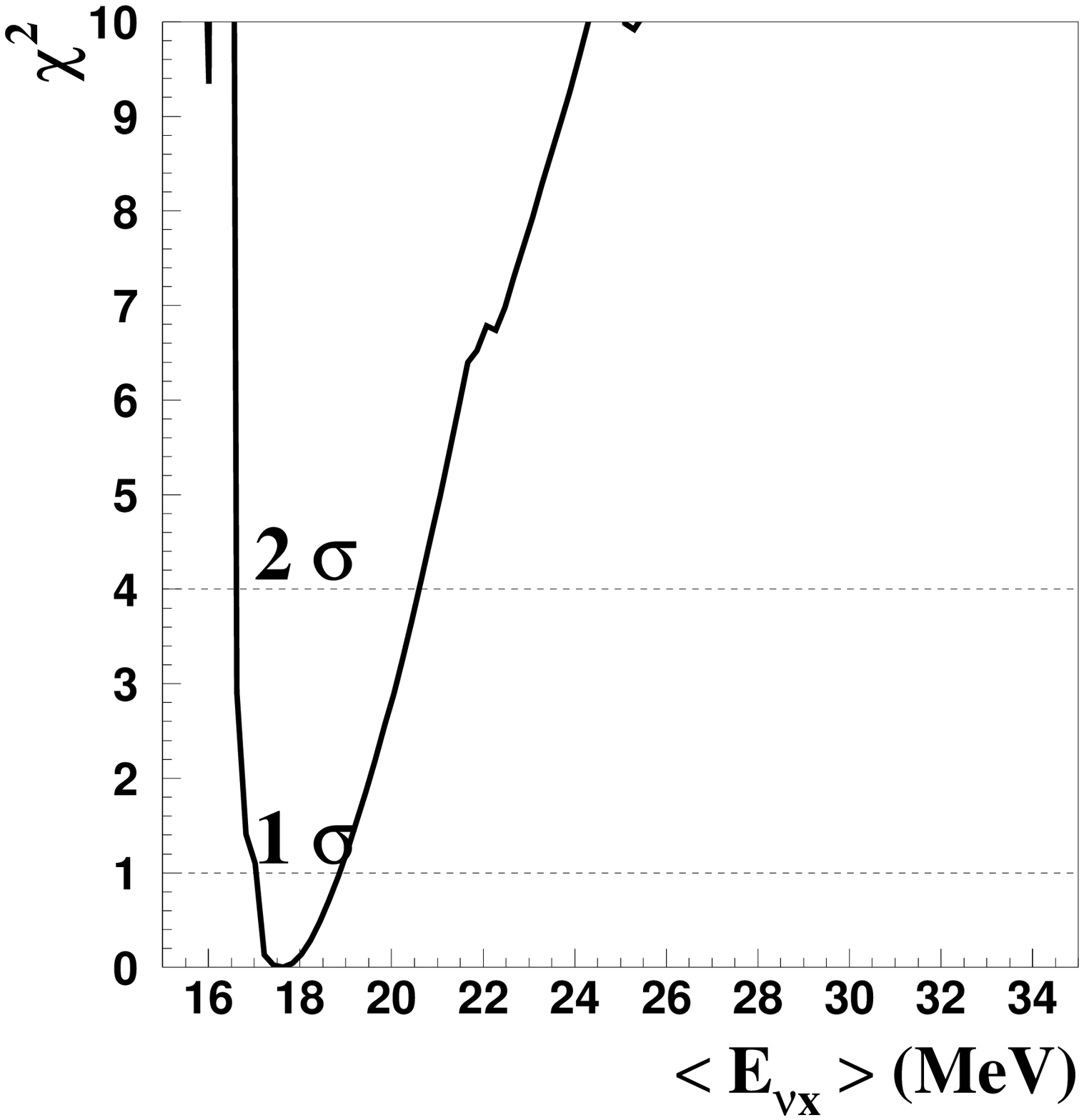,width=0.3\linewidth}
\end{center}
\vspace{-1cm}
\begin{center}
\epsfig{file=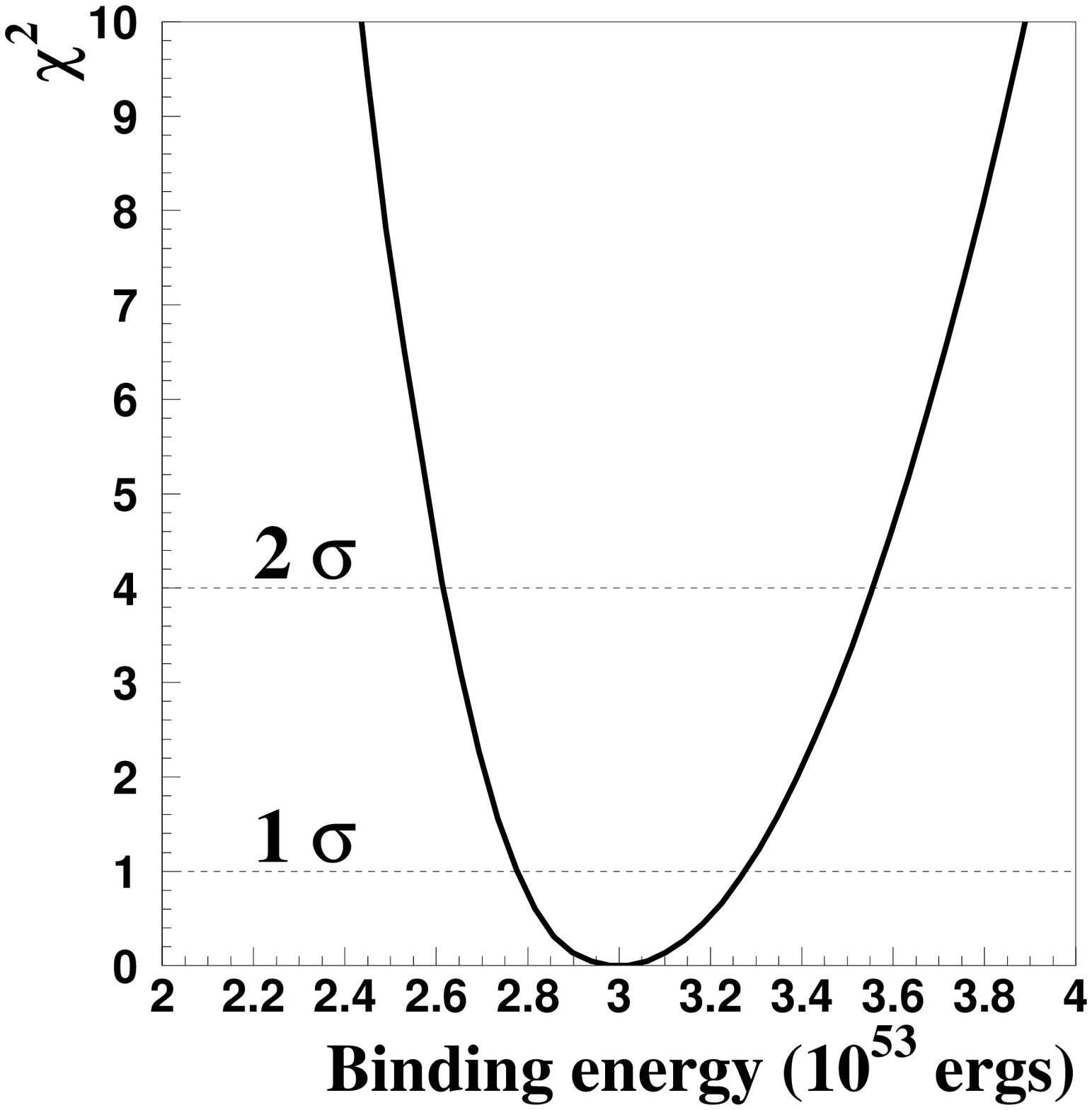,width=0.3\linewidth} \hspace{-0.5cm}
\epsfig{file=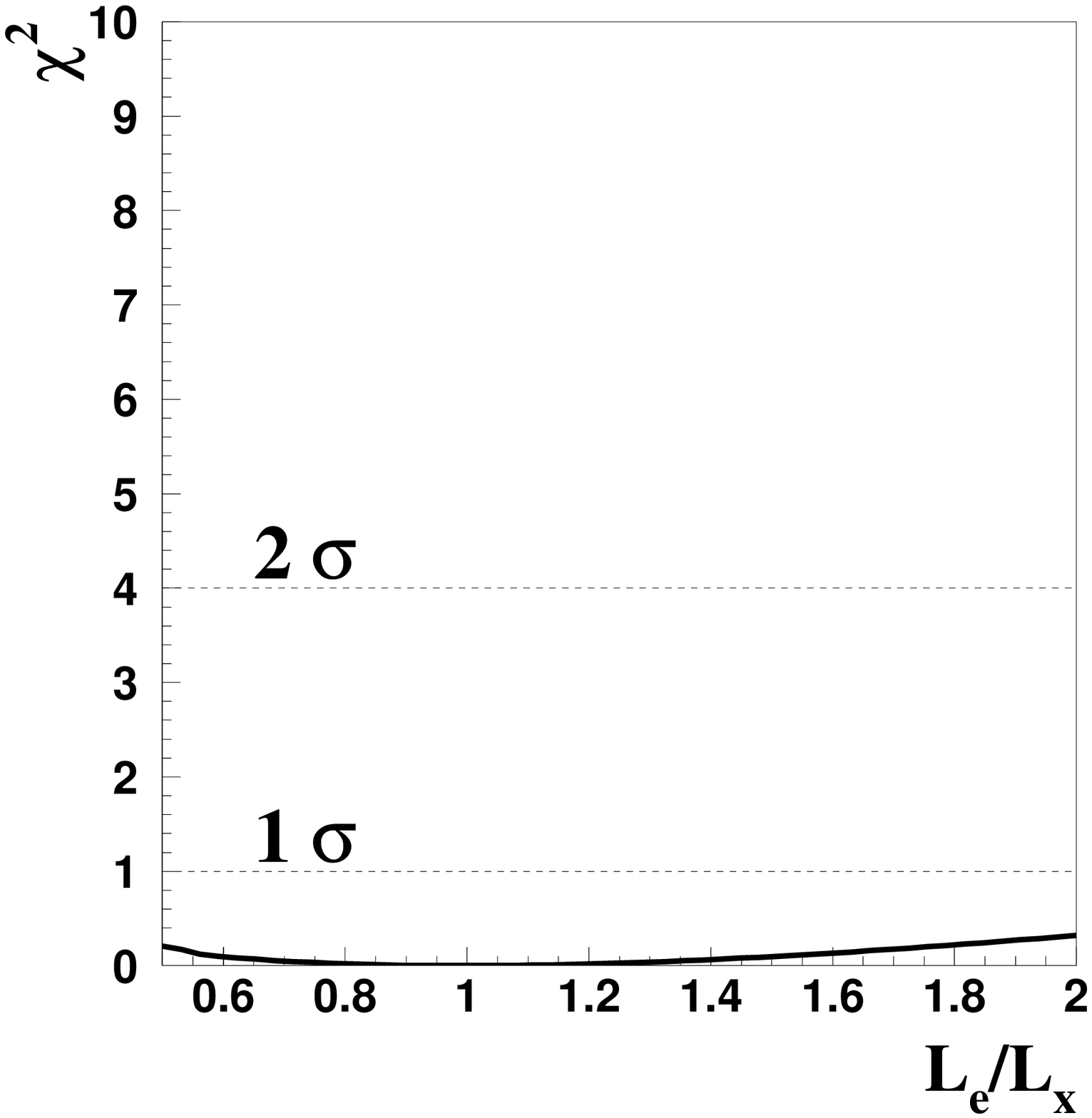,width=0.3\linewidth}
\end{center}

\begin{center}
\Large \bf 100 kton LAr
\end{center}
\vspace{-0.7cm}
\epsfig{file=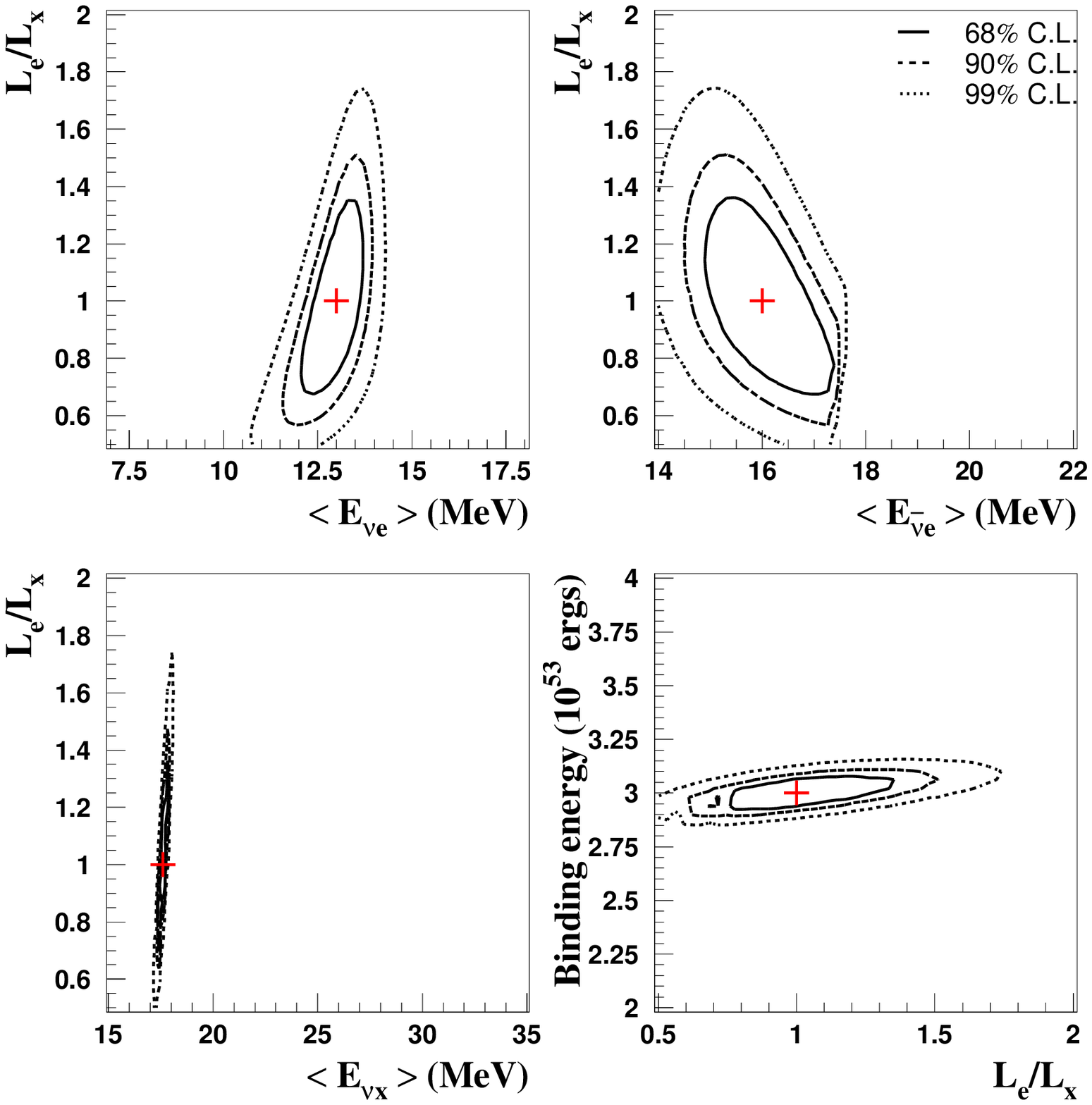,width=0.55\linewidth}
\hspace{-1cm}
\epsfig{file=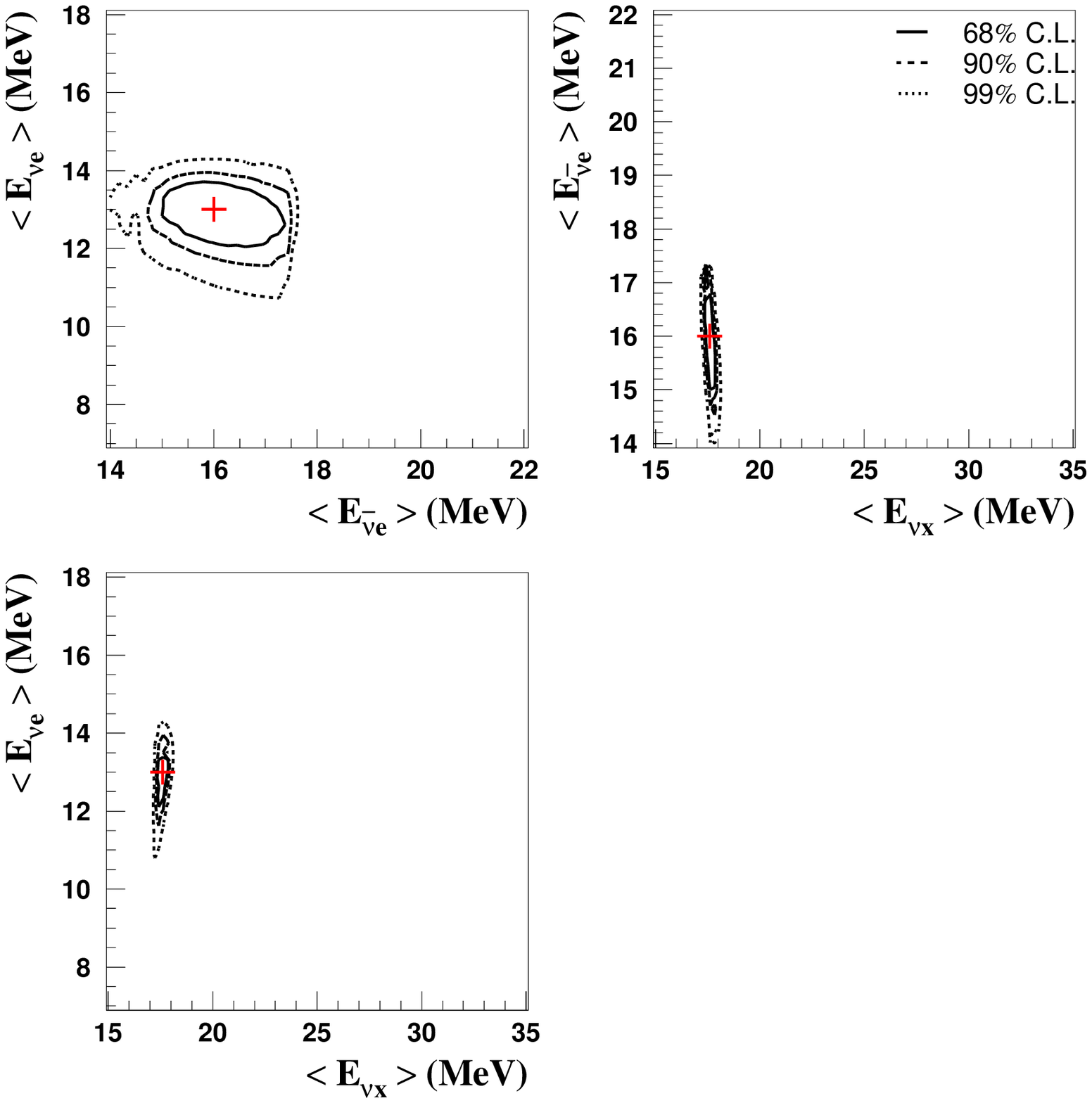,width=0.55\linewidth}
\caption{(Top) $\chi^2$ value of the fit as a function of the
supernova parameters for a 3 kton detector, assuming that the \th13
mixing angle has been measured with a precision of 10\% (\s2t13 =
10$^{-3}$ $\pm$ 10$^{-4}$) and the mass hierarchy is inverted ($\dm31$
$<$ 0). The reference values taken for the neutrino average energies
are the ones of scenario II (see table
\ref{tab:sncoolscenario}). (Bottom) 68\%, 90\% and 99\% C.L. allowed
regions for the supernova parameters with a 100 kton detector. Crosses
indicate the value of the parameters for the best fits.}
\label{fig:2dfitcase1ihscen2}
\end{figure}

If we consider true values those given by scenario II (see table
\ref{tab:sncoolscenario}), the corresponding results for the $\chi^2$
fit are presented in figures \ref{fig:2dfitcase1scen2} (n.h.) and
\ref{fig:2dfitcase1ihscen2} (i.h.). 
For the normal hierarchy case we see that results are quite similar to
those obtained for scenario I. Only the \avenux ~parameter can be
measured with a 3 kton detector. However, the inverted hierarchy case
shows a worse situation. The \nue ~and \nux ~average energies are
closer for scenario II and it makes more difficult the determination
of a region in the plane (\avenue, \avenux). 

For a 100 kton detector, scenarios I and II give quite similar results
except for the \lelx ~parameter in the case of i.h., which is worse
determined for scenario II ($\sim$ 32\%) than for scenario I ($\sim$
9\%) (see table \ref{tab:accurasnpar}).  

%The expected accuracies at 90\% C.L. in the determination of the
%supernova parameters are shown in table \ref{tab:accurasnpar} for different
%LAr detector masses (3 and 100 kton) and supernova scenarios.

\subsection{If an upper bound on \th13 angle is set by long-baseline experiments}
%%%%%%%%%%%%%%%%%%%%%%%%%%%%%%%%%%%%%%%%%%%%%%%%%%%%%%%%%%%%%%%%%%%%%%%%%%%%%%%%%%%%%%%%%%

Another possibility is that future long-baseline neutrino experiments
will not be sensitive enough to measure \s2t13 and will place an upper limit on its value.
Assuming the limit \s2t13 $<$ 10$^{-4}$, we study
the possible determination of the supernova parameters.

In this case the large mixing angle region for both hierarchies is
excluded and we only have the possibility that the \th13 angle be
in the intermediate (n.h.-i or i.h.-i) or in the small (n.h.-S or
i.h.-S) mixing angle regions.

Putting the condition of \s2t13 $<$ 10$^{-4}$ and fixing the mass
hierarchy, we perform a $\chi^2$ minimization letting at each step
all variables free, included the \th13 angle. The upper bound on the 
\th13 angle is inserted as an appropriate term in the $\chi^2$ function.
We study two cases for
the ``true'' data: normal and inverted hierarchy. 

Table
\ref{tab:accurasnparcase2} shows the expected accuracies at 90\%
C.L. in the determination of the supernova parameters for 3 and 100
kton detectors and supernova scenarios I and II. 

\begin{table}[htbp]
\centering
\begin{tabular}{|c|c|c||c|c|c|c|c|} 
\hline
\multicolumn{8}{|c|}{With constraint \s2t13 $<$ 10$^{-4}$ from terrestrial experiment} \\
\hline 
Detector mass & True hierarchy & SN scen. & $\frac{\Delta\favenue}{\favenue}$ &
$\frac{\Delta\faveanue}{\faveanue}$ &
$\frac{\Delta\favenux}{\favenux}$ & $\frac{\Delta\fEB}{\fEB}$ &
$\frac{\Delta(\flelx)}{(\flelx)}$ \\ \hline \hline 
3 kton & n.h. & I & $\sim$ 30\% & $\sim$ 30\% & $\sim$ 7\% & $\sim$ 26\% & -- \\
3 kton & i.h. & I & $\sim$ 52\% & $\sim$ 30\% & $\sim$ 8\% & $\sim$ 17\% & -- \\
\hline
3 kton & n.h. & II & -- & $\sim$ 25\% & $\sim$ 16\% & $\sim$ 27\% & -- \\
3 kton & i.h. & II & -- & $\sim$ 25\%& $\sim$ 23\% & $\sim$ 17\% &
-- \\ \hline \hline
100 kton & n.h. & I & $\sim$ 6\% & $\sim$ 4\% & $<$ 1\% & $\sim$
1\% & $\sim$ 11\% \\
100 kton & i.h. & I & $\sim$ 5\% & $\sim$ 4\% & $<$ 1\% & $\sim$ 2\% &
$\sim$ 9\% \\ \hline 
100 kton & n.h. & II & $\sim$ 9\% & $\sim$ 5\% & $<$ 1\% & $\sim$
2\% & $\sim$ 35\% \\
100 kton & i.h. & II & $\sim$ 8\% & $\sim$ 4\% & $\sim$ 1\% &
$\sim$ 2\% & $\sim$ 37\% \\
\hline
\end{tabular}
\caption{Expected accuracies at 90\% C.L. in the determination of the
supernova parameters using the neutrinos measured with a 3 kton and a
100 kton detector. We have assumed that the mass hierarchy is
known and an upper limit on the mixing angle of \s2t13 $<$ 10$^{-4}$
has been set by long-baseline neutrino experiments. Supernova
scenarios I and II are tested.}      
\label{tab:accurasnparcase2}
\end{table}

\begin{figure}[htbp]
\begin{center}
\fbox{\LARGE \sf \s2t13 $<$ 10$^{-4}$ and n.h.}
\end{center}
\begin{center}
\Large \bf 3 kton LAr
\end{center}
\vspace{-0.7cm}
%\epsfig{file=EPS/2dfitcase2p_2rebin.eps,width=0.55\linewidth}
%\hspace{-1cm}
%\epsfig{file=EPS/2dfitcase2p_3rebin.eps,width=0.55\linewidth}
%\vspace{-1cm}
\begin{center}
\epsfig{file=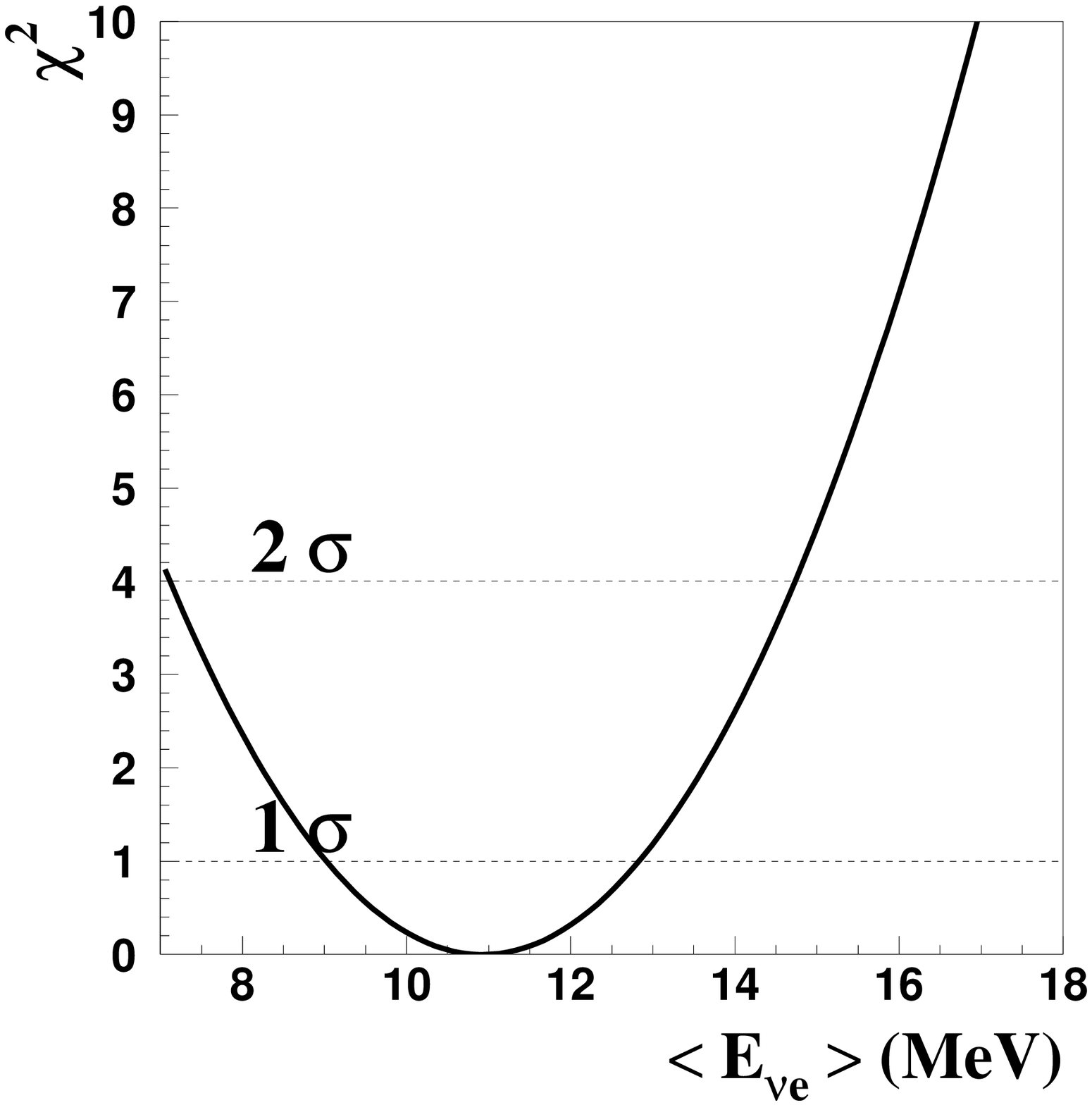,width=0.3\linewidth} \hspace{-0.5cm}
\epsfig{file=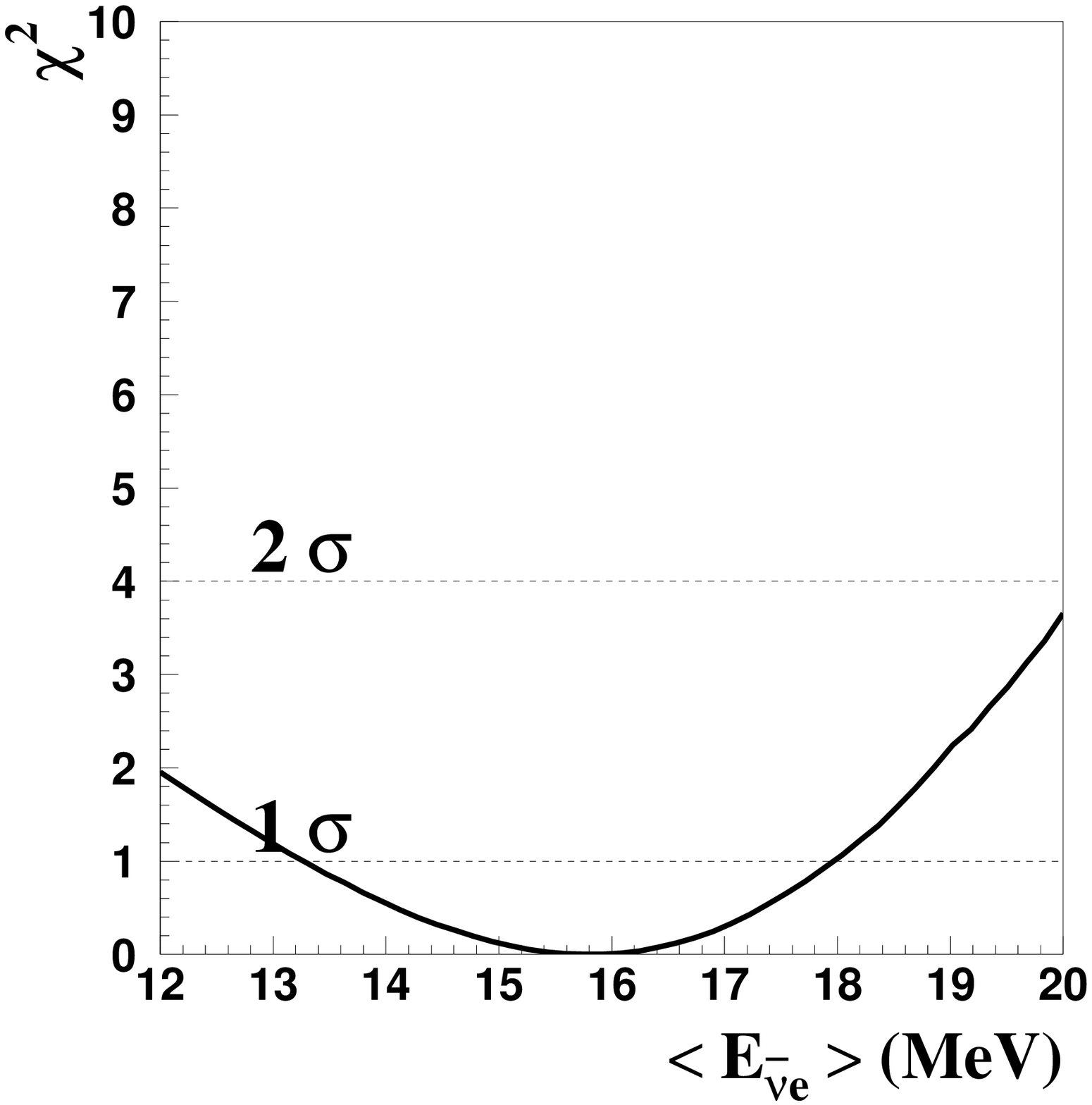,width=0.3\linewidth} \hspace{-0.5cm}
\epsfig{file=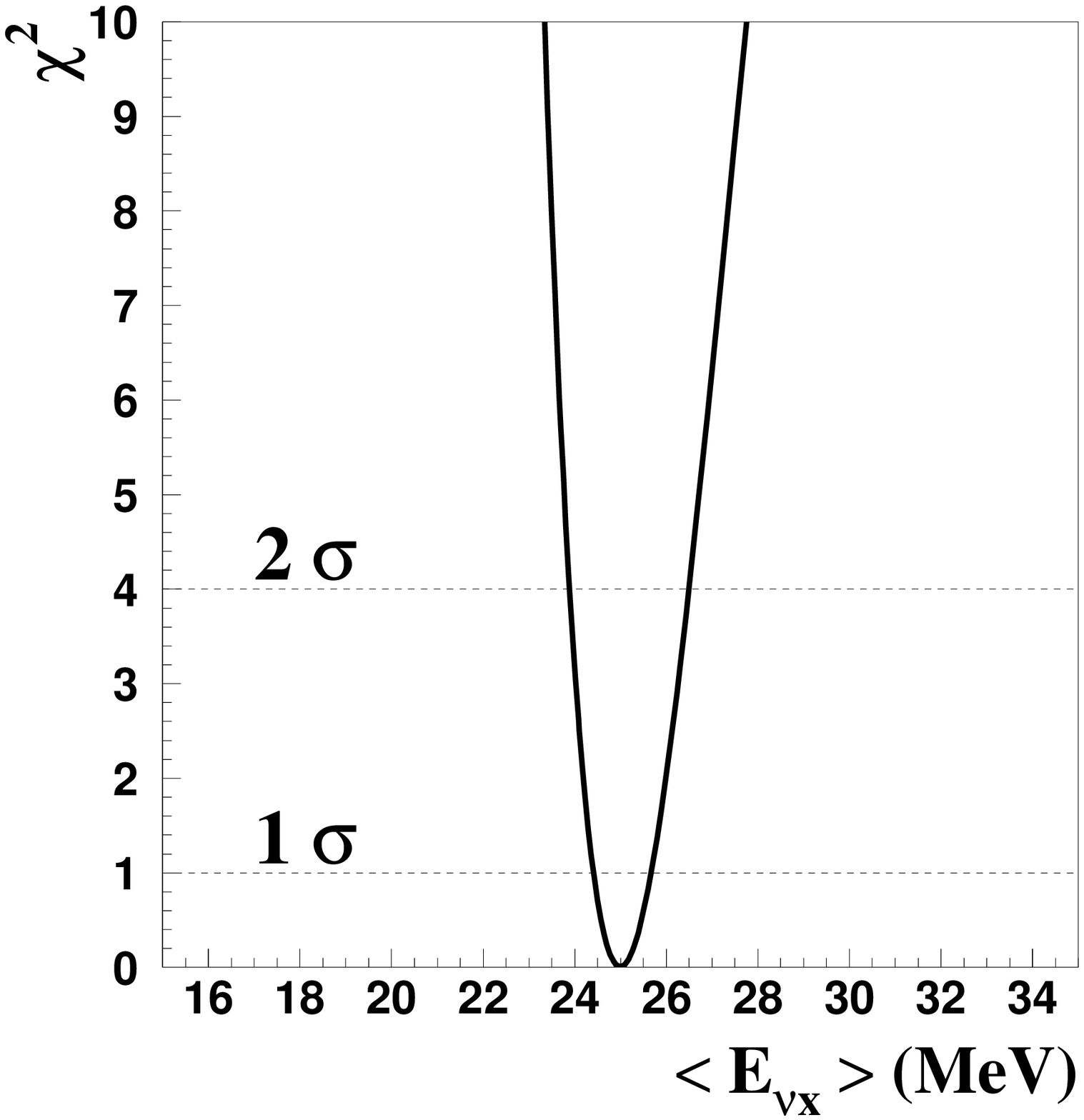,width=0.3\linewidth}
\end{center}
\vspace{-1cm}
\begin{center}
\epsfig{file=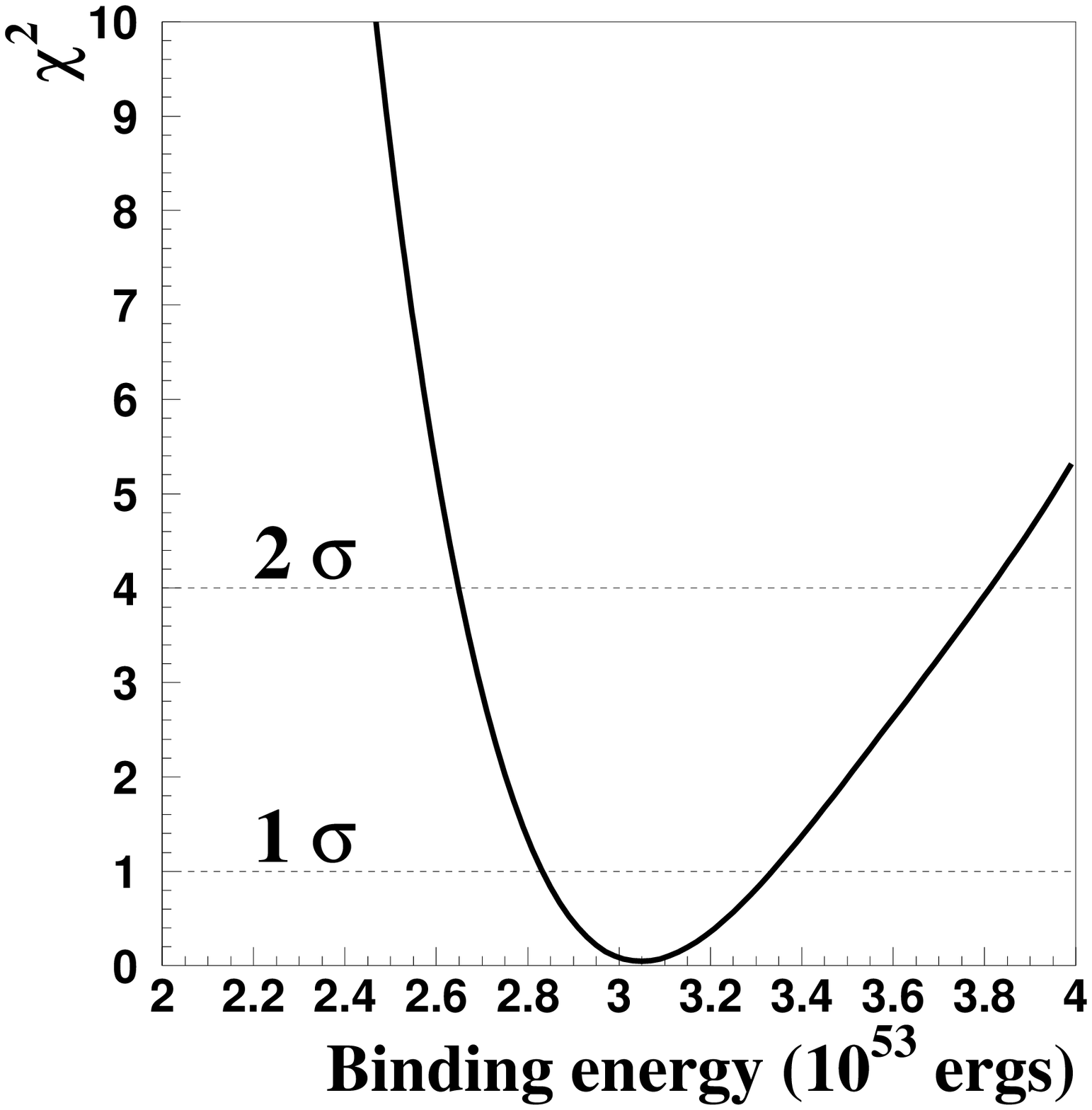,width=0.3\linewidth} \hspace{-0.5cm}
\epsfig{file=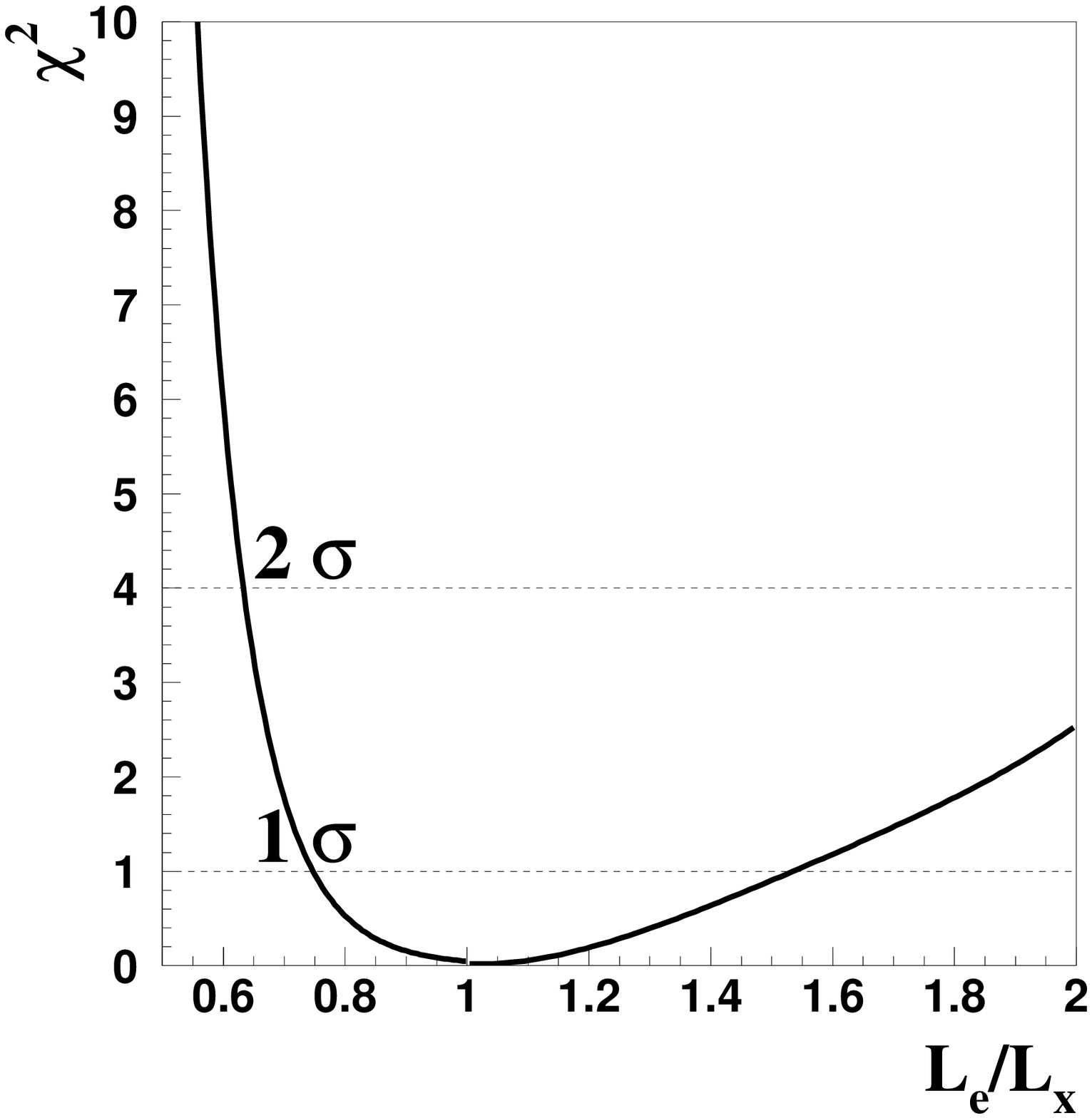,width=0.3\linewidth}
\end{center}

\begin{center}
\Large \bf 100 kton LAr
\end{center}
\vspace{-0.7cm}
\epsfig{file=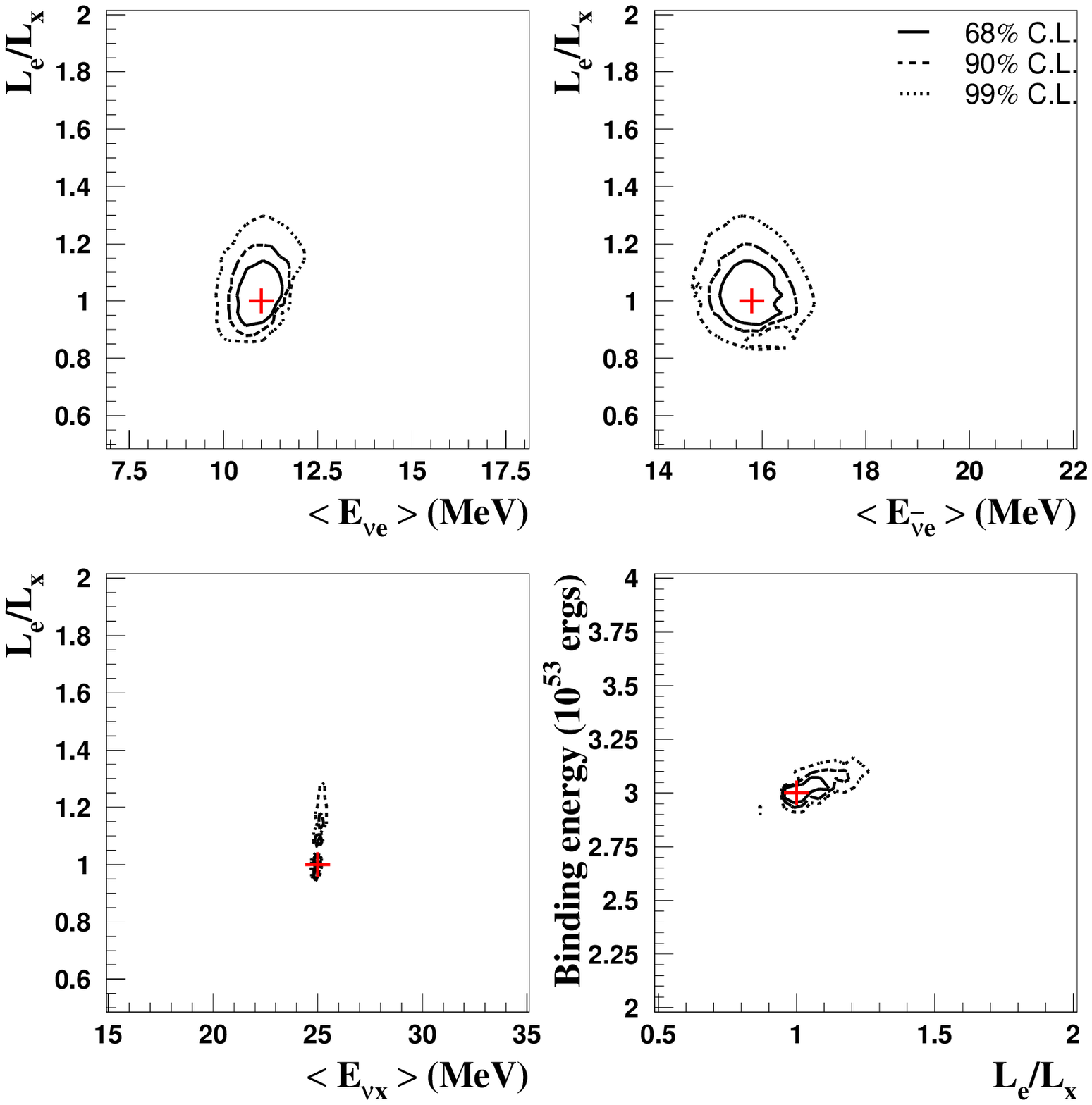,width=0.55\linewidth}
\hspace{-1cm}
\epsfig{file=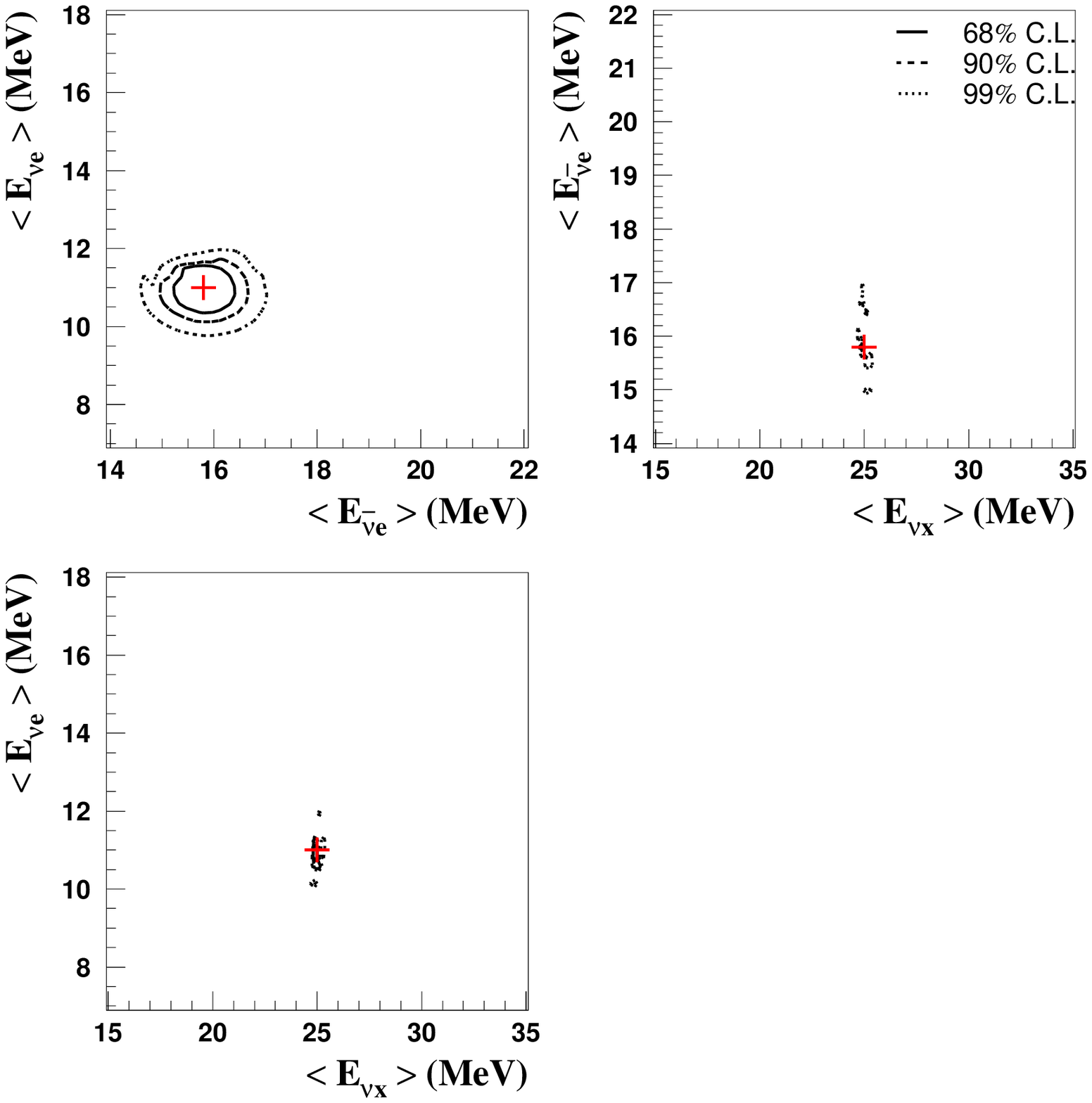,width=0.55\linewidth}
\caption{(Top) $\chi^2$ value of the fit as a function of the
supernova parameters for a 3 kton detector, assuming that an upper
limit on the value of the \th13 mixing angle has been set (\s2t13 $<$
10$^{-4}$) and the mass hierarchy is normal ($\dm31$ $>$ 0). (Bottom)
68\%, 90\% and 99\% C.L. allowed regions for the supernova parameters
with a 100 kton detector. Crosses indicate the value of the parameters
for the best fits.}
\label{fig:2dfitcase2}
\end{figure}

\begin{figure}[htbp]
\begin{center}
\fbox{\LARGE \sf \s2t13 $<$ 10$^{-4}$ and i.h.}
\end{center}
\begin{center}
\Large \bf 3 kton LAr
\end{center}
\vspace{-0.7cm}
%\epsfig{file=EPS/2dfitcase2ihp_2.eps,width=0.55\linewidth}
%\hspace{-1cm}
%\epsfig{file=EPS/2dfitcase2ihp_3.eps,width=0.55\linewidth}
%\vspace{-1cm}
\begin{center}
\epsfig{file=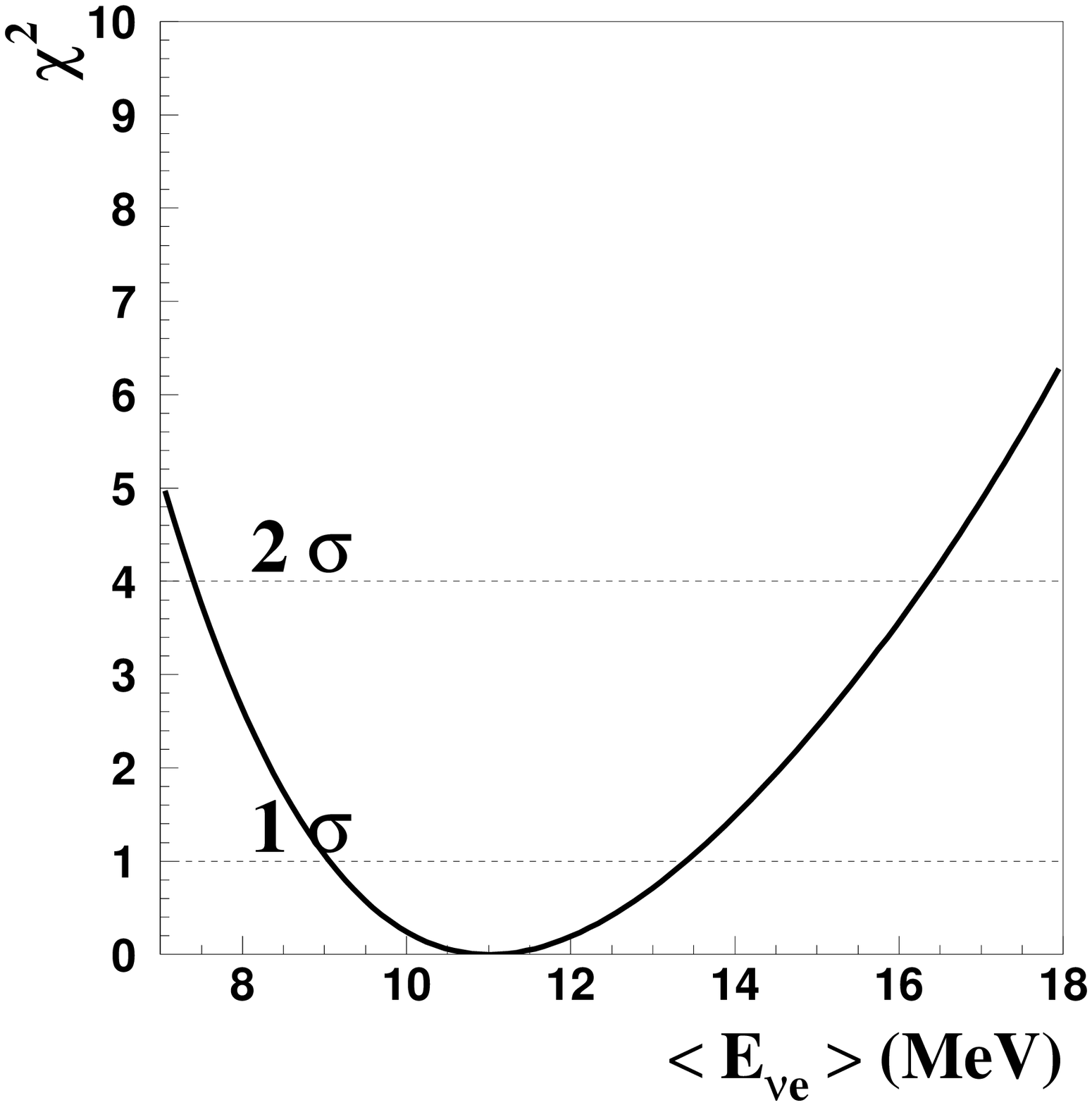,width=0.3\linewidth} \hspace{-0.5cm}
\epsfig{file=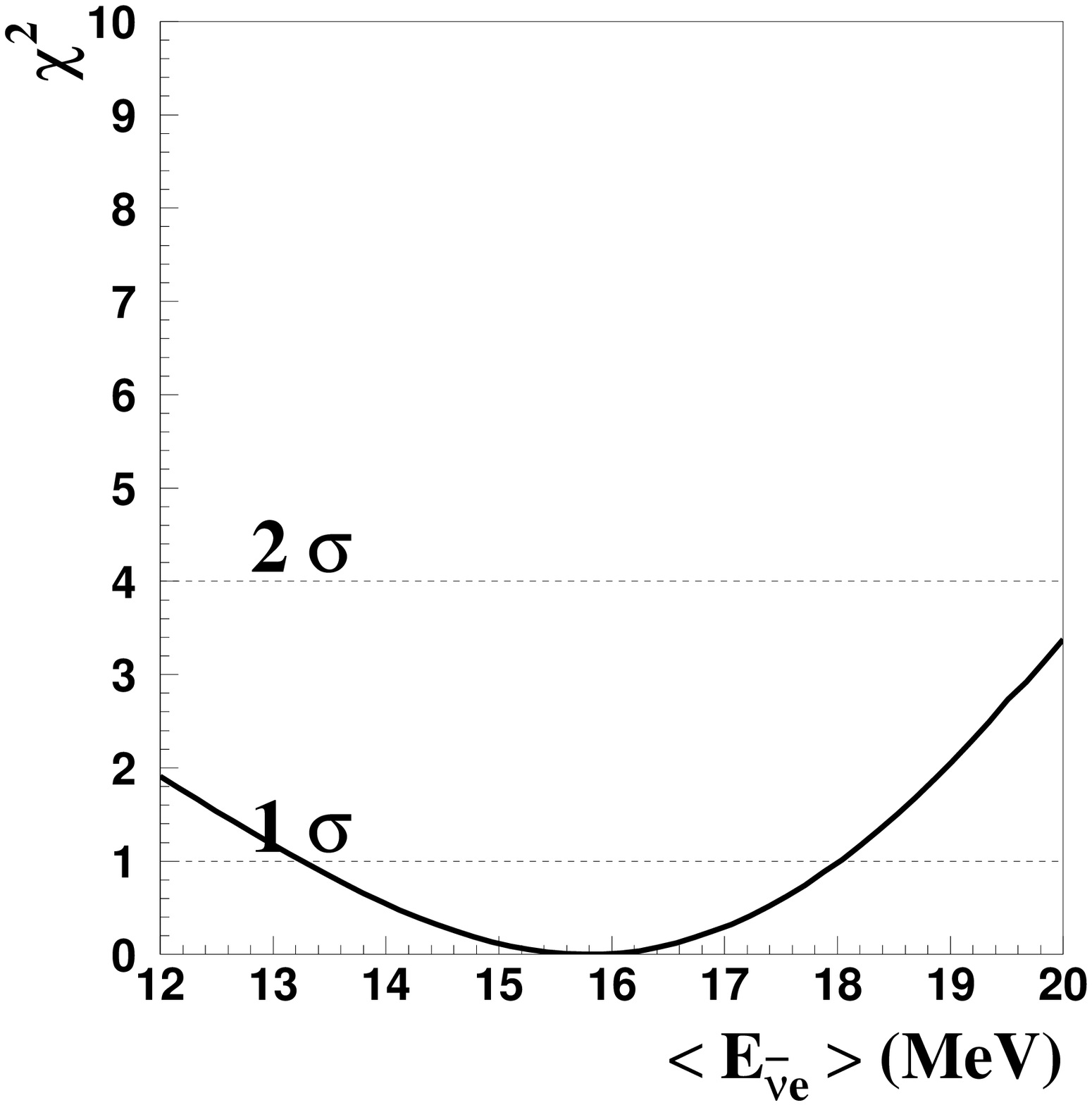,width=0.3\linewidth} \hspace{-0.5cm}
\epsfig{file=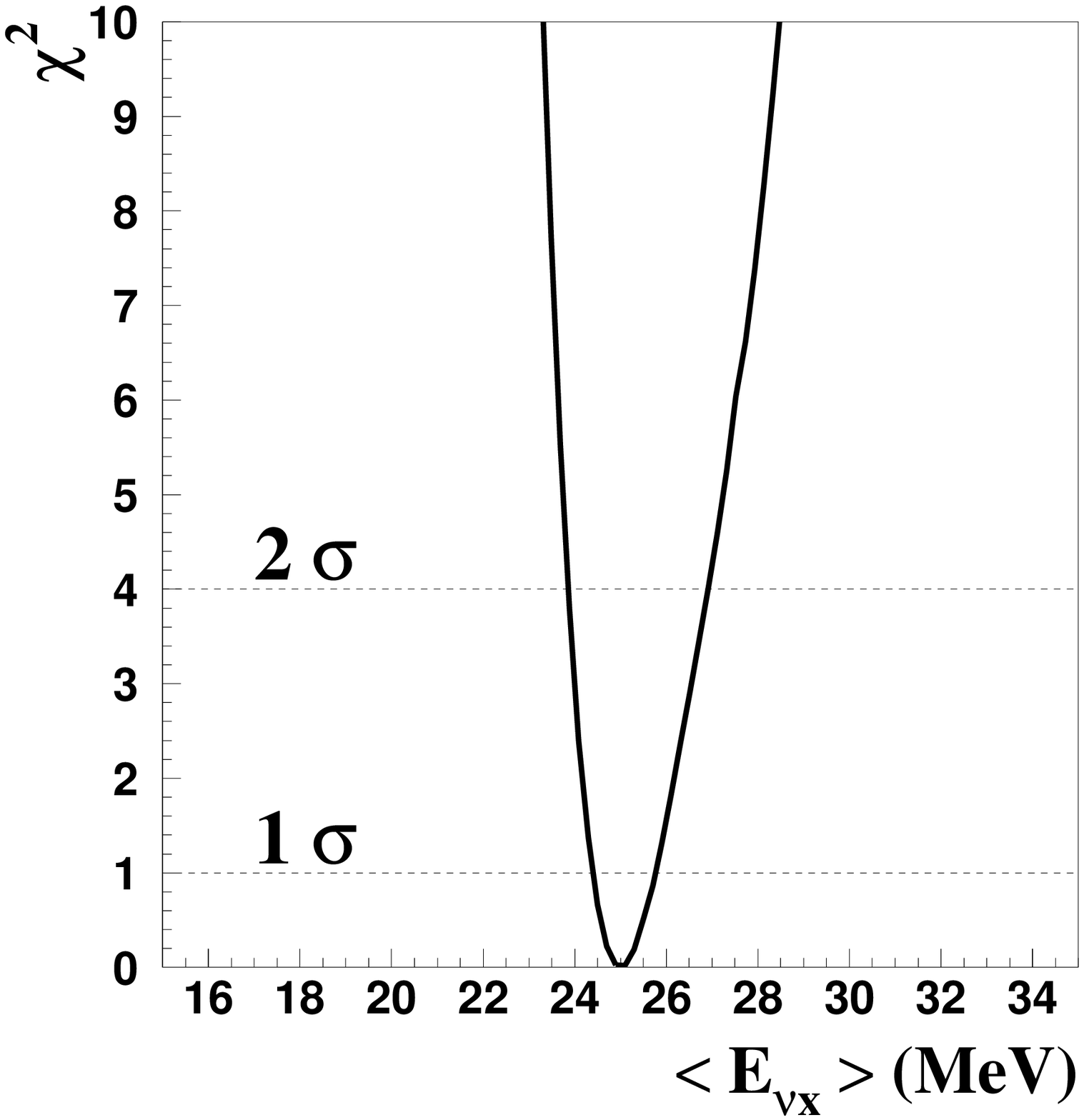,width=0.3\linewidth}
\end{center}
\vspace{-1cm}
\begin{center}
\epsfig{file=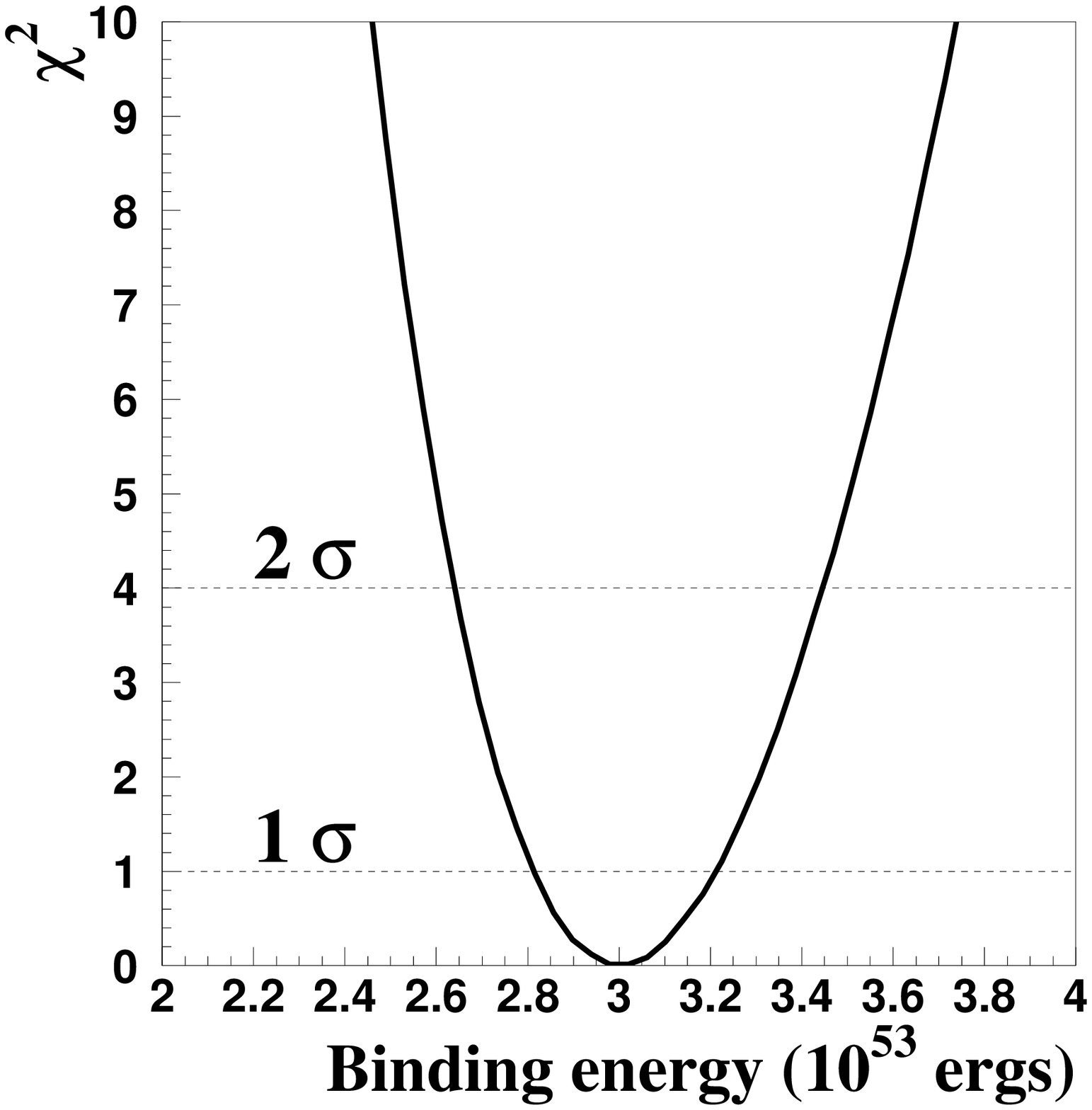,width=0.3\linewidth} \hspace{-0.5cm}
\epsfig{file=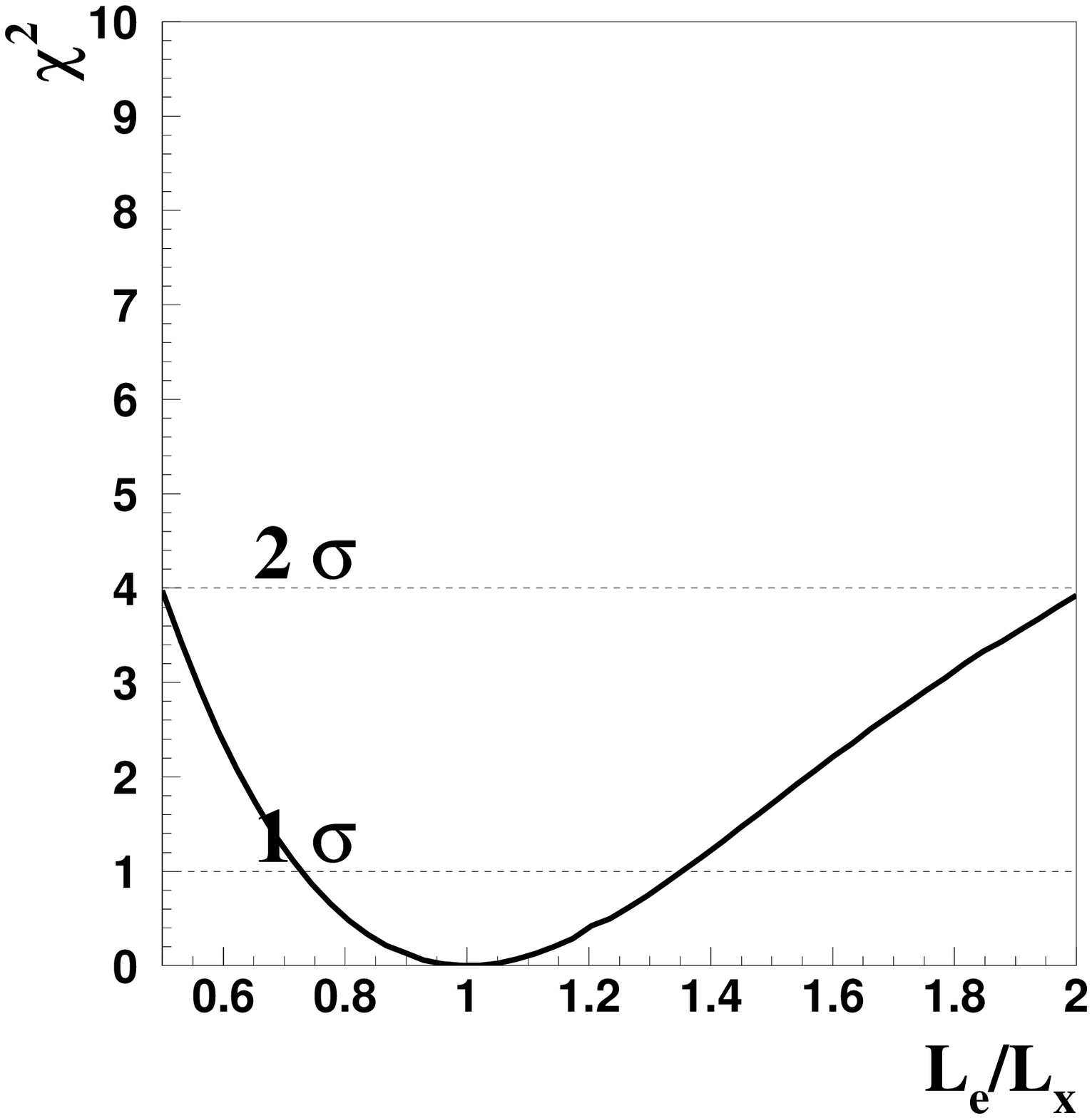,width=0.3\linewidth}
\end{center}

\begin{center}
\Large \bf 100 kton LAr
\end{center}
\vspace{-0.7cm}
\epsfig{file=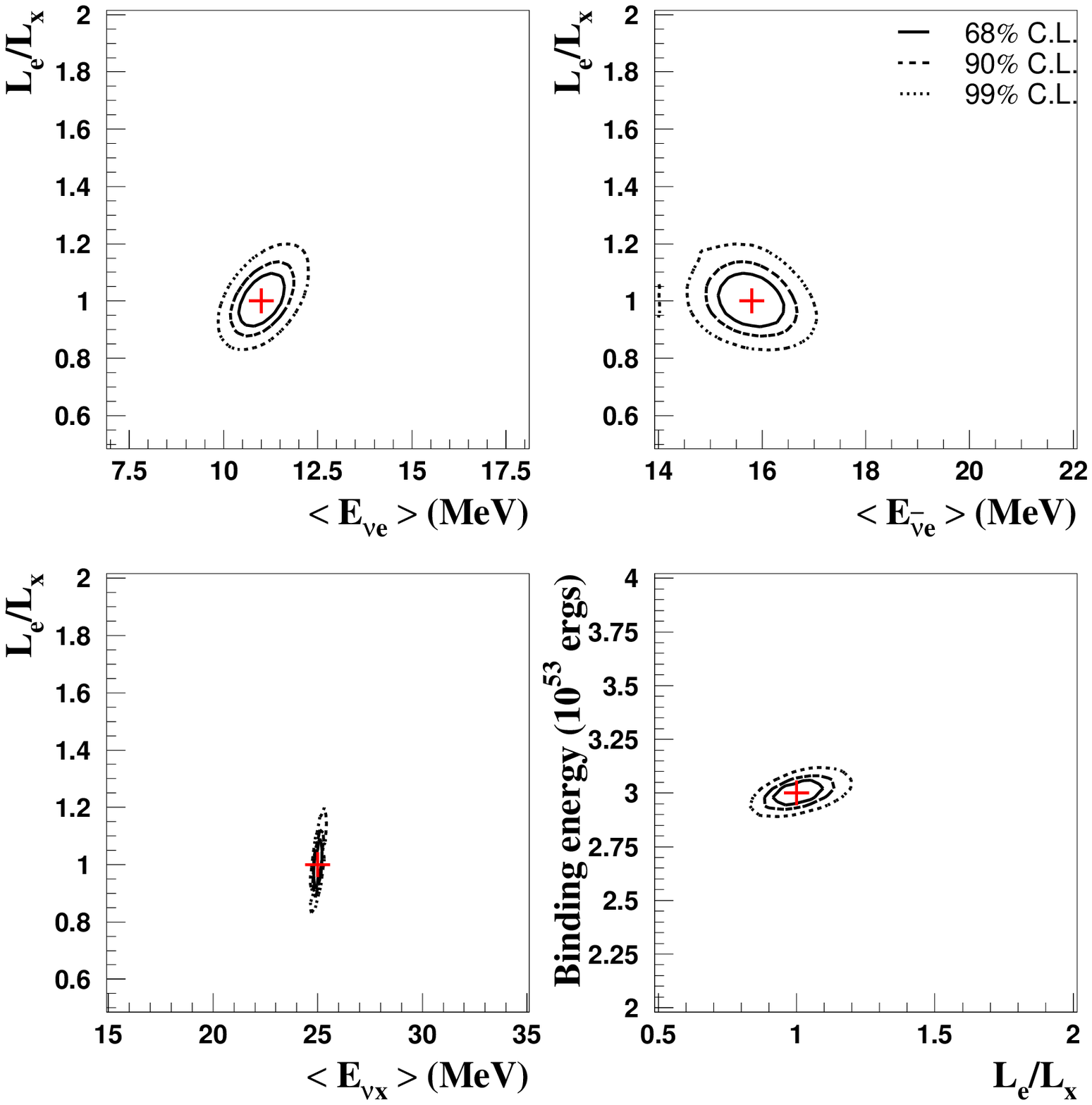,width=0.55\linewidth}
\hspace{-1cm}
\epsfig{file=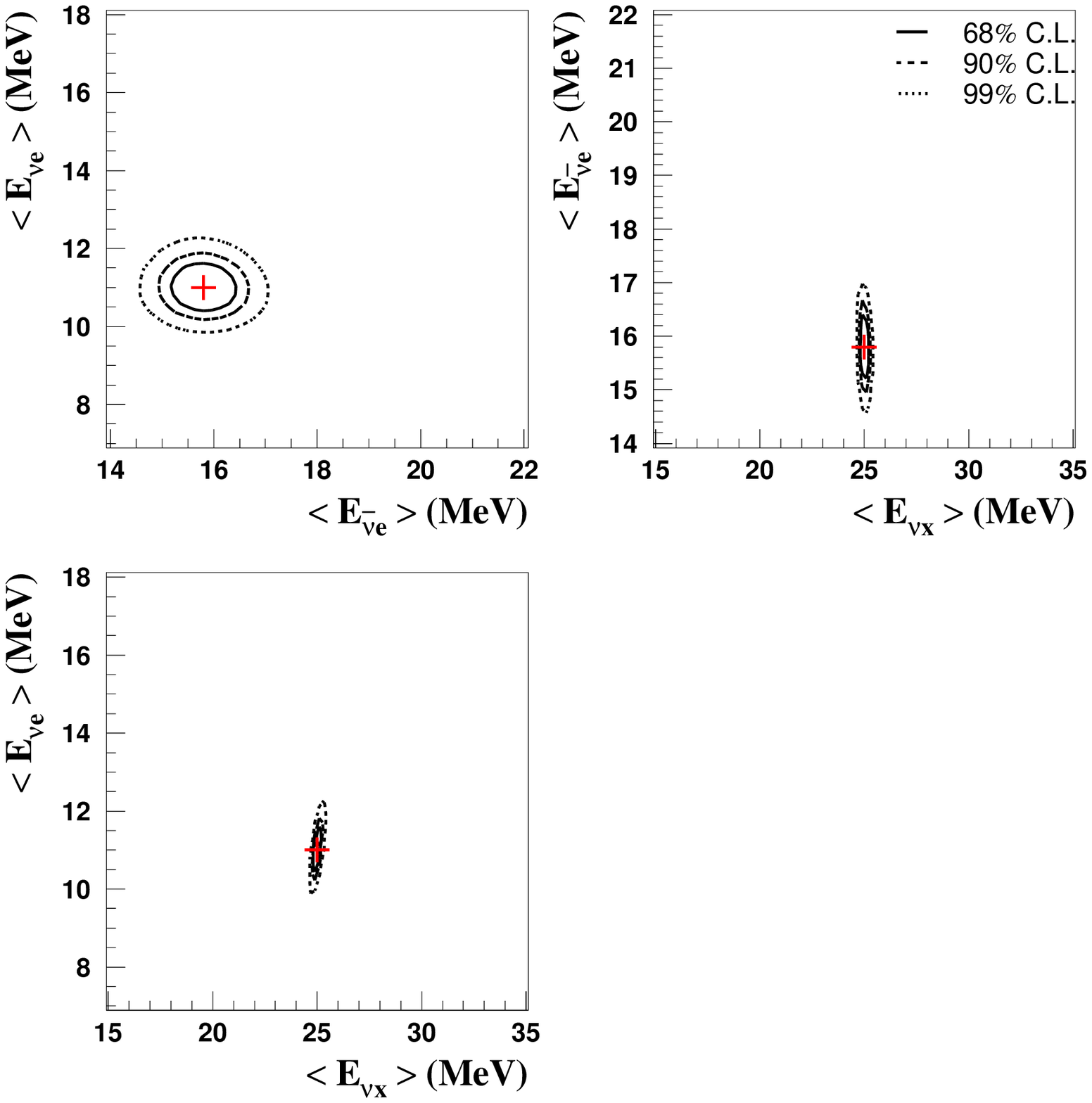,width=0.55\linewidth}
\caption{(Top) $\chi^2$ value of the fit as a function of the
supernova parameters for a 3 kton detector, assuming that an upper
limit on the value of the \th13 mixing angle has been set (\s2t13 $<$
10$^{-4}$) and the mass hierarchy is inverted ($\dm31$ $<$ 0). (Bottom)
68\%, 90\% and 99\% C.L. allowed regions for the supernova parameters
with a 100 kton detector. Crosses indicate the value of the parameters
for the best fits.}
\label{fig:2dfitcase2ih}
\end{figure}

Figures \ref{fig:2dfitcase2} and \ref{fig:2dfitcase2ih} show on the
top the $\chi^2$ value of the fit as a function of the supernova
parameters for a 3 kton detector considering that an upper limit on
the value of \th13 has been set (\s2t13 $<$ 10$^{-4}$) and for normal
or inverted mass hierarchy, respectively. The bottom plots are the
allowed regions at 68\%, 90\% and 99\% C.L. on different supernova
parameters planes for a 100 kton detector. A value of \s2t13 =
10$^{-7}$ was taken as reference for the fit, i.e. in the small mixing
angle region. However, the fitted angle can be in both intermediate
and small mixing angle regions. We have checked that if the ``true''
value of the angle is \s2t13 = 10$^{-5}$, the results of the $\chi^2$
fit are only slightly modified.

In this case the poor determination of the parameters obtained with 3
ktons indicates that statistics are not enough for providing good
results. However, a 100 kton detector will solve these problems.

Normal and inverted hierarchy cases give different results
because they depend on the value of the angle. For \s2t13 $\lesssim$
10$^{-6}$ it is impossible to distinguish between both
hierarchies. But if the angle is the intermediate region (2 $\times$
10$^{-6}$ $<$ \s2t13 $<$ 3 $\times$ 10$^{-4}$) there is a dependence
with the neutrino energy and the results depend on the particular
value of \th13. This is specially important in the n.h.~case,
where the survival probability changes with energy in the neutrino
channel, which is the most sensitive for LAr detectors. 

With a 100 kton detector, the expected accuracies are similar for
normal and inverted hierarchies, as shown in table
\ref{tab:accurasnparcase2}. Thanks to the high statistics available
with a very massive detector, we can fit very precisely the data and
the results are essentially equal for both mass hierarchies, as expected in the
small \th13 region.  
 
\begin{figure}[htbp]
\begin{center}
\fbox{\LARGE \sf \s2t13 $<$ 10$^{-4}$ and n.h. (scen II)}
\end{center}
\begin{center}
\Large \bf 3 kton LAr
\end{center}
\vspace{-0.7cm}
%\epsfig{file=EPS/2dfitcase2scen2p_2bis.eps,width=0.55\linewidth}
%\hspace{-1cm}
%\epsfig{file=EPS/2dfitcase2scen2p_3bis.eps,width=0.55\linewidth}
%\vspace{-1cm}
\begin{center}
\epsfig{file=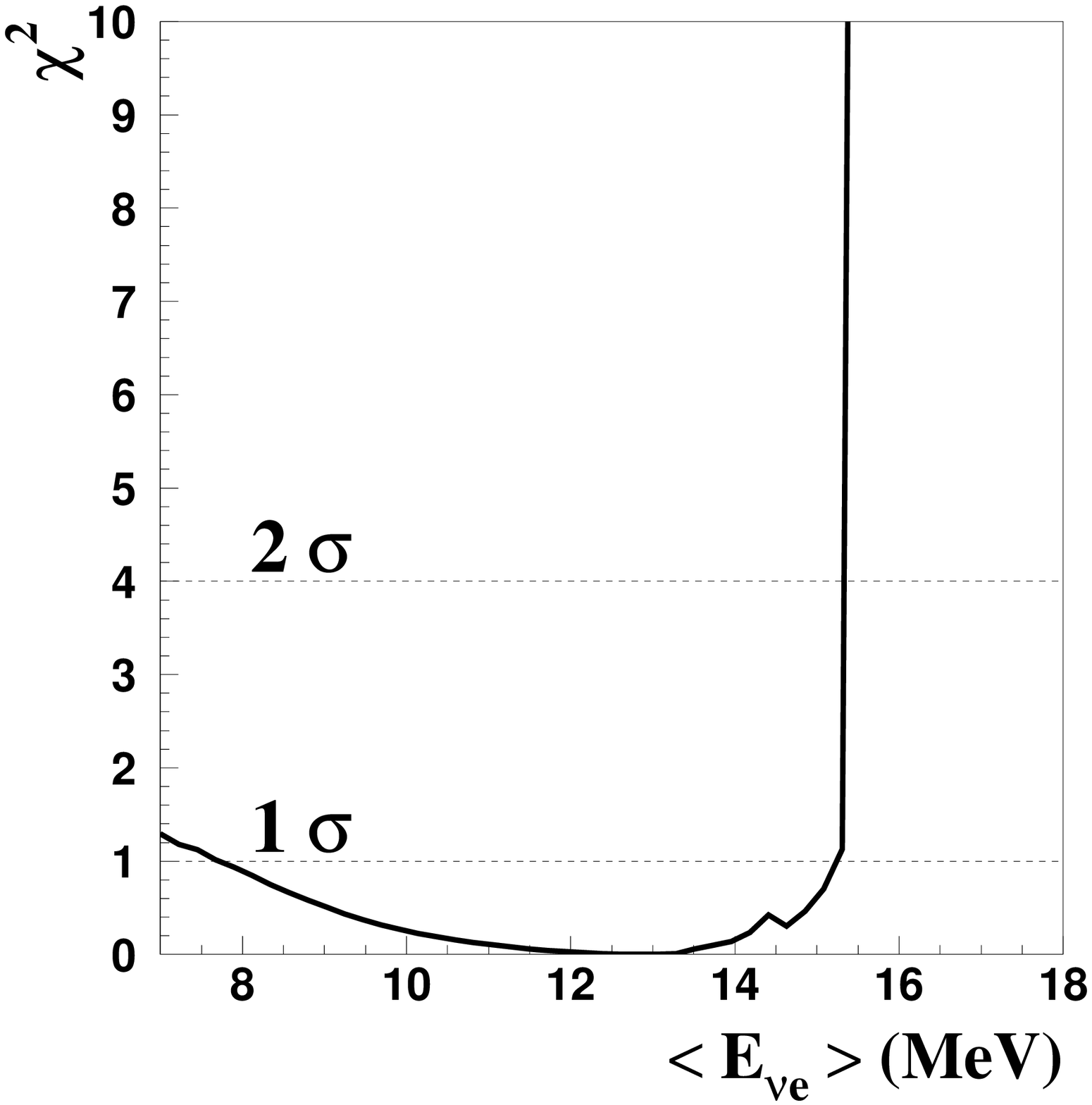,width=0.3\linewidth} \hspace{-0.5cm}
\epsfig{file=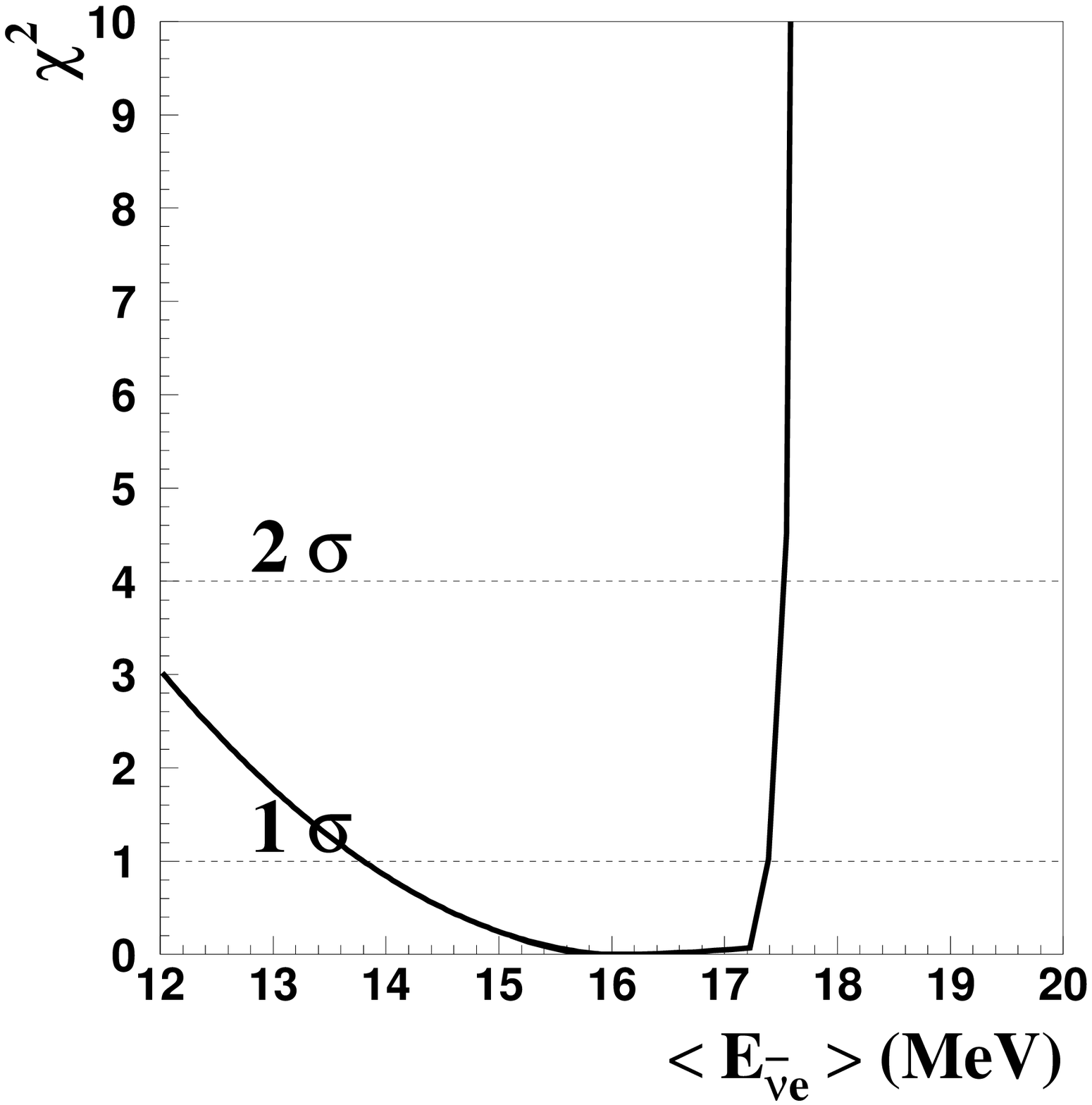,width=0.3\linewidth} \hspace{-0.5cm}
\epsfig{file=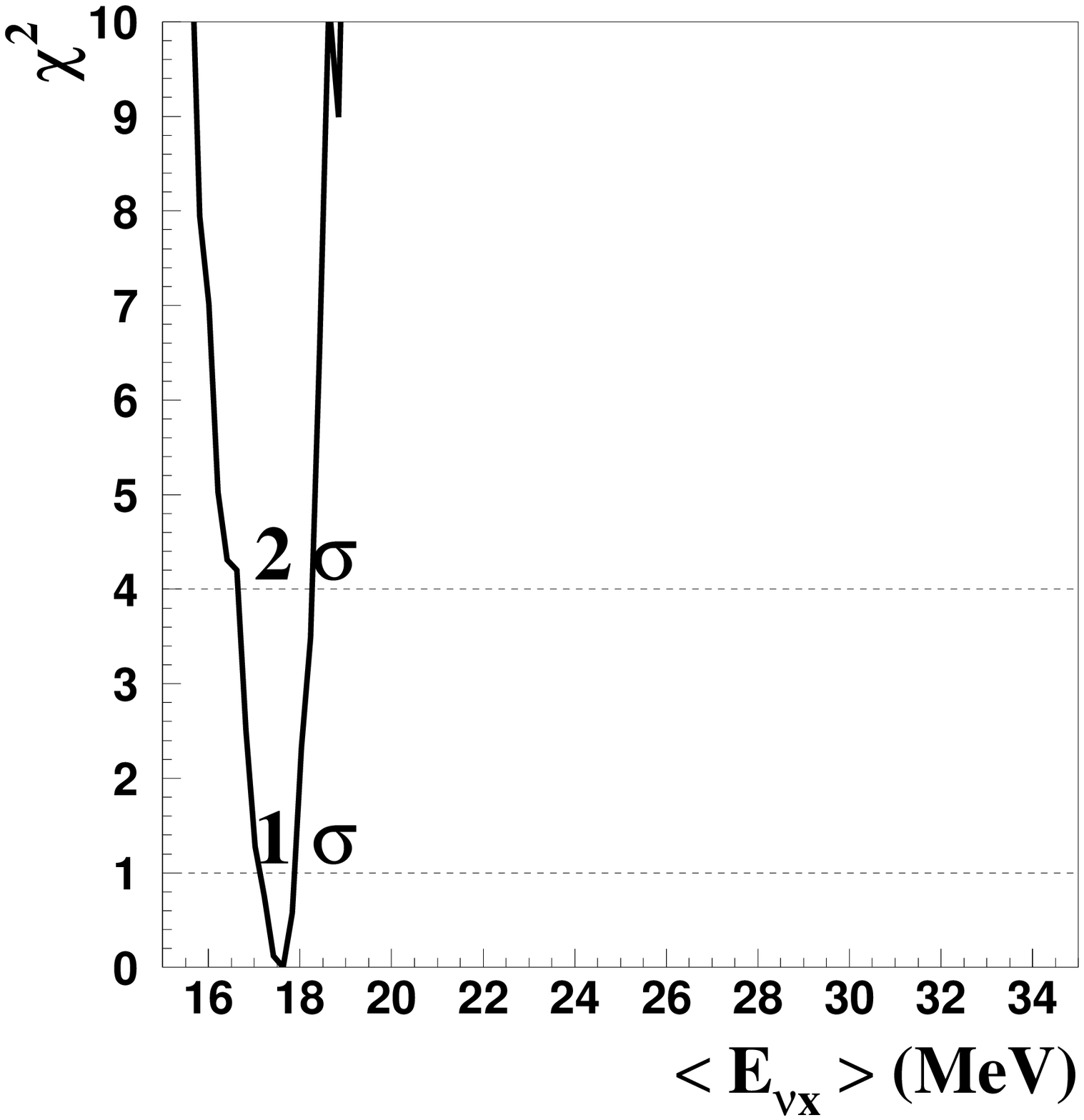,width=0.3\linewidth}
\end{center}
\vspace{-1cm}
\begin{center}
\epsfig{file=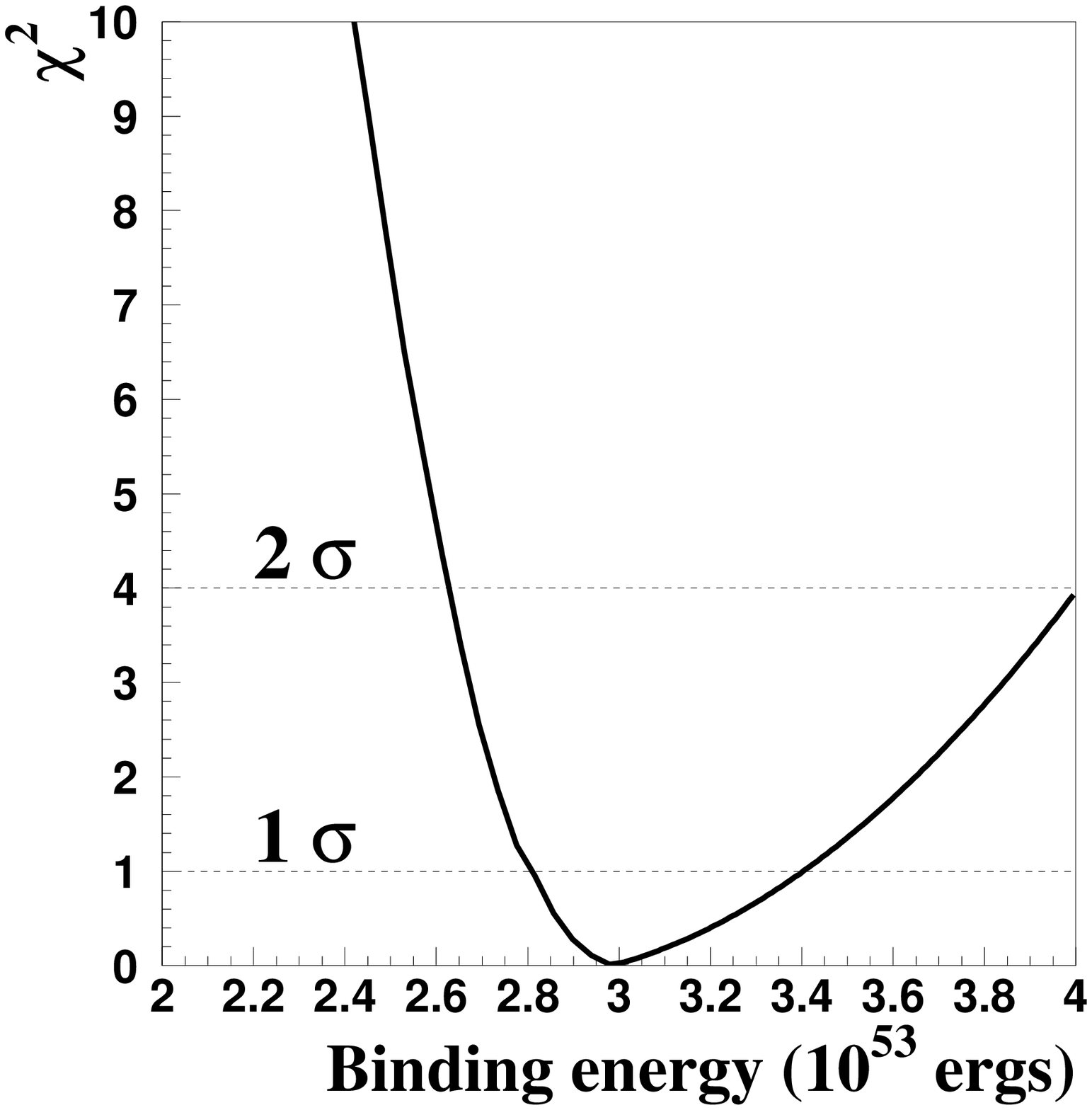,width=0.3\linewidth} \hspace{-0.5cm}
\epsfig{file=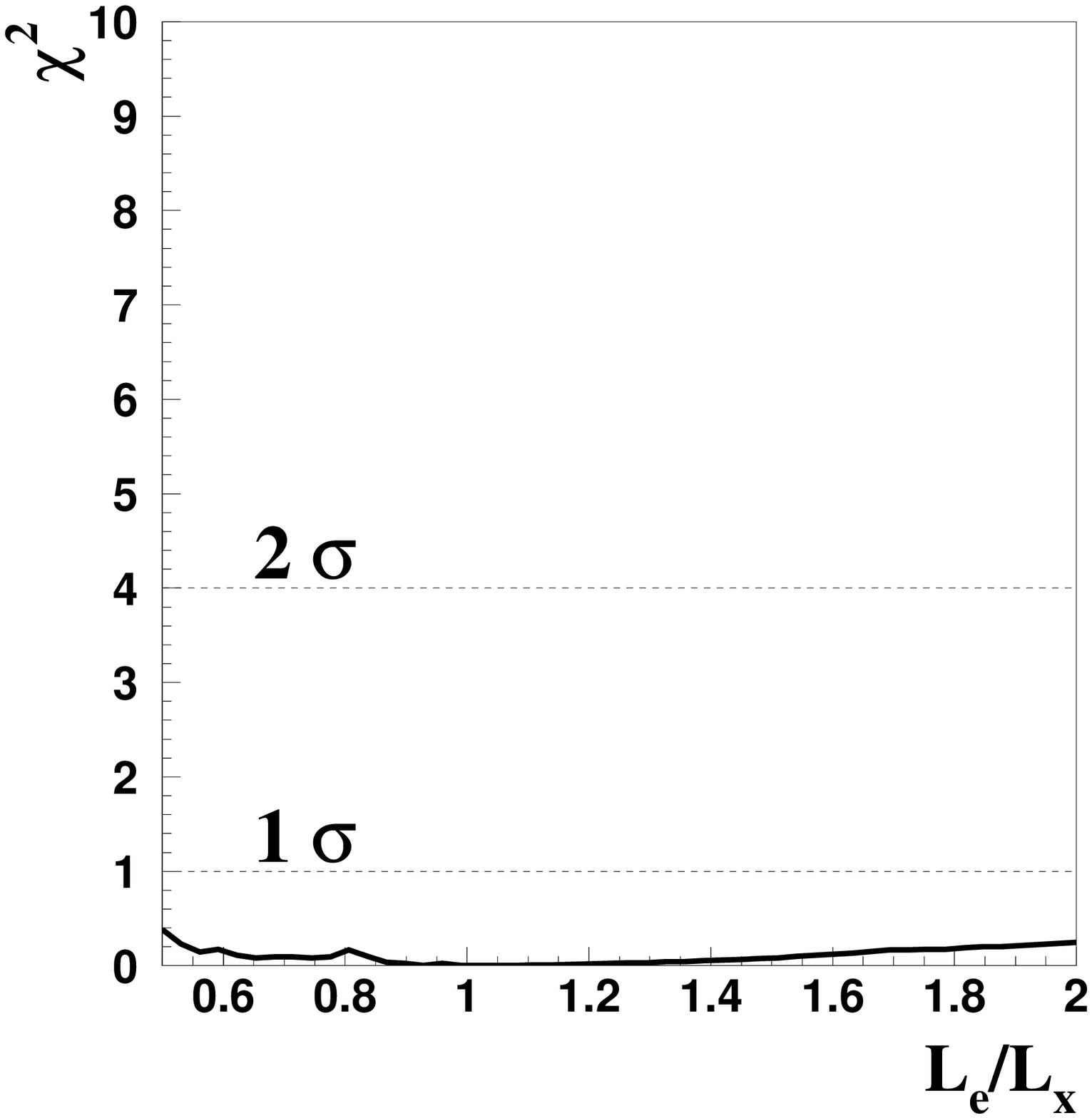,width=0.3\linewidth}
\end{center}

\begin{center}
\Large \bf 100 kton LAr
\end{center}
%\vspace{-0.7cm}
\epsfig{file=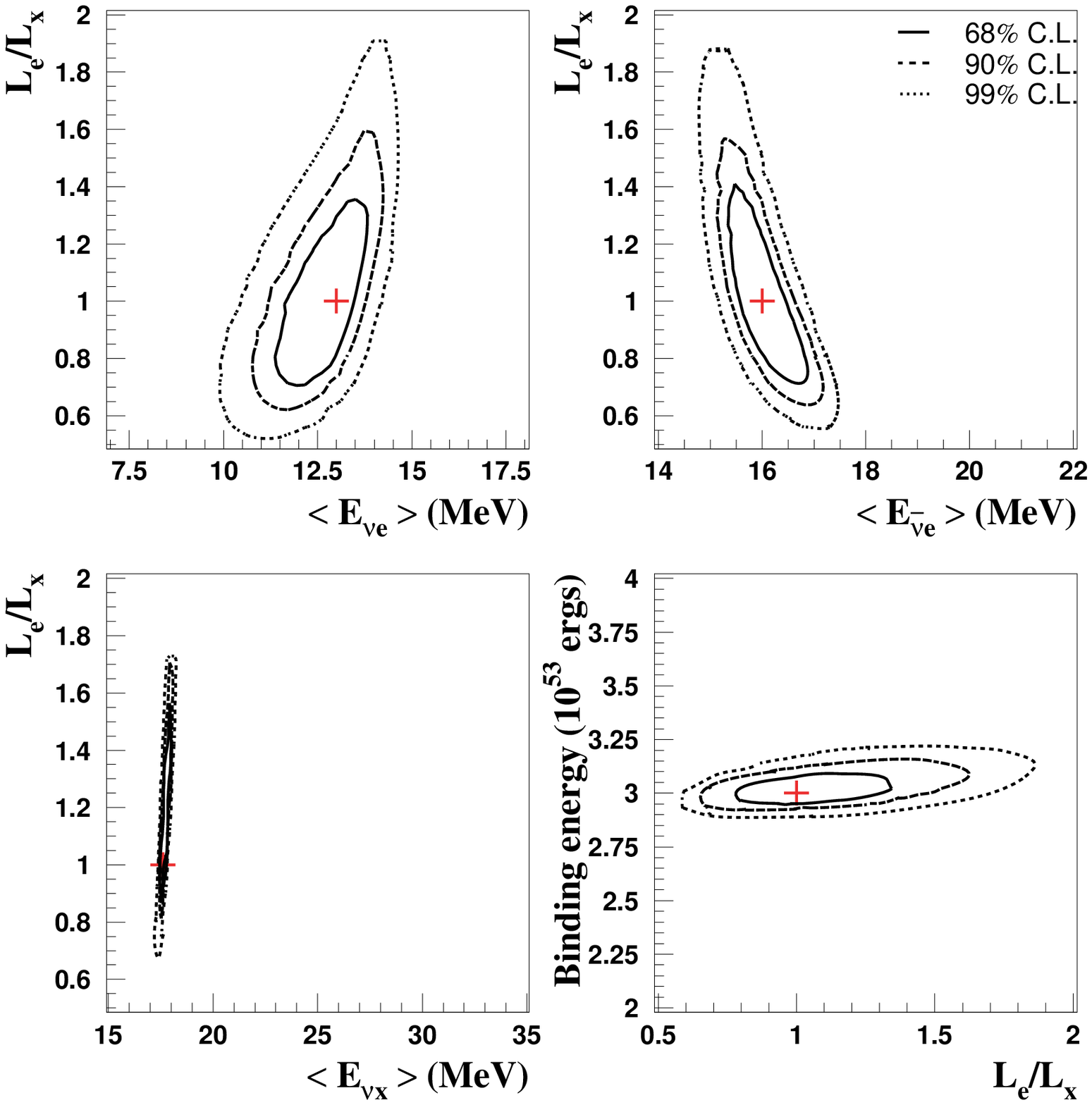,width=0.49\linewidth}
%\hspace{-1cm}
\epsfig{file=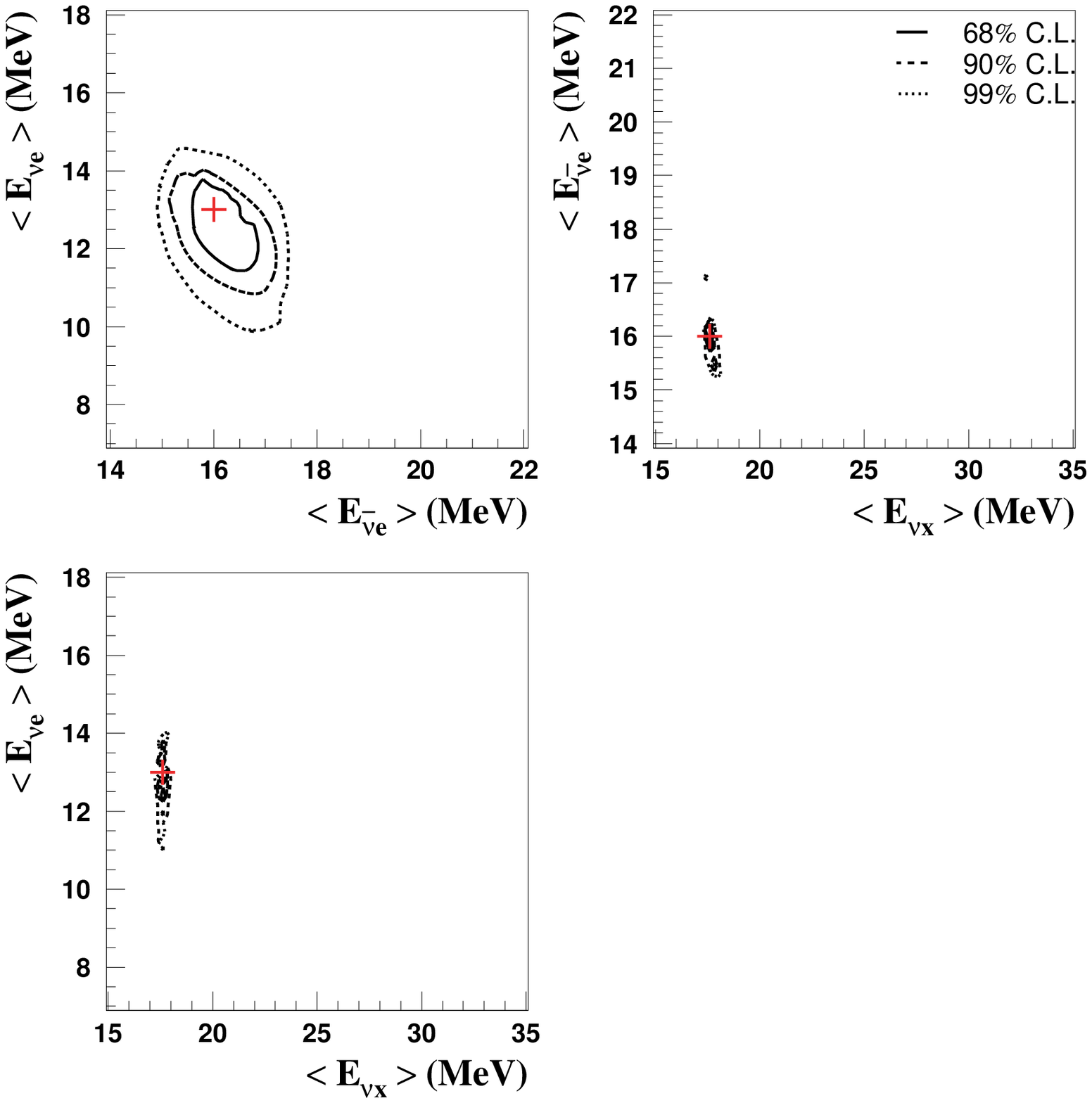,width=0.49\linewidth}
\caption{(Top) $\chi^2$ value of the fit as a function of the
supernova parameters for a 3 kton detector, assuming that an upper
limit on the value of the \th13 mixing angle has been set (\s2t13 $<$
10$^{-4}$) and the mass hierarchy is normal ($\dm31$ $>$ 0). The
reference values taken for the neutrino average energies are the ones
of scenario II (see table \ref{tab:sncoolscenario}). (Bottom) 68\%,
90\% and 99\% C.L. allowed regions for the supernova parameters with a
100 kton detector. Crosses indicate the value of the parameters for
the best fits.}
\label{fig:2dfitcase2scen2}
\end{figure}

\begin{figure}[htbp]
\begin{center}
\fbox{\LARGE \sf \s2t13 $<$ 10$^{-4}$ and i.h. (scen II)}
\end{center}
\begin{center}
\Large \bf 3 kton LAr
\end{center}
\vspace{-0.7cm}
%\epsfig{file=EPS/2dfitcase2ihscen2p_2bis.eps,width=0.55\linewidth}
%\hspace{-1cm}
%\epsfig{file=EPS/2dfitcase2ihscen2p_3tris.eps,width=0.55\linewidth}
%\vspace{-1cm}
\begin{center}
\epsfig{file=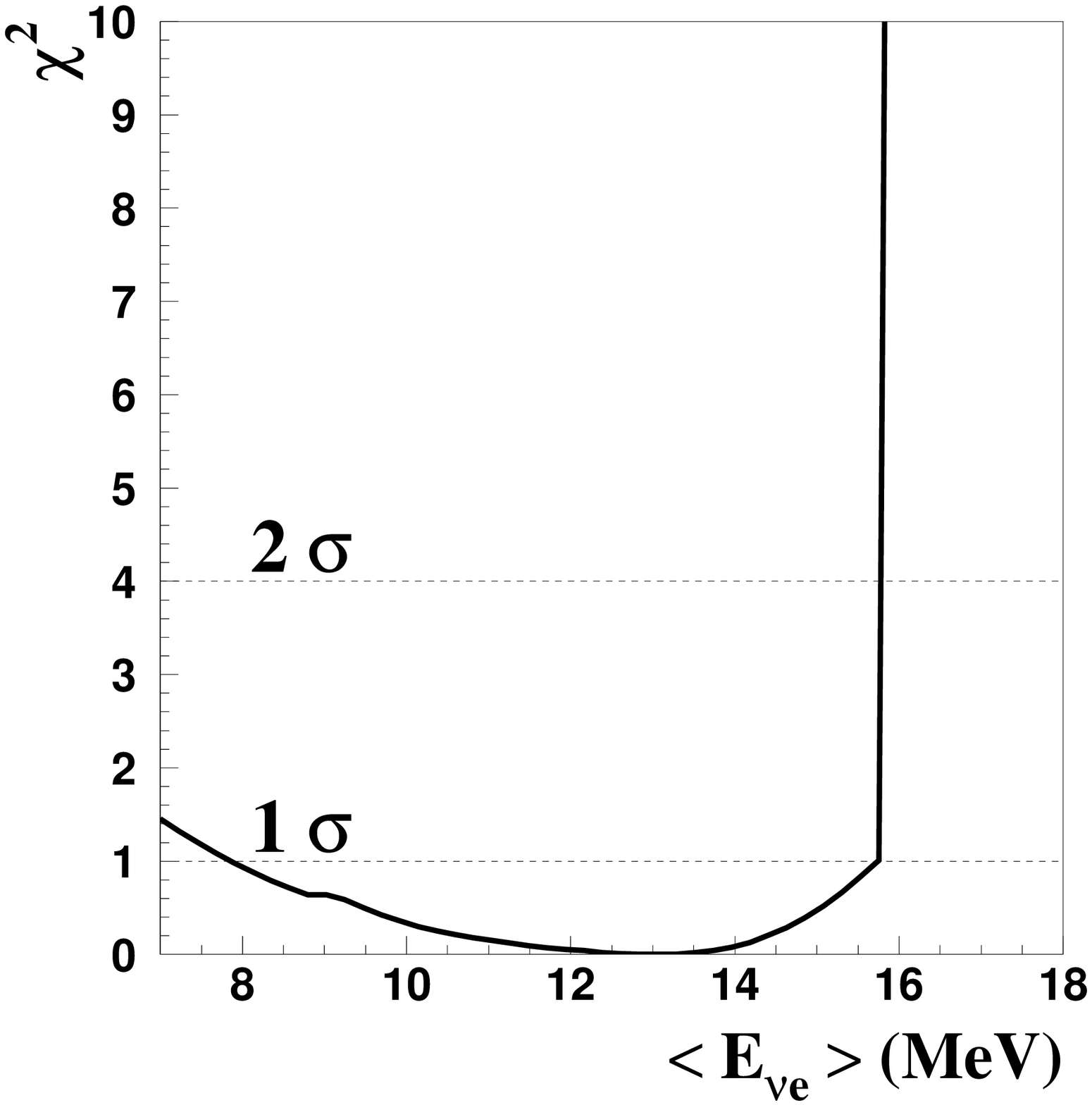,width=0.3\linewidth} \hspace{-0.5cm}
\epsfig{file=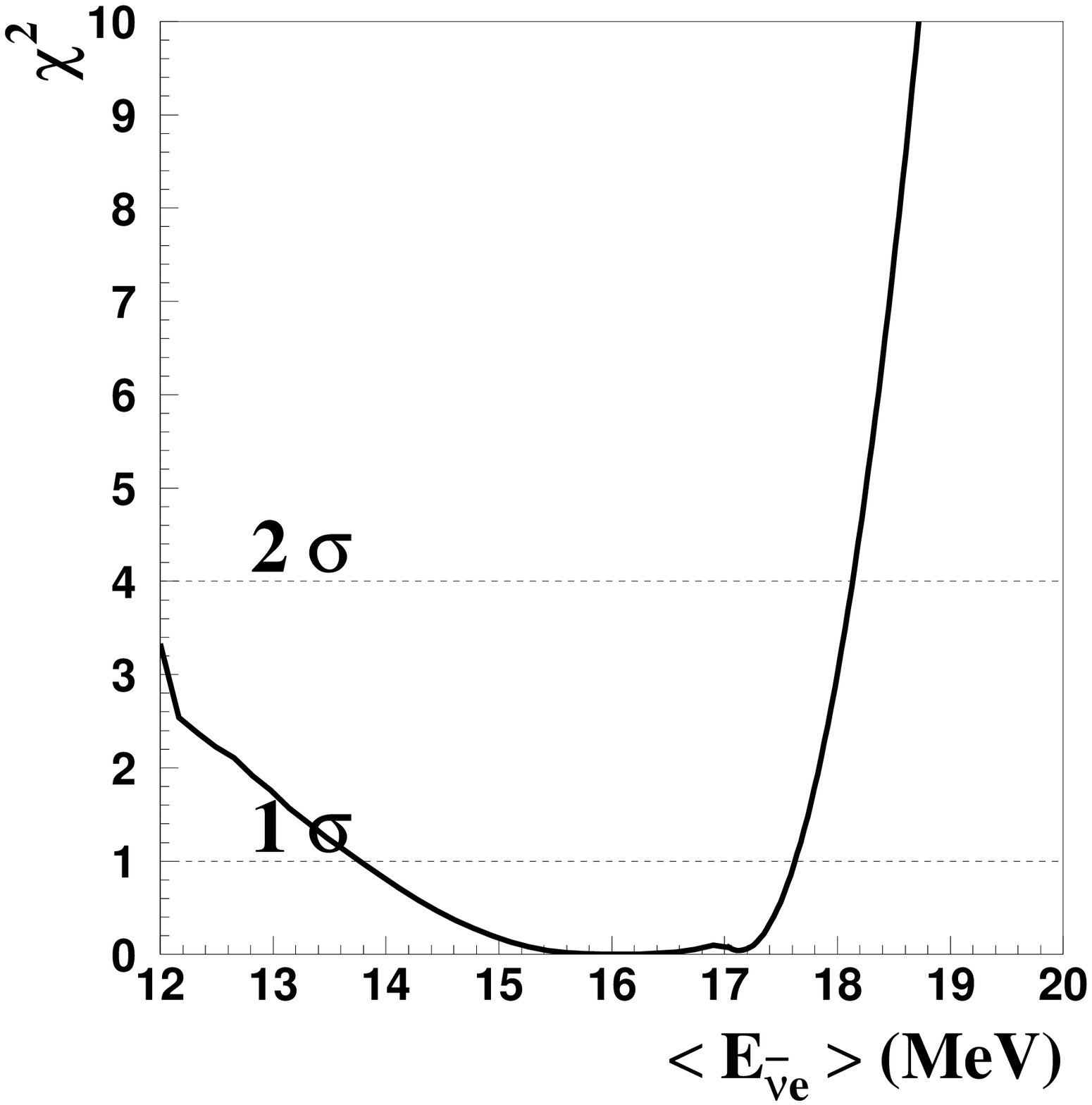,width=0.3\linewidth} \hspace{-0.5cm}
\epsfig{file=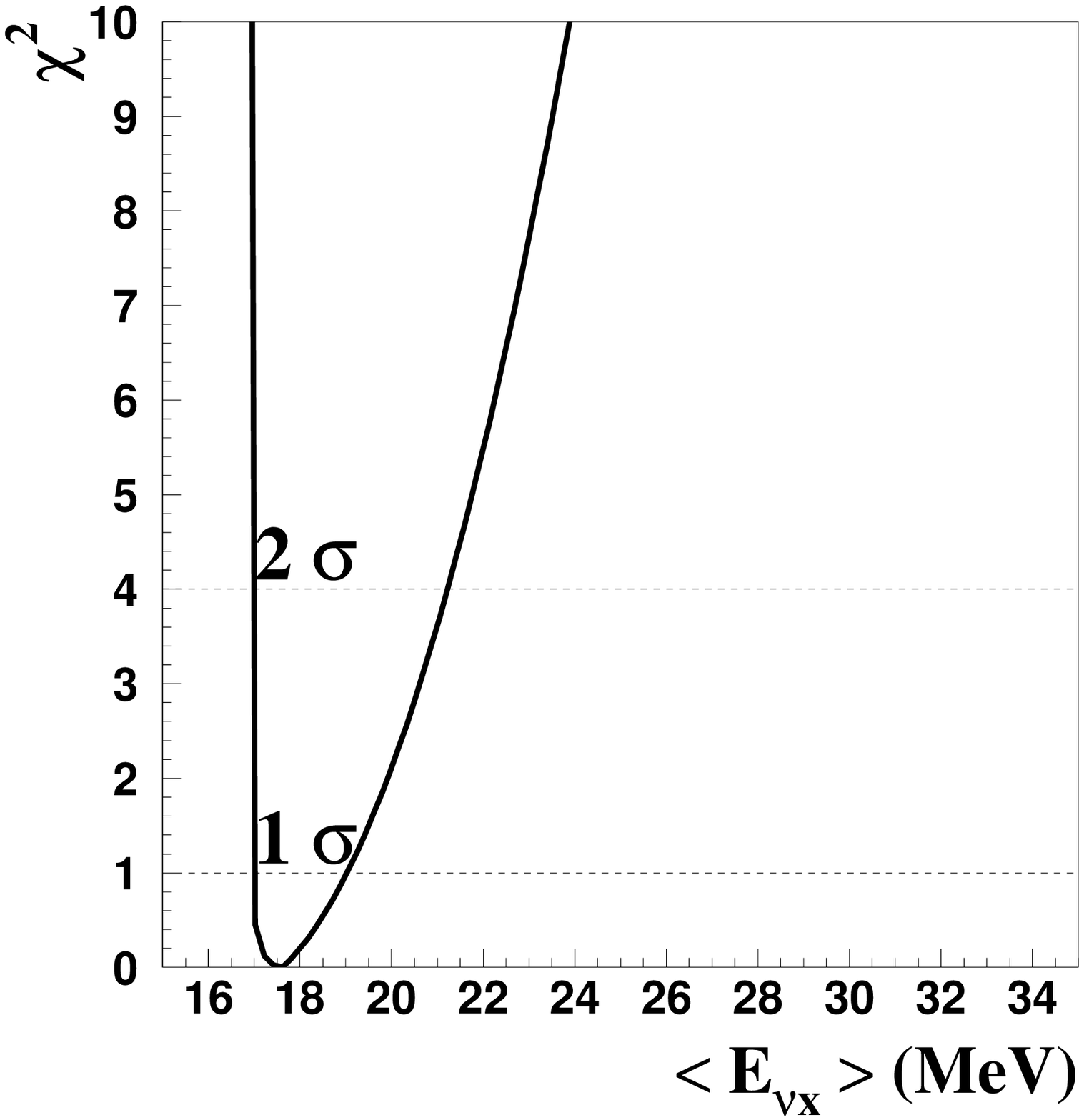,width=0.3\linewidth}
\end{center}
\vspace{-1cm}
\begin{center}
\epsfig{file=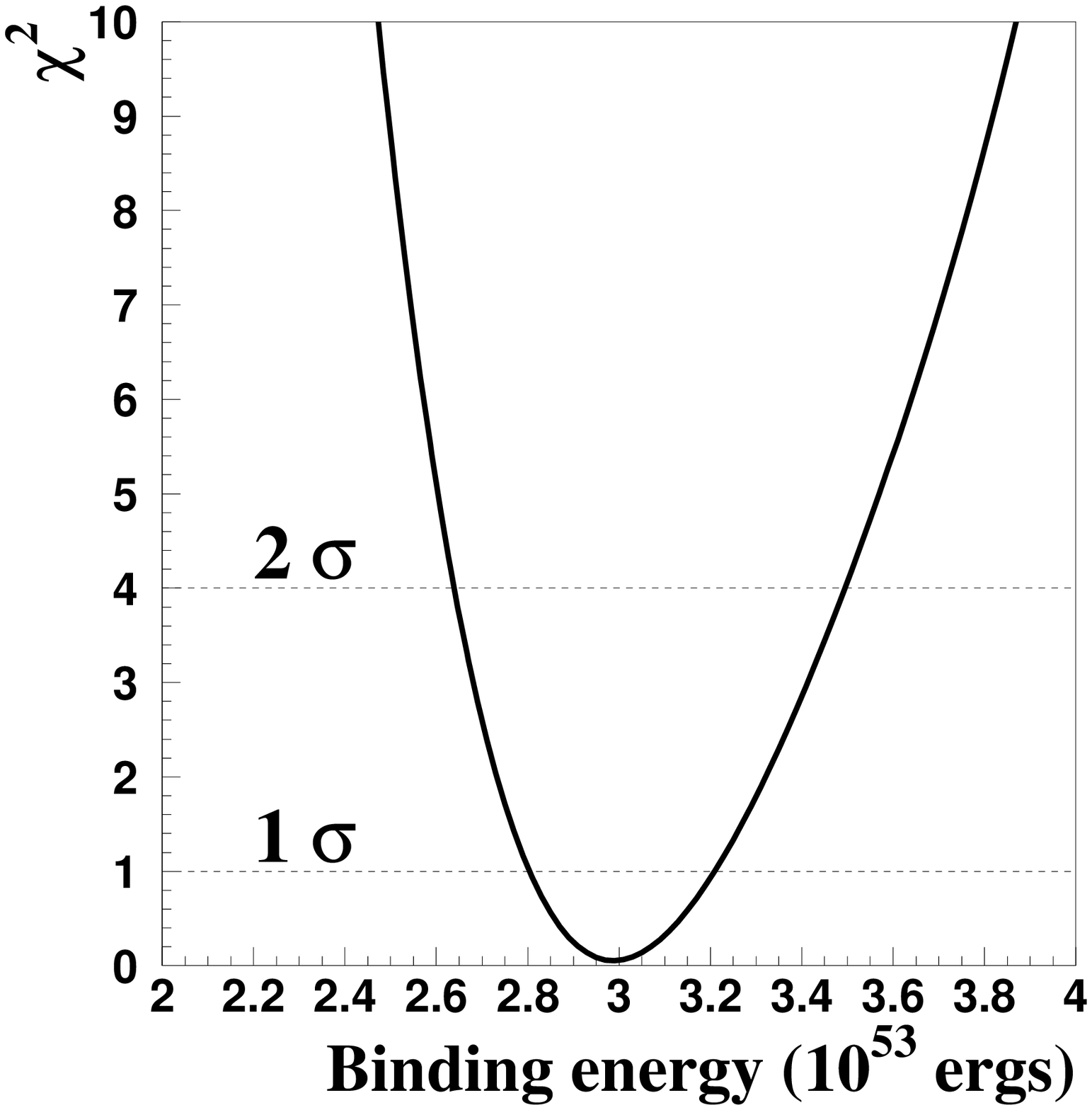,width=0.3\linewidth} \hspace{-0.5cm}
\epsfig{file=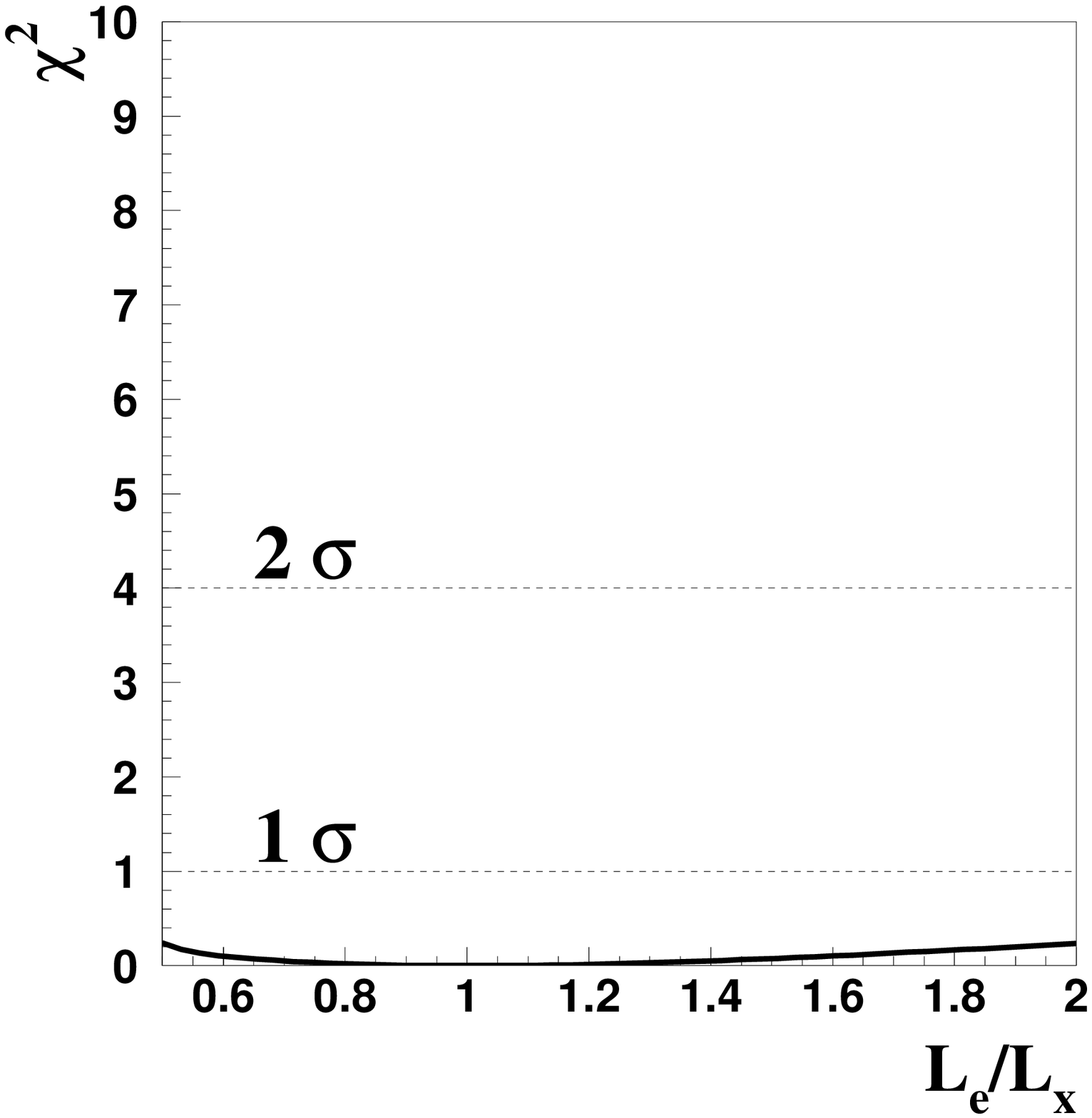,width=0.3\linewidth}
\end{center}

\begin{center}
\Large \bf 100 kton LAr
\end{center}
\vspace{-0.7cm}
\epsfig{file=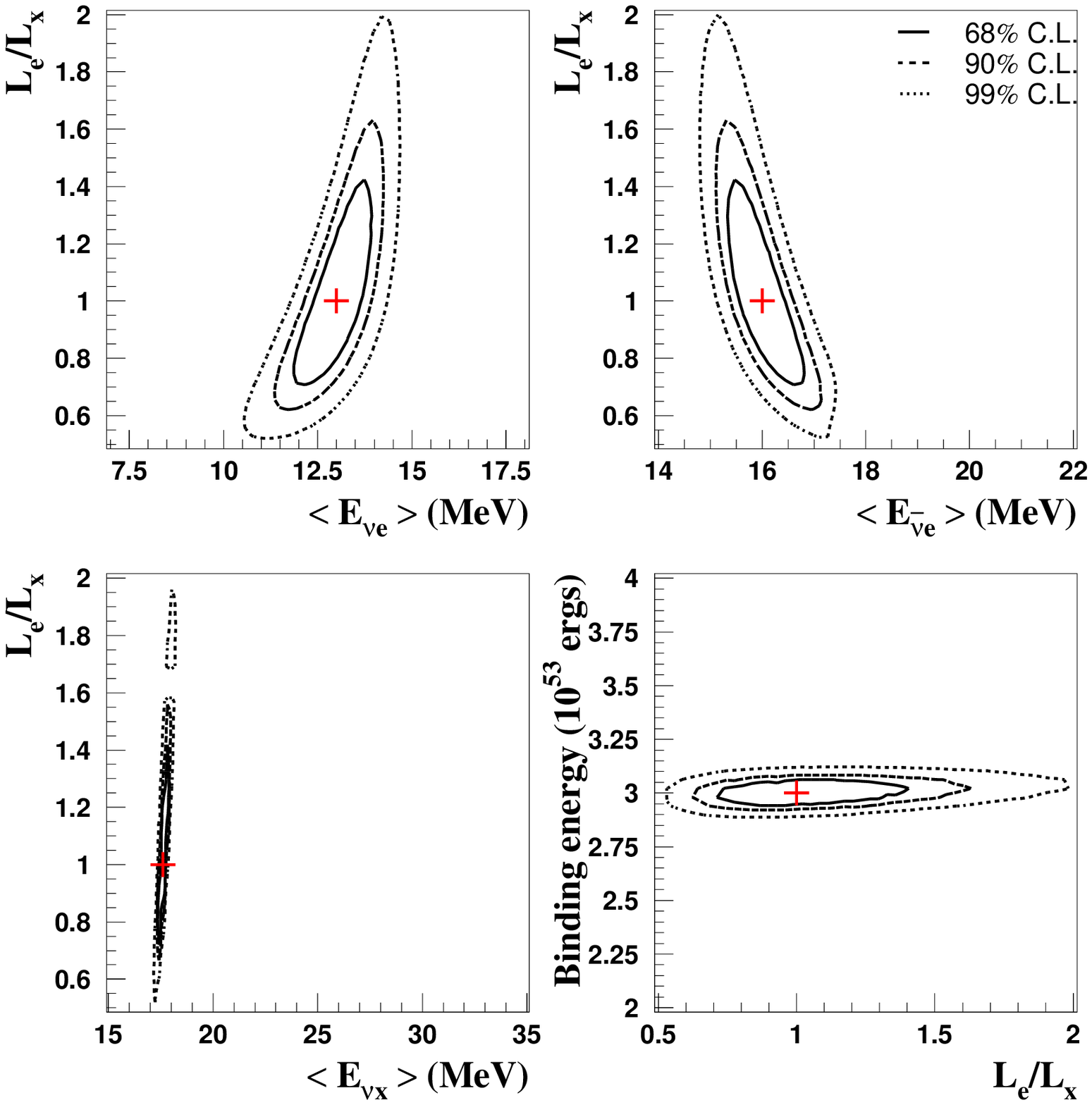,width=0.55\linewidth}
\hspace{-1cm}
\epsfig{file=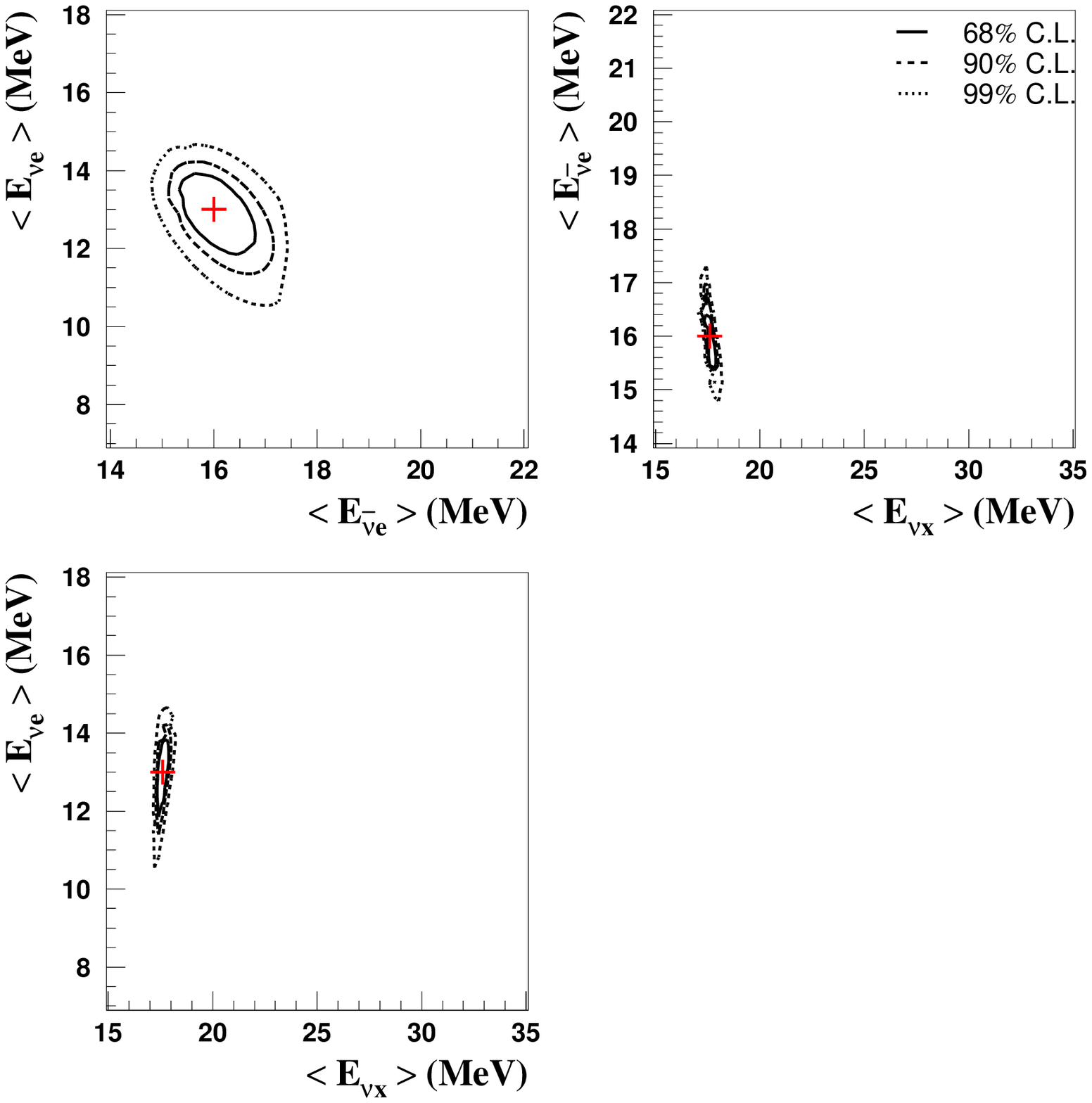,width=0.55\linewidth}
\caption{(Top) $\chi^2$ value of the fit as a function of the
supernova parameters for a 3 kton detector, assuming that an upper
limit on the value of the \th13 mixing angle has been set (\s2t13 $<$
10$^{-4}$) and the mass hierarchy is inverted ($\dm31$ $<$ 0). The
reference values taken for the neutrino average energies are the ones
of scenario II (see table \ref{tab:sncoolscenario}). (Bottom) 68\%,
90\% and 99\% C.L. allowed regions for the supernova parameters with a
100 kton detector. Crosses indicate the value of the parameters for
the best fits.}
\label{fig:2dfitcase2ihscen2}
\end{figure}

Considering the parameters of scenario II as reference values, the
corresponding 2D allowed regions are presented in figures
\ref{fig:2dfitcase2scen2} and \ref{fig:2dfitcase2ihscen2}, for both
mass hierarchies. In this case the determination of \avenue ~and
\avenux ~is more complicated due to the non-hierarchical
scenario. Likewise, \lelx ~presents a big uncertainty even with a
100 kton detector.

%%%%%%%%%%%%%%%%%%%%%%%%%%%%%%%%%%%%%%%%%%%%%%%%%%%%%%%%%%%%%%%%%%%%%%%
\section{Study of the supernova parameters without any knowledge on the
neutrino oscillation parameters}
\label{sec:third}
%%%%%%%%%%%%%%%%%%%%%%%%%%%%%%%%

We consider four cases as ``true'' values of the oscillation
parameters: n.h.-L, i.h.-L, n.h.-S, i.h.-S. Without assuming any
value of the mixing angle and mass hierarchy, we perform a $\chi^2$
minimization letting all parameters \{\EB, \avenue, \aveanue, \avenux, \lelx,
\s2t13, sign[$\D32$]\} freely vary and we study how well we can
determine them in the case of a 100~kton detector. For the 3~kton detector,
the results can be rescaled accordingly.

\begin{figure}[htbp]
\begin{center}
\fbox{\LARGE \sf n.h.-L and n.h.-S (all parameters free)}
\end{center}
%\begin{center}
%\Large \bf 3 kton LAr
%\end{center}
\vspace{-1cm}
\begin{center}
\epsfig{file=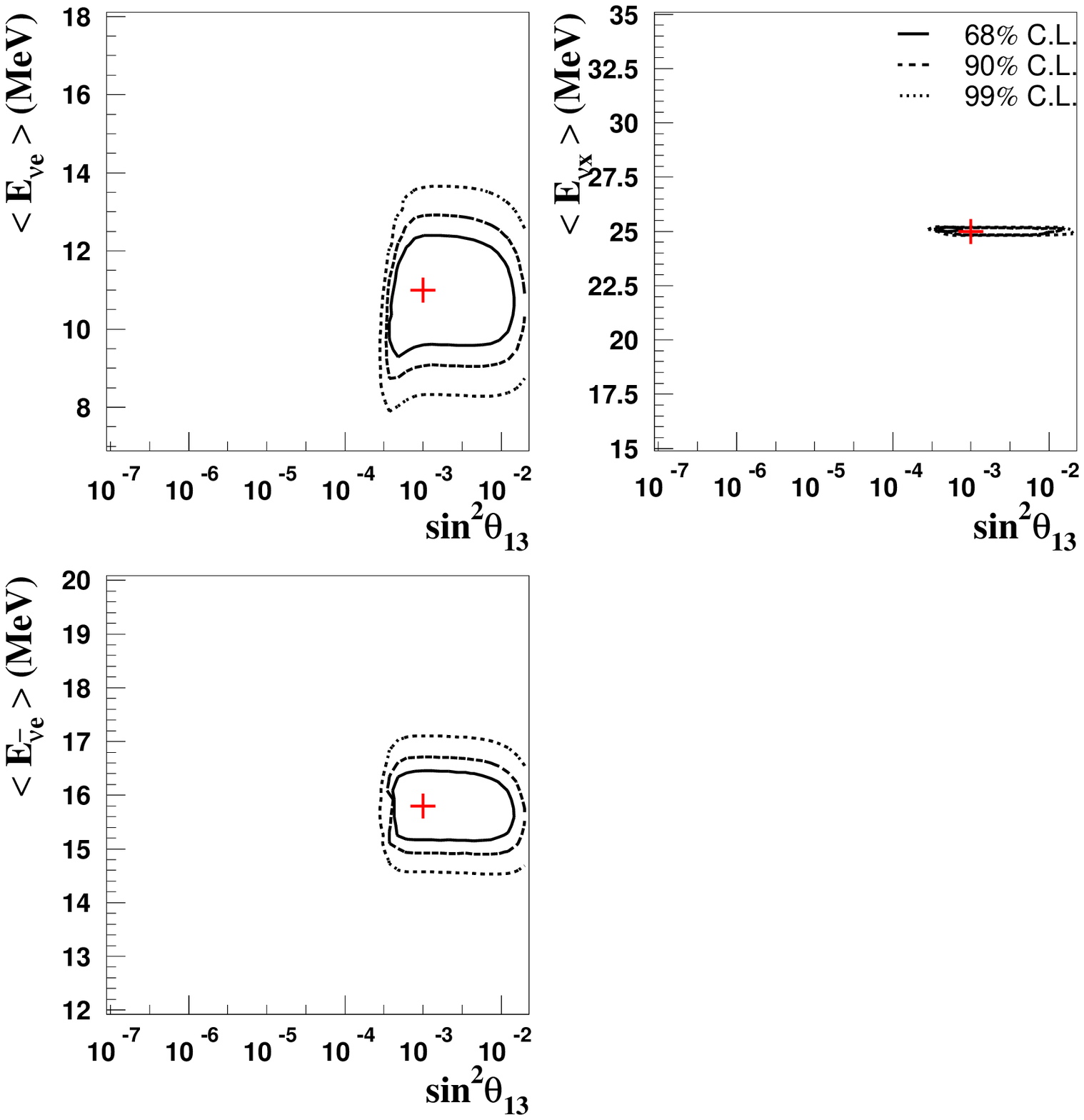,width=0.45\linewidth}
\epsfig{file=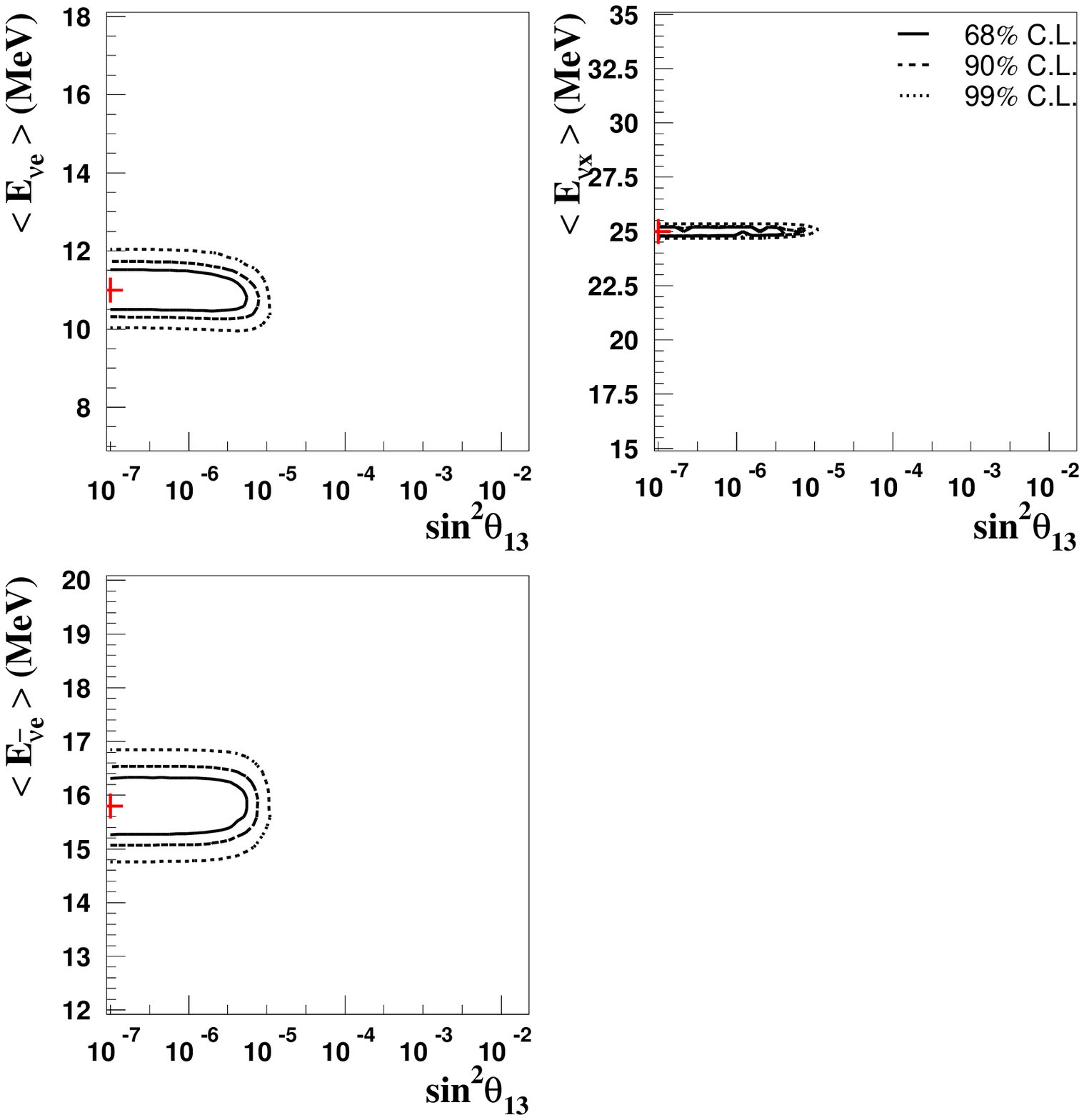,width=0.45\linewidth}
\end{center}
\begin{center}
\vspace{-1cm}
\epsfig{file=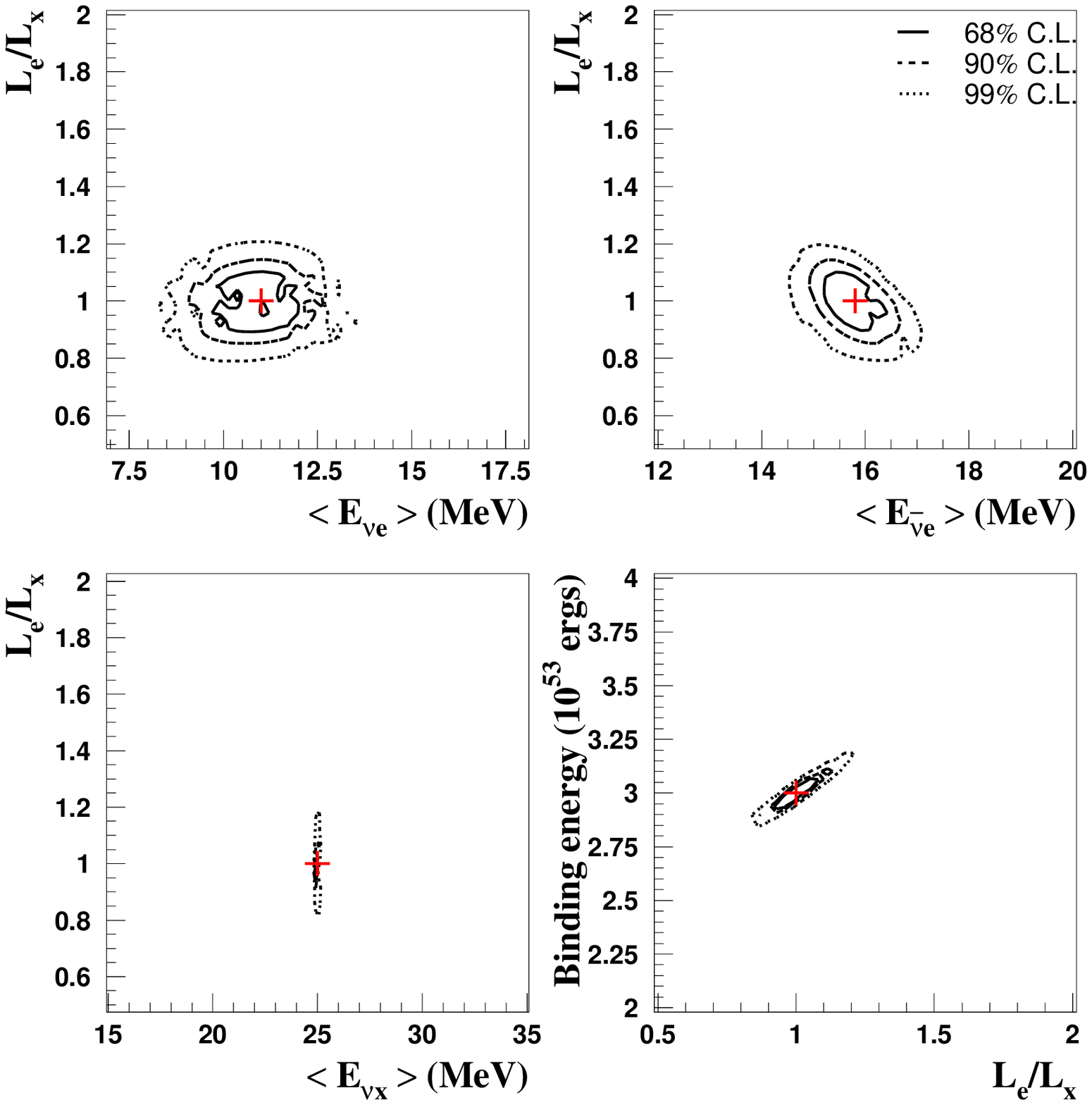,width=0.45\linewidth}
\epsfig{file=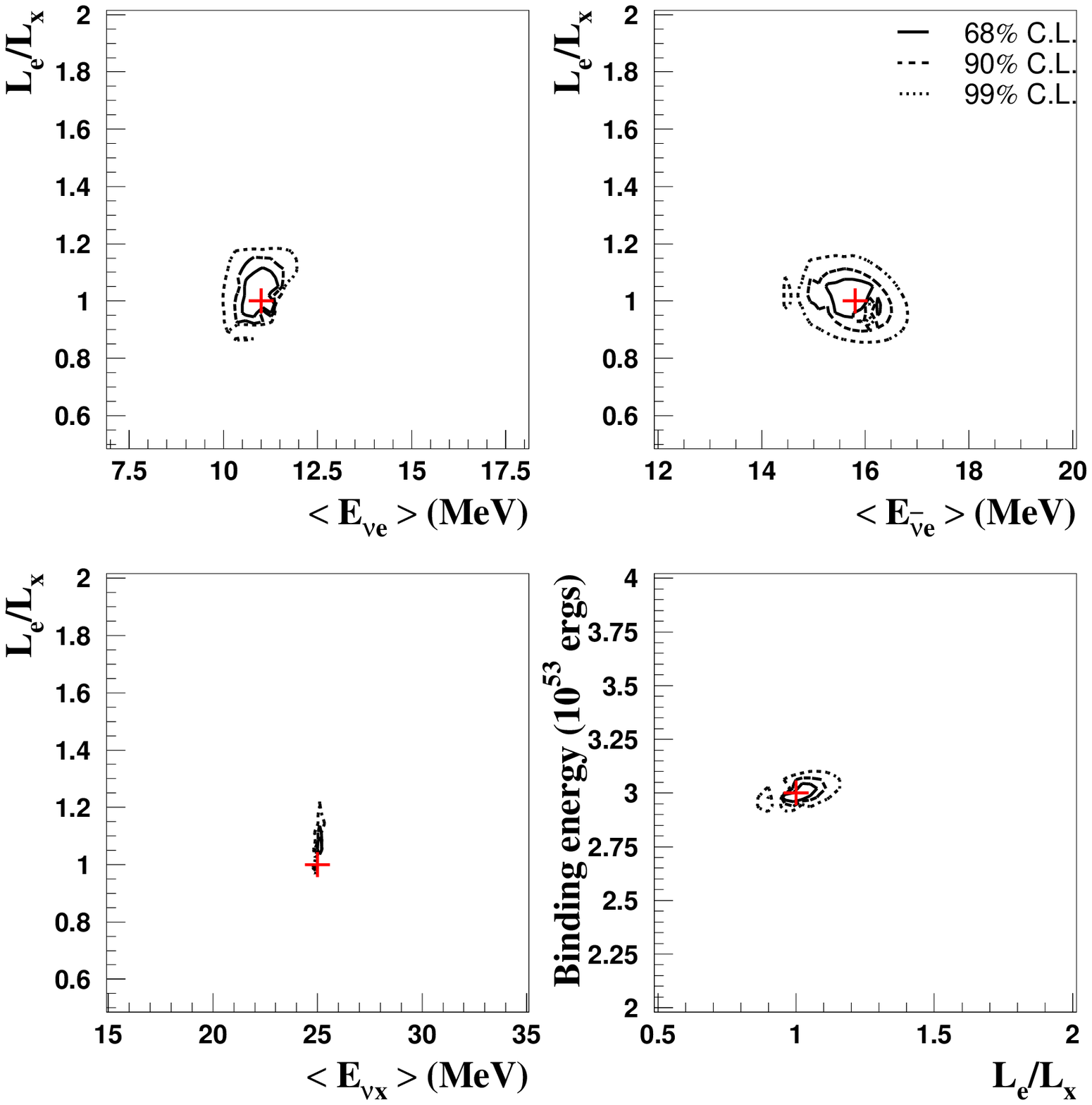,width=0.45\linewidth}
\end{center}
\begin{center}
\vspace{-1cm}
\epsfig{file=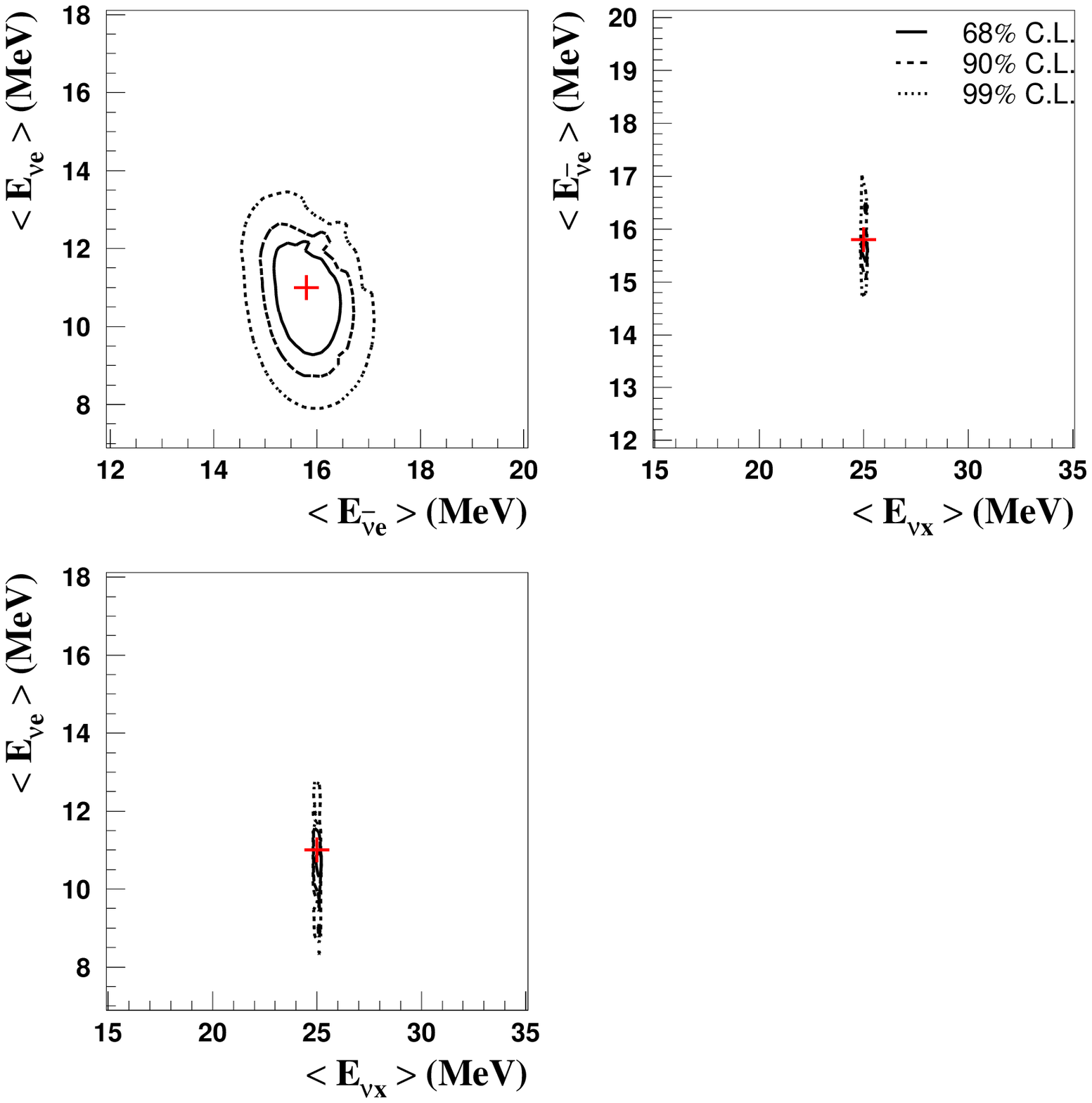,width=0.45\linewidth}
\epsfig{file=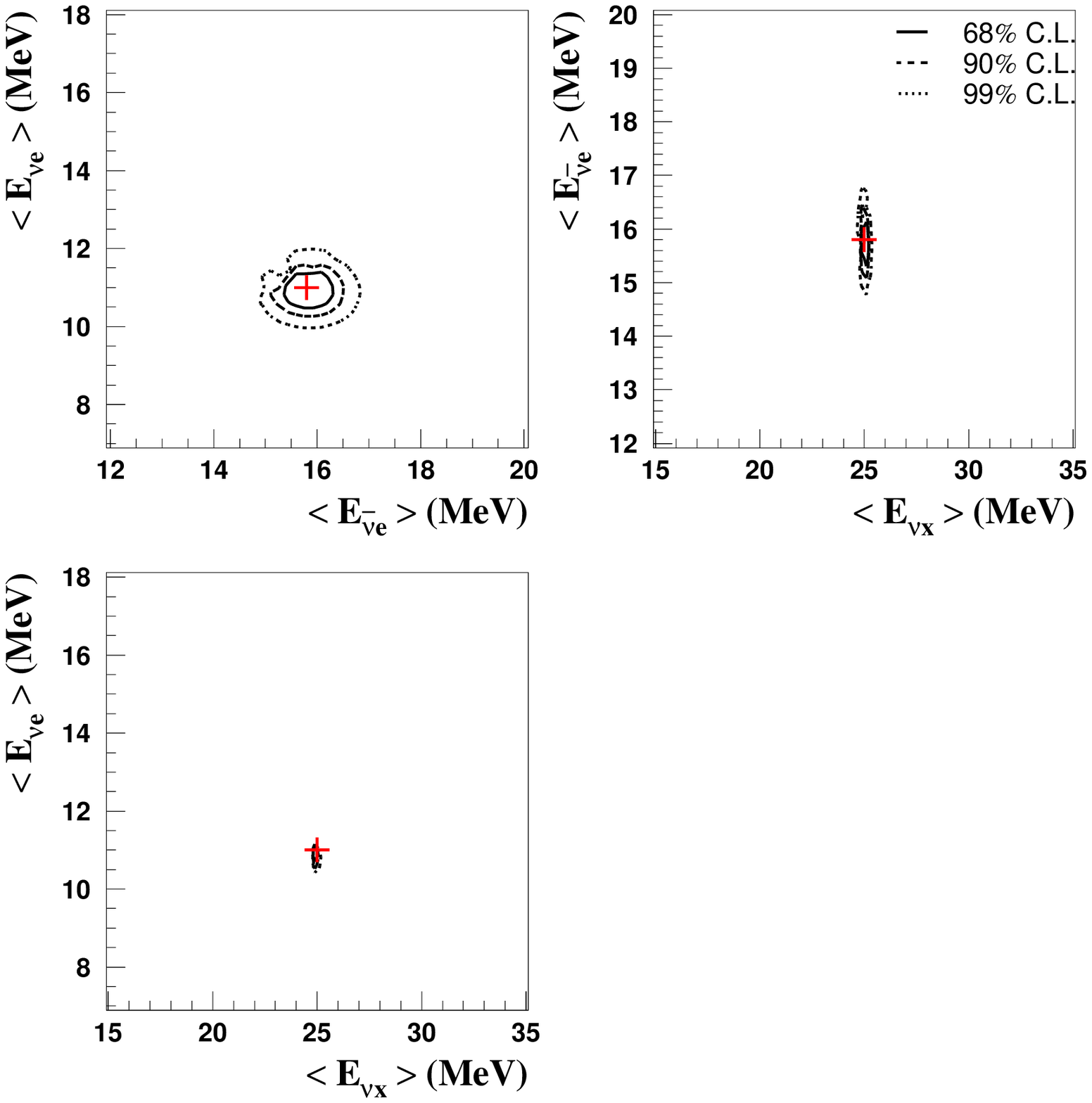,width=0.45\linewidth}
\end{center}
\caption{Correlation between oscillation and supernova parameters: 
68\%, 90\% and 99\% C.L. allowed regions for the supernova
parameters without any assumption on the oscillation parameters for a
100 kton detector. The reference values correspond to the
normal mass hierarchy, large (left) and small (right) \th13 mixing
angle cases. Crosses indicate the value of the parameters for the best fits.} 
\label{fig:allfreenhl}
\end{figure}

\begin{figure}[htbp]
\begin{center}
\fbox{\LARGE \sf i.h.-L and i.h.-S (all parameters free)}
\end{center}
%\begin{center}
%\Large \bf 3 kton LAr
%\end{center}
\vspace{-1cm}
\begin{center}
\epsfig{file=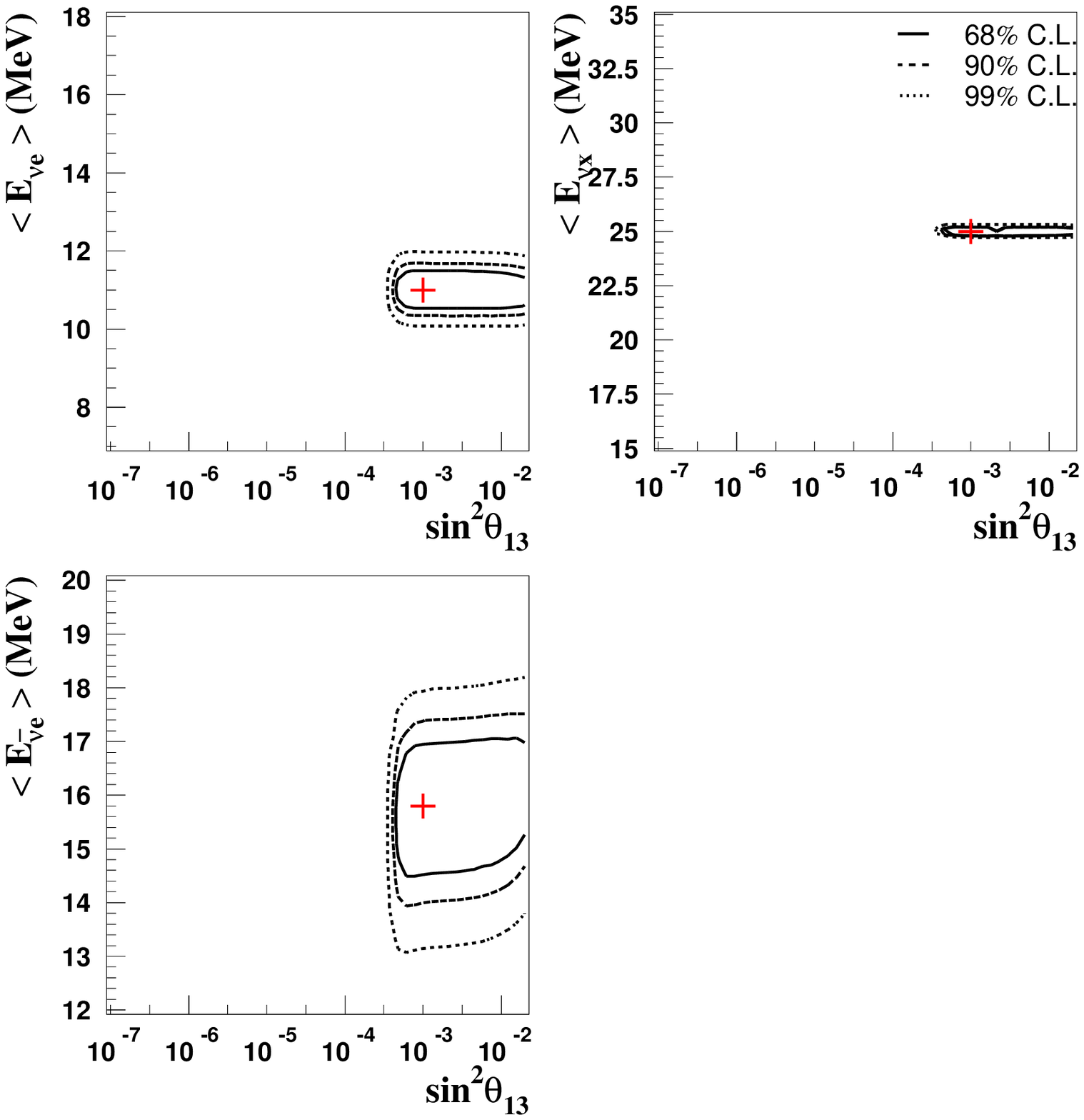,width=0.45\linewidth}
\epsfig{file=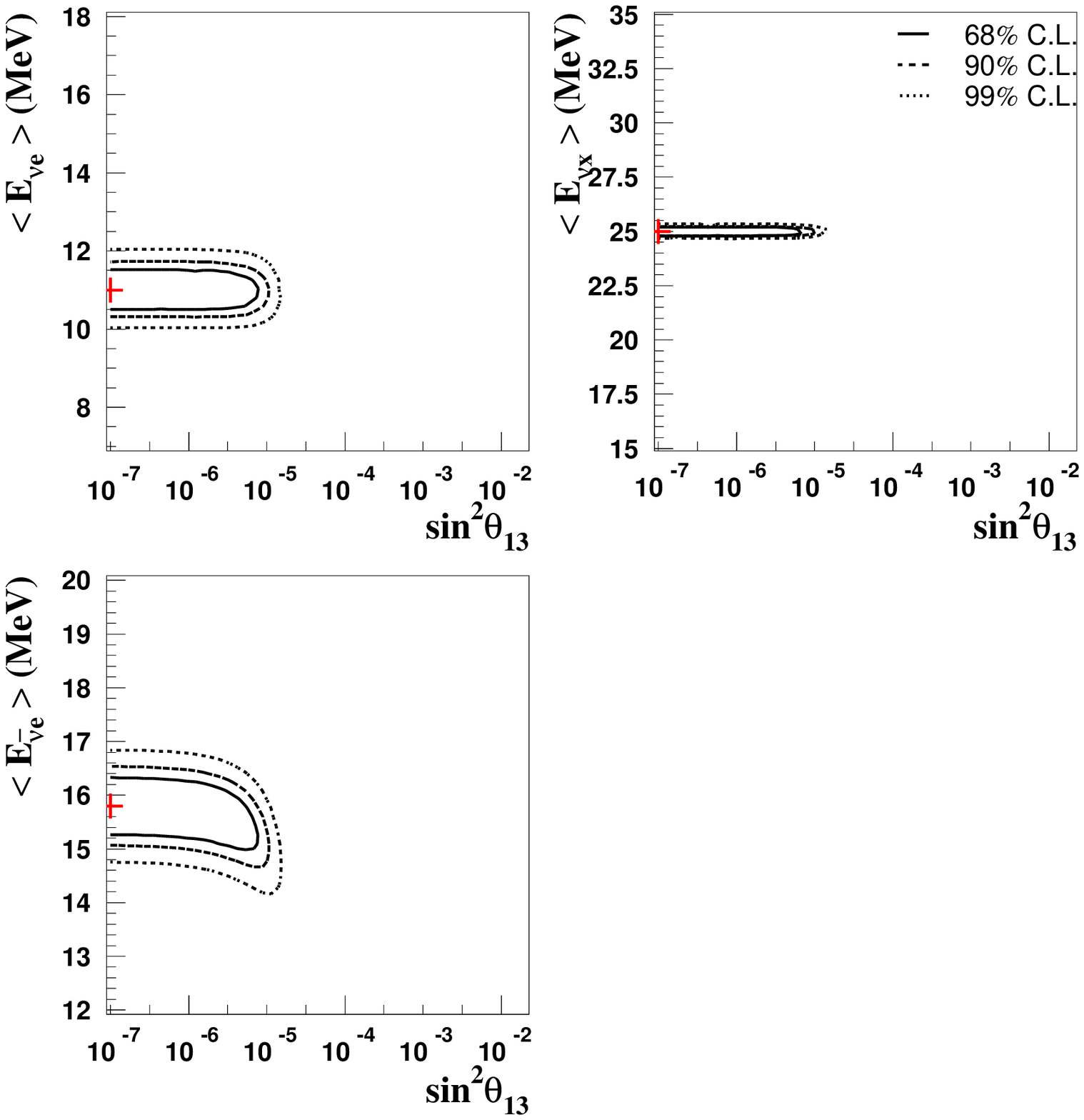,width=0.45\linewidth}
\end{center}
\begin{center}
\vspace{-1cm}
\epsfig{file=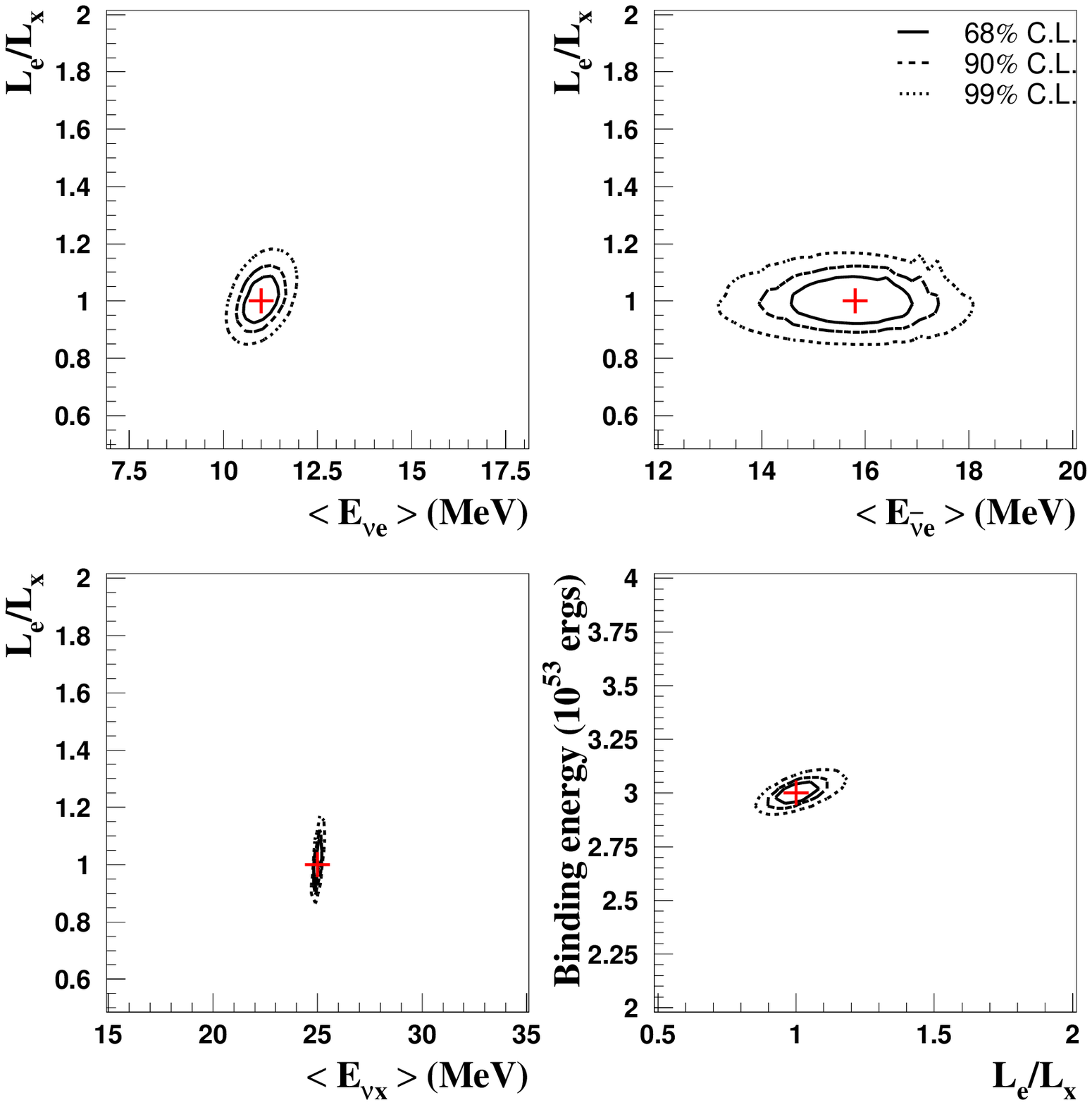,width=0.45\linewidth}
\epsfig{file=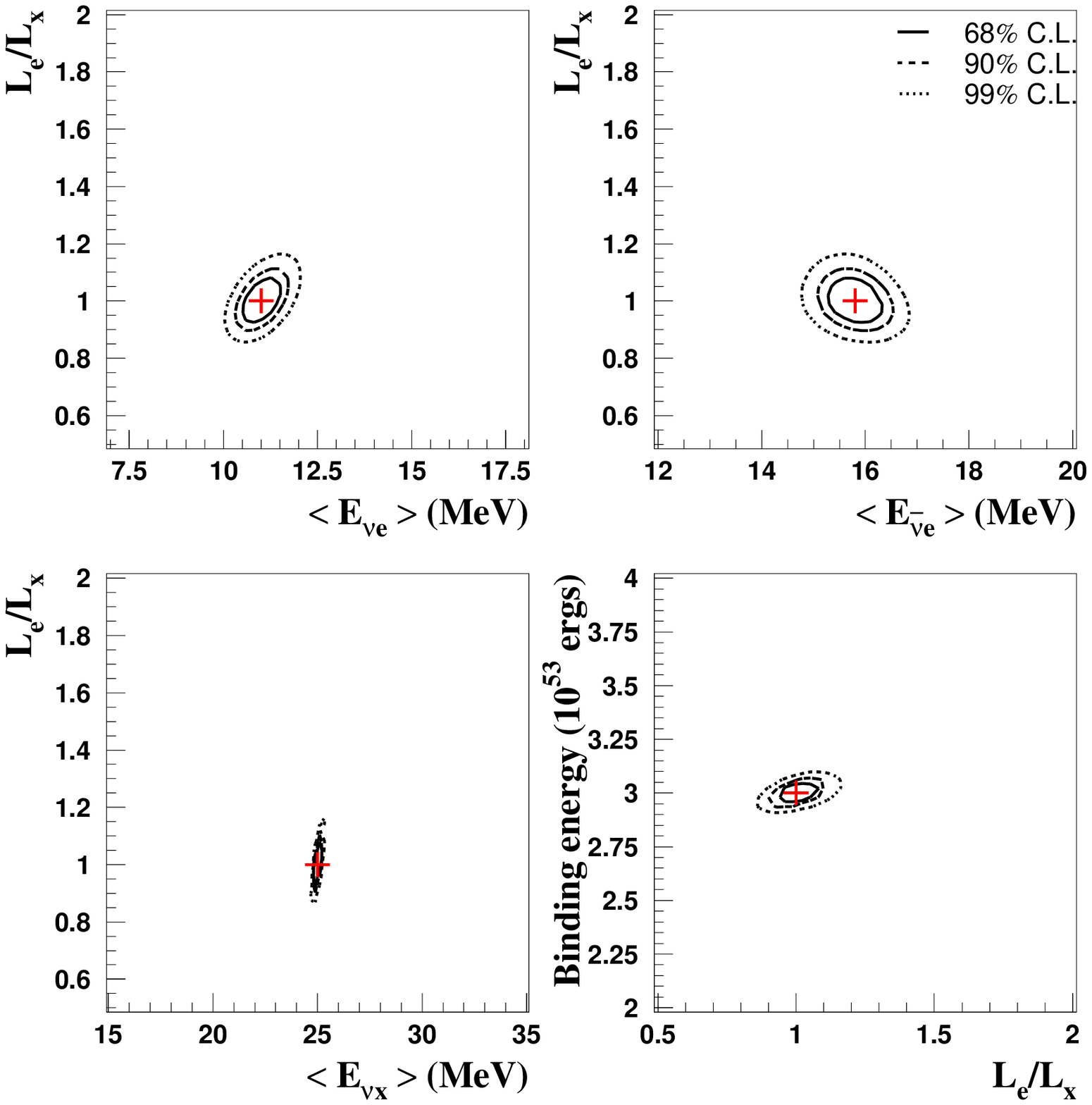,width=0.45\linewidth}
\end{center}
\begin{center}
\vspace{-1cm}
\epsfig{file=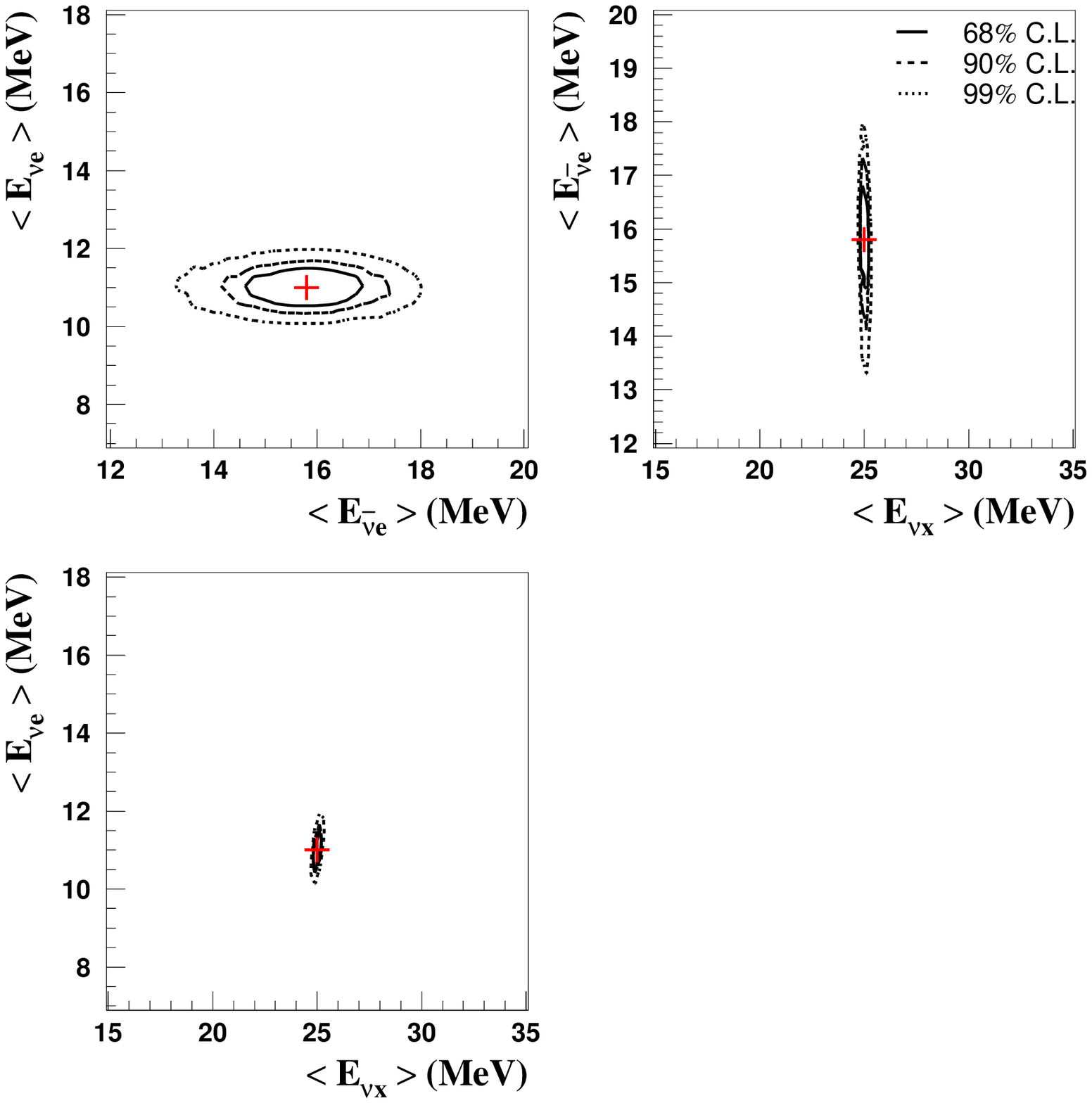,width=0.45\linewidth}
\epsfig{file=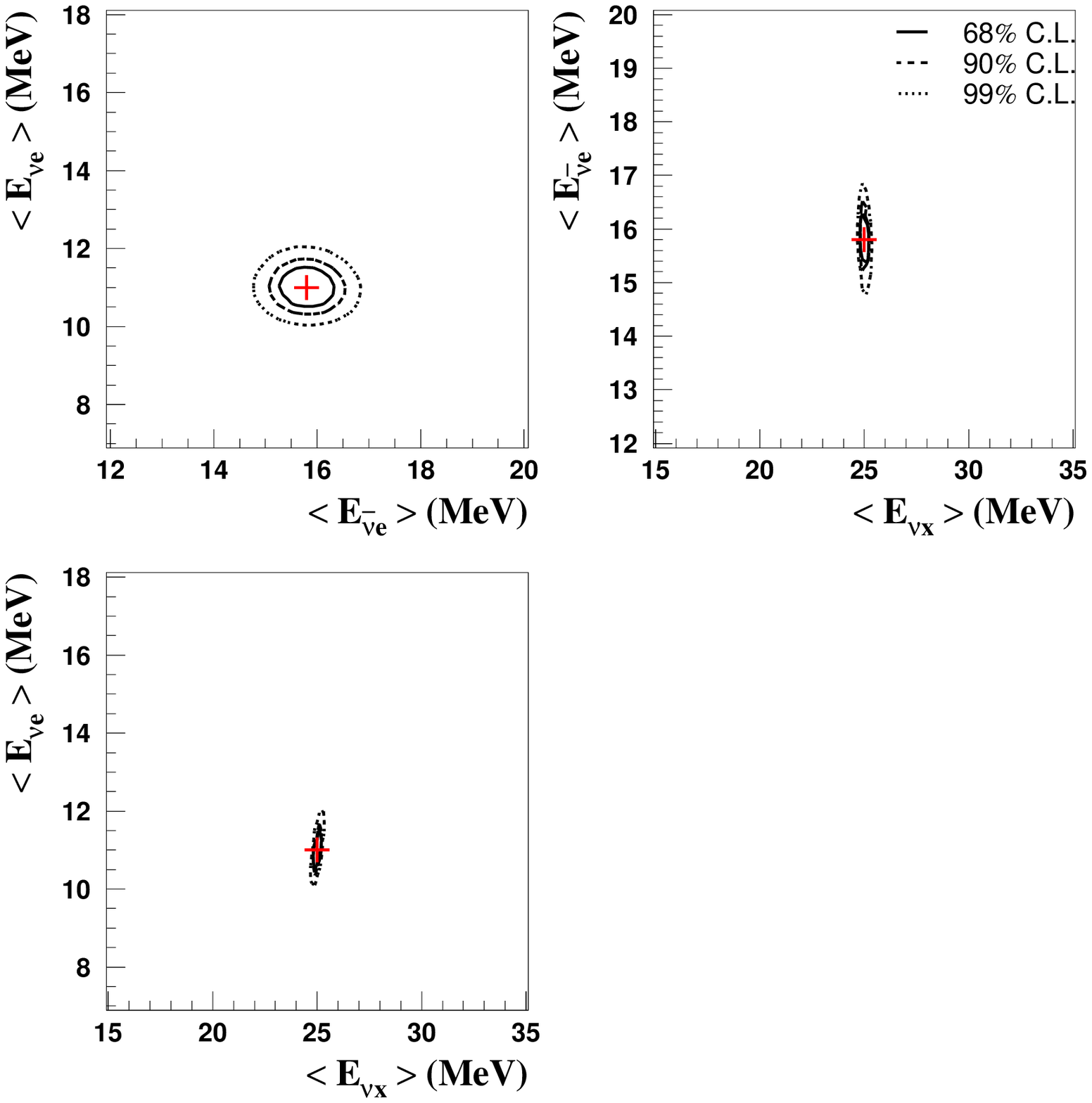,width=0.45\linewidth}
\end{center}
\caption{Correlation between oscillation and supernova parameters: 
68\%, 90\% and 99\% C.L. allowed regions for the supernova
parameters without any assumption on the oscillation parameters for a
100 kton detector. The reference values correspond to the
inverted mass hierarchy, large (left) and small (right) \th13 mixing
angle cases. Crosses indicate the value of the parameters for the best fits.} 
\label{fig:allfreeihl}
\end{figure}

Figures \ref{fig:allfreenhl} and
%, \ref{fig:allfreenhs},
\ref{fig:allfreeihl} 
% and \ref{fig:allfreeihs} 
show the allowed regions
at 68\%, 90\% and 99\% C.L. of the oscillation and supernova parameters. 
The reference values of the parameters
correspond to scenario I. 

The results show that with large statistics it is possible
to decouple the supernova and oscillations parameters and
determine the parameters with high precision, as
presented in Table~\ref{tab:allfree}. The expected accuracies at 90\%
C.L. are listed for every astrophysical parameter, supernova scenario
and ``true'' oscillation case considered.
The total binding energy (\EB) has an error
1--4\%, 
the average energy of electron
neutrinos \avenue ~at the core is determined with an error
5--21\%,  
the average energy of electron
antineutrinos \aveanue ~at the core, with an error
4--9\%,  
the average energy of other
(anti)neutrinos \avenux ~at the core is determined with an error
$\sim 1\%$,  
and the relative luminosities of the
electron and non-electron flavor neutrinos \lelx,
with an error 9--37\%. The errors with a 3~kton
detector are roughly 6 times larger than those with a 100~kton
detector, hence, many of the parameters would be poorly
determined in this case.

\begin{table}[htbp]
\centering
\begin{tabular}{|c|c||c|c|c|c|c|} 
\hline
\multicolumn{7}{|c|}{100~kton detector. All parameters free.} \\
\hline 
``True'' osc. case & SN scen. & $\frac{\Delta\favenue}{\favenue}$ &
$\frac{\Delta\faveanue}{\faveanue}$ &
$\frac{\Delta\favenux}{\favenux}$ & $\frac{\Delta\fEB}{\fEB}$ &
$\frac{\Delta(\flelx)}{(\flelx)}$ \\ \hline \hline 
\bf n.h.-L & I & $\sim$ 17\% & $\sim$ 4\% & $<1 \%$ & $\sim$ 2\% &
$\sim$ 11\% \\
\bf i.h.-L & I & $\sim$ 5\% & $\sim$ 9\% & $<1 \%$ & $\sim$ 2\% & $\sim$
9\% \\ 
\hline
\bf n.h.-S & I & $\sim$ 6\% & $\sim$ 4\% & $<1\%$ & $\sim$
1\% & $\sim$ 11\% \\
\bf i.h.-S & I & $\sim$ 5\% & $\sim$ 4\% & $<1\%$ & $\sim$ 2\% &
$\sim$ 9\% \\ \hline \hline
\bf n.h.-L & II & $\sim$ 21\% & $\sim$ 3\% & $<$ 1\% & $\sim$ 4\% &
$\sim$ 14\% \\ 
\bf i.h.-L & II & $\sim$ 6\% & $\sim$ 9\% & $\sim$ 1\% & $\sim$ 2\% & $\sim$
34\% \\ 
\hline
\bf n.h.-S & II & $\sim$ 11\% & $\sim$ 5\% & $<$ 1\% & $\sim$
2\% & $\sim$ 35\% \\
\bf i.h.-S & II & $\sim$ 8\% & $\sim$ 4\% & $\sim$ 1\% & $\sim$ 2\% &
$\sim$ 37\% \\ \hline
\end{tabular}
\caption{Expected accuracies at 90\% C.L. in the determination of the
supernova parameters using the neutrinos measured with a 100 kton
detector. No conditions on the \th13 angle and mass hierarchy have
been considered. Four oscillation scenarios have been studied as 
possible ``true'' cases. Supernova scenarios I and II are tested.}     
\label{tab:allfree}
\end{table}

%
%\begin{figure}[htb]
%\begin{center}
%\fbox{\LARGE \sf n.h.-S (all parameters free)}
%\end{center}
%%\begin{center}
%%\Large \bf 3 kton LAr
%%\end{center}
%%\vspace{-0.7cm}
%%\epsfig{file=EPS/snfitallfreenhs_100kton_2.eps,width=0.55\linewidth}
%%\hspace{-1cm}
%%\epsfig{file=EPS/snfitallfreenhs_100kton_3.eps,width=0.55\linewidth}
%\begin{center}
%\epsfig{file=EPS/snfitallfreenhs_100kton_4.eps,width=0.55\linewidth}
%\end{center}
%\caption{Correlation between oscillation and supernova parameters: 
%68\%, 90\% and 99\% C.L. allowed regions for the supernova
%parameters without any assumption on the oscillation parameters for a
%100 kton detector. The reference values correspond to the
%normal mass hierarchy and small \th13 mixing angle case. Crosses
%indicate the value of the parameters for the best fits.} 
%\label{fig:allfreenhs}
%\end{figure}

%\begin{figure}[htb]
%\begin{center}
%\fbox{\LARGE \sf i.h.-S (all parameters free)}
%\end{center}
%%\begin{center}
%%\Large \bf 3 kton LAr
%%\end{center}
%%\vspace{-0.7cm}
%%\epsfig{file=EPS/snfitallfreeihs_100kton_2.eps,width=0.55\linewidth}
%%\hspace{-1cm}
%%\epsfig{file=EPS/snfitallfreeihs_100kton_3.eps,width=0.55\linewidth}
%\begin{center}
%\epsfig{file=EPS/snfitallfreeihs_100kton_4.eps,width=0.55\linewidth}
%\end{center}
%\caption{Correlation between oscillation and supernova parameters: 
%68\%, 90\% and 99\% C.L. allowed regions for the supernova
%parameters without any assumption on the oscillation parameters for a
%100 kton detector. The reference values correspond to the
%inverted mass hierarchy and small \th13 mixing angle case. Crosses
%indicate the value of the parameters for the best fits.} 
%\label{fig:allfreeihs}
%\end{figure}

\begin{figure}[htbp]
\centering
\epsfig{file=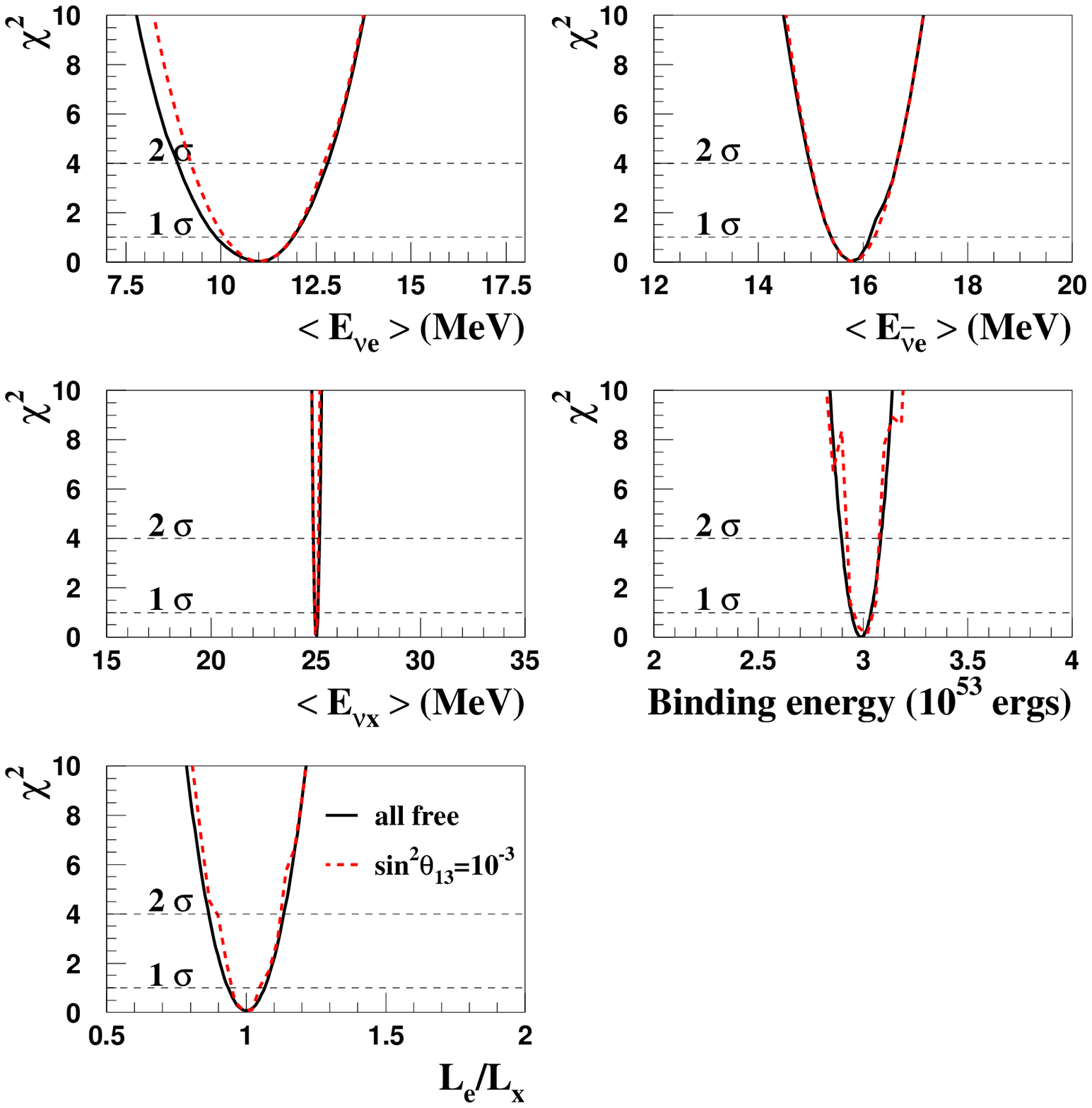,width=0.52\linewidth}
\hspace{-1cm}
\epsfig{file=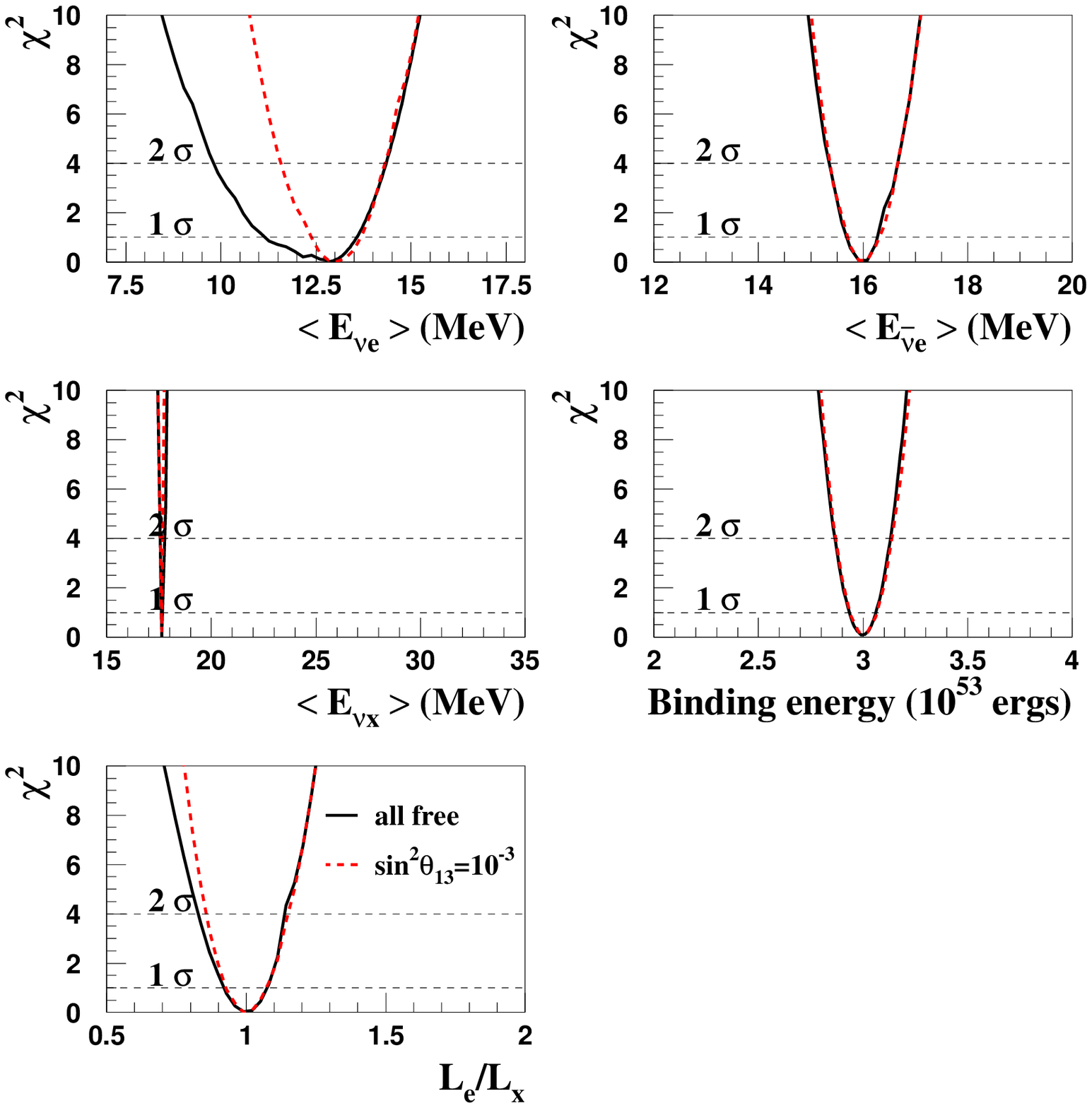,width=0.52\linewidth}
\caption{{\bf Determination of the supernova parameters:}
$\chi^2$ value of the fit as a function of the supernova
parameters for a 100 kton detector. We compare the results obtained
fixing the angle to \s2t13 = 10$^{-3}$ $\pm$ 10$^{-4}$ and normal mass
hierarchy with the case of leaving all parameters free. The ``true''
supernova parameters correspond to scenario I (left) and to scenario
II (right). The ``true'' oscillation parameters are normal mass
hierarchy and large mixing angle.} 
\label{fig:comp_nhl_100kton}
\end{figure}

Figure~\ref{fig:comp_nhl_100kton} 
shows the $\chi^2$ value from the minimization as a function of the supernova
parameters for a 100 kton detector, considering scenario I (left) and
II (right) as ``true'' supernova parameters. We compare the
results obtained fixing the mixing angle to \s2t13 = 10$^{-3}$ $\pm$
10$^{-4}$ and normal mass hierarchy with the case of
leaving the mixing angle free. We have chosen the case where
the effect of the neutrino mixing is largest, namely when the 
``true'' oscillation parameters
correspond to normal mass hierarchy and large mixing angle. 
For other oscillation parameters, in particular for small mixing,
the effect is invisible.

We see that  the expected accuracies are similar
to the ones obtained constraining the mixing angle and the mass
hierarchy, except for the \avenue ~parameter and ``true'' oscillation
case n.h.-L. In this scenario, the knowledge of the \th13 value
improves the determination of the \avenue ~energy from 17\% to 14\%
for scenario I and from 21\% to 9\% for scenario II.

Hence, the observation of different channels which have different sensitivities to the
neutrino flavors allows to decouple the supernova and neutrino oscillation physics,
effectively allowing these two sectors to be studied independently. With the statistics
provided by a 100~kton detector, a single supernova explosion would allow to
determine the parameters of the supernova cooling phase quite precisely.

%%%%%%%%%%%%%%%%%%%%%
\section{Conclusions}
%%%%%%%%%%%%%%%%%%%%%

In this paper we have investigated the capabilities of liquid argon TPC
detectors to study and decouple supernova and neutrino oscillation physics
in the event of the detection of a single core collapse supernova at an assumed
distance of 10~kpc. 

The neutrino physics was determined by the oscillation angle \th13 and the mass
hierarchy. The supernova properties were summarized in
five astrophysical parameters: the average energies of the
neutrinos emitted from the supernova (\avenue, \aveanue, \avenux),
the total binding energy (\EB) and the relative luminosities of the
electron and non-electron flavor neutrinos (\lelx).
We considered two specific scenarios (I \& II) in order
to understand the effects of a hierarchical versus non-hierarchical 
distribution of energies of supernova neutrinos, since this issue is
still debated. For the supernova scenario II, the
differences between the average energies of electron and non-electron
neutrinos is of the order of 10\%. This is an important issue since the 
interplay between hierarchy between neutrino flavors and neutrino
mixing is expected to be a striking feature of matter enhanced oscillations
in the supernova. In the case of degenerate neutrinos, the effect of
oscillation is more difficult to observe. To contemplate two scenarios allowed us
to address this in a quantitative way.

We have considered the four neutrino detection channels available in
argon: elastic scattering on atomic electrons, charged-current
interactions on argon (independently sensitive to \nue ~and \anue ~neutrinos) and
neutral-current interactions on argon. 

In the first part of the work, we have studied the sensitivity 
to the \th13 and mass hierarchy parameters under the assumption
of the supernova scenario I. 
For the true large mixing angle ($>$ 3 $\times$ 
10$^{-4}$), a mass hierarchy identification is possible and lower
limits on the \th13 value are set. We have shown that the detection
of the NC process is important in this context.
For small mixing angle ($<$ 2 $\times$ 10$^{-6}$), it is not possible
to distinguish the mass hierarchy but upper limits on \th13 can be obtained. 
If the mixing angle is in an intermediate range (2 $\times$ 10$^{-6}$
$<$ \s2t13 $<$ 3 $\times$ 10$^{-4}$), measurements of its value are
possible.

For the supernova scenario II, we found with the statistics provided
by a 3 kton detector the possibility to determine the \th13 mixing angle is limited.
More statistics, such as the one obtained with a 100~kton detector,
will be mandatory to determine quantitatively the
oscillation parameter.
  
In the second part of our study, we have studied the ability to determine the supernova
parameters considering two different cases for the \th13 angle: (a) the angle
has been determined by long-baseline experiments
with $\approx$ 10\% precision, (b) an upper
limit on the mixing angle has been set by long-baseline experiments. 
For a 3 (100) kton detector, the supernova parameters are measured
with such accuracies: the total binding energy (\EB) has an error
20--30\% (2--4\%), 
the average energy of electron
neutrinos \avenue ~at the core is determined with an error
30--120\% (5--14\%),  
the average energy of electron
antineutrinos \aveanue ~at the core, with an error
30--80\% (3--9\%),  
the average energy of other
(anti)neutrinos \avenux ~at the core is determined with an error
3--23\% ($<1\%$),  
and the relative luminosities of the
electron and non-electron flavor neutrinos \lelx,
with an error $>100\%$ (10--40\%).

In the last part of our study, we have considered the case where both
supernova and oscillation parameters are free in the minimization. 
%The possibility to measure different channels which have different
%sensitivities to the neutrino flavors, one can disentangle the
%correlation between supernova and neutrino mixing
%parameters, effectively allowing them to be decoupled. 
The precision with which the supernova parameters can be determined
without knowledge on the neutrino mixing parameter is essentially the
same as when information from terrestrial experiments would be available,
except in the supernova scenario II and in large mixing angle case where
the determination of the \avenue ~energy profits largely from the terrestrial
knowledge on \th13.

In conclusion, the possibility provided by liquid Argon TPCs
to observe elastic, charged and neutral current channels with different sensitivities to the
neutrino flavors allows to disentangle the supernova and neutrino oscillation physics,
allowing these two sectors to be effectively decoupled and
studied independently. With the statistics
provided by a 100~kton detector, a single supernova explosion would allow to
determine the supernova cooling phase quite precisely. This is true even in the case that the
energies of the neutrinos of different flavors are almost degenerate, as some recent supernova
simulations indicate.

%%%%%%%%%%%%%%%%%%%%%%%%%%%%%%%%%%%%%%%%%%%%%%%%%%%%%%%%%%%%%%%%%%%%%%%%%%%%%%%
%
%   REFERENCES
%
%%%%%%%%%%%%%%%%%%%%%%%%%%%%%%%%%%%%%%%%%%%%%%%%%%%%%%%%%%%%%%%%%%%%%%%%%%%%%%%


\begin{thebibliography}{00}

\bibitem{SN1987}
K. Hirata \etal, \PRD ~{\bf 38} (1988) 448; \PRL ~{\bf 58} (1987) 1490. \\
R. Bionta \etal, \PRL ~{\bf 58} (1987) 1494.

%\cite{Fukuda:2000fq}
\bibitem{Fukuda:2000fq}
Y.~Fukuda,
``Observation of supernova neutrino burst at Super-Kamiokande,''
%\href{http://www.slac.stanford.edu/spires/find/hep/www?irn=5646154}{SPIRES entry}
{\it Prepared for International Symposium on Origin of Matter and Evolution of Galaxies 2000, Tokyo, Japan, 19-21 Jan 2000}

%\cite{Tanner:zr}
\bibitem{Tanner:zr}
N.~W.~Tanner and G.~Doucas,
``Supernova Detection By SNO,''
%\href{http://www.slac.stanford.edu/spires/find/hep/www?irn=4089529}{SPIRES entry}
{\it Prepared for 3rd NESTOR Workshop, Pylos, Greece, 19-21 Oct 1993}

\bibitem{nuobserv}
S.Fukuda et al., \PRL ~{\bf 86} (2001) 5656.\\
SNO Collaboration, \PRL ~{\bf 87} (2001) 071301. \\
Y. Fukuda et al., \PRL ~{\bf 82} (1999) 2644. \\
KamLAND Collaboration, \PRL ~{\bf 90} (2003) 021802. \\
K2K Collaboration, \PRL ~{\bf 90} (2003) 041801.

\bibitem{reactors}
%\cite{Apollonio:1999ae}
%\bibitem{Apollonio:1999ae}
M.~Apollonio {\it et al.}  [CHOOZ Collaboration],
%``Limits on neutrino oscillations from the CHOOZ experiment,''
Phys.\ Lett.\ B {\bf 466} (1999) 415, {\hepex{9907037}}.
%%CITATION = HEP-EX 9907037;%%
%\cite{Boehm:2001ik}
%\bibitem{Boehm:2001ik}
F.~Boehm {\it et al.},
%``Final results from the Palo Verde neutrino oscillation experiment,''
Phys.\ Rev.\ D {\bf 64} (2001) 112001, {\hepex{0107009}}.
%%CITATION = HEP-EX 0107009;%%

\bibitem{msw}
L. Wolfenstein, \Journal{\PRD}{17}{1978}{2369} 
\Journal{\PRD}{20}{1979}{2634} \\
S. P. Mikheyev and A. Yu. Smirnov, {\em Sov. J. Nucl. Phys.} {\bf 42}
(1986) 913.

\bibitem{Barger}
V. Barger, D. Marfatia and B.P. Wood, \PLB ~{\bf 547} (2002) 37.

\bibitem{Valle}
H. Minakata, H. Nunokawa, R. Tomas and J.W.F. Valle, \PLB ~{\bf 542}
(2002) 239.

\bibitem{Lunardini}
C. Lunardini and A. Y. Smirnov, {\it Probing the neutrino mass
hierarchy and the 13-mixing with supernovae}, {\it
J. Cosmol. Astropart. Phys.} {\bf 6} (2003) 009.

\bibitem{Choubey}
A.~Bandyopadhyay, S.~Choubey, S.~Goswami and K.~Kar, {\it Prospects of
probing $\theta_{13}$ and neutrino mass hierarchy by supernova neutrinos
in KamLAND}, {\hepph{0312315}}. 

\bibitem{IApaper}
I. Gil-Botella and A. Rubbia, {\it Oscillation effects on supernova
neutrino rates and spectra and detection of the shock breakout in a
liquid argon TPC}, {\it J. Cosmol. Astropart. Phys.} {\bf 10}
(2003) 009, {\hepph{0307244}}.

\bibitem{icarus3000}
%\cite{Aprili:2002wx}
%\bibitem{Aprili:2002wx}
P.~Aprili {\it et al.}  [ICARUS Collaboration],
``The ICARUS experiment: A second-generation proton decay experiment and
neutrino observatory at the Gran Sasso laboratory. Cloning of T600 modules to
reach the design sensitive mass. (Addendum),'' LNGS-EXP 13/89 add.2/01,
and CERN-SPSC-2002-027.
%\href{http://www.slac.stanford.edu/spires/find/hep/www?r=cern-spsc-2002-027}{SPIRES entry}
All proposals are available at {\it http://www.cern.ch/icarus}.

%\cite{Cline:2001pt}
\bibitem{Cline:2001pt}
D.~B.~Cline, F.~Sergiampietri, J.~G.~Learned and K.~McDonald,
{\it LANNDD: A massive liquid argon detector for proton decay, supernova
and  solar neutrino studies, and a neutrino factory detector}, \NIMA
~{\bf 503} (2003) 136, {\astroph{0105442}}.

\bibitem{Rubbia:2004tz}
A.~Rubbia,
{\it Experiments for CP-violation: A giant liquid argon scintillation, Cerenkov
and charge imaging experiment?}, {\hepph{0402110}}. To appear in {\it Proceedings of the II
International Workshop on Neutrino Oscillations in Venice, Italy}. 

\bibitem{Raffelt}
M.~T.~Keil, G.~G.~Raffelt and H.~T.~Janka,
{\it Monte Carlo study of supernova neutrino spectra formation},
\APJ ~{\bf 590} (2003) 971, {\astroph{0208035}}. \\
G.~G.~Raffelt, M.~T.~Keil, R.~Buras, H.~T.~Janka and M.~Rampp,
{\it Supernova neutrinos: Flavor-dependent fluxes and spectra},
{\astroph{0303226}}. 

\bibitem{Langanke}
K. Langanke, P. Vogel and E. Kolbe,
\PRL ~{\bf 76} (1996) 2629.

\bibitem{Dighe}
A. S. Dighe and A. Y. Smirnov, \PRD ~{\bf 62} (2000) 033007.

\bibitem{martinez}
E. Kolbe, K. Langanke, G. Mart{\'i}nez Pinedo and P. Vogel, {\it
Neutrino-nucleus reactions and nuclear structure}, {\it J. Phys. G:
Nucl. Part. Phys.} {\bf 29} (2003) 2569.

\bibitem{minuit} F.James, MINUIT manual, CERN program libraries p. 43

%\cite{Apollonio:2002en}
\bibitem{Apollonio:2002en}
M.~Apollonio {\it et al.},
{\it Oscillation physics with a neutrino factory}, CERN-TH-2002-208,
{\hepph{0210192}}. \\ 
%%CITATION = HEP-PH 0210192;%%
%\cite{Huber:2002mx}
%\bibitem{Huber:2002mx}
P.~Huber, M.~Lindner and W.~Winter,
{\it Superbeams versus neutrino factories},
\NPB ~{\bf 645} (2002) 3, 
{\hepph{0204352}}.
%%CITATION = HEP-PH 0204352;%%

\end{thebibliography}
\end{document}